\documentclass{article}
\usepackage[english]{babel}
\usepackage{natbib}
\usepackage{amssymb}
\usepackage{booktabs}

\usepackage[letterpaper,top=2cm,bottom=2cm,left=3cm,right=3cm,marginparwidth=1.75cm]{geometry}

\usepackage{amsmath}
\usepackage{graphicx}
\usepackage{mhchem}
\usepackage[colorlinks=true, allcolors=blue]{hyperref}

\title{Star-Planet Interactions}

\usepackage[dvipsnames]{xcolor}
\usepackage{authblk}

\author[1]{Arghyadeep Paul}
\author[2]{Kristina Kislyakova\thanks{Corresponding author: Kristina Kislyakova (kristina.kislyakova@univie.ac.at)}}
\author[2]{Manuel Güdel}
\author[3]{Rim Fares}
\author[1]{Judy Chebly}
\author[9]{Sergio Joya}
\author[4,10]{Miljenko \v{Cemelji\'c}}
\author[5,6]{Katja Poppenhäger}
\author[5]{Julian Alvarado-Gomez}
\author[5]{Silva Järvinen}
\author[6]{Cesar Bertucci}
\author[7,8]{Dibyendu Nandy}
\author[1]{Antonio Garc\'ia Mu\~noz}
\author[1]{Antoine Strugarek}
\author[11]{Mayank Narang}
\author[12]{Shyama Narendranath}

\affil[1]{Université Paris Cité, Université Paris-Saclay, CEA, CNRS, AIM, F-91191, Gif-sur-Yvette, France}
\affil[2]{Department of Astrophysics, University of Vienna, T\"urkenschanzstrasse 17, A-1180 Vienna, Austria}
\affil[3]{Physics Department, United Arab Emirates University, P.O. Box 15551, Al-Ain, United Arab Emirates}
\affil[4]{Nicolaus Copernicus Superior School, College of Astronomy and Natural Sciences, ul. Nowogrodzka 47a, 00-695, Warsaw, Poland}
\affil[5]{Leibniz-Institut für Astrophysik Potsdam (AIP), An der Sternwarte 16, 14482 Potsdam, Germany}
\affil[6]{Institut f\"ur Physik und Astronomie, Universit\"at Potsdam, Karl-Liebknecht-Str.\ 24/25, 14476 Potsdam, Germany}
\affil[7]{Center of Excellence in Space Sciences India, Indian Institute of Science Education and Research Kolkata, Mohanpur 741246, West Bengal, India}
\affil[8]{Department of Physical Sciences, Indian Institute of Science Education and Research Kolkata, Mohanpur 741246, West Bengal, India}
\affil[9]{Physikalisches Institut, University of Bern, Sidlerstrasse 5, 3012 Bern, Switzerland}
\affil[10]{Nicolaus Copernicus Astronomical Center, Polish Academy of Sciences, Bartycka 18, 00-716 Warsaw, Poland}
\affil[11]{Academia Sinica Institute of Astronomy and Astrophysics}
\affil[12]{U R Rao Satellite Centre, Indian Space Research Organisation}

\begin{document}
\maketitle

\tableofcontents

% \author{
% \begin{tabular}{p{0.9\textwidth}}
% \centering
% Arghyadeep Paul$^{5}$, Kristina Kislyakova$^{5}$, Manuel Güdel$^{5}$, Rim Fares$^{1}$,\\
% Judy Chebly$^{5}$, Sergio Joya$^{2}$, Miljenko \v{C}emelji\'{c}$^{2}$,\\
% Katja Poppenhäger$^{3,4}$, Julian Alvarado-Gomez$^{3}$, Silva Järvinen$^{3}$,\\
% Cesar Bertrucci$^{5}$, Dibyendu Nandy$^{5,6}$, Antonio Garcia Munoz$^{5}$,\\
% Antoine Strugarek$^{5}$, Mayank Narang$^{5}$, Shyama Narendranath$^{5}$\\[1ex]

% $^{1}$ Physics Department, United Arab Emirates University, P.O. Box 15551, Al-Ain, United Arab Emirates\\
% $^{2}$ Nicolaus Copernicus Superior School, College of Astronomy and Natural Sciences, ul. Nowogrodzka 47a, 00-695, Warsaw, Poland\\
% $^{3}$ Leibniz-Institut für Astrophysik Potsdam (AIP), An der Sternwarte 16, 14482 Potsdam, Germany\\
% $^{4}$ Institut f\"u Physik und Astronomie, Universit\"at Potsdam, Karl-Liebknecht-Str.\ 24/25, 14476 Potsdam, Germany\\

% \end{tabular}
% }

\begin{abstract}
Star-planet interactions (SPIs) encompass the coupled exchange of energy, momentum, and mass between exoplanets and their host stars, mediated through radiative, tidal, magnetic, and particle-driven processes. These mechanisms operate in a strongly interconnected manner, jointly shaping the physical and dynamical evolution of planetary systems across a wide range of orbital architectures and stellar environments. In this review, we synthesise the current theoretical and observational understanding of SPIs, with emphasis on the physical processes that link stellar radiation, winds, and magnetic activity to planetary atmospheres, interiors, and orbital evolution, while drawing on Solar-System analogues to contextualise key mechanisms. Stellar radiation across infrared to X-ray wavelengths provides a primary driver of atmospheric heating, photochemistry, and escape, with far and extreme-ultraviolet photons initiating ionisation and dissociation processes that regulate upper-atmospheric structure and composition. The resulting thermal and non-thermal escape processes are further modulated by stellar wind interactions and magnetic coupling, leading to complex feedbacks that can either stabilise or erode planetary atmospheres over time. In parallel, tidal forcing redistributes energy and angular momentum within star-planet systems, producing internal heating, altering the spin rates of both stars and planets, and driving orbital migration and circularisation. Magnetic interactions introduce additional channels of energy and momentum transfer through star–planet magnetic coupling, reconnection, and current systems, which can enhance atmospheric escape, deposit energy in planetary ionospheres and interiors, and generate observable signatures such as radio emission and enhanced stellar surface activities. The combined action of these processes governs the long-term evolution and observable properties of exoplanetary systems, with outcomes that depend sensitively on stellar activity evolution, planetary mass and composition, atmospheric structure, and magnetic field strength and configurations. By integrating radiative, tidal, and magnetic interactions within a unified framework, and by leveraging insights from well-characterised Solar-System analogues, this review highlights the key physical pathways through which SPIs shape planetary environments and identifies the primary observational diagnostics that will inform future studies of exoplanet system evolution and habitability.
\end{abstract}
% \manuel{\textbf{can we use another citation format? The one used now is pretty strange. A reference to [EKL+07] is not very useful.  I  temporarily introduced bibliographystyle\{aasjournal-3\} at the end.}
% } 

\section{Introduction}\label{sec:intro}
%%%%%%%%%%%%%%%%%%%%%%%%%%%%%%%%%  Description  %%%%%%%%%%%%%%%%%%%%%%%%%%%%%%%%%%%%%%
% \begin{itemize}
%     \item Concept: Define SPI as the exchange of mass, momentum, and energy.
%     \item Driver: Briefly introduce the "Stellar Driver" -- XUV, stellar winds, and tides, leading into the following sections.
% \end{itemize}

\textit{Star-planet interactions (SPIs)} manifest in a wide variety of ways and involve the exchange of mass, momentum, and energy between exoplanets and their host stars. In this review, we summarize the contemporary state-of-the-art understanding of SPIs and discuss tidal, magnetic, radiative, and particle interactions operating in exoplanetary systems. We also address the Solar-System analogues of these mechanisms and explore how decades of studies of our own planetary system can inform the much younger field of exoplanet science. 

SPIs can produce a broad range of observable phenomena. For example, they may drive powerful radio emissions from exoplanetary systems, offering insight into planetary magnetic moments and the surrounding plasma environment, which is often dominated by the stellar wind. SPIs may also manifest through atmospheric escape, enhanced stellar activity, orbital migration, planetary heating, or localized emission features on host stars. These phenomena can be investigated from both theoretical and observational perspectives, and this review presents a synthesis of the current understanding of SPIs from both approaches.

\textit{Stellar radiation} plays a crucial role in determining both the present state and the long-term evolution of planetary atmospheres. Infrared radiation with wavelengths of roughly 1--20$~\mu$m is absorbed especially by molecules such as CO$_2$, H$_2$O, and CH$_4$ as they become vibrationally excited (e.g. \citealp{johnstone2018}). Their radiative excitation followed by collisional de-excitation effectively deposits thermal energy into the atmospheric gas when photon absorption occurs sufficiently deep within the atmosphere. Such heating mechanisms therefore depend fundamentally on atmospheric composition, which determines the chemical, collisional, and radiative properties of the atoms and molecules interacting with the incident photons.

The broad range of short-wavelength and high-energy radiation, extending from the far-ultraviolet (FUV, 912~\AA--2000~\AA\ = 91~nm--200~nm), through the extreme-ultraviolet (EUV, $\sim$100~\AA--912~\AA\ = 10~nm--91~nm or 0.014~keV--0.1~keV), and into the X-ray regime ($\sim$0.1~\AA--100~\AA\ = 0.01~nm--10~nm or $\sim 0.1$~keV--100~keV), collectively referred to as XUV radiation, is also of central importance for planetary evolution. FUV radiation is particularly important for the photodissociation of molecules and the ionization of metals with low ionization potentials, while shorter-wavelength radiation additionally ionizes atoms and molecules, depositing further thermal energy through the production of photoelectrons. These photoelectrons may subsequently drive additional atmospheric chemistry \citep{Gillet_2023,garciamunozbataille2024}.

The altitude at which stellar radiation is absorbed depends sensitively on both the wavelength of the radiation and the atmospheric composition in different layers as a consequence of radiative transfer through the upper atmosphere. Ionization, photodissociation, and heating processes can therefore initiate complex chemical networks. The coupled system of atmospheric chemistry, radiative heating, radiative and conductive cooling, and transport processes may, under stable irradiation conditions, evolve toward a quasi-steady-state atmosphere (e.g. \citealt{roble1995, ridley2006} for Solar-System planets, or \citealt{tian2008a} for exoplanets).

Such a steady state, however, may only be temporarily maintained through a balance between continuous atmospheric replenishment from the planetary interior via outgassing or volcanism (e.g. \citealt{noack2014}) and atmospheric escape to space \citep{watson1981, erkaev2007, garciamunoz2007, lundin2007, tian2008a, kislyakova2013}. Atmospheric escape processes are conventionally divided into \textit{thermal mass loss}, resulting from thermal particles in the upper atmosphere reaching escape velocity, and \textit{non-thermal mass loss}, driven by interactions between atmospheric particles and the ionized stellar wind, including its magnetic and electric fields.

Over evolutionary timescales, steady-state conditions may break down because stellar XUV emission, stellar winds, and planetary outgassing rates evolve with time \citep{tu2015, johnstone2015, johnstone2021}. During the early stages of stellar evolution in particular, XUV irradiation can become sufficiently intense to exceed atmospheric replenishment rates even within the habitable zone, potentially causing irreversible atmospheric erosion \citep{garciamunozetal2021}. Stellar radiation is also a major driver of photochemistry. In addition to the irradiation level and stellar wind strength, the long-term evolution of planetary atmospheres depends strongly on planetary mass, magnetic fields, and atmospheric composition, which itself evolves over time. Consequently, the atmospheric state at any given epoch is intimately connected to the evolutionary history of the host star, especially the evolution of stellar magnetic activity, rotation, and magnetized winds (e.g. \citealp{Gronoff2020}).

\textit{Tidal interactions} shape the evolution of exoplanets and Solar-System bodies in a variety of ways. Within planetary interiors, tidal dissipation can act as a powerful heat source and, in some cases, drive volcanic activity far more intense than that currently observed on Earth (e.g. \citealp{Nicholls2025}). The energy powering these interactions originates from the gravitational potential energy of the coupled star-planet system and, in some cases, from interactions among multiple planets within the same system \citep{Beuthe2013,Ogilvie_2014}. Such effects are particularly important in systems where planets or moons participate in orbital resonance chains that maintain non-zero orbital eccentricities. In the Solar System, this process is observed among the Galilean satellites. Comparable resonant configurations also exist among exoplanets, for example in the TRAPPIST-1 system, which hosts seven closely orbiting terrestrial-sized planets around a very late-type low-mass M dwarf \citep{Luger2017}.

Tidal interactions are not restricted to planetary interiors alone; they can substantially reshape the architecture and rotational evolution of entire star-planet systems. Tidal dissipation acts to circularize planetary orbits and can drive orbital migration either inward or outward depending on the exchange of angular momentum. Inward migration occurs when orbital angular momentum is transferred from the planet to the stellar rotation, causing the host star to spin up \citep{Revol2023}. Conversely, when the host star initially rotates faster than the planetary orbit, angular momentum can instead be transferred from the stellar rotation to the planet, causing the star to spin down while the planet migrates outward. A well-known example of this process in the Solar System is the Earth-Moon system, where the Moon is gradually receding from Earth over time \citep{Tyler2021}.

In addition to gravitational tides, \textit{thermal tides} may also influence planetary evolution. Thermal tides arise from periodic atmospheric heating caused by stellar irradiation, which generates density asymmetries in the atmosphere. These asymmetries can exert torques on the planet and modify its rotational and orbital evolution. Thermal tides are thought to play an important role in the asynchronous rotation of Venus (e.g. \citealp{Revol2023}) and may also affect the spin states and atmospheric dynamics of strongly irradiated exoplanets, particularly hot Jupiters and terrestrial planets orbiting close to their host stars. In some cases, thermal tides may compete with gravitational tides and alter the efficiency of tidal synchronization and orbital migration.

\textit{Magnetic interactions} constitute another major class of SPIs. In this case, the interaction is mediated through magnetic fields connecting the host star and the planet. Similar to tidal interactions, magnetic SPIs can induce orbital migration by transferring angular momentum between the planetary orbit and stellar rotation \citep{Bromley2022}. Depending on the direction of angular momentum exchange, the planet may either lose or gain orbital angular momentum while the stellar rotation correspondingly accelerates or decelerates.

In addition to angular momentum exchange, magnetic SPIs can generate a range of phenomena that arise specifically from magnetic coupling between the planet and the host star. For example, magnetic flux tubes connecting a planet and its host star may act as efficient particle accelerators, while magnetic reconnection events due to SPI can accelerate particles to significantly higher energies \citep{Vidotto2025}. Such interactions may also produce observable radio emissions from energised particles, induce hot spots on stellar surfaces, and generate a variety of additional electromagnetic signatures.

Magnetic interactions can furthermore lead to heat generation within planetary interiors and atmospheres. Similar to tidal dissipation, magnetic heating may enhance planetary volcanism \citep{Kislyakova2017}. Whereas tidal dissipation primarily deposits energy within planetary interiors, magnetic heating can also occur directly within the ionosphere, the ionized upper layer of a planetary atmosphere \citep{Strugarek_2025}. Under certain conditions, this heating can become highly efficient and drive strong atmospheric escape, potentially leading to substantial atmospheric erosion and the formation of observable plasma tori along planetary orbits. As with tidal interactions, the Galilean satellites  (especially Io, Europa and Ganymede) provide important Solar-System analogues for these magnetic SPIs \citep{Murakami2016}.

Stellar activity is discussed in \citet{Isik2026} and \citet{AlvaradoGomez2026}, while stellar evolutionary processes and their implications for planetary habitability are addressed in \citet{Segura2026}. This review focuses on the physical mechanisms through which host stars interact with their planets and vice versa. In particular, we examine how radiative, tidal and magnetic forcing shape planetary atmospheres, interiors, orbital evolution, and the long-term fate of exoplanets, and how these interactions give rise to potentially observable signatures across exoplanetary systems.

\section{Physical Foundations: The Three Pillars of Star-Planet Interactions}
The physical regimes of interactions between stars and their planetary companions is underpinned by three fundamental mechanisms that dictate the long-term evolution, dynamical stability, and observability of these systems. In the following sections, we characterize these as the three pillars of Star-Planet Interactions: gravitational (tidal), magnetic (electrodynamic), and radiative coupling. We commence this analysis with a description of tidal interactions, which serve as the primary channel for the secular redistribution of angular momentum and the synchronization of spin-orbit architectures, especially in close-in exoplanetary systems.

\subsection{Tidal Interactions}
\label{sec:tidal_interactions}

Tidal interactions have been a known effect in close stellar binaries \citep{Zahn_1977, Mazeh2008}, and they have also been proposed to operate similarly in close star-planet systems. For binaries, tidal locking leads to the synchronization of the rotation periods of the stellar components to the orbital period of the binary. Standard magnetic braking of cool stars (see \citealt{AlvaradoGomez2026}) causes the stars to lose angular momentum, but through the locking the vast angular momentum reservoir of the stellar orbits around each other can be tapped, which allows the stars to maintain rapid rotation (and hence high activity levels) for much longer than single cool stars of the same spectral type, which spin down efficiently through magnetic braking. Analogously, in exoplanets, tidal interactions with the host star can lead to a variety of effects, such as planetary migration, formation of resonant chains, spin-orbit alignment, stellar and planetary spin synchronization, and radius inflation for gaseous exoplanets \citep{Deeg2018HandbookOfExoplanets}. Stellar rotation can also be accelerated due to tidal interactions with its planet \citep{2023MNRAS.520.3749L}, and if a star engulfs a planet, this leaves long term observational fingerprints of the past interaction process in the stellar rotation \citep{Benbakoura2019,Lazovik2026}.

\subsubsection{Mechanisms of Tidal Dissipation and Secular Evolution}\label{sec:tidal_dissipation}

The efficiency of tidal interactions is ultimately governed by the dissipation of mechanical energy within the interiors of the planetary and stellar components. Two distinct physical regimes contribute to this process: \emph{equilibrium tides} and \emph{dynamical tides} \citep{Zahn_1977, Ogilvie_2014}.

The equilibrium tide corresponds to the quasi-hydrostatic, large-scale deformation of a body in response to the companion’s gravitational potential, commonly described as the tidal bulge and the associated large-scale flow that carries the bulge around the body. By contrast, the dynamical tide arises from the excitation of internal oscillation modes, including inertial waves in convective regions and internal gravity waves in radiative zones of stars. The subsequent dissipation of these waves is highly sensitive to the respective internal structures, stratification, and rotation of the body, and in short-period systems, often dominates the long-term tidal torque budget \citep{Fuller2012,Ogilvie_2014}.

This dissipation is commonly parameterized using the dimensionless tidal quality factor $Q$, which characterises how efficiently the kinetic energy of the tidal flows (or, in solid bodies, the elastic energy of the tidal deformation) is converted into heat ($Q^{-1}$ measures the energy dissipated per tidal cycle relative to the peak energy stored in the tide), and through the second-order (degree-2) Love number $k_2$, which measures the deformational response of the body to the tidal potential. As written here and used in Eq.~(\ref{eq:tidal_torque}), $k_2$ denotes the real part of the complex Love number at the relevant forcing frequency, usually approximated by its elastic or hydrostatic value, and is treated as independent of the tidal frequency and amplitude. In contrast, $Q$ is an effective parameter that depends on the internal structure of the body and, in more realistic treatments, varies with tidal frequency and possibly with forcing amplitude. In a more general formalism, both effects are combined in the complex Love number $\tilde{k}_2$, whose real part corresponds to the $k_2$ used here and whose imaginary part quantifies the dissipation, $\mathrm{Im}(\tilde{k}_2) \simeq -k_2/Q$ \citep[e.g.,][]{Ogilvie_2014, Mathis2018, Barker2026}; it is $\mathrm{Im}(\tilde{k}_2)$ that in general depends on the tidal forcing frequency (and possibly amplitude). In a dissipative (non-elastic) body, internal friction introduces a phase lag $\epsilon$ between the tidal potential and the induced deformation, such that $Q^{-1} \approx \sin(2\epsilon)$, where $\epsilon$ is typically small ($\epsilon \ll 1$, since $Q \gg 1$ for stars and giant planets). This phase lag displaces the tidal bulge away from the instantaneous line of centers, producing a non-zero gravitational torque $\mathcal{T}$.

For a planetary body of radius $R_p$ and mass $M_p$, perturbed by a companion (typically a star for exoplanetary systems) of mass $M_{\star}$ at orbital separation $a$, the leading-order tidal torque is given by:
\begin{equation}
\mathcal{T} = \frac{3}{2} \frac{G M_{\star}^2 R_p^5}{a^6} \frac{k_2}{Q}.
\label{eq:tidal_torque}
\end{equation}
This expression represents the leading-order magnitude of the tidal torque: its dependence on the tidal forcing frequency is contained implicitly in $Q$ (and, in general, in the Love number), while its sign is not explicitly specified here. This frequency is typically defined as the difference between the body's rotational angular velocity, $\omega$, and its orbital mean motion, $n$ (e.g., the synodic frequency $\omega - n$), which determines the sign and detailed magnitude of the torque in more complete tidal models; the torque on the body's spin changes sign with $\omega - n$ and acts to drive $\omega$ towards $n$. This torque drives the secular evolution of the system by redistributing angular momentum while monotonically reducing the total mechanical energy. The resulting evolution naturally separates into a hierarchy of characteristic timescales,
\[
\tau_{\text{sync}} \lesssim \tau_{\text{obl}} \ll \tau_{\text{circ}} \ll \tau_{\text{decay}},
\]
 wherein the terms $\tau_{\text{sync}}, \tau_{\text{obl}}, \tau_{\text{circ}} \& \tau_{\text{decay}}$ corresponds to the synchronization, obliquity damping, circularization and orbital decay timescales respectively, reflecting the progressively larger energy reservoirs involved in each process \citep{Goldreich_1966, Rasio_1996}. This ordering holds as a general trend in equilibrium tide theory for close-in planets with weak dynamical tidal effects, but can be altered in systems with strong resonances, high eccentricities, or frequency-dependent dissipation. This hierarchy must, however, be interpreted with care. Here, $\tau_{\text{sync}}$ and $\tau_{\text{obl}}$ refer to the synchronization and obliquity damping of the spin of the \textit{planet}, driven by tides raised on the planet. Because the rotational angular momentum of a planet is much smaller than its orbital angular momentum, the planetary spin is the first to respond to tides and tends to synchronize first; if $Q$ is similar for the different tidal components involved, the planetary obliquity evolves on timescales comparable to $\tau_{\text{sync}}$. This does not apply to the \textit{stellar} spin: the evolution of the stellar obliquity, which is the quantity most commonly discussed in the literature in connection with the spin--orbit angles of hot Jupiters, is driven by tides raised on the star and typically proceeds on timescales similar to those of stellar spin synchronization or of orbital migration and decay. Likewise, whether circularization is faster than orbital migration depends on where tidal dissipation primarily occurs. Circularization is usually attributed to tides raised in the planet, whereas the orbital decay of close-in planets is driven by tides raised in the star, so that the ordering $\tau_{\text{circ}} \ll \tau_{\text{decay}}$ ultimately depends on the relative tidal dissipation (i.e., the relative values of $Q/k_2$) in the star and in the planet, and may not apply to all systems \citep[e.g.,][Sect.~6]{BarkerOgilvie2009}. We describe them in further detail below.

\paragraph{Synchronization ($\tau_{\text{sync}}$) and Obliquity Damping ($\tau_{\text{obl}}$).}
The synchronization timescale $\tau_{\text{sync}}$ describes the alignment of the rotational period of a body with the orbital period. Because it depends primarily on the body's moment of inertia and spin angular momentum, it is typically the fastest tidal process. For a planet, it scales approximately as:
\begin{equation}
\tau_{\text{sync}} \sim \frac{2\alpha_p}{3} \frac{Q}{k_2} \left( \frac{\omega}{n} \right) \left( \frac{M_p}{M_\star} \right) \left( \frac{a}{R_p} \right)^{3} n^{-1},
\end{equation}
where $\omega$ is the initial spin frequency of the planet, $n$ is the mean motion, $Q$ and $k_2$ refer to the planet, and $\alpha_p = C_p/(M_p R_p^2)$ is the normalised moment of inertia of the planet ($\alpha_p \approx 0.25$ for Jupiter). For close-in gas giants, this yields a planetary spin synchronization timescale of $\tau_{\text{sync}} \sim 10^3$–$10^6$ years, assuming $Q/k_2 \sim 10^5$--$10^6$, orbital periods of $\sim$1--4~days, and an initial planetary rotation period of $\sim$10~h, implying rapid tidal locking of the planet (this timescale does not refer to the spin of the host star).

In parallel, the dissipation of the obliquity-related tidal component damps the spin-axis inclination relative to the orbital normal on a comparable or slightly longer timescale $\tau_{\text{obl}}$, leading to low-obliquity, tidally locked configurations \citep{Heller_2011}. However, the coupling between spin synchronization and obliquity damping depends on the tidal model and internal dissipation mechanisms; in some cases, obliquity can persist in Cassini states (equilibrium states of the spin axis of a body when its orbit is perturbed) \citep{Gladman1996} or resonant configurations.

\paragraph{Circularization ($\tau_{\text{circ}}$) and Orbital Decay ($\tau_{\text{decay}}$).}
Eccentricity damping requires dissipation of orbital energy, which represents a substantially larger reservoir than rotational energy. The leading-order eccentricity evolution driven by tides raised on the planet is:
\begin{equation}
\tau_{\rm circ}^{-1} = \frac{1}{e} \frac{de}{dt} = -\frac{21}{2} \frac{k_2}{Q} \frac{M_\star}{M_p} \left( \frac{R_p}{a} \right)^5 n.
\end{equation}
This expression is valid to leading order in $e$ (i.e., for low eccentricities), assumes that the planet rotates synchronously, and assumes that all components of the tide are characterised by the same $Q$.

The strong dependence on $a$ ($\tau_{\rm circ}^{-1} \propto a^{-13/2}$ once the dependence of $n$ on $a$ is included) implies that very close-in planets circularize rapidly, while wider systems may retain measurable eccentricities over gigayear timescales. The exact semi-major axis dependence depends on whether a constant-Q or constant time-lag tidal prescription is used and the treatment is more rigorous than what has been presented here. A detailed discussion on the various prescriptions has been elaborated in \citet{Mathis2018}. It should be emphasised, however, that both the constant-$Q$ and the constant time-lag models are idealised prescriptions that are unlikely to apply in practice; they are best regarded as crude illustrative models of tidal evolution, and they may not even predict the correct qualitative evolution in all cases (e.g., at high eccentricities; \citealt{Barker2026}).

At still longer timescales, tides raised on the host star by the planet extract orbital angular momentum, producing secular orbital decay ($\tau_{\text{decay}}$). Because typical stellar dissipation efficiencies are low ($Q_\star \sim 10^5$–$10^7$) and $M_\star \gg M_p$, this process is generally the slowest, often exceeding stellar lifetimes except in extreme ultra-short-period systems such as WASP-12b \citep{2016A&A...588L...6M}. The efficiency of tidal dissipation in stars remains highly uncertain, with $Q_\star$ estimates spanning several orders of magnitude depending on stellar type, rotation, and evolutionary state.

\subsubsection{Planetary Interiors and Climate}
\begin{figure}
\begin{center}
%\hbox{\hspace{-0.5cm}
\includegraphics[width=0.5\textwidth]{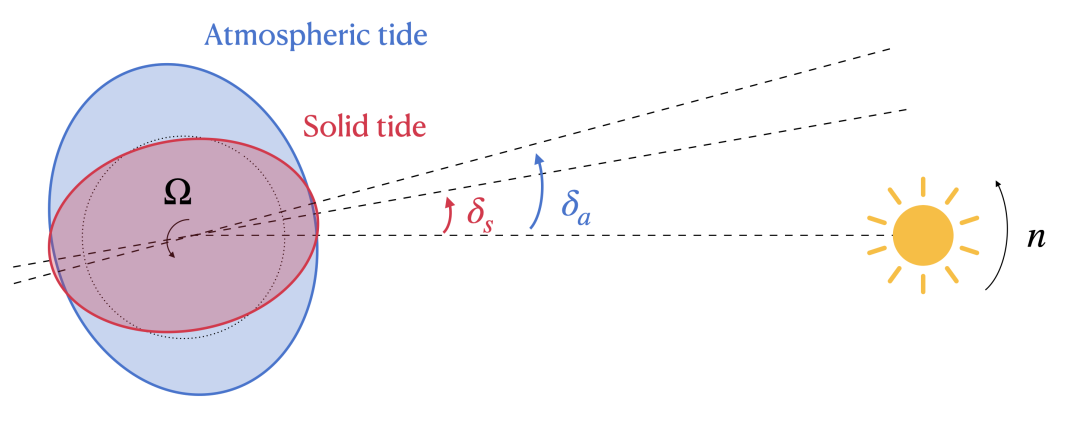} 
%%\hspace{-0.3cm}
%}
%\vspace*{-0.2 cm}
\caption{Schematic illustration of tidal influence on a rocky planet with an atmosphere. In this example, two tidal contributions, gravitational and thermal (atmospheric tides caused by variations in day-night insolation), are acting in opposition on each other \citep{Revol2023}.  }
\label{fig:TidalBulge}
\end{center}
\end{figure}

Tidal interaction between a star and an exoplanet leads to tides raised on the star by its planet, as well as tidal distortion of a planet by its star. For the latter, substantial heating of the planetary interior can result (see \citealt{Jackson2008, Driscoll2015}). This may lead to increased outgassing and volcanism of rocky exoplanets (see for example \citealt{Bello-Arufe2025, Gkouvelis2025, Nicholls2025}) and has been investigated as a factor in exomoon heating \citep{Heller2013}. Furthermore, the spin-orbit coupling of the planetary spin to the planetary orbit leads to constant day and night sides with large temperature contrasts. For planets with substantial atmospheres, the atmosphere and the solid core of the planet interact with the host star in different ways. In this case, multiple evolutionary outcomes for the planetary rotation are possible; an example of a planet which rotational evolution has likely been substantially altered by tidal interactions is Venus \citep{Revol2023}. Fig.~\ref{fig:TidalBulge} presents a schematic illustration of the influence of tidal interactions on a rocky planet with an atmosphere; as one can see, the atmospheric and solid tide can influence the planet in different manners. The spatial distribution of tidal heating can vary depending on the system's parameters. The total energy release in planetary interiors can be described by the relation from \citet{Beuthe2013}:
\begin{equation}
\dot{E} = - \mathrm{Im}(\tilde{k}_2) \frac{(n R_p)^5}{G} \left( \frac{21}{2}e^2 + \frac{3}{2} \sin^2 I \right).
\label{eq:tidal_heat}
\end{equation}

Here, $e$ is the orbital eccentricity, $I$ is the obliquity, $\tilde{k}_2$ is the complex tidal gravity Love number of degree two, whose imaginary part is related to the tidal quality factor by $\mathrm{Im}(\tilde{k}_2) \simeq -k_2/Q$ (Sect.~\ref{sec:tidal_dissipation}), $G$ is the gravitational constant, $R_p$ is the radius of the planet, and $n$ is the orbital mean motion. This expression assumes a synchronously rotating planet, so that the eccentricity and obliquity tides are both forced at the frequency $n$, and is valid to leading order in $e$ and $\sin I$ \citep{Segatz1988, Beuthe2013}. The Love number $k_2$ depends on the interior structure and can change during evolution due to phase changes or melting. From Eq.~\ref{eq:tidal_heat}, planets on eccentric orbits are clearly subject to strong tidal heating, which drives volcanic activity and orbital circularization. Due to the $(n R_p)^5$ dependence, close-in planets are the most affected. Furthermore, the second term shows that a non-zero obliquity also leads to tidal dissipation. Since tidal dissipation drives the system towards its minimum-energy state at fixed angular momentum, which for the planetary spin corresponds to alignment with the orbital normal, the associated tidal torque acts to reduce the obliquity. Unless the obliquity is maintained by other mechanisms (e.g., in Cassini states; \citealt{Gladman1996}), the spin axis of a close-in planet is therefore driven towards alignment with the orbital normal on a timescale comparable to that of spin synchronization. Together, spin synchronization and obliquity damping lead to tidally locked, low-obliquity configurations. This tidal locking fundamentally alters circulation patterns; for tidally locked gas giants, 3D climate simulations predict the formation of fast equatorial eastward wind jets \citep{Carone2020}. In rocky exoplanets, tidal locking can lead to a variety of climate regimes, including `eye-ball' states \citep{Angerhausen2014} where only the substellar point can sustain liquid water \citep{Godolt2015, Macdonald2025}. Tidal interactions can also lead to formation of resonant chains of exoplanets, similar to the resonant chain formed by the Galilean satellites in the Solar System. In the Solar System, Io is the most volcanically active body due to constant energy release due to tidal interactions with Jupiter and other Galilean moons (e.g. \citealp{Segatz1988}). The same process can take place in exoplanetary systems, potentially supporting long-lived volcanic activity, as suspected for the TRAPPIST-1 system \citep{Luger2017,Barr2018}.

\subsubsection{Orbital Evolution and the Darwin Stability Limit}
For star-planet systems, one may in principle expect spin–orbit coupling between the planetary orbit and the stellar rotation. A frequently cited example where the stellar rotation period appears close to the planetary orbital period is the $\tau$~Boo~b system \citep{Catala2007, Donati2008}, suggesting strong tidal coupling in that case. However, a key difference between star-planet systems and stellar binaries is the extreme mass ratio. Because $M_p/M_\star \lesssim 10^{-3}$ for typical hot Jupiters, the tidal torque exerted on the star is much weaker than in binary systems, so that the stellar spin is only weakly affected. The orbit, in contrast, can still evolve significantly: because the orbital angular momentum of a planet is small compared to that of a stellar binary, even this comparatively weak torque can drive substantial orbital decay in the closest systems, as observed for WASP-12b (Sect.~\ref{sec:tidal_migration}). Whether such systems can reach a true spin–orbit equilibrium is further constrained by the Darwin stability criterion \citep{Hut_1980, Damiani2015}. A stable synchronous configuration exists only if the total angular momentum exceeds a critical threshold, $L_{\text{tot}} > 4 S_{\text{sync}}$, where $S_{\text{sync}}$ denotes the stellar spin angular momentum in the hypothetical synchronous state. In most hot Jupiter systems, this condition is not satisfied because the orbital angular momentum dominates but remains insufficient to support synchronous equilibrium when combined with the required stellar spin. As a result, these systems are Darwin unstable: instead of evolving toward a stable synchronous state, they undergo continuous orbital decay driven by tidal dissipation. Moreover, magnetic braking continuously removes angular momentum from the system, so that a true equilibrium does not exist on long timescales and classical Darwin stability arguments may not be relevant in practice \citep{Damiani2015, Barker2026}. Consistent with this picture, most observed systems hosting close-in giant planets do not exhibit matched stellar rotation and orbital periods, indicating that full tidal synchronization is rarely achieved in practice (note that this is different from `tidal locking' wherein the orbital period of the planet is equal to the rotation period of the \textit{planet} rather than the star).

\subsubsection{Tidal migration and stellar spin-up}\label{sec:tidal_migration}
\begin{figure}[ht]
\begin{center}
\includegraphics[width=0.5\textwidth]{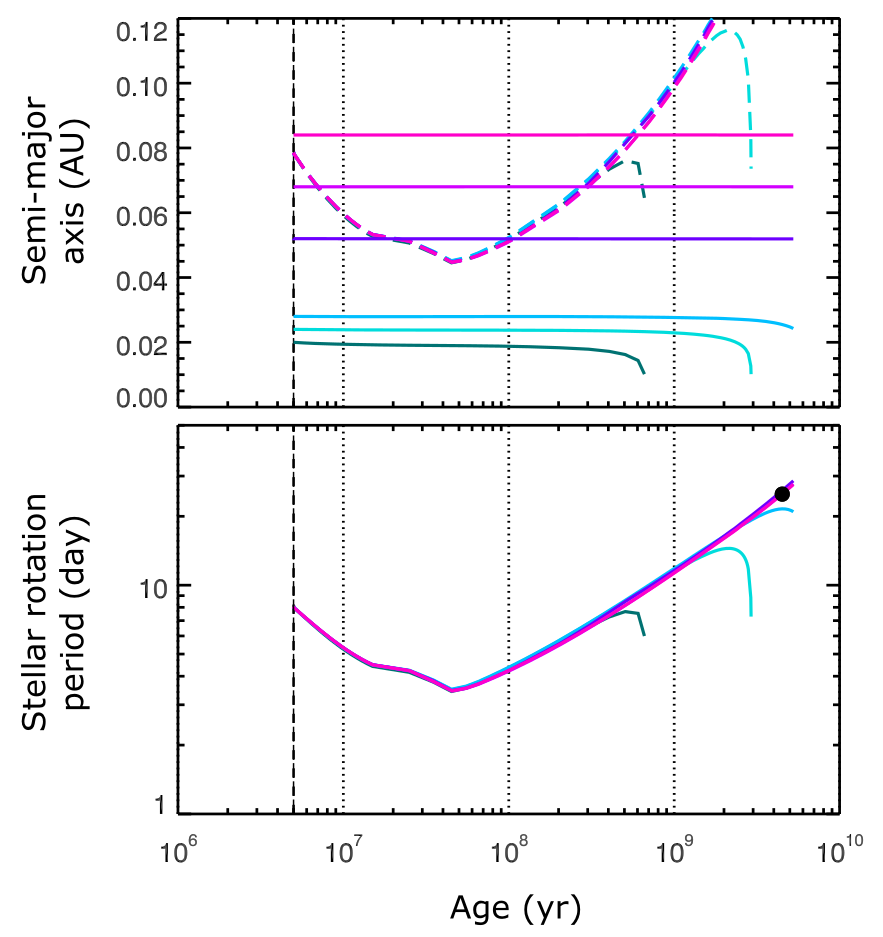}
\caption{Tidal evolution of a planet with a mass equal to the mass of Jupiter around an initially slowly rotating star with a mass equal to the solar mass. Top panel: evolution of the semi-major axis of the planet (solid lines) and the corresponding corotation radius (dashed lines). Bottom panel: evolution of the rotation period of the star. Adapted from \cite{Bolmont2016}.}
\label{fig:TidalMigration}
\end{center}
\end{figure}

At typical moderate ages beyond the first few hundred Myr, the stellar rotation period is usually longer than the orbital period of a close-in planet ($P_{\text{rot}} > P_{\text{orb}}$). In this regime, tidal dissipation transfers angular momentum from the planetary orbit to the stellar spin. However, this transfer does not lead to long-term synchronization because cool stars undergo continuous spin-down due to magnetic braking. As angular momentum is lost through magnetized stellar winds, the system cannot maintain a steady spin-up balance. Instead, tidal interactions result in a net extraction of orbital angular momentum, leading to gradual orbital decay and, ultimately, potential engulfment of the planet. The rate of this decay is highly sensitive to orbital separation, scaling approximately as:
\begin{equation}
\frac{1}{a} \frac{da}{dt} \propto \frac{M_p}{M_\star} \left( \frac{R_\star}{a} \right)^8
\end{equation}
(this scaling corresponds to a constant time-lag prescription for tides raised on a slowly rotating star; a constant-$Q_\star$ prescription instead gives $a^{-1}\,da/dt \propto (M_p/M_\star)(R_\star/a)^5 n$, i.e., a migration timescale $\propto a^{13/2}$; see also Sect.~\ref{sec:magnetic_migration}).
The steep dependence on $a$ implies that tidal migration is significant primarily for ultra-short-period systems. Observationally, evidence for ongoing tidal orbital decay is typically sought through precise long-term timing of transit or radial velocity measurements. Despite extensive searches, only the WASP-12b system is widely accepted as showing measurable orbital decay \citep{2016A&A...588L...6M, 2021AJ....161...72T, 2022AJ....163..175W}. Fig.~\ref{fig:TidalMigration} illustrates an example of the coupled evolution of the semi-major axis and stellar rotation period under tidal interaction and magnetic braking. As one can see, the outcome of the interaction depends on the initial separation between the star and the planet, with the planets being on closest orbits undergoing the most efficient tidal migration with the subsequent engulfment by their host star \citep{Benbakoura2019}. The star gains angular momentum as the result of this interaction and spins up, which has direct influence on its activity level \citep{johnstone2021}. The initial stellar rotation period also matters, as it determines the initial co-rotation radius (at which $n = \Omega_\star$) and its subsequent evolution. %although only one case is shown in Fig.~\ref{fig:TidalMigration}. 
On the stellar side, tidal star-planet interactions lead to a spin-up of the host star relative to a planet-free evolutionary track. This enhanced rotation can in turn amplify magnetic dynamo activity, producing elevated levels of flaring, starspots, and increased X-ray and chromospheric emission. Identifying these effects requires a well-defined non-interacting baseline for stellar evolution. Such reference states are typically constructed using wide binaries, open cluster populations, or asteroseismic constraints.

Tidal star--planet interactions can also manifest themselves as planet-induced stellar pulsations. In the highly eccentric HAT-P-2 system, \citet{deWit2017} detected in $\sim$350~h of \textit{Spitzer} 4.5~$\mu$m photometry stellar pulsations with amplitudes of $\sim$40~ppm at exact harmonics of the planetary orbital frequency, indicative of a tidal origin; high-frequency pulsations consistent with this signal were later also retrieved in independent \textit{HST} observations \citep{Jacobs2025}. \citet{Bryan2024} interpreted such behaviour in terms of episodic resonance locking: in their models of eccentric giant planets around F-type main-sequence stars such as HAT-P-2, dynamical tides raised by the planet excite chains of resonance locks with multiple stellar oscillation modes as the star evolves, yielding orders-of-magnitude larger changes in orbital eccentricity and in the amplitudes of harmonic pulsations than a single resonance-locking episode or non-resonant tidal interactions. This framework predicts episodes of enhanced tidal dissipation, alternating inward and outward migration, and corresponding stellar spin-up and spin-down. Series of harmonic pulsations may therefore provide both a diagnostic of close-in eccentric companions, including non-transiting or highly inclined systems that are otherwise difficult to identify, and a probe of stellar interiors through tidal asteroseismology. Because such resonances become more frequent as the star approaches the end of the main sequence, episodic resonance locking also offers a pathway for the late-stage migration of eccentric giant planets, potentially revealing a population of ``dormant migrating giants'' \citep{Bryan2024}. 

\subsection{Magnetic interactions}\label{sec:magnetic_interactions}
As discussed in Section \ref{sec:tidal_interactions}, tidal interactions can prevent or delay the standard magnetic braking of cool stars, maintaining high levels of magnetic activity over long timescales. This sets the stage for the second major component of star-planet interactions: the electromagnetic coupling between the stellar magnetic field and the planet. While the fundamental plasma processes and specific configurations of our own neighborhood have been detailed as a baseline in Section \ref{sec:solar_system_laboratory} (The Solar System Laboratory), we focus here on the generalized theoretical framework of star-planet magnetic interactions. 

Exoplanets, as well as all the bodies of our Solar System, orbit within the magnetic fields of their respective host stars. For satellites of large planets, the intrinsic magnetic field of the host planet also plays a role in the overall magnetic interactions. In the Solar System, this is the case for the Galilean moons of Jupiter (e.g. \citealp{Kivelson2004,Roth2017,Bagenal2020}). External magnetic fields influence planets in various ways. If a planet or a moon has its own magnetic field, then the external magnetic field compresses it and can induce a response from the planet's mantle and, in some cases, the core (the latter case is especially interesting for Mercury, where the external magnetic field can temporarily amplify the dynamo; \citealp{Glassmeier2007,Zomerdijk-Russell2021}). If the external magnetic field is variable, then it can also drive the planet's magnetospheric configuration (e.g. \citealp{Baumjohann2010, Baumjohann2012}). A time varying field can also generate currents in the planet's atmosphere or interiors, which drive Ohmic heating (Sec.~\ref{sec:OhmicHeating}). Even if the external magnetic field can be approximated as static, the motion of a planet or moon through the field generates an electromotive force (EMF), which can drive currents connecting the primary (the star or the host planet) and the secondary (the planet or the moon). These currents are expected to be carried by Alfv\'en waves.

Before turning to the formalism, it is useful to summarise where the different magnetic interaction regimes are expected to matter. A key quantity is the stellar Alfv\'en surface, i.e., the surface on which the stellar wind speed equals the local Alfv\'en speed, $v_A = B/\sqrt{4\pi\rho}$ (in Gaussian units). Inside this surface the wind is sub-Alfv\'enic, and perturbations generated by the planet can propagate back towards the star along Alfv\'en characteristics; outside it, the wind is super-Alfv\'enic and the interaction is essentially one-way (Sect.~\ref{sec:faroutsystems}). For Sun-like stars, the Alfv\'en surface lies at roughly 10--20 stellar radii (i.e., $\lesssim 0.1$~au), so that only close-in planets such as hot Jupiters can be in the sub-Alfv\'enic regime, whereas for young, rapidly rotating stars and active M dwarfs with strong (hundreds of G to kG) large-scale fields, the Alfv\'en surface can extend much further and may even encompass the habitable zone \citep[e.g.,][]{Garraffo2017}. Close-in planets around strongly magnetised stars are therefore the most favourable cases for star--planet magnetic coupling through Alfv\'en wings, with consequences such as stellar hotspots, coherent radio emission, and magnetic migration (Sects.~\ref{sec:stellarhotspots}--\ref{sec:magnetic_migration} and \ref{sec:emissionnearplanet}). The nature of the planet then determines the type of obstacle: unmagnetised, conducting bodies (e.g., rocky planets without an active dynamo) are described by the unipolar inductor model, while magnetised planets (e.g., gas giants or rocky planets with an active dynamo) present a larger magnetospheric obstacle described by the dipolar inductor model (Sects.~\ref{sec:unipolar_inductor} and \ref{sec:dipolar_inductor}). Induction (Ohmic) heating additionally requires the planet to experience a time-varying field, which is favoured by inclined stellar dipoles, eccentric or inclined orbits, and compact planetary systems around M dwarfs (Sect.~\ref{sec:OhmicHeating}). Finally, far-out planets, including Earth-like planets in the habitable zones of Sun-like stars, interact with the super-Alfv\'enic stellar wind mainly through magnetospheric compression, reconnection, and non-thermal escape (Sects.~\ref{sec:faroutsystems} and \ref{sec:nonthermalescape}).

The interaction between a conducting body and an external magnetic field is governed by the magnetic induction equation, derived from Maxwell’s equations under the assumptions of non-relativistic magnetohydrodynamics (neglecting displacement current) and Ohm’s law $\mathbf{J} = \sigma (\mathbf{E} + \mathbf{v} \times \mathbf{B}/c)$ (e.g. \citealp{Laine2008OhmicDiffusion, Saur_2010, Laine2012UnipolarInductor}):
\begin{equation}
    \frac{\partial \mathbf{B}}{\partial t} = \nabla \times ( \mathbf{v} \times \mathbf{B}) - \nabla \times (\eta \nabla \times \mathbf{B}),
    \label{eq:induction}
\end{equation}
where $\eta = c^2/(4\pi \sigma)$ is the magnetic diffusivity in Gaussian units. If $\eta$ is constant, the second term becomes $\eta \nabla^2 \mathbf{B}$ and describes magnetic diffusion. The relative importance of induction and diffusion is quantified by the magnetic Reynolds number $\mathrm{Rm} = vL/\eta$. For $\mathrm{Rm} \gg 1$, magnetic field lines are effectively frozen into the flow, whereas for $\mathrm{Rm} \ll 1$, Ohmic diffusion dominates over induction, so that magnetic perturbations are rapidly smoothed out. Note that this does not imply that Ohmic dissipation is efficient: in this regime, diffusion can damp the magnetic perturbations to such an extent that the associated Ohmic heating becomes negligible.

\begin{figure}
\begin{center}
\includegraphics[width=1.0\textwidth]{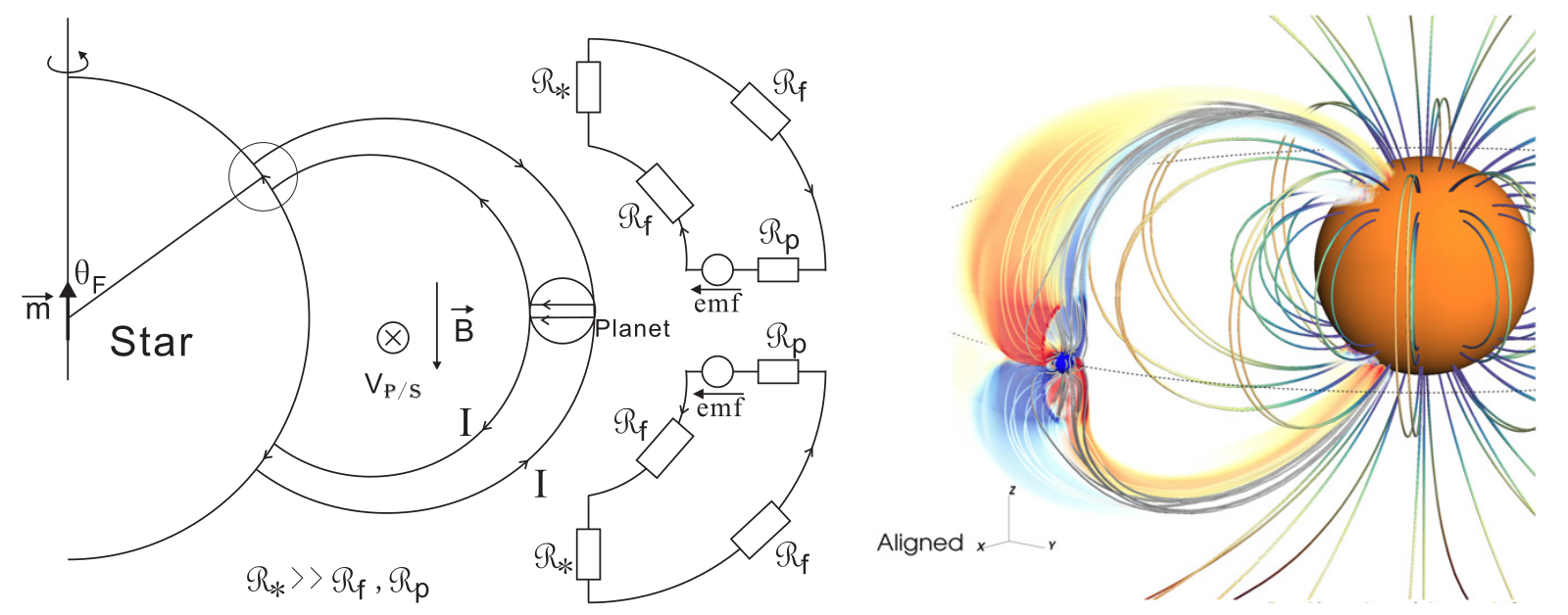} 
\caption{The unipolar inductor mechanism. Left panel: schematic of a conducting body moving with a velocity denoted by $v_{P/S}$ through an external magnetic field denoted by $\vec{B}$ and generating currents that connect it to the central object. The currents are expected to form a loop and impacted by the electrical resistance posed by the planet's ionosphere (denoted by $\mathcal{R}_P$), the magnetic field lines themselves (denoted by $\mathcal{R}_f$) as well as the stellar atmosphere (denoted by $\mathcal{R}_*$). Right panel: Associated current systems in star-planet interactions obtained from a numerical simulation of a planet exhibiting a dipolar magnetic field orbiting around the star in a sub-Alfv\'enic regime. The left and right panels are adapted from \citet{Laine2012UnipolarInductor} and \citet{Strugarek_2025} respectively.}
\label{fig:UnipolarInductor}
\end{center}
\end{figure}

For an exoplanet experiencing a quasi-static ambient stellar magnetic field, particularly in the sub-Alfvénic regime where the relative velocity between the planet and the surrounding plasma is less than the local Alfvén speed, magnetic interactions can lead to a two-way communication between the star and the planet. These interactions can be categorized into two theoretical regimes based on the nature of the planetary obstacle: the unipolar inductor model for non-magnetized, conducting bodies, and the dipolar inductor model for planets possessing an intrinsic magnetic field \citep{Zarka2007}.
Both regimes are governed by the motional electric field $\mathbf{E} = -(\mathbf{v}_{rel} \times \mathbf{B})/c$ established by the planet's velocity relative to the magnetized stellar plasma. This steady-state coupling is physically distinct from the time-varying induction that occurs when a planet encounters fluctuations in the stellar magnetic flux ($\partial \mathbf{B}/\partial t$), a process leading to atmospheric and interior Ohmic heating discussed in Section \ref{sec:OhmicHeating}.

\subsubsection{Unipolar Inductor Model}
\label{sec:unipolar_inductor}

The unipolar induction mechanism describes the interaction of a conducting, non-magnetized body (such as a terrestrial planet or a moon) as it moves through a magnetized plasma. This interaction generates a motional electromotive force, a concept originally and historically applied to the Io-Jupiter system by \citet{Goldreich1969}.

As illustrated in the left panel of Figure \ref{fig:UnipolarInductor}, a conducting body moving with a relative velocity $\mathbf{v}_{P/S}$ through an external magnetic field $\mathbf{B}$ establishes an internal electric field. In the reference frame of the planet, this field is given by:
\begin{equation}
\mathbf{E} = -\frac{1}{c}(\mathbf{v} \times \mathbf{B})
\end{equation}

For a perfectly conducting sphere of radius $R$ rotating with angular frequency $\Omega$ in a uniform magnetic field $B_0$ aligned with the rotation axis, the resulting potential difference between the pole and the equator is derived by integrating the motional field:
\begin{equation}
\mathcal{E} = \frac{1}{c} \int \limits_{0}^{R} B_{0}\Omega r \, dr = \frac{B_{0}\Omega R^{2}}{2c}
\end{equation}

To sustain this internal field, a volume charge density $\rho_{\text{total}}$ must exist within the conductor, a concept mathematically analogous to the Goldreich-Julian density found in pulsar magnetospheres \citep{Goldreich}:
\begin{equation}
\rho_{\text{total}} = \frac{\nabla \cdot \mathbf{E}}{4\pi} = -\frac{\Omega B_{0}}{2\pi c}
\end{equation}

In the case of a sphere that is electrically neutral as a whole, the external potential $V(r>R)$ must satisfy Laplace's equation and match the internal potential at the boundary. This leads to a quadrupolar potential distribution:
\begin{equation}
V(r>R) = -\frac{B_{0}\Omega R^{5}}{3cr^{3}}P_{2}(\cos\theta)
\end{equation}
where $P_{2}(\cos\theta) = \frac{1}{2}(3\cos^2\theta - 1)$. The corresponding external electric field is:
\begin{equation}
\mathbf{E} (r>R) = -\frac{B_{0}\Omega R^{5}}{cr^{4}}\left( P_{2}(\cos\theta)\hat{\mathbf{r}} + \frac{\sin{2\theta}}{2}\hat{\mathbf{\theta}}\right)
\end{equation}

In a plasma-filled environment, this potential drives a global current system. Following the circuit analogy presented in \citet{Laine2012UnipolarInductor} and Figure \ref{fig:UnipolarInductor}, the total current is regulated by the series combination of several resistive components:
\begin{itemize}
    \item $\mathcal{R}_P$: The electrical resistance of the planetary ionosphere or mantle.
    \item $\mathcal{R}_f$: The resistance associated with the magnetic field lines (The Alfvénic conductance of the plasma).
    \item $\mathcal{R}_*$: The resistance of the stellar atmosphere.
\end{itemize}

\subsubsection{Dipolar Inductor Model }
\label{sec:dipolar_inductor}

When a planet possesses an intrinsic dipolar magnetic field $\mathbf{B}_p$, the primary interaction shifts from the planetary surface to the magnetopause. This regime is defined as the dipolar inductor or the reconnection-driven model \citep{Zarka2007, Lanza2012, Saur2013}. In this configuration, the planet presents a magnetic obstacle significantly larger than its physical radius $R_p$. The effective interaction radius $R_{obs}$ (or magnetospheric radius $R_{mag}$) is determined by the balance between the planetary magnetic pressure and the total pressure (dynamic and magnetic) of the stellar wind:
\begin{equation}
R_{obs} \approx R_p \left( \frac{B_p^2}{8\pi P_{sw}} \right)^{1/6}
\end{equation} 
In this regime, energy transfer is primarily mediated by magnetic reconnection and pressure balance at the magnetopause. The planetary magnetosphere supports its own system of field-aligned currents, which locally govern the current structure near the planet. On larger scales, the overall current topology is shaped by additional field-aligned currents that connect the planet to the host star. This configuration is illustrated in the right panel of Fig.~\ref{fig:UnipolarInductor}. In the immediate vicinity of the planet, the currents are controlled by the intrinsic planetary magnetic field. In contrast, the large-scale currents linking the planet to the star are determined by a combination of magnetospheric current systems and the motional EMF generated by the magnetospheric obstacle as it interacts with the surrounding plasma flow.

While the above models work even under a quasi-static field, electromagnetic induction effects arise when a conducting body is embedded in a varying external magnetic field. In the Solar System, these induction effects are most prominent in the Galilean satellites of Jupiter and in Mercury, providing a vital empirical baseline for the `induction heating' regime. For the Galilean satellites, the time variation of the magnetic field ($\partial \mathbf{B}/\partial t$) in the frame of the moons is driven by the $\sim 10^\circ$ tilt of Jupiter’s magnetic dipole and its rapid rotation. For Io, although tidal dissipation is the primary heat source, electromagnetic induction contributes significantly to the heating of its interior \citep{Roth2017}. Such variations for the case of Europa also drives large-scale eddy currents in Europa's saline oceans, creating a secondary magnetic field that was instrumental in the ocean's discovery \citep{Kivelson2000, Gissinger2019}. In the case of Mercury, the dynamic interaction with the solar wind drives electric currents in the planet's core. These currents produce a transient magnetic response that can temporarily amplify the planet's internal dynamo (e.g., \citealp{Glassmeier2007, Johnson2016, Zomerdijk-Russell2021}). In the exoplanetary context, this time-varying magnetic flux affects the evolution of a planet in several ways: it can lead to the heating of interiors and atmospheres or trigger extensive volcanic activity. This time-varying magnetic flux ($\partial \mathbf{B}/\partial t$) typically arises from three primary configurations: an inclined stellar magnetic dipole relative to the rotation axis \citep[as often observed in M-dwarfs;][]{Morin2010, See2025}, a high orbital inclination that causes the planet to sweep through varying magnetic latitudes \citep{Albrecht2021, Bourrier2023}, or orbital eccentricity where the ambient field strength fluctuates as a function of the planet's varying distance from the host star \citep{Kislyakova2017}. Additionally it could also be simply due to the irregular magnetic topology of the star for realistic stellar magnetic fields. 

\subsubsection{Ionospheric and Interior Ohmic heating by Induction}
\label{sec:OhmicHeating}

\begin{figure}
\begin{center}
\includegraphics[width=1.0\textwidth]{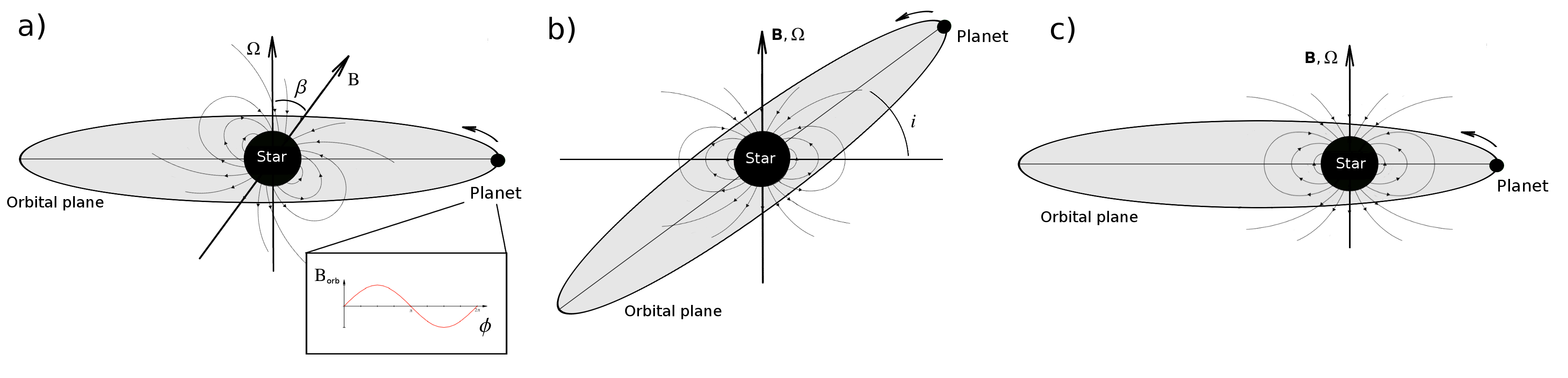} 
\caption{Sketch of the induction heating mechanism. In every case, the planet continuously experiences magnetic flux changes in its interior along its orbital motion. The change of the magnetic flux penetrating the planet generates eddy currents which dissipate and heat the planetary mantle. Panel a: the origin of the magnetic field variation is an inclined stellar magnetic dipole. Panel b: the star exhibits a dipolar field aligned with the rotation axis (roughly like the Sun during activity minima), and the planet is on an inclined orbit. Panel c: the planet is on an eccentric orbit, and the variation of the magnetic field is due to varying distance to the host star (the left panel adapted from \citet{Kislyakova2017}).}  \label{fig:schematic}
\end{center}
\end{figure}

\paragraph{The Physics of Dissipation}
The upper layers of a planet's atmosphere are ionized by stellar XUV radiation, energetic particle precipitation, and the stellar wind, forming a plasma composed of neutrals ($n$), ions ($i$), and electrons ($e$). In typical exoplanetary upper atmospheres, ions are often predominantly singly ionized, so that quasi-neutrality ($n_e = \sum_i Z_i n_i$) reduces to $n_e \approx n_i$, where $Z_i$ is the charge state of ion species $i$.

To model the resulting dissipation, we introduce the magnetic vector potential $\mathbf{A}$ (defined such that $\mathbf{B} = \nabla \times \mathbf{A}$) and adopt the Coulomb gauge ($\nabla \cdot \mathbf{A} = 0$). Combining Faraday's Law, Ampère's Law, and the generalized Ohm's law, the master equation for the evolution of the potential is:
\begin{equation}
\label{eq:MasterEq}
\nabla^2 {\bf A} - \frac{4\pi}{c^2}\sum_\alpha \underline{\sigma}_\alpha \cdot  \frac{\partial {\bf A}}{\partial t} = 0
\end{equation}
where $\underline{\sigma}_\alpha$ is the conductivity tensor of species $\alpha$, describing the electrodynamic response of each plasma component.

In the ionosphere, conductivity is highly anisotropic. The three primary components are the parallel (field-aligned, or Spitzer) conductivity $\sigma_\parallel$, the Pedersen conductivity $\sigma_P$ (perpendicular to $\mathbf{B}$ and parallel to $\mathbf{E}$), and the Hall conductivity $\sigma_H$ (perpendicular to both $\mathbf{E}$ and $\mathbf{B}$) \citep{landau2003electrodynamics, SchunkNagy2009}. In terms of these components, the generalised Ohm's law for the current density in the ionosphere reads
\begin{equation}
\mathbf{J} = \sigma_\parallel \mathbf{E}'_\parallel + \sigma_P \mathbf{E}'_\perp + \sigma_H \, \hat{\mathbf{b}} \times \mathbf{E}'_\perp ,
\label{eq:generalised_ohm}
\end{equation}
where $\hat{\mathbf{b}} = \mathbf{B}/|\mathbf{B}|$, $\mathbf{E}' = \mathbf{E} + \mathbf{u}_n \times \mathbf{B}/c$ is the electric field in the rest frame of the neutral gas moving with velocity $\mathbf{u}_n$, and the subscripts $\parallel$ and $\perp$ denote the components parallel and perpendicular to $\mathbf{B}$. The total conductivity tensor $\sum_\alpha \underline{\sigma}_\alpha$ in Eq.~(\ref{eq:MasterEq}) is the tensor representation of these three components. For ionospheric Ohmic heating, the Pedersen conductivity typically dominates the total energy dissipation \citep{Strugarek_2025}.

Eq.~\ref{eq:MasterEq} provides a general formalism for any conducting medium and has been solved analytically for piece-wise constant conductivities \citep{Parkinson1983, Bidinosti2007, Saur_2010} and numerically for arbitrary conductivity profiles \citep{Kislyakova2017, Strugarek_2025}. Solving this equation allows one to determine where energy is deposited and the degree to which the time-varying field penetrates the planetary interior.

\paragraph{Atmospheric vs. Interior Heating}

The location and nature of induction heating depend primarily on the presence of an intrinsic planetary magnetic field, which controls how externally time-varying magnetic perturbations couple into conducting layers.

\begin{figure}[htpb]
\begin{center}
\includegraphics[width=0.9\textwidth]{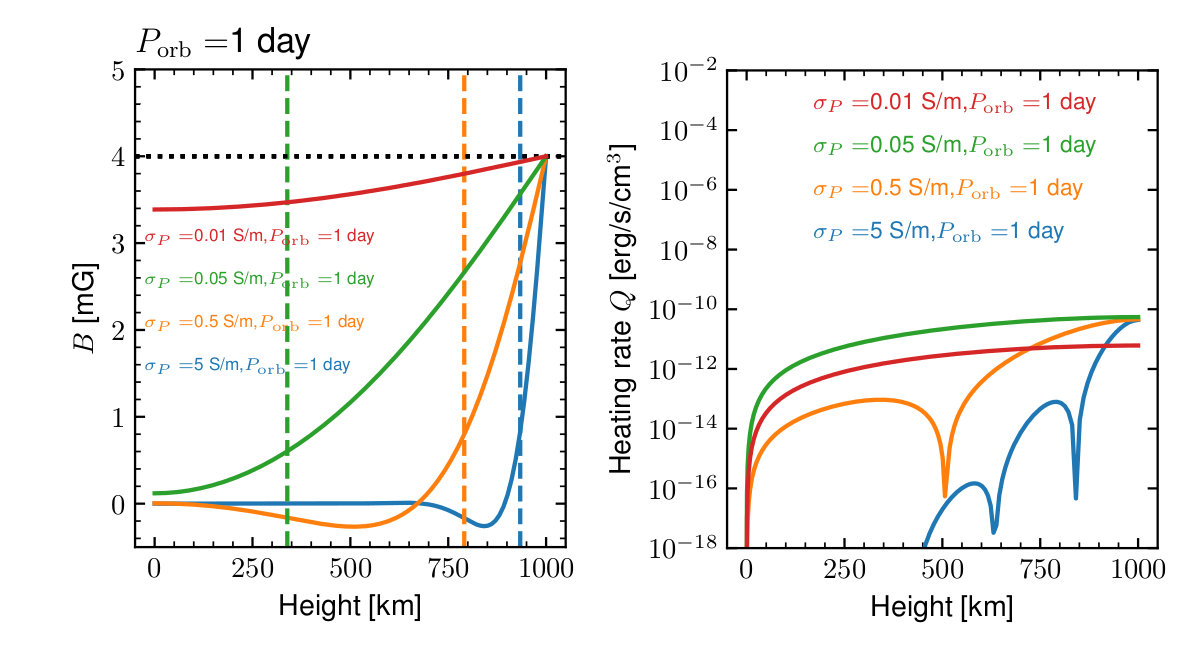}
\caption{Left: Penetration of the time-varying magnetic field into the atmosphere for varying Pedersen conductivities. As conductivity increases, the field is more efficiently screened and decays over a shorter distance from the upper boundary (where $z$ is the height above the surface). Right: Associated volumetric Ohmic heating rate $Q$. Colored lines show that as the atmosphere becomes a more efficient screen, most of the heat is deposited at the top of the atmosphere. Adapted from \citet{Strugarek_2025}.}
\label{fig:Strugarek2025_OhmicHeating}
\end{center}
\end{figure}

\textit{Unmagnetized Planets:} In the absence of an intrinsic magnetic field, the external time-varying stellar magnetic field directly drives electric currents in the upper atmosphere/ionosphere. As shown by \citet{Strugarek_2025}, Ohmic dissipation is therefore concentrated in the ionosphere and thermosphere (Fig.~\ref{fig:Strugarek2025_OhmicHeating}). The spatial distribution and magnitude of this heating are governed by the magnetic screening effect. The left panel of Fig.~\ref{fig:Strugarek2025_OhmicHeating}  illustrates how the external magnetic field $B_{\rm ext}$ penetrates the atmosphere. As the Pedersen conductivity $\sigma_P$ increases, the induced currents become stronger, more effectively shielding the deeper atmospheric layers. This causes the time-varying field to decay rapidly within a narrow layer at the top of the atmosphere, a scale defined by the Pedersen skin depth, $\delta_P = c/\sqrt{2\pi\Omega\sigma_P}$. The right panel shows the resulting volumetric heating rate. In the regime where the skin depth $\delta_P$ is much smaller than the vertical extent of the atmosphere, the system approaches a saturated response. In this limit, any further increase in conductivity merely restricts the dissipation to an even thinner layer without increasing the total energy deposition. Consequently, the heating rate becomes only weakly dependent on the local conductivity and is bounded by a maximum value: \begin{equation} \label{eq:maxQ_simplified} Q_{\rm max} = \frac{\Omega B_{\rm ext}^2}{8\pi}, \end{equation} where $\Omega$ is the characteristic angular frequency of the large-scale magnetic forcing (e.g. orbital motion or stellar magnetic variability), and $B_{\rm ext}$ is the external stellar magnetic field strength at the planet. In this limit, energy deposition is controlled primarily by the imposed electromagnetic forcing rather than the detailed ionospheric conductivity profile

\textit{Magnetized Planets:} When a planet possesses an intrinsic magnetic field, the external time-varying stellar field first interacts with the magnetosphere and drives time-dependent currents in the ionospheric closure region. These currents can then induce further magnetic field perturbations that can propagate downward and couple to the planetary interior. This scenario is expected for many rocky exoplanets \citep{Zhang2022} and hot Jupiters \citep{Christensen2006}.

In this case, the response of the interior is governed by electromagnetic diffusion rather than direct atmospheric dissipation. The conductivity $\sigma$ is treated as an effective scalar conductivity in the deep interior, and the characteristic depth of energy deposition is set by the electromagnetic skin depth,
\begin{equation}
\delta = \sqrt{\frac{2 c^2}{4 \pi \sigma \omega}},
\end{equation}
where $\omega$ is the angular frequency of the time-varying magnetic field experienced locally by the planet.

This scale determines whether induction heating is confined to shallow layers or can penetrate into the mantle and core, depending on the conductivity structure and forcing frequency.

In the strong skin effect regime ($\delta \ll R_p$), the total energy release is concentrated in the upper layers:
\begin{equation}
Q = \Delta B^2 \frac{3 c^2 R_p^2}{16 \pi \sigma \delta},
\label{eq_strong_skin}
\end{equation}
where $\Delta B$ is the characteristic amplitude of magnetic field variations in the planetary environment, and $R_p$ is the planetary radius. Conversely, for the weak skin effect ($\delta \gg R_p$), typical of Moon-sized objects or asteroids, the total energy release is:
\begin{equation}
Q = \Delta B^2 \frac{3 c^2 R_p}{20 \pi \delta (\mu +2)^2} \left( \frac{R_p}{\delta} \right)^4,
\label{eq_weak_skin}
\end{equation}
where $\mu$ is the relative magnetic permeability of the medium \textcolor{red}{\citep{Kislyakova2023}}.

Late-type M dwarfs are particularly favorable hosts for induction heating because they often possess strong, large-scale magnetic fields \citep{Shulyak2017, Shulyak2019}, maintain high rotation rates over gigayear timescales \citep{johnstone2021, See2025}, and frequently host compact planetary systems \citep{Deeg2018HandbookOfExoplanets, Sabotta2021}. While earlier studies suggested that induction heating might be relevant for super-Earths orbiting G-type stars \citep{KislyakovaNoack2020}, more recent work indicates that the process is significantly more efficient around M dwarfs \citep{Peng2025}.

\begin{figure}[h]
\begin{center}
\includegraphics[width=0.9\textwidth]{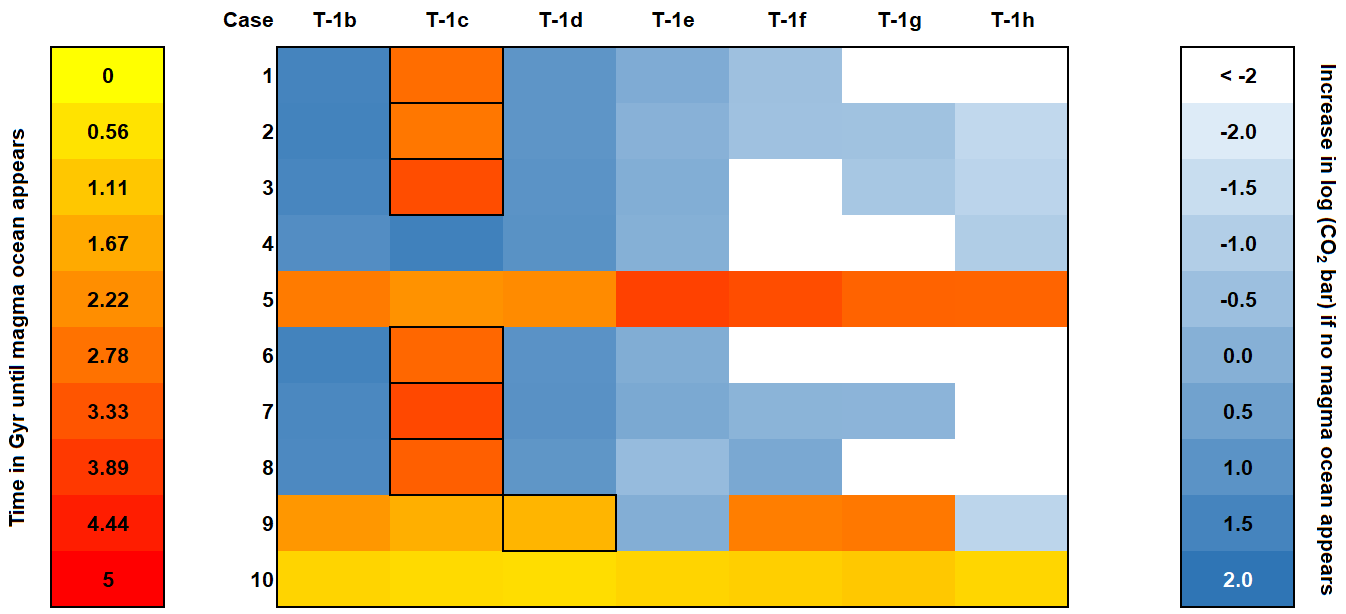}
\caption{Induction heating effects on TRAPPIST-1 planets. Yellow–red colors indicate regions where mantle temperatures exceed the liquidus, forming magma oceans. Thick boundaries highlight cases where melting only occurs when induction heating is included. The blue scale shows the logarithmic increase in outgassed CO$_2$. From \citet{Kislyakova2017}.}
\label{fig:InductionTrappist1}
\end{center}
\end{figure}

For rocky worlds like those in the TRAPPIST-1 system, electromagnetic induction heating, driven by the star’s rotation and the planets' orbital motion, can serve as a primary internal energy source. As illustrated in Figure \ref{fig:InductionTrappist1}, modeling across ten different parameter scenarios reveals that the three innermost planets, TRAPPIST-1b, c, and d, are particularly susceptible to extreme heating. In many cases, this process pushes mantle temperatures beyond the liquidus (yellow–red regions), creating global magma oceans. Notably, the thick boundaries in the figure highlight `tipping point' cases for TRAPPIST-1c and -d, where melting occurs exclusively because of induction heating. Even when a magma ocean is absent, induction heating can enhance volcanic outgassing by up to several tens of bars of CO$_2$ (blue scale) over 5 billion years of evolution. Such persistent geodynamic activity has profound implications for atmospheric observability; it suggests that even if primary atmospheres are lost to stellar activity, volcanic degassing can sustain transient secondary atmospheres or produce detectable plasma tori around the star, analogous to the Io plasma torus \citep{Kislyakova2018, Bouma2025}. In gas giants, Ohmic dissipation provides an internal heat source that contributes to radius inflation and affects early contraction phases \citep{HuangCumming2012, Knierim2022}.

\subsubsection{Stellar Hotspots Caused by Exoplanets} \label{sec:stellarhotspots}

Early observational studies provided some of the first evidence for magnetic interactions between close-in exoplanets and their host stars. In particular, \citet{Shkolnik_2005, Shkolnik2008} reported periodic enhancements in chromospheric activity in systems such as HD 179949 and $\upsilon$~And. These variations were observed to be \textbf{phase-locked} to the orbital periods of their close-in giant planets, suggesting a direct magnetic link between the star and the planet. The alignment between the chromospheric emission and the planetary orbital phase supports the hypothesis that the planet induces localized energy deposition in the stellar chromosphere, generating observable activity signatures (hotspots). Subsequent work, including that of \citet{Cauley_2018} and \citet{Cauley_2019}, reinforced this interpretation by identifying similar chromospheric variability in additional systems.

These studies sought to quantify the energetics of the interaction, estimating the excess power emitted by the star due to its interaction with the planet relative to the baseline stellar emission. However, it is important to note that such interaction signatures are not consistently persistent. Long-term monitoring has revealed substantial variability in their detectability; for example, \citet{Cauley_2018} found that HD 189733 exhibited planetary-induced activity during some epochs but not during others. This variability may be linked to the changing magnetic environment of the star or the transition between different plasma interaction regimes, as modeled for non-magnetized rocky planets by \citet{Cohen2015}, and characterized by the Alfvénic Mach number, $M_{A}$, defined as the ratio of the relative plasma flow velocity in the obstacle's frame, $v_{\text{rel}}$, to the local Alfvén speed, $v_{A}$:
\begin{equation}
    M_{A} =\frac{v_{\text{rel}}}{v_{A}}.
\end{equation}
The flow is super-Alfvénic when $(M_{A}>1)$ and sub-Alfvénic when $(M_{A}<1)$.

\paragraph{Alfvén Wings}
The theoretical framework for these interactions is generally based on the unipolar inductor model (see Section \ref{sec:unipolar_inductor}), which describes the generation of an electromotive force across the orbiting body. The actual transport of energy along the circuit, however, is primarily governed by magnetohydrodynamic (MHD) wave dynamics. As the conducting obstacle (the exoplanet) moves through the magnetized stellar wind in the sub-Alfvénic regime, it excites Alfvén waves that propagate along the magnetic field lines. Because the Alfvén velocity ($\mathbf{v}_{A}$) varies along the field lines and the planet moves with a relative orbital velocity ($\mathbf{v}_{\mathrm{rel}}$), these disturbances are not emitted isotropically. Instead, they organize into two time-stationary structures, known as Alfvén wings, aligned with the Alfvén characteristics ($\mathbf{c}_{A}^{\pm} = \mathbf{v}_{\mathrm{rel}} \pm \mathbf{v}_A$) in the planet’s reference frame, where (+) denotes propagation toward one hemisphere of the planet and (-) toward the other hemisphere. These wings act as the “transmission lines” of the unipolar inductor circuit, carrying field-aligned currents (FACs) that connect the planetary generator to the stellar surface \citep[e.g.,][]{Saur2013, Paul_2025}.

\begin{figure*}
    \centering
    \includegraphics[width=\linewidth]{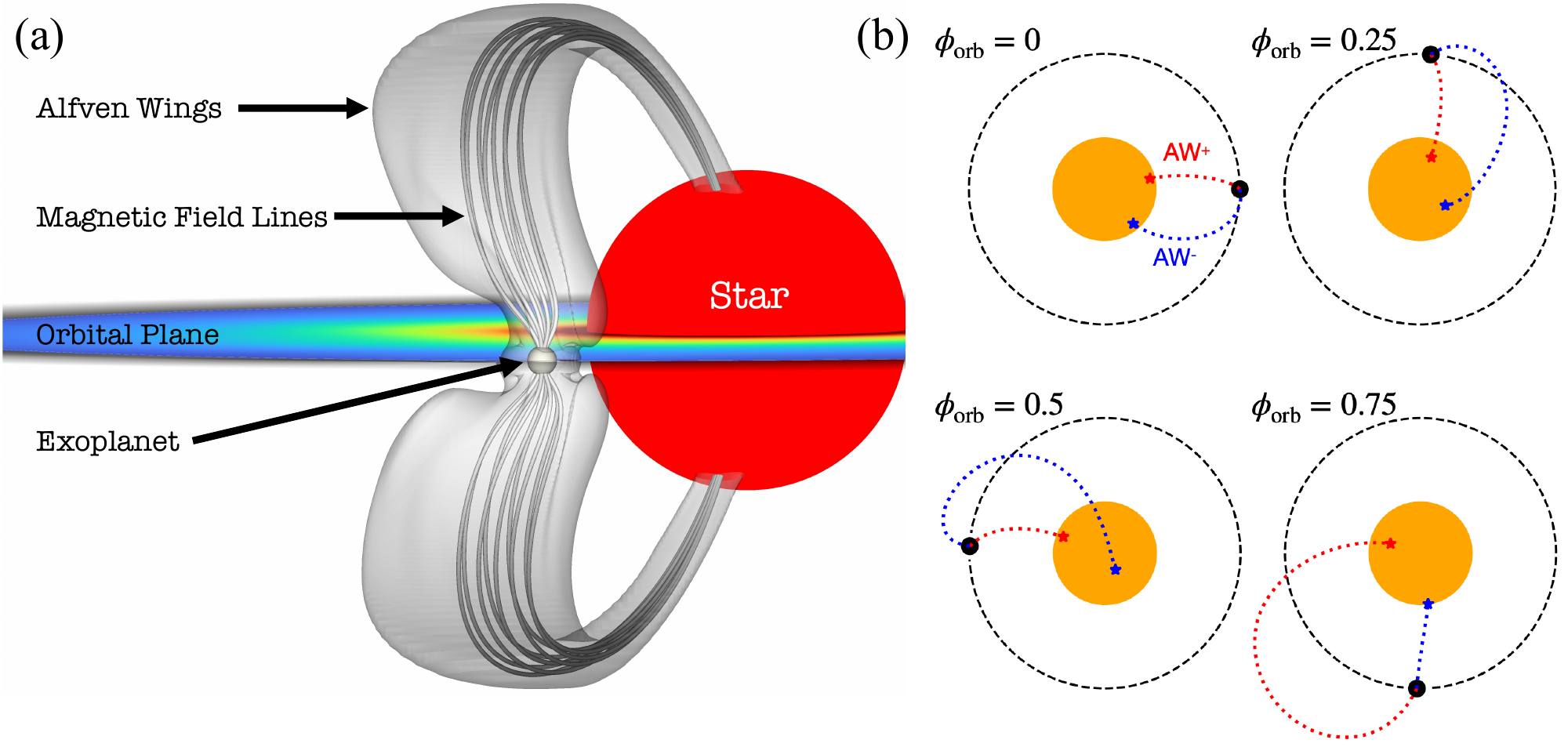}
    \caption{Panel (a) presents a three-dimensional rendering of a close-in exoplanet moving through a magnetized, sub-Alfvénic stellar wind. The Alfvén wings and the associated magnetic field lines within them are highlighted. Panel (b) displays the projection of the two Alfvén wings onto the orbital plane of the exoplanet at four different orbital phases around the star Kepler-78 \citep{Strugarek2019}. The projection is based on a 3D MHD simulation using the observed magnetic map of the star.}
    \label{fig:3DAW_2dproj}
\end{figure*}
The morphology of Alfvén wings is governed by the Alfvén characteristics ($\mathbf{c}_{A}^{\pm}$), with their geometry primarily determined by the ambient plasma flow and the local magnetic field configuration, rather than by the detailed properties of the planetary obstacle. Panel (a) of Figure \ref{fig:3DAW_2dproj} illustrates the canonical structure of the wings in an idealized system where the magnetic dipole moments of both the planet and the star are oriented perpendicular to the orbital plane. In this configuration, the magnetic field lines within the wings act as a `magnetic tether,' carrying field-aligned currents (FACs) that close the unipolar inductor circuit. This connectivity enables the transport of energy from the planet to the stellar surface, potentially generating localized heating at the magnetic footpoints of the Alfvén wings.

The Alfvén characteristics define the boundaries within which the Alfvén mode can exert its influence. In contrast to fast MHD disturbances, which propagate approximately isotropically, slow MHD disturbances are confined by the geometry of the slow mode characteristics  $\mathbf{c}_{s}^{\pm}$. The Alfvén characteristics form an angle $\theta_{A}^{\pm}$ relative to the background magnetic field, defined as:
\begin{equation}
    \sin \theta_{A}^{\pm}= \frac{M_{A}\cos \theta}{(1+M_{A}^{2}\pm 2M_{A}\sin \theta)^{1/2}},
\end{equation}
where $\theta$ measures the deviation of the flow from the direction perpendicular to the magnetic field. In the particular case where the background magnetic field is oriented perpendicular to the flow direction, the resulting Alfvén waves make an angle:
\begin{equation}
    \theta_{A}=\tan^{-1}M_{A}.
\end{equation}

Note that, following the original works, the expressions from here onwards in Sect.~\ref{sec:magnetic_interactions} are written in SI units, whereas Gaussian units are used in Sects.~\ref{sec:unipolar_inductor}--\ref{sec:OhmicHeating}. The Alfvén conductance controls the maximum current that can be carried by the Alfvén wave. It is (\citealt{Neubauer1980})
\begin{equation}
    \Sigma_{A}=\frac{1}{\mu_{0}v_{A}(1+M_{A}^{2}\pm 2M_{A}\sin \theta)^{1/2}}
\end{equation}
and the magnetic field in an Alfvén wing is constrained to:
\begin{align}
    \mathbf{B}_{\perp}&=\mu_{0}\Sigma_{A}\frac{\mathbf{c}_{A}^{\pm}}{|\mathbf{c}_{A}^{\pm}|}\times  \mathbf{E}\\
    B_{||}&=\pm (B_{0}^{2}-B_{\perp}^{2})^{1/2}
\end{align}
where $\mathbf{B}_{\perp}$ and $\mathbf{B}_{||}$ are the perpendicular and parallel decomposition of the magnetic field perturbation relative to the background magnetic field $\mathbf{B}_{0}$.   The perturbations are perpendicular to the background magnetic field and the wave generates no perturbation of the magnetic field strength. 

In realistic systems, however, stellar magnetic fields are far more topologically complex than simple dipoles. Consequently, the morphology of the Alfvén wings is highly sensitive to variations in stellar wind conditions, such as plasma density, velocity, and magnetic field orientation, as well as to the evolving magnetic connectivity between the star and the planet. Panel (b) of Figure \ref{fig:3DAW_2dproj} illustrates this complexity for a planet orbiting Kepler-78, using a realistic stellar magnetic field extrapolation. The projected pattern of the Alfvén wings (blue dotted and dashed curves) changes continuously with the orbital phase $\phi$, highlighting that these wings are dynamic structures shaped by both orbital motion and, in turn, the time-varying stellar magnetic environment.

\subsubsection{Energy Dissipation Mechanisms for Sub Alfv\'enic Magnetic Interactions}
Building on the work of \citet{Neubauer_1980} on the interaction between Jupiter and its moon Io, \citet{Zarka_2001} extended this framework to describe energy dissipation in sub-Alfvénic star–planet interactions by analogy. In this picture, the interaction behaves like an electrical circuit. Perturbation electric fields generated at the planetary obstacle in the ambient plasma flow, together with the Alfvén conductance of the connecting magnetic field lines, define the effective source and loads within this circuit. For the case in which the ambient plasma flow is perpendicular to the ambient stellar magnetic field, \citet{Zarka_2001} estimated the total power dissipated within this current loop as:
\begin{equation}
    \mathcal{P}_{loop} = \frac{B_{\star}^{2}}{2\mu_{0}}\pi R_p^{2} v_{\mathrm{rel}} 
    \left(\frac{2 M_{A}}{1+M_{A}^{2}}\right)^{1/2}
    \label{eq:zarka2001_power}
\end{equation}
where $B_\star$ is the stellar magnetic field strength at the orbit of the planet, $\mu_0$ is the permeability of free space, $R_p$ is the planetary radius, $v_{\mathrm{rel}}$ is the relative speed between the planet and the ambient plasma flow, and $M_A = v_{\mathrm{rel}}/v_A$ is the Alfvén Mach number with $v_A$ being the Alfvén speed. The dependence of the power on the load imposed by the Alfvén conductance, which itself depends on $v$ and $M_A$, has been absorbed into the corresponding terms in the formula for simplification.

This circuit analogy established an initial conceptual framework for understanding the total energy budget in SPMI, particularly in terms of planetary radio emissions. More recently, however, star–planet magnetic interactions have been inferred from excess stellar emissions, commonly referred to as stellar `hotspots.' These hotspots represent enhanced emissions from the stellar surface at visible wavelengths (generally Ca~{\sc ii} H\&K) and are believed to be modulated by the energetics of SPMI. Current understanding suggests that, in sub-Alfvénic interactions, energy from SPMI is ultimately transferred to the star and dissipated near its surface, giving rise to the observed emission enhancements. Several theoretical mechanisms have been proposed to explain how this transfer occurs, likely acting in combination depending on the plasma environment. Below, we outline the premises and provide analytical estimates for the three primary mechanisms: magnetic reconnection, Alfvén wave propagation, and magnetic stress accumulation.

\subparagraph{Magnetic Reconnection}

The release of magnetic energy via magnetic reconnection in star–exoplanet systems was first proposed by \citet{Rubenstein_2000} as a potential explanation for stellar superflares (highly energetic events with energies approximately $10^2$–$10^7$ times greater than those of typical solar flares) observed on slowly rotating F- and G-type dwarf stars that would otherwise be expected to exhibit low levels of magnetic activity. In their scenario, magnetic field lines associated with the star and its close-in planet become increasingly tangled, ultimately undergoing reconnection and triggering the eruption of these superflares. Magnetic reconnection can be very powerful pathway for energy conversion in the Solar System and drives multiple important physical phenomena such as particle acceleration, planetary aurorae, and so on, and can happen for both magnetized and non-magnetized bodies, though at different locations \citep{Baumjohann2010}. \citet{Cuntz_2000} provided preliminary scalings indicating the magnetic energy released during star–planet reconnection in the form of a proportionality. Thereafter,
% \begin{equation}
%     P \propto 
%     \frac{\epsilon B_{\langle \star,\text{surface}\rangle} B_{\langle p,\text{surface}\rangle}^{1/3} 
%     v_{\rm rel}}{d^2 R_p^2 F_X^{1/6}}
%     \label{eq:cuntz_scaling}
% \end{equation}
% where $\epsilon$ is an efficiency factor, $B_{\langle \star,\text{surface}\rangle}$ and $B_{\langle p,\text{surface}\rangle}$ are the mean surface magnetic field strengths of the star and planet, $d$ is the orbital distance, and $F_X$ is the stellar surface X-ray flux.
\citet{Zarka2007} derived a general analytical estimate for the power released through magnetic reconnection between planetary and stellar magnetic fields:
\begin{equation}
\mathcal{P}_{Za07} = \frac{\epsilon\, v_{\rm rel} B_{\star}^2 \pi R_{\rm obs}^2}{\mu_0}
= \epsilon \, \mathcal{P}_{obs} ,
\label{eq:zarka2007_reconnection}
\end{equation}
where this scaling physically corresponds to the stellar-wind Poynting flux density incident on the exoplanet, $S_w \sim v_{\rm rel} B_{\star}^2 / \mu_0$, multiplied by the effective cross-sectional area of the planetary obstacle, $\pi R_{\rm obs}^2$. Accordingly, the total electromagnetic power intercepted by the obstacle is $\mathcal{P}_{obs} \sim (v_{\rm rel} B_{\star}^2 / \mu_0)\,\pi R_{\rm obs}^2$. The dimensionless coefficient $\epsilon$ then parameterizes the fraction of the incident power that is ultimately dissipated through magnetic reconnection. Its value depends on both the magnetic field topology, specifically, how favorable the stellar and planetary field orientations are for reconnection, and on the intrinsic efficiency of the reconnection process itself. The quantity $R_{\rm obs}$ denotes the effective radius of the planetary obstacle interacting with the stellar wind, which may correspond to the planetary magnetosphere or, in the absence of a strong intrinsic field, the ionospheric size. For a magnetized planet, the effective obstacle size is determined by pressure balance (both thermal and magnetic) between the planetary magnetosphere and the surrounding stellar wind. Building on this framework, \citet{Lanza2012} further simplified the expected energy release by explicitly incorporating the exoplanetary magnetic field strength $B_p$ and the planetary radius $R_p$ into the scaling law in equation \ref{eq:zarka2007_reconnection}. Under the assumptions of a magnetostatic stellar corona and a favorable magnetic field topology, the reconnection power in the formulation of \citet{Lanza2012} is given by
\begin{equation}
\mathcal{P}_{La12} = \frac{\epsilon \pi R_p^2 B_{\star}^{4/3} B_{p}^{2/3} v_{\rm rel}}{\mu_0},
\label{eq:lanza2012_reconnection}
\end{equation}

 Applying \eqref{eq:zarka2007_reconnection} and \eqref{eq:lanza2012_reconnection} to a realistic system, say, $\tau$~Boo b yields powers released during magnetic reconnection of $10^{16}$–$10^{17}$ W considering all efficiency factors to be equal to unity, i.e., the most optimistic estimate. \citet{Lanza2012} also noted that adopting a force-free approximation for the stellar corona can increase these estimates by up to two orders of magnitude, highlighting the sensitivity of predictions to the assumed stellar magnetic environment.

% \begin{figure}
%     \centering
%     \includegraphics[width=0.5\linewidth]{figures/Paul_power_vs_Bp.pdf}
%     \caption{SPMI power estimates from the Magnetic Reconnection, Saur, and Lanza models. Colors indicate two exoplanetary systems; line styles correspond to the semi-analytical models.}
%     \label{fig:Bp_vs_P}
% \end{figure}

\subparagraph{The Alfvén-wing Poynting Flux Model}

\citet{Saur2013} proposed a unified framework for calculating the electromagnetic energy flux in sub-Alfvénic star–planet interactions. By ``unified,'' we mean that the framework is derived for general sub-Alfvénic interactions under certain approximations and is applicable to any system that satisfies these conditions. In the limit $M_A \rightarrow 0$, the integrated Poynting flux, or the total power directed toward the host star by an exoplanet is given in this model by:
\begin{equation}
    \mathcal{P}_{Sa13} = \frac{2 \pi R_{obs}^2 v_A (\alpha M_A B_\star \cos\theta)^2}{\mu_0} = \mathcal{P}_{obs} (\, 2 \alpha^2 M_A \cos^2\theta) \label{eq:saur_poynting}
\end{equation}
Here, $R_{\rm obs}$ again represents the effective obstacle size presented by the exoplanet. The key difference in this case is that the obstacle is not defined by the planetary magnetosphere itself, but rather by the size of the Alfvén wing generated by the planet. The quantity $B_\star$ denotes the stellar magnetic field, while $\theta$ is the angle between $B_\star$ and the relative flow direction $v_{\rm rel}$. The dimensionless factor $\alpha$ characterizes the efficiency of the magnetic interaction. \citet{Saur2013} further characterized the interaction efficiency $\alpha$ as dependent on the Pedersen conductance $\Sigma_P$ of the planetary ionosphere and the Alfvénic conductance $\Sigma_A$ of the connecting magnetic field lines:
\begin{equation}
\alpha = \frac{\Sigma_P}{\Sigma_P + 2 \Sigma_A}
\label{eq:alpha_definition}
\end{equation}
where the Alfvén conductance is given by:
\begin{equation}
\Sigma_A = 
\frac{1}{\mu_0 v_A \sqrt{1 + M_A^2 - 2 M_A \sin \theta}}
\label{eq:alfven_conductance}
\end{equation}

The effective obstacle size posed by the planet generally depends on the pressure balance between the planetary magnetic field and the ambient flow induced by the stellar wind. In scenarios where the planet generates Alfvén wings, the obstacle can also be effectively defined by the size of the Alfvén wings near the planet rather than the planetary body or the magnetosphere itself. In this context, \citet{Saur2013} found that the `effective' obstacle size posed by the Alfvén wings depend on the planetary magnetic field strength and orientation as:
\begin{equation}
R_{obs} = R_p 
\left(\frac{B_p}{B_\star}\right)^{1/3}
\left(3 \cos \frac{\delta}{2}\right)^{1/2}
\label{eq:obstacle_radius}
\end{equation}
where $B_p$ is the planetary magnetic field and $\delta$ is the angle between the planetary and stellar dipole axes. This framework successfully accounts for the power budget observed in sub-Alfvénic interactions in the Solar System. As an example, for Io, the model predicts powers of $\sim 10^{12}$ W, consistent with observed Jovian auroral enhancements ($\sim 10^{11}$ W), allowing for dissipation losses. Similar consistency is found for several Jovian and Saturnian moons orbiting in sub-Alfvénic conditions. However, when applied to exoplanets, this model often underestimates the power required to explain chromospheric hotspots. For HD 179949, where the excess power is $\sim 10^{20}$ W, equation \ref{eq:saur_poynting} estimates a Poynting flux of only $\sim 10^{17}$ W (for $\alpha \sim 1$, i.e., the most optimistic estimate).

\subparagraph{The Magnetic Stress Model}
\citet{Lanza_2013} proposed an alternative approach relying on the accumulation of magnetic stress rather than pure wave propagation. In this framework, the stellar corona is assumed to be in a force-free state. A magnetic field loop connecting the star and planet is steadily stretched by the orbital motion of the planet leading to an accumulation of energy. The accumulation persists until a critical threshold in energy is reached. Thereafter an impulsive energy release occurs relaxing the magnetic field lines and the cycle continues. Such a stretch-and-break mechanism is considered analogous in its cyclic nature to solar Coronal Mass Ejections (CME) which exhibit similar cycles of gradual energy accumulation and impulsive release. By equating the average energy dissipation to the available Poynting flux at the base of the loop (ideal efficiency) near the planet, \citet{Lanza_2013} derived the average power dissipated due to such an interaction as :
\begin{equation}
\mathcal{P}_{La13} \approx 
\frac{2 \pi f_{AP} R_p^2 B_p^2 v_{\rm rel}}{\mu_0} = 2 f_{AP} \left(\frac{B_p}{B_\star}\right)^{4/3} \mathcal{P}_{obs}
\label{eq:lanza_power}
\end{equation}
where $f_{AP}$ represents another efficiency factor quantifying the fraction of one hemisphere of the planet ($2\pi R_p^2$) open to the connecting field lines. Assuming that the motional electric field just above the planetary surface is $E \sim -v_{rel} \times B_p$, with $B_p$ denoting the surface magnetic field of the exoplanet. The functional form of $f_{AP}$, following \citet{Adams_2011} is given as
\begin{equation}
f_{AP} = 
1 - \left(1 - \frac{3 \zeta^{1/3}}{2 + \zeta} \right)^{1/2}
\label{eq:fAP_definition}
\end{equation}
where
\begin{equation}
\zeta = \frac{B_\star}{B_p}
\label{eq:zeta_definition}
\end{equation}
Note that, as derived by \citet{Lanza_2013}, the exoplanetary magnetic field $B_p$ used here corresponds to the polar magnetic field strength of the exoplanet, which is twice the equatorial surface magnetic field for a planet with a simple dipolar magnetic field configuration. In the context of star–planet magnetic interactions, \citet{Lanza_2013} provided an order-of-magnitude estimate assuming $B_p$ and $B_\star$ to be on the order of 10 G each, $v_{\rm rel} \sim 10^4$ m/s, $R_p \sim 10^8$ m, and $f_{AP} \sim 0.1$. With these values, the estimated power is of the order of $10^{20}$ W, comparable to the tentative stellar emission enhancements observed by \citet{Shkolnik_2005, Shkolnik2008}. They further suggest that the dissipation of magnetic stresses driving the Poynting flux is likely not confined to a single region, potentially occurring throughout much of the volume of the interconnecting loop and regions with stronger magnetic fields, such as those near the planet or the star, contribute most significantly to the total dissipated energy. Nevertheless, the multi-scale nature of the proposed stretch-and-break mechanism complicates its realization in global models, rendering its direct verification in simulations an area requiring significant further investigation.

\paragraph{Transition to Numerical Models}
Models describing sub-Alfvénic star–planet interactions as shown above have, until recently, been predominantly analytical, relying on simplified geometries and assumptions about the stellar environment and available energy budgets. While these formulations provide an essential basis for the scaling laws, they are constrained by idealized conditions: for example, the Saur model assumes spatially homogeneous plasma flow on the scale of the planetary obstacle and often adopts the limit of small Alfvénic Mach numbers ($M_A \rightarrow 0$). Almost all the existing scaling laws have efficiency factors baked in that are arbitrarily constrained. To move beyond these approximations and evaluate the robustness of the proposed mechanisms in realistic, time-dependent systems, recent studies \citep[e.g.,][]{Strugarek2015} have employed global numerical simulations of sub-Alfvénic star–planet magnetic interactions. It should be kept in mind, however, that such simulations are themselves idealised. For numerical reasons, they adopt dissipation coefficients (resistivity and viscosity) that are many orders of magnitude larger than in real stellar coronae and winds, and hence magnetic Reynolds and Lundquist numbers that are far smaller than realistic values; the planetary obstacle must be resolved by the numerical grid, which often requires placing the planet at small orbital distances and/or adopting simplified stellar magnetic field topologies and wind properties. The resulting scaling laws must therefore be extrapolated over several orders of magnitude in some parameters when applied to real systems.

The first attempt at modelling systematically the energetics of star-planet magnetic interactions was carried out by \citet{Strugarek2016}. In a series of 18 3D global simulations varying the orbital distance of the planet, its intrinsic magnetic field strength and directionality, and the Ohmic resistivity in the model, a scaling law was derived to characterise the energy transfer going through each of the Alfvén wings. This scaling law was pushed even further by encompassing more inclinations of the planetary field with respect to the ambient magnetic field in \citet{2017ewas.confE...6S}, that lead to the following scaling law
\begin{equation}
    \mathcal{P}_{\rm St17} = 4.55 \; \mathcal{P}_{pl} \ f_{1.81}(\Theta_M) \left( c_d  M_A^{1.11f_{-0.29}(\Theta_M)} \right) \Lambda_P^{0.23} \, , \label{eq:S17_PF}
\end{equation}
where function $f_x(\theta)=1+(x-1)\cos(\theta)$ encodes the dependencies on the angle between the planetary magnetic field and the ambient field magnetic denoted $\Theta_M$ (the interaction is maximized for $\Theta_M=0$, in which case $f_x(0)=x$), and where the drag coefficient is taken after \citet{Zarka2007} as $c_d = M_A/\sqrt{1+M_A^2}$. 

The quantity $\mathcal{P}_{pl}$ plays a role analogous to that of $\mathcal{P}_{obs}$ in the analytical scaling laws, with a subtle but important distinction: $\mathcal{P}_{pl}$ represents the stellar-wind Poynting flux intercepted by the physical cross-section of the planetary body itself, rather than by the full planetary obstacle, which may also include the magnetosphere. Mathematically, this quantity is given by $\mathcal{P}_{pl} = \pi R_p^2 \frac{v_{\rm rel} B^2}{\mu_0}.$ The Poynting flux (Eq. \ref{eq:S17_PF}) was obtained in numerical models integrating the net signed Poynting flux passing through the Alfvén wings just outside of the planet magnetosphere. It was shown in \citet{Strugarek2016} that the models predict that this net Poynting flux is directed towards the star, and its amplitude is compatible with the predictions of the classical Alfvén wings model of \citet{Saur2013}.

In order to obtain an upper limit on the Poynting flux that can be carried through Alfvén wings, \citet{Paul_2026} re-analyzed and extended this set of 3D global simulations, now computing the Poynting flux all along the Alfvén wing. In the classical Alfvén wing model, the limit $M_A\rightarrow 0$ is taken and the interaction of the wing itself with the surrounding environment is generally neglected except for regions very close to the planet. In addition, in \citet{Paul_2026} the maximal Poynting flux was computed by considering only the flux directed towards the star, avoiding some compensation made by the return Poynting flux as in the calculations of \citet{Saur2013} and \citet{Strugarek2016}. Putting those together, \citet{Paul_2026} arrived at the following scaling law for the maximal Poynting flux reached in 3D global interactions:
\begin{equation}
    \mathcal{P}_{\rm PS26} = 0.857 \mathcal{P}_{pl} \left(\frac{B_p}{1\rm T}\right)^{0.5} \left(\frac{1\rm T}{B_\star}\right)^{0.85} \, [\rm W] \label{eq:power_ps25}.
\end{equation}
Although, it should be noted that the simulations therein assumed dipolar magnetic field topologies for both the star and the planet, with both dipole moments along the 
($\pm$)z-axis. Deviations from this idealised configuration, either due to non-dipolar field geometries or a variation in the relative orientations of the magnetic fields, may lead to variations in the estimate. Therefore, this expression should be regarded as an upper limit, corresponding to an optimistic scenario in which the maximum amount of energy is channelled through the magnetic interaction.
\begin{figure}
    \centering
    \includegraphics[width=0.7\linewidth]{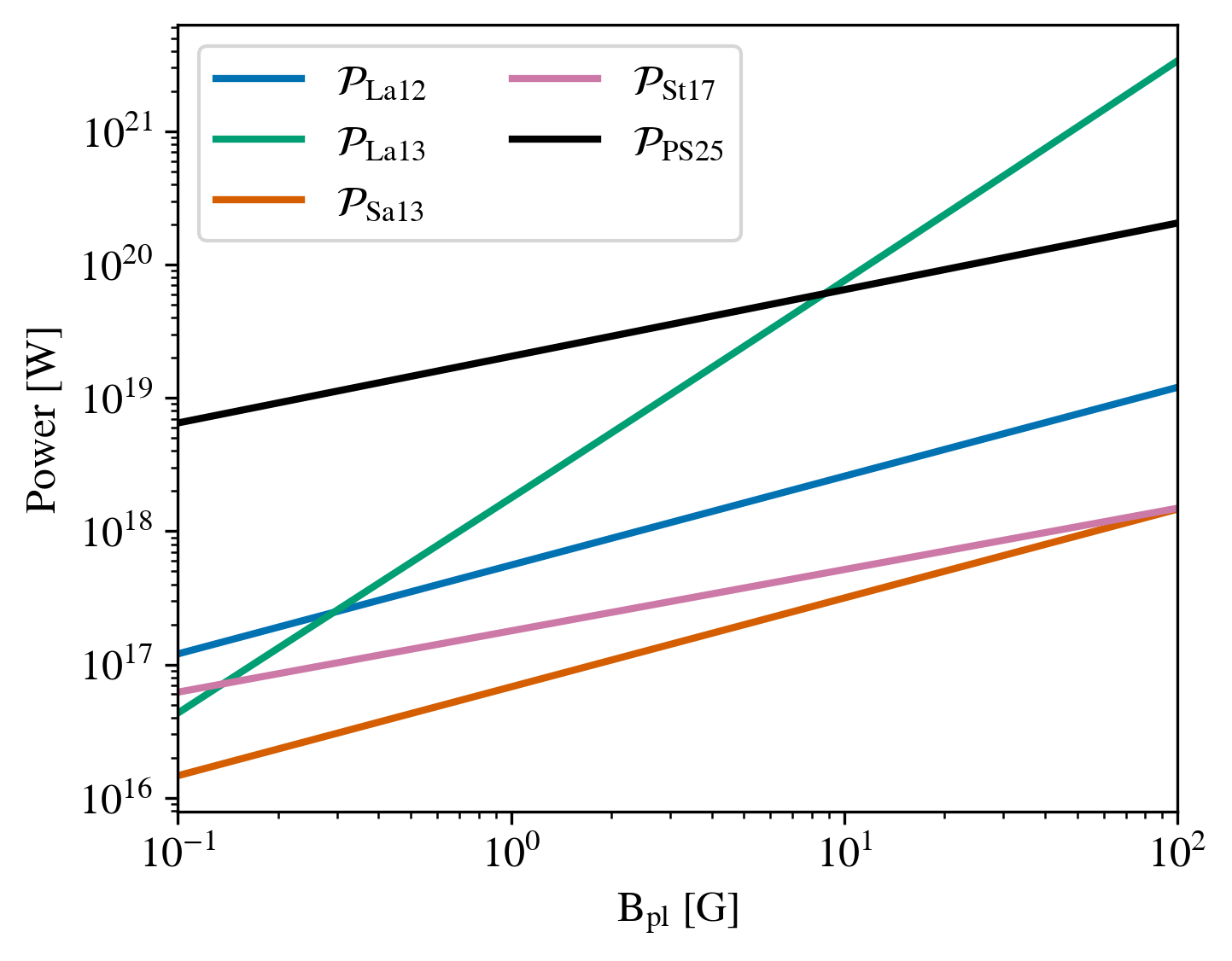}
    \caption{Analytical and numerical scaling laws for SPMI energetics evaluated for the parameters of the $\tau$~Boo system.}
    \label{fig:scaling_powers}
\end{figure}

Figure \ref{fig:scaling_powers} presents an overview of the SPMI power predicted by the scaling laws described above, evaluated using parameters appropriate to the $\tau$~Boo system. We assume a stellar surface magnetic field strength of 2.6 G, a stellar radius of 1.48 $R_{\odot}$, and a stellar mass of 1.3 $M_{\odot}$. The planet is assumed to have a radius of 1.13 $R_J$ and an orbital period of 3.3 days, while the stellar rotation period is taken to be 3.7 days. The orbital separation is 0.048 AU. The relative velocity between the planet and the surrounding plasma is computed as the vector sum of the planetary orbital velocity and the local stellar wind velocity. We assume the stellar wind speed is 169 km s$^{-1}$, as obtained from wind modelling of Sun-like stars at this orbital distance. To estimate the theoretical upper bound on the SPMI power, we adopt a number of simplifying assumptions. In particular, all efficiency factors are set to unity, namely $\epsilon$ in Equation \ref{eq:lanza2012_reconnection} and $\alpha$ in Equation \ref{eq:saur_poynting}. We also take $\cos\theta$ in Equation \ref{eq:saur_poynting} to be equal to 1 and set $\theta_M$ in Equation \ref{eq:S17_PF} to 0. These choices are intended to address the substantial uncertainties associated with constraining efficiency factors for individual systems. For instance, the parameter $\epsilon$ depends on the reconnection efficiency, which is typically expected to lie between 0.1 and 0.2 \cite{Zarka2007}, as well as on the local magnetic topology, both of which are currently difficult to constrain. Likewise, the parameter $\alpha$ depends on the exoplanetary ionospheric conductance, which remains poorly determined. In the absence of such constraints, Equations \ref{eq:zarka2007_reconnection} and \ref{eq:lanza2012_reconnection} should be interpreted as estimates of the total stellar wind Poynting flux intercepted by the planet rather than of the power that is ultimately channelled back to the star due to SPMI. Along similar lines, using  Equation \ref{eq:lanza_power} for SPMI power represents an idealized case in which the available Poynting flux near the planet is fully converted into SPMI power and directed toward the host star. Numerical scaling laws such as Equation \ref{eq:S17_PF} also depend on parameters including $\Theta_M$ and $c_d$, which can only be reliably constrained through detailed 3D stellar wind modeling. Finally, Equation \ref{eq:power_ps25} is inherently formulated to describe the maximum possible inflowing power as it only takes into account the negative radial component of the Poynting flux on concentric surfaces centered on the star. Consequently, Figure \ref{fig:scaling_powers} presents upper-limit estimates derived from the different scaling laws, each of which already carries significant intrinsic uncertainties. The fact that these upper limits differ from one another further as can be seen in figure \ref{fig:scaling_powers}, illustrates the level of uncertainty involved and emphasizes the sensitivity of the predicted SPMI power to the underlying assumptions of each formulation. It has recently been shown that atmospheric escape from exoplanets can modulate the power associated with SPMI. In particular, \citet{Presa2026} demonstrated that, in the presence of ongoing atmospheric mass loss, the Poynting flux channelled from the planet towards the star is modulated by a factor proportional to $\dot{M}_{\mathrm{d}}^{0.5}$, where $\dot{M}_{\mathrm{d}}$ denotes the dayside mass-loss rate.

In addition to the interaction mechanisms discussed above, a distinct form of magnetic coupling has recently been identified. \citet{Ilin2025} reported observations that highlight that, in sub-Alfvénic star–planet systems, an orbiting planet may directly trigger flaring activity on the stellar surface. In this scenario, pre-existing, metastable magnetic loops in the stellar corona are driven toward instability by the additional energy and perturbations deposited through SPMI. The planetary interaction thus acts as a catalyst, initiating the release of stored magnetic energy in the form of stellar flares. A more detailed discussion of the observational signatures and diagnostics of such planet-induced flaring is presented in Section \ref{sec:xray_whitelight_obs}.

\paragraph{Efficiency factors}
\begin{figure}
    \centering
    \includegraphics[width=1.\linewidth]{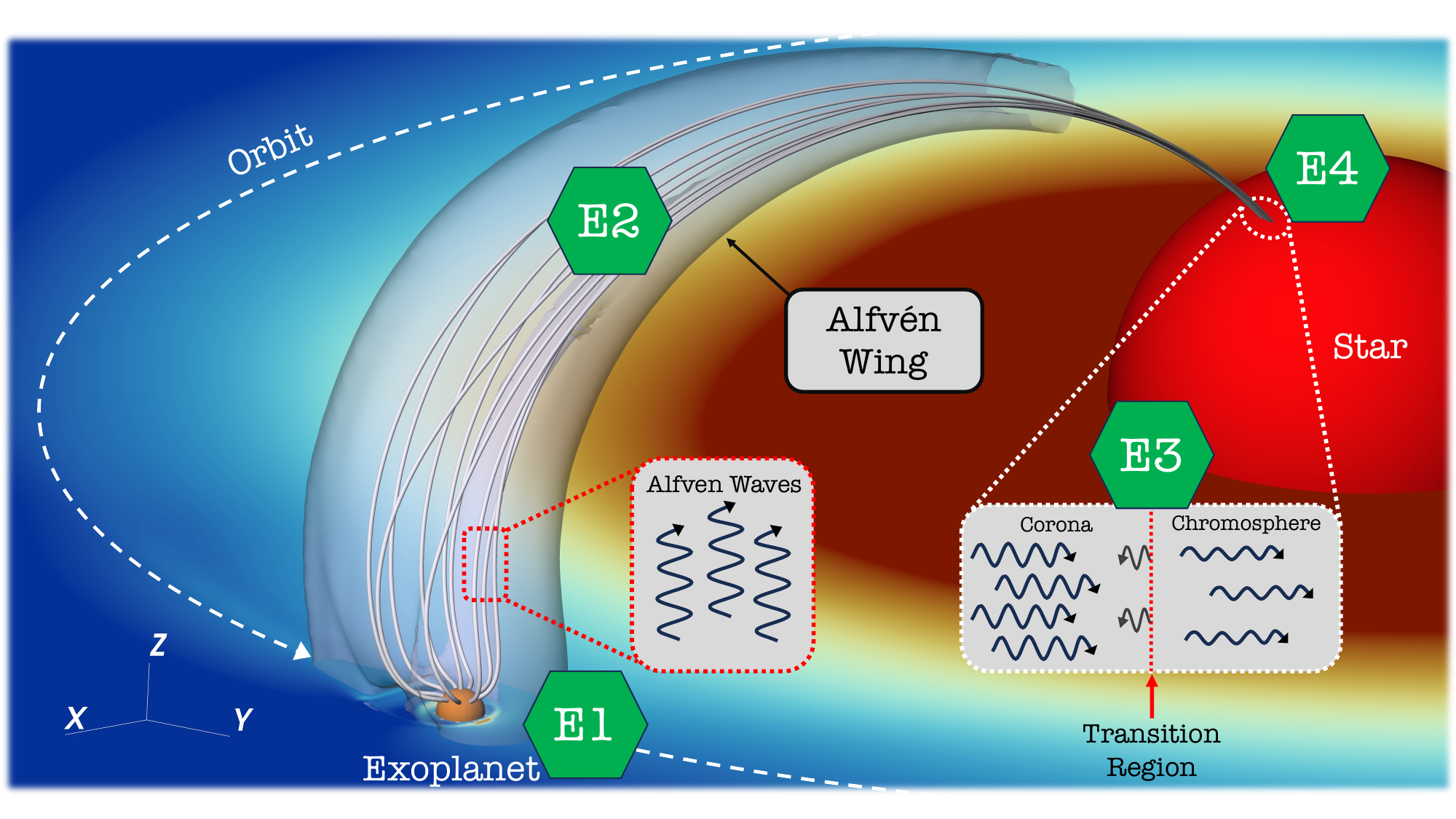}
    \caption{A schematic illustrating the current understanding of star–planet magnetic interactions (SPMI) in sub-Alfvénic systems, driven by the formation of Alfvén wings that carry Alfvén waves from the exoplanet back towards the host star. The schematic also highlights several efficiency factors (E1, E2, E3 and E4), which play a key role in determining the energetics of SPMI.}
    \label{fig:efficiency_factors}
\end{figure}

Numerous studies in the literature have reached a general consensus that periodic enhancements in stellar chromospheric emission arise from the energy transfer via star–planet magnetic interactions. Localized perturbations generated by the planetary obstacle channel a fraction of the interaction power into a Poynting flux at the planet, which is transported toward the star along Alfvén wings. This Poynting flux may be described either as energy carried by Alfvénic perturbations or as a quasi-steady flux associated with enhanced bulk plasma motions that produce a motional electric field, with both descriptions representing equivalent manifestations of electromagnetic energy transport.

The dissipation of this SPMI-driven energy in the stellar chromosphere leads to localized heating and the formation of chromospheric hotspots, resulting in enhanced emission. It is immediately apparent that the efficiency of this entire process depends on multiple factors, several of which are explicitly incorporated into the analytical formulations presented above. In the following section, we present a comprehensive overview of the efficiency factors governing the full SPMI process, from energy generation and transport to its ultimate dissipation in the chromosphere and the resulting observable hotspots. Figure \ref{fig:efficiency_factors} presents a schematic view of one Alfvén wing in a numerically simulated star–planet magnetic interaction system. The power redirected toward the star by the interaction constitutes only a fraction of the incident power impinging on the obstacle. Depending on the physical configuration, this obstacle may correspond to the planetary body itself, the planetary magnetosphere, or the extended Alfvén wing structure that magnetically connects the planet to the star. Consequently, the first efficiency factor is determined by the fraction of the incident stellar wind Poynting flux that can be intercepted and redirected toward the star by the obstacle. This efficiency is denoted by E1 in Figure \ref{fig:efficiency_factors}. This efficiency factor is implicitly incorporated into the coefficients $\epsilon$ in the scaling laws given by Equations \ref{eq:zarka2007_reconnection} and \ref{eq:lanza2012_reconnection}, as well as the parameters $\alpha$, $\theta$, and $\delta$ in Equations \ref{eq:saur_poynting} and \ref{eq:obstacle_radius}, and the factor $f_{AP}$ in Equation \ref{eq:lanza_power}. Collectively, these efficiency parameters depend on a range of physical properties, including the relative orientation of the stellar wind magnetic field and the bulk plasma flow, the strength and topology of the planetary magnetic field, and the specific mechanism responsible for generating and redirecting the interaction power, among other factors. 

 The second efficiency factor, denoted by E2 in figure \ref{fig:efficiency_factors}, is associated with the propagation of SPMI power from the planet toward the star. Alfvén wings represent the primary channel through which this power is transported; however, recent studies have shown that the surrounding plasma outside the Alfvén wings can also contribute to the channelling of energy \citep{Paul_2026}. Regardless of the specific propagation pathway, it is expected that a fraction of the power will be dissipated during transit from the planet to the stellar atmosphere. Such dissipation can arise from a variety of mechanisms, including inhomogeneities in the ambient stellar wind, spatial variations in the Alfvénic conductance along the wing, or localized heating induced by processes such as wave–particle interactions or kinetic turbulence. This efficiency factor E2, therefore quantifies the fraction of the initially redirected power that successfully reaches the vicinity of the star without being lost to intermediate dissipation processes and remains largely unconstrained at present.

 The stellar transition region introduces a sharp variation in the Alfvén speed along the direction of Poynting flux propagation due to the abrupt changes in stellar properties between the corona and the chromosphere. Analogous to the partial reflection of light at the interface of a medium with a reduced speed of light, Alfvén waves propagating along the Alfvén wings are expected to be partially reflected at the transition region as a result of this Alfvén speed gradient. This reflection constitutes an additional efficiency factor, denoted as E3 in Figure \ref{fig:efficiency_factors}, which accounts for the fraction of energy reflected rather than transmitted to the lower stellar atmosphere. \citet{Paul_2025} quantified this efficiency using a simplified one-dimensional stellar wind model and found that, for realistic sub-Alfvénic star–planet systems, approximately 90\% of the Alfvén wave energy is reflected at the stellar transition region. However, the role of the stellar transition region in reflecting other types of waves, such as magnetosonic waves, or in influencing Poynting flux carried by bulk plasma flows and associated currents which may also contribute to the net Poynting flux, remains largely uncharacterized.

Finally, the last step in the SPMI process is the enhanced emission from the stellar chromosphere, which we interpret as localized stellar hotspots resulting from the dissipation of energy that reaches the chromosphere via the transition region. These enhanced emissions are most commonly observed in the Ca~{\sc ii} H\&K lines. At present, only very weak constraints exist on the fraction of the incident energy that is ultimately dissipated and radiated in these specific wavebands, giving rise to the efficiency factor $E_4$. \citet{Cauley_2019} provided an initial estimate of this factor by drawing a simplified analogy to solar flares, considering the fraction of the total flare energy emitted in the Ca~{\sc ii} lines. They found that approximately 0.2\% of the total energy budget of a solar flare is emitted in these wavebands. While this offers a useful starting point for characterizing $E_4$, it should be interpreted with caution, as the physical mechanisms underlying solar flares and stellar hotspots induced by SPMI are fundamentally different. Improved constraints on $E_4$ will require dedicated 3D numerical simulations capable of capturing the detailed energy transport and dissipation processes in the stellar atmosphere.

% All these scenarios involve energy transfer via Alfvénic perturbations, as they provide a viable pathway for magnetic coupling between the planet and the star. For Alfvén waves to effectively transport energy from the planet to the star, the planet must reside within the Alfvén surface of the star. The Alfvén surface is a three-dimensional boundary within the stellar wind, beyond which the stellar wind becomes super-Alfvénic. Inside this surface, the stellar wind speed is slower than the Alfvén speed, allowing Alfvénic perturbations generated by the planet to propagate upstream toward the star. Outside the Alfvén surface, such perturbations are carried away from the star by the faster-moving stellar wind. When a planet orbits within the stellar Alfvén surface, magnetic field lines connect the two bodies through structures known as Alfvén wings. These magnetic tethers enable the transmission of energy and angular momentum between the planet and star. Alfvén wings have been described in theoretical and simulation-based studies, including those by Strugarek et al. (2015, 2019) and Fischer et al. (2022). 

\subsubsection{Magnetic Migration of Exoplanets}
\label{sec:magnetic_migration}

Planets can theoretically migrate due to magnetic drag forces applied to their orbital motion. This is especially true for exoplanets in close-in orbits around strongly magnetized stars, where the magnetic field they encounter can reach values of several hundred Gauss. This effect was first estimated by \citet{Strugarek2015}, who found that magnetic drag suffered by close-in planets around T-Tauri stars could lead to migration on timescales of a few Myrs. 

\citet{Bromley2022} established a general theoretical framework for these interactions, characterizing the instantaneous magnetic forces when both the host star and the orbiting body are treated within the dipole approximation. For permanently magnetized bodies, the steep radial dependence of the dipole-dipole interaction can lead to measurable orbital precession. In the case of conducting bodies, the stellar field induces a time-varying magnetic dipole moment, creating the potential for eccentricity pumping and resonance trapping. Furthermore, the cumulative effect of ohmic losses in induced eddy currents can drain orbital energy, driving an inward inspiral; conversely, a rapidly spinning stellar host can transfer energy to the orbiting body, potentially driving it outward.

While similar electromagnetic couplings occur within the Solar System, studies of planet-satellite magnetic interactions have typically found these torques to be too small to induce secular effects on satellite migration \citep[e.g.,][]{Saur2013}. This realization has prompted systematic comparisons between magnetic and tidal torque theories \citep{2017ApJ...847L..16S}. Although classical planetary migration driven by gaseous or planetesimal disks is often efficient enough to remove planets as quickly as they form \citep{Bromley2011}, population synthesis models suggest a distinct mass-dependency in migration \citep{2021A&A...650A.126A, 2023MNRAS.520.3749L, 2023A&A...679L..12G}. The consensus is that magnetic drag may dominate over tidal migration for low-mass planets orbiting low-mass stars, whereas tidal migration remains the primary driver for more massive close-in companions.

It must finally be noted that both tidal and magnetic torques exhibit a strong dependency on the orbital distance. For tides raised on the star, the migration timescale scales as $\tau_{\rm mig}^{\rm tide} \propto a^{13/2}$ for a constant tidal quality factor, and even more steeply, $\tau_{\rm mig}^{\rm tide} \propto a^{21/2}$, if the tide is dissipated as fully damped internal gravity waves in the radiative core, as proposed for the orbital decay of some hot Jupiters \citep{Barker2020}. For magnetic torques, \citet{Strugarek1} derived a migration timescale scaling as $\tau_{\rm mig}^{\rm mag} \propto a^{4.5}$ from 2.5D simulations; accounting for the three-dimensional nature of the interaction yields an approximate scaling of $\tau_{\rm mig}^{\rm mag} \propto a^{5.5}$. This scaling depends on the stellar magnetic field and wind properties, as well as on the magnetic interaction regime, and should therefore be regarded as indicative only. Consequently, these forces are most potent for exoplanets with short orbital periods, typically $P_{\rm orb} < 10$ days \citep{2021A&A...650A.126A}. As with tidal interactions, the orbital angular momentum exchanged magnetically is generally transferred to or from the stellar rotation, leading to a corresponding spin-up or spin-down of the central star depending on the direction of the planet's migration.

\subsubsection{Far out exoplanetary systems}\label{sec:faroutsystems}
Beyond the regime of mutual magnetic and tidal coupling characteristic of close-in systems, a distinct physical paradigm emerges for planets at larger orbital separations: the so-called ``far-out'' star–planet systems. While close-in interactions have attracted considerable attention due to the observational accessibility of hot Jupiters, the interactions in far-out systems are equally consequential but fundamentally different in nature. Specifically, the magnetic star–planet interaction in these systems is largely one-sided: while the host star affects the planet, the planetary magnetism has negligible feedback on the star. Our own Earth, along with the majority of planets residing in the habitable zone, belongs to this category, making the exploration of far-out SPI essential as well for understanding planetary evolution and habitability.

In this regime, the stellar wind has typically accelerated past the host star’s Alfvén surface. Consequently, the planet’s interaction with its environment is governed by a (typically) super-Alfvénic stellar wind, ionizing radiation, and transient events such as flares and coronal mass ejections (CMEs) with their associated energetic particles \citep{Georgoulis2024}. The nature of this interaction is dictated by the interplay between the planet's intrinsic attributes, most notably its magnetism, as well as the activity of the host star. Stellar magnetism still remains a central driver, similar to close in systems, providing the causal pathway for SPI \citep{Nandy2021, Daglis2021, Nandy2023}; its strength and secular evolution \citep{Nandy2004,Ribas2005,Gudel2007,Nandy2007} determine the wind conditions a planet encounters throughout its history \citep{Vidotto2021}. Collectively, these drivers shape the planetary magnetosphere \citep{Das2019,Carolan2019}, modulate atmospheric erosion rates, and ultimately determine the long-term maintenance of habitable conditions. Insights from our Solar System—particularly the divergent histories of Earth and Mars—provide compelling case studies that are now being generalized to exoplanetary contexts through increasingly sophisticated numerical modeling.

The response of a planet to magnetized stellar wind forcing depends critically on the presence of an intrinsic global magnetic field, typically generated through an internal dynamo. The stand-off distance of the magnetopause is determined by the global pressure balance between the incoming stellar wind and the planetary magnetosphere-atmosphere system, thereby regulating the extent to which the upper atmosphere is shielded from direct wind exposure (see Section \ref{sec:pressurebalance}). In far-out systems this interaction generally occurs under super-Alfv\'enic conditions, naturally producing a bow shock upstream of the planetary obstacle. However, this interaction is inherently dynamic: turbulence, magnetic reconnection, and transient stellar disturbances continuously restructure the magnetospheric topology and alter the efficiency with which stellar plasma couples to the planetary environment.

Utilizing global 3D MHD simulations, \cite{Das2019} showed that magnetic reconnection in the nightside magnetotail acts as a key driver of magnetospheric dynamics, enabling stellar wind plasma to penetrate into the inner magnetosphere. These processes become particularly complex for planets possessing highly tilted dipole fields, where reconnection geometries and current systems evolve substantially over an orbital cycle. Such configurations may resemble Uranus- or Neptune-like exoplanets, or planets undergoing magnetic reversals. These dynamics have direct implications for auroral morphology, the chemical enrichment of atmospheres with interplanetary ions, and the detectability of magnetospheres via radio observations.

\begin{figure*}
    \centering
    \includegraphics[width=0.8\textwidth]{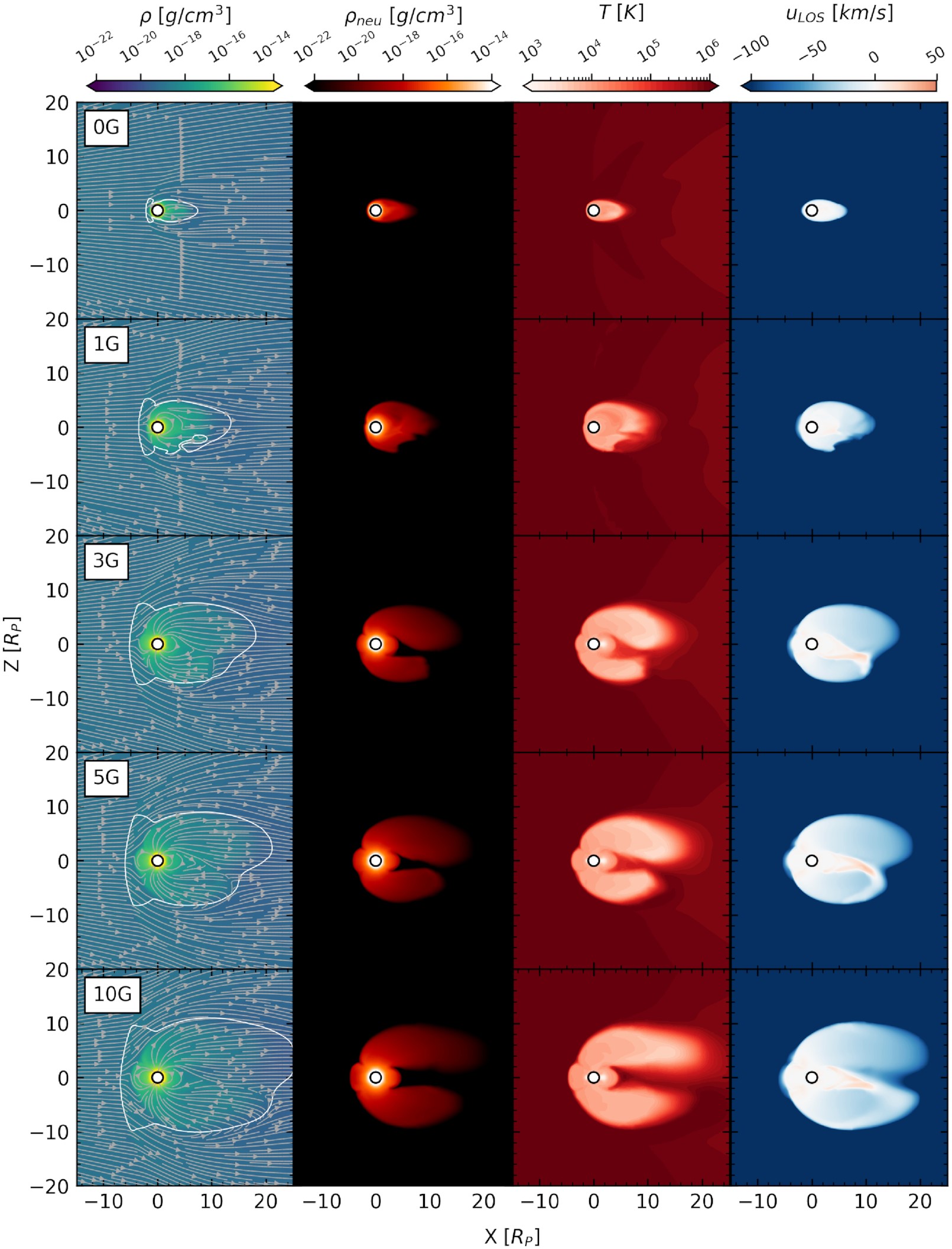}
    \caption{Global radiative-MHD simulations of atmospheric escape from a magnetized exoplanet adapted from \citet{Carolan2021}. The rows correspond to increasing intrinsic planetary magnetic field strength, progressing from weakly magnetized to strongly magnetized configurations. The columns display different physical diagnostics of the escaping atmosphere. As the planetary magnetic field increases, the outflow transitions from a single comet-like tail to a magnetically controlled double-tail morphology separated by an equatorial ``dead zone'' of confined plasma. These simulations demonstrate how intrinsic magnetism can strongly restructure atmospheric escape pathways and modify observable transit signatures.}
    \label{fig:doubletail}
\end{figure*}

Recent radiative-MHD simulations have further revealed that planetary magnetic fields can fundamentally reorganize atmospheric escape itself. Figure \ref{fig:doubletail} presents a sequence of simulations from \citet{Carolan2021} illustrating how the large-scale morphology of escaping plasma changes as the planetary magnetic field strength increases. In the weakly magnetized regime, the escaping atmosphere forms an approximately comet-like tail shaped primarily by the stellar wind. As the intrinsic dipole field strengthens, however, magnetic confinement creates an equatorial ``dead zone'' of trapped plasma while simultaneously channeling atmospheric escape along open polar magnetic field lines. This transition produces the characteristic double-tail morphology visible in the strongly magnetized cases. The four rows of the figure correspond to different magnetic field strengths, while the columns display complementary diagnostics of total density, neutral density, temperature, and line of sight velocity, respectively. These simulations demonstrate that planetary magnetic fields do not merely suppress atmospheric escape, but instead redistribute and reorganize the outflow geometry in a highly anisotropic manner. Such magnetic structuring has direct consequences for observational diagnostics including Ly$\alpha$ transit absorption, auroral emission, and planetary radio signatures.

The situation is qualitatively more severe for planets with weak or absent intrinsic fields, such as Mars or rocky exoplanets whose dynamos ceased early in their history. In these cases, the stellar wind interacts directly with the conductive ionosphere, inducing an imposed magnetosphere through the ``piling-up'' of magnetized plasma and associated stellar wind magnetic fields lines on the dayside. This induced shield provides far less protection, as stellar plasma and fields continuously slip past the body. This leads to direct coupling between the stellar wind and the (exo)planetary atmosphere, having the potential to drive significant erosion \citep{Basak2021}. This process likely explains the current state of Mars; after its dynamo ceased $\sim$4.0--4.1~Gyr ago, the loss of its global magnetosphere exposed the atmosphere to the active young Sun, characterized by harsher winds and frequent CMEs, triggering the irreversible loss of its volatiles.

\begin{figure*}[t]
    \centering
    \includegraphics[width=0.7\columnwidth]{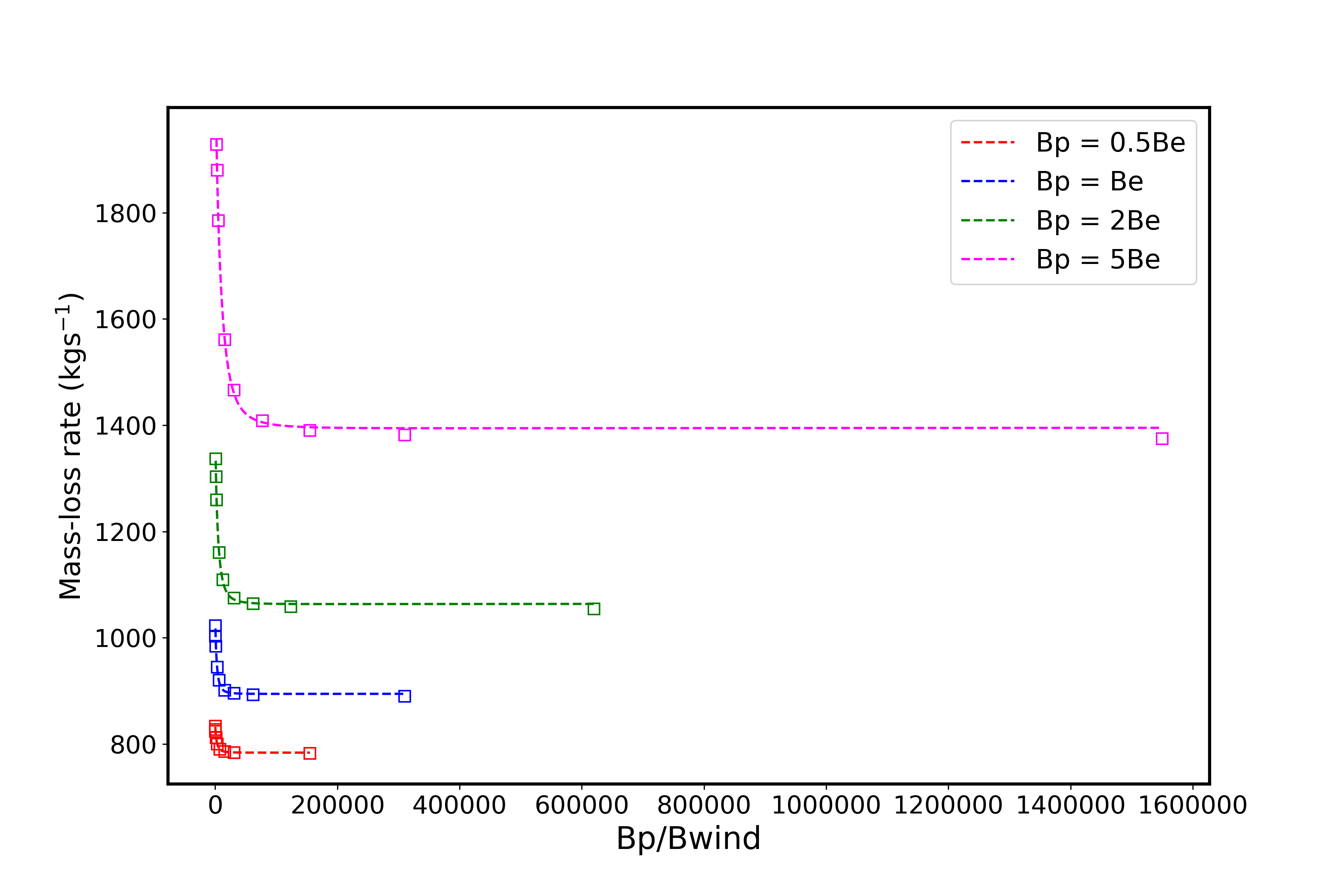}
    \caption{Dependence of atmospheric mass loss rate on the ratio of the planetary ($B_p$) to the stellar wind ($B_{wind}$) magnetic field strength. Simulation results are represented by square boxes, while the fit from the analytic expression derived in \cite{Gupta2023} is shown by dotted lines. Different curves correspond to planets with different intrinsic dipole field strength relative to the Earth ($B_e$). The figure is from \cite{Gupta2023}.}
    \label{fig:massloss}
\end{figure*}

A central unresolved question in far-out SPI is how the relative strengths of stellar and planetary magnetic fields modulate erosion efficiency. As stars age, their magnetic activity, wind speeds, and high-energy radiation decline. Concurrently, exoplanets are expected to span a wide range of magnetic field strengths, although these remain observationally unconstrained for nearly all exoplanets (Sect.~\ref{sec:emissionnearplanet}). \cite{Gupta2023} performed a systematic parameter space exploration using 3D MHD simulations in an attempt to decouple these effects. Their findings indicate that atmospheric mass loss is determined not by either field in isolation, but by their relative magnitudes. This relationship is non-linear: for a fixed planetary field, a stronger stellar field increases mass loss; however, if the ratio remains constant but the absolute strengths increase, the mass loss rate also rises (see Fig.~\ref{fig:massloss}). Based on their simulations, the semi-analytic relationship provided by \cite{Gupta2023} serves as a tool for predicting which exoplanetary systems are most capable of retaining their atmospheres.

\subsection{Radiative Interactions}

% \noindent\manuel{I am adding below text on thermal escape and some model results for terrestrial planets; I imagine we should also somewhere at least mention core-powered mass loss although this is not SPI; the other major subsection here could focus on the photochemistry of atmospheres that again comes up in Chapter 4. }

In addition to magnetic interactions, the deposition of stellar electromagnetic radiation also serves as the fundamental driver for the thermal and chemical state of planetary upper atmospheres and thereby forms the third pillar of star-planet interactions.
\subsubsection{Stellar radiation and planetary thermospheres}\label{thermosphere}

The combined stellar radiation in the far ultraviolet, the extreme ultraviolet, and the X-ray ranges (``XUV'', see \citealt{AlvaradoGomez2026, Segura2026}) is the main driver of chemical processing, ionization, and heating in planetary upper atmospheres. Other significant contributors to heating include stellar infrared radiation, collisions between neutrals and charged particles, interactions with the stellar wind, exothermic chemical reactions, or Joule heating.
We briefly summarize the key physical and chemical phenomena occurring in exoplanets as a result of radiative interactions. These processes must be accounted for in models of the upper atmospheres of planets, in particular of planets with secondary atmospheres, i.e., atmospheres formed after the loss of the primordial hydrogen--helium envelope (e.g., through outgassing) and dominated by heavier species such as N$_2$, CO$_2$, or H$_2$O. More detailed descriptions, including methodologies, can be found, for example, in 
\citet{garciamunoz2007}, 
\citet{johnstone2018}, \citet{tian2008a}, and the references therein.

\begin{itemize}
\item\textbf{Hydrodynamics:} Hydrodynamic equations are the basis for a model of a thermal, collisionally dominated atmosphere. The equations must include all source and sink terms for mass, momentum, and energy, which comprise diffusion, chemical source and sink terms related to chemical reactions, all heating and cooling terms including external irradiation, line and continuum radiation, heat exchange between ions, electrons, and neutrals, conductivity terms. The gas is assumed to be thermal, the particles velocities following a Maxwell-Boltzmann distribution for a temperature $T$ and a ratio $\gamma$ of the specific heats (typically $\gamma = 5/3$). An optional, special case is the assumption of the absence of velocity fields, $v= 0$, which makes the atmosphere \textit{hydrostatic.} A standard way to solve the equations is to require \textit{steady-state}, i.e., the time derivatives of the mass and momentum densities are set to zero.

\item\textbf{Mass loss} to space is possible if particles gain sufficient energy to overcome the gravitational potential of the planet, or interact with the surrounding stellar wind. \textit{Thermal}
loss processes include single-particle Jeans loss at the exobase, or a hydrodynamic wind driven by high heating rates (see below). \textit{Non-thermal} loss includes various interactions between particles of the stellar wind and particles in the atmosphere.

\item\textbf{Stellar high-energy irradiation} is one of the key heating terms via absorption in the gas, but also drives chemical reactions in the upper atmosphere. Knowledge of the XUV spectrum typically in the range  0.1-200~nm covering X-rays, the extreme ultraviolet, and the far ultraviolet is important; the spectral flux and the spectral shape depend on the stellar activity level (see \citealt{AlvaradoGomez2026}). To calculate the interaction between radiation and gas at all relevant altitudes, photon-energy dependent radiative transport mechanisms must be included, with appropriate optical depth expressions describing photoreactions (ionization, dissociation, chemical reactions). Diffuse radiation may be important where the corresponding continuum and line opacities become large, thereby affecting excited state populations and the overall gas energy budget.

\item\textbf{Stellar infrared (IR) irradiation} can be of similar importance for atmospheric heating, depending on the presence of IR absorbing molecules in the upper atmosphere, such as CO$_2$ or H$_2$O. The heating is achieved via photoabsorption and excitation of molecules that then collisionally de-excite, heating the gas but also partially re-emit radiation back into space. Again, the wavelength-dependent radiative transport must be calculated, including individual lines.

\item\textbf{Ionization and photoelectrons:}
The ionization process by incoming stellar photons normally involves energies that largely exceed the ionization threshold for individual  particles. The excess energy ends up in the ejected electron as kinetic energy. Such ``photoelectrons'' can continue to ionize another particle, or eventually thermalize via elastic collisions between thermal and non-thermal electrons. For sufficiently low-energy photoelectrons, the heat transfer can be assumed to be local. For very high-energy electrons (e.g., ``cosmic rays''), electron transport through the atmosphere must be considered.

\item\textbf{Chemical networks:} The chemical composition is a fundamental driver of the atmospheric state, as it determines the local opacity, the degree of ionization, and the efficiency of heating and cooling. A chemical network tracks the abundances of various species, including neutrals, ions, and electrons, through a system of coupled reactions. In modelling, this network must be solved self-consistently with the physical mechanisms because of a continuous feedback loop: stellar irradiation drives photodissociation and ionization, potentially producing species like NO or CO$_2$, that may act as primary coolants to regulate the temperature, while the resulting temperature and vertical transport (diffusion) determine the chemical reaction rates  as a feedback.

\item\textbf{Diffusion:} We distinguish two diffusion mechanisms. Turbulent diffusion mixes volumes containing different concentrations of some species with each other, via eddies stirring the fluid. Eddy diffusion is a model-approach to the description of the turbulent and large-scale mixing that occurs in atmospheres.
In eddy diffusion, the density profiles follow the pressure scale height of the entire gas. Eddy diffusion is responsible for the mixing of species in the lower atmosphere, making the mixing ratios independent of altitude (``homosphere''). In contrast, molecular diffusion is due to the individual thermal motion of particles, producing a net flux of a species from higher to lower concentrations; the species follow their own pressure scale heights. This process is important in the upper atmosphere where the scale heights of different particle types are different (``heterosphere''). Heavy species have a shorter scale height and therefore accumulate at lower altitudes, while light elements expand to higher altitudes. Ambipolar diffusion, which arises from the constraint of gas neutrality and the higher mobility of electrons, must be included for some long-lived ion species \citep{
garciamunoz2007_ambipolar,tian2008a}.

\item\textbf{Heating mechanisms:} We already mentioned the important radiative XUV and IR heating mechanisms above. Not all energy of the incoming photons ends up in thermal energy; some XUV energy may be used to photodissociate molecules, in which case the energy will be transferred to kinetic energy of the products that eventually heats the gas. Some XUV energy may be used to ionize atoms or molecules; in this case, energy is transferred to the free photoelectron as kinetic energy, producing a non-thermal spectrum of electrons; a fraction of these will further collide elastically with the thermal electron population at which point the energy is thermalized. But another fraction will further ionize other particles or excite atoms and molecules; in the latter case, de-excitation will usually be radiated away without making a heating contribution.

Stellar IR photons can be absorbed by neutral molecules, exciting them after which collisional de-excitation extracts the energy as heat.
Chemical photoreactions may transfer energy to chemical potential energy that can be released as thermal energy in exothermic reactions.

In the presence of a magnetic field, Joule heating may be relevant in the presence of a finite electrical conductivity and an electric field. Also, high-energy particles may precipitate along magnetic field lines into the atmosphere, further contributing to heating.

\item\textbf{Cooling mechanisms:} The dominant cooling mechanism in the upper atmosphere is optical and IR radiation from atoms and/or molecules into space after collisions. Therefore, the chemical composition and the photoreactions in the upper atmosphere play a pivotal role. Some molecules such as CO$_2$ or atoms like O are very efficient radiators and therefore coolants. Other species like N$_2$ hardly radiate and do not counteract heating terms. Efficient coolants include transitions of CO$_2$ (in particular line at 15~$\mu$m), H$_2$O (many mid-IR lines), NO (vibrational band at 5.3~$\mu$m), and O (630~nm, 63~$\mu$m, and 147~$\mu$m). Excitation occurs via inelastic molecular collisions, the absorption of line radiation emitted by adjacent particles, and the absorption of stellar (solar) photons; additionally, radiative pumping via upwelling terrestrial thermal infrared radiation modulates the vibrational populations of specific non-LTE species like $\text{CO}_2$ and $\text{H}_2\text{O}$ in the mesosphere, though it becomes negligible for thermospheric $\text{NO}$ and high-altitude cooling budgets. De-excitation occurs through collisional quenching or radiative relaxation (spontaneous or stimulated emission).

\item\textbf{Conduction:} In the rarefied environment of the planetary thermosphere, thermal conduction is a key mechanism for transporting deposited energy from regions of peak heating toward cooler, denser layers near the lower boundary (e.g., the mesopause). This downward heat flux plays an important role in the overall energy balance, particularly where radiative cooling becomes inefficient at low densities. Conduction must be treated separately for neutrals, ions, and electrons, as these populations are often thermally decoupled ($T_e > T_i > T_n$). Neutral thermal conductivity typically scales as $\sqrt{T}$, whereas electron conductivity is much more efficient, following the $T_e^{5/2}$ Spitzer relation. In magnetized environments, thermal conduction for charged species is highly anisotropic, occurring primarily along magnetic field lines and being strongly suppressed perpendicular to them, which can significantly influence the global temperature structure.

\item\textbf{Collisional energy exchange between ions, electrons, and neutrals:} We also need to consider energy exchange via collisions between electrons, ions, and neutrals as these three groups of particles maintain different temperature profiles in the thermosphere. Electrons transfer energy to ions and neutrals via Coulomb collisions. The ions then further transfer energy to neutrals via collisions, to be determined for all pairs of ion vs. neutral species. The latter dominates neutral heating in the upper atmosphere. It involves collisions of neutrals with their ionized equivalents involving charge exchange; and collisions between non-equivalent ion vs. neutral pairs. Finally electron-neutral collisions 
importantly include inelastic collisions that produce fine-structure transitions, in particular in oxygen; furthermore, collisional excitation of rotational and vibrational modes in molecules are important.

\end{itemize}

\subsubsection{Radiatively driven thermal escape}

A crucial consequence of radiative heating of upper planetary atmospheres is atmospheric mass loss, also known as ``thermal escape'' or ``evaporation''. This mechanism can remove significant fractions if not the entirety of a planetary atmosphere over millions to billions of years if the stellar XUV flux above the atmosphere is sufficiently high and the atmospheric replenishment rate is smaller than the loss rate. Atmospheric loss has far-reaching consequences for the survivability of atmospheres and their habitability (see \citealt{Segura2026}). We briefly summarize the main mechanisms. The presentation of the formalism below largely follows \citet{bauer2004} where more details can be found.

\textbf{Basic formalism for thermal atmospheric escape:} Thermal escape is to be distinguished from non-thermal escape processes that are described in section \ref{sec:nonthermalescape}. For thermal escape, we consider particles following a Maxwell-Boltzmann velocity distribution 
for a given temperature $T$. Microscopically, the velocity distribution is isotropic. However, if the radially upward directed velocity component of a particle exceeds the local escape velocity,
\begin{equation}
    v_{\infty} = \left(2GM\over r\right)^{1/2}
\end{equation}
(where $G$ = gravitational constant, $M$ = planetary mass, $r$ = distance of the considered particle from the planetary center)
the particle could in principle escape to space. However, collisions with other particles may prevent most particles from escaping. The mean free path of a particle is
\begin{equation}
    \lambda = \frac{1}{n\sigma}
\end{equation}
where $n$ is the total particle density and $\sigma$ is the collisional cross section. The probability that a given particle 
undergoes a collision over a distance $z$ is
\begin{equation}
    P(z) = e^{-z/\lambda}.
\end{equation}
We define the \textit{exobase} as the upper boundary of the thermosphere by the probability $P(z) = 1/e$, i.e.
\begin{equation}
    z = \frac{kT}{mg} \equiv H 
\end{equation}
where $H$ is called the (local) scale height, $k$ is Boltzmann's constant, $m$ the mass of the particle (atom, ion, electron, molecule), and $g$ is local gravitational acceleration ($GM/r^2$).

This means that at the exobase level and above, the mean free path of a particle is equal to or exceeds the local scale height, making escape of particles likely. While the thermosphere is normally considered to be a thermal gas with a Maxwell-Boltzmann velocity distribution, the \textit{exosphere} above the exobase is essentially collisionless, and the particle distribution is not Maxwellian (because the higher-energy particles are missing due to their escape). It should be clear that the concept of an exobase is a simplification. The two regimes gradually change into one another.

We define the \textit{Jeans escape flux} $F_{\rm J}$ as the outward flux of particles with an upward velocity component exceeding the escape velocity. The escape flux at the exobase level is given by the Jeans formula,
\begin{equation}
    F_{\rm J} = \frac{v_0}{2\pi^{1/2}}n_{j,e}(1+X_c)e^{-X_c} \quad[{\rm particles~cm^{-2}~s^{-1}}] \label{eq:jeans}
\end{equation}
and therefore 
\begin{equation}
    \dot{M} = 4\pi R_{\rm e}^2m_{j} F_{\rm J}
    \label{eq:energylimitedmdot}
\end{equation}
where $n_{j, e}$ is the exobase density of the considered species $j$,
$m_{j}$ is the mass of one particle of the species,
$R_{\rm e}$ is the radius of the exobase, the velocity
\begin{equation}
    v_0 = \left(\frac{2kT}{m_{j}}\right)^{1/2}
\end{equation}
($T$ = exobase/exosphere temperature) is the most probable velocity of the Maxwellian distribution, and  
\begin{equation}\label{Xc}
    X_c = X(R_{\rm e}) = \frac{GMm_j}{R_{\rm e}kT} \equiv \left(\frac{v_{\infty}}{v_0}\right)^2
\end{equation}
is the \textit{escape parameter} determined at $R_{\rm e}$.
Escape is important for $X < 15$, while the atmosphere is stable for $X\sim 30$. However, for $X \leq 1.5$ escape becomes very large as the value of 1.5 (for monatomic gas, see \citealt{oepik1963} and \citet{volkovetal2011}) means that the energy of the thermal motion, $(3/2)kT$, is equal to or exceeds the gravitational binding energy. In this case, also called the ``blow-off'' state \citep{oepik1963, tian2008a} the entire gas flows. A more orderly but hydrodynamic flux sets in for $ X < 1.5$. Hydrodynamic simulations show that the atmosphere can sometimes enter the ``blow-off'' state even for $ X < 15$ \citep{Fossati2017}.
For a hydrodynamically flowing atmosphere, it is important to consider adiabatic cooling due to expansion; adiabatic cooling can substantially suppress mass loss and prevent blow-off conditions \citep{tian2008a}. 

\textbf{Energy-limited escape:} Expansion and flow begins at $X_c = 1.5-3$ where this value refers to the main constituent of the atmosphere. For accurate mass-loss estimates, complex atmospheric models including chemical reaction networks must be calculated numerically. However, an upper limit to the mass-loss rate $\dot{M}$ is often estimated using the formula for \textit{energy-limited escape} that was developed for pure hydrogen atmospheres \citep{watson1981, erkaev2007}
\begin{equation}
    \dot{M} = \pi \frac{\nu \Phi_{\rm X}r_{\rm X}^2r_0}{KGM}
    \label{eq:energylimited}
\end{equation}
where $\nu$ is the efficiency of XUV heating (typically a few tens of percent), $\Phi_{\rm X}$ is the stellar XUV flux incident on the planetary atmosphere (in erg~cm$^{-2}$~s$^{-1}$), $r_{\rm X}$ is
the (idealized, sharp) radius where XUV is absorbed, $r_0$ is a ``planetary radius'' typically defined at the photospheric level in the atmosphere, and $M$ is again the planetary mass. The constant $K$, $0 \leq K \leq 1$, corrects for the presence of a Roche lobe, 
\begin{equation}
    K = 1 - \frac{3}{2}\left(\frac{R}{R_{\rm R}}\right) + \frac{1}{2}\left(\frac{R}{R_{\rm R}}\right
)^3
\end{equation}
where $R$ is the planetary radius and $R_{\rm R}$ is the Roche radius
\begin{equation}
    R_{\rm R} \approx d \left(\frac{M}{3M_*}\right)^{1/3}
\end{equation}
\citep{krenn2021}, $M$ and $M_*$ being the planetary and the stellar mass, respectively, and $d$ the orbital separation between star and planet. 

The energy-limited formula assumes that a fraction $\nu$ of the incident energy is used to lift particles in the upper atmosphere to infinity without remaining excess energy in the particles. The formula is often used to estimate the mass-loss rate itself. This however is not appropriate because the framework in which it was developed aims at estimating an upper limit. But even then, a number of conditions apply (see \citealt{krenn2021} for an in-depth description): i) The XUV energy is absorbed in a narrow layer of the atmosphere; ii) the minimum temperature in the thermosphere must be 0~K at some point below the XUV absorption layer to minimize conductive losses to even lower layers; iii) the estimate holds only for hydrogen atmospheres; as we will motivate below, it is inappropriate to apply it to secondary atmospheres; iv) the gas is non-viscous and has a constant mean molecular weight everywhere; v) escape mechanisms other than hydrodynamic escape are assumed not to play a role; vi) the pressure decreases outward to zero at infinity; 
vii) the atmosphere below the lower boundary $r_0$ is tightly bound, i.e., is quasi-static and expands sub-sonically (thermal velocity is smaller than the escape velocity).

A number of studies have tested the applicability of energy-limited escape estimates by comparing results with more sophisticated simulations \citep{garciamunoz2007}. The direction and magnitude of the discrepancy depend not only on the irradiation level but also on other planetary parameters, in particular on the gravitational potential of the planet (see the escape parameter $\Lambda$ below). Energy-limited escape was found to underestimate mass loss for strongly irradiated planets with a low gravitational potential (low $\Lambda$), for which escape is driven largely by the thermal energy of the atmosphere itself, but to overestimate loss from weakly irradiated or near-hydrostatic atmospheres \citep{erkaev2013, kubyshkina2018a}. For massive, strongly irradiated hot Jupiters with a high gravitational potential, in contrast, it can overestimate the loss by up to two orders of magnitude, because at very high irradiation levels the outflow is limited by radiative recombination rather than by the available energy \citep{murrayclay2009}.

\begin{figure}
\begin{center}
\hbox{
\includegraphics[width=0.47\textwidth]{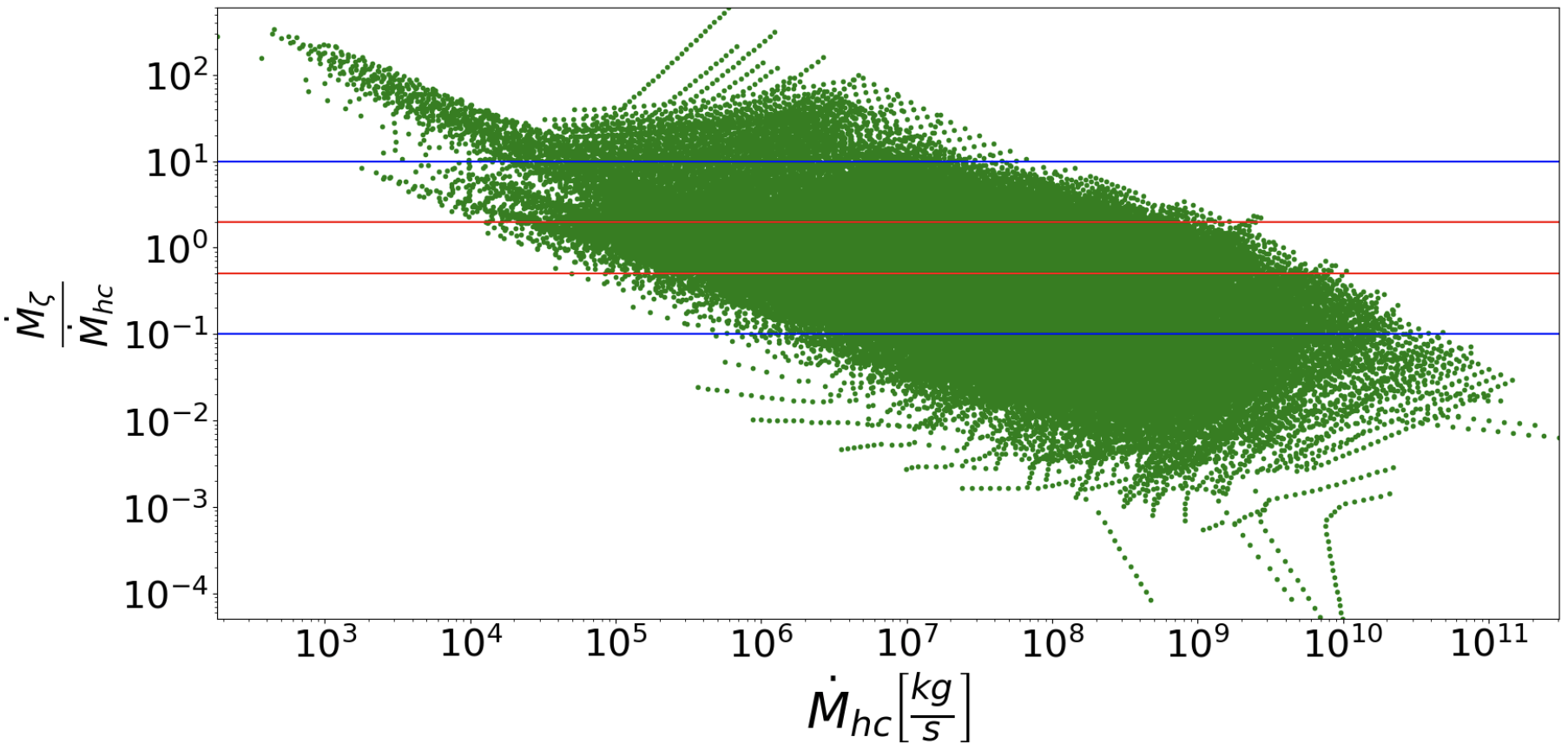} 
\hskip 0.3cm
\includegraphics[width=0.49\textwidth]{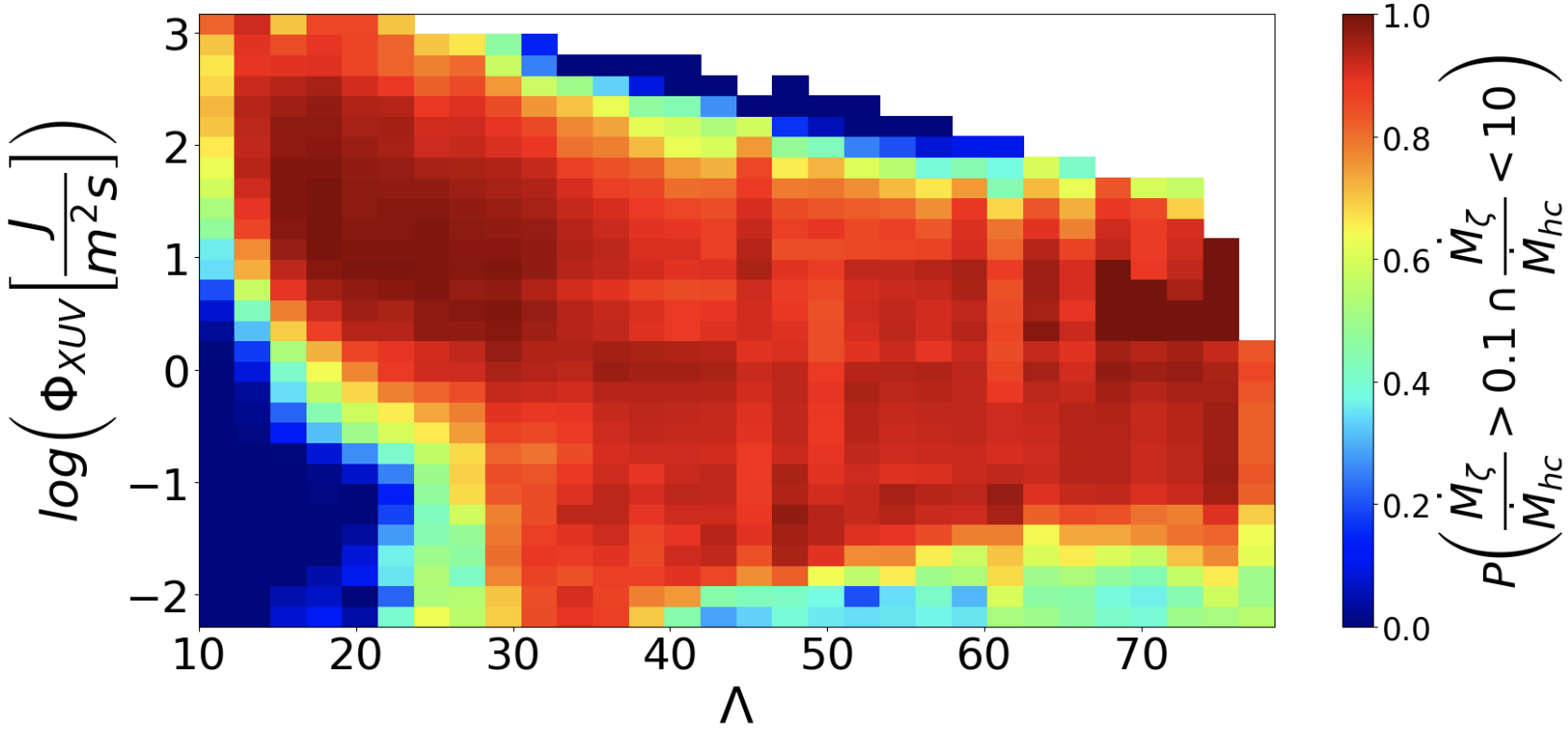} 
}
%\vspace*{-0.2 cm}
\caption{\textit{Left:} Ratio of mass-loss rates obtained from the energy-limited approach (here denoted by $\dot{M_{\zeta}}$) and hydrodynamic simulations (here denoted by $\dot{M}_{hc}$) as a function of $\dot{M}_{hc}$, for a wide range of planetary and irradiation parameters (see text; from \citealt{krenn2021}). -- \textit{Right:}
Probability for agreement between the energy-limited estimate of mass loss with hydrodynamically derived values (color), as a function of the escape parameter $\Lambda$ (Eq.~\ref{escapeparameter2}) and the incident stellar XUV flux. The colors comprise discrepancies of up to one order of magnitude (over- or underestimates; note 1~J~m$^{-2}$~s$^{-1} =  10^3$~erg~cm$^{-2}$~s$^{-1}$; from \citealt{krenn2021}).
}\label{fig:energylimited}
\end{center}
\end{figure}
A thorough analysis of the validity of this approach was given by \citet{krenn2021} based on previous work by \citet{kubyshkina2018a} and \citet{kubyshkina2018b}. The grid of investigated planets comprises the following parameter ranges: i) Planet mass from 0.5~$M_{\oplus}$ to 318~$M_{\oplus} \equiv M_{\rm Jupiter}$; ii) planet radius from 1$~R_{\oplus}$ to 11.2$~R_{\oplus} \equiv R_{\rm Jupiter}$; iii) equilibrium temperature from 300~K to 2000~K; iv) $L_{\rm XUV}$ from $4.75\times 10^{27}-1.64\times 10^{30}$~erg~s$^{-1}$.
Their major findings comparing the energy-limited escape rate $\dot{M}_{\rm el}$ with those from hydrodynamic simulations, $\dot{M}_{\rm hd}$, are the following (see Fig.~\ref{fig:energylimited}):
\begin{itemize}
    \item $\dot{M}_{\rm el}$ overestimates loss rates by up to two orders of magnitude if the rates are small, $\dot{M}_{\rm hd} < 10^5$~kg~s$^{-1}$, and underestimates the loss rates by up $\sim$three orders of magnitude if the loss rates are large, $\dot{M}_{\rm hd} > 10^7$~kg~s$^{-1}$.  The reason is that for the former, the assumptions for hydrodynamic loss are violated and for the latter that the boil-off process takes over as the dominant loss mechanism.
    \item For the intermediate range of loss rates, say, $10^5-10^7$~kg~s$^{-1}$, discrepancies between one and two orders of magnitude above or below the hydrodynamic estimate are common.
    \item Even though for some parameter combinations approximate agreement is reached,  this agreement is often by chance or because two ignored or violated parameters cancel their influence.
    \item The energy-limited approach does not provide reliable upper-limit estimates for $\Lambda < 30$, where 
    \begin{equation}\label{escapeparameter2}
     \Lambda = \frac{KGMm}{R_pkT_{\rm eq}} \approx KX_c, 
    \end{equation}
    is an approximation of the Jeans parameter $X_c$ (Eq.~\ref{Xc}), including the correction $K$ for the Roche lobe, $m$ being the particle mass, $R_p$ the planetary radius, and $T_{\rm eq}$ the equilibrium temperature.
    \item For $\Lambda > 30$, the method can be used to obtain upper limits (rather than estimates of $\dot{M}$), but these upper limits may be significant overestimates of the actual mass-loss rates.
    \item The best parameter range of application is for planets with intermediate gravitational potential, low to intermediate  equilibrium temperatures, and low to intermediate X-ray irradiation levels; in these cases, the mass-loss estimates are still only order-of-magnitude (see Fig.~6-bottom in \citealt{krenn2021}).
    For evolutionary studies in which planets necessarily move through a large parameter range, the formalism cannot be used.
    \item Judging the validity of the energy-limited approach requires external information from more sophisticated hydrodynamic simulations, making its apriori application unreliable if not meaningless. 
\end{itemize}
Based on these, we conclude that reliable hydrodynamic loss rates require comprehensive hydrodynamic simulations for most cases of primordial atmospheres.

\textbf{Secondary atmospheres:} Energy-limited escape estimates are not meaningful for \textit{secondary atmospheres} because often a multitude of atmospheric constituents (atoms, molecules) are present in the lower atmosphere that undergo ionization, chemical reactions, and photodissociation in the upper atmosphere. Some molecules that are expected to be abundant on low-mass exoplanets, such as H$_2$O or CO$_2$, are strong emitters in the infrared, and some atoms also emit strongly in the ultraviolet, optical and infrared (e.g., O).

The cooling behavior and therefore an assessment of the temperature profile in the upper atmosphere requires complex models addressing radiative transport, thermal conduction and chemical networks, as summarized above in Sect.~\ref{thermosphere}. Recent cooling and heating functions for H$_2$O have been demonstrated by \citet{garciamunozetal2024}, showing that this molecule's cooling and heating occur over a broad of wavelengths and involve collisional and radiative excitation of the rotational and vibrational modes of molecular motion.

\begin{figure}
\begin{center}
\hbox{
\includegraphics[width=0.49\textwidth]{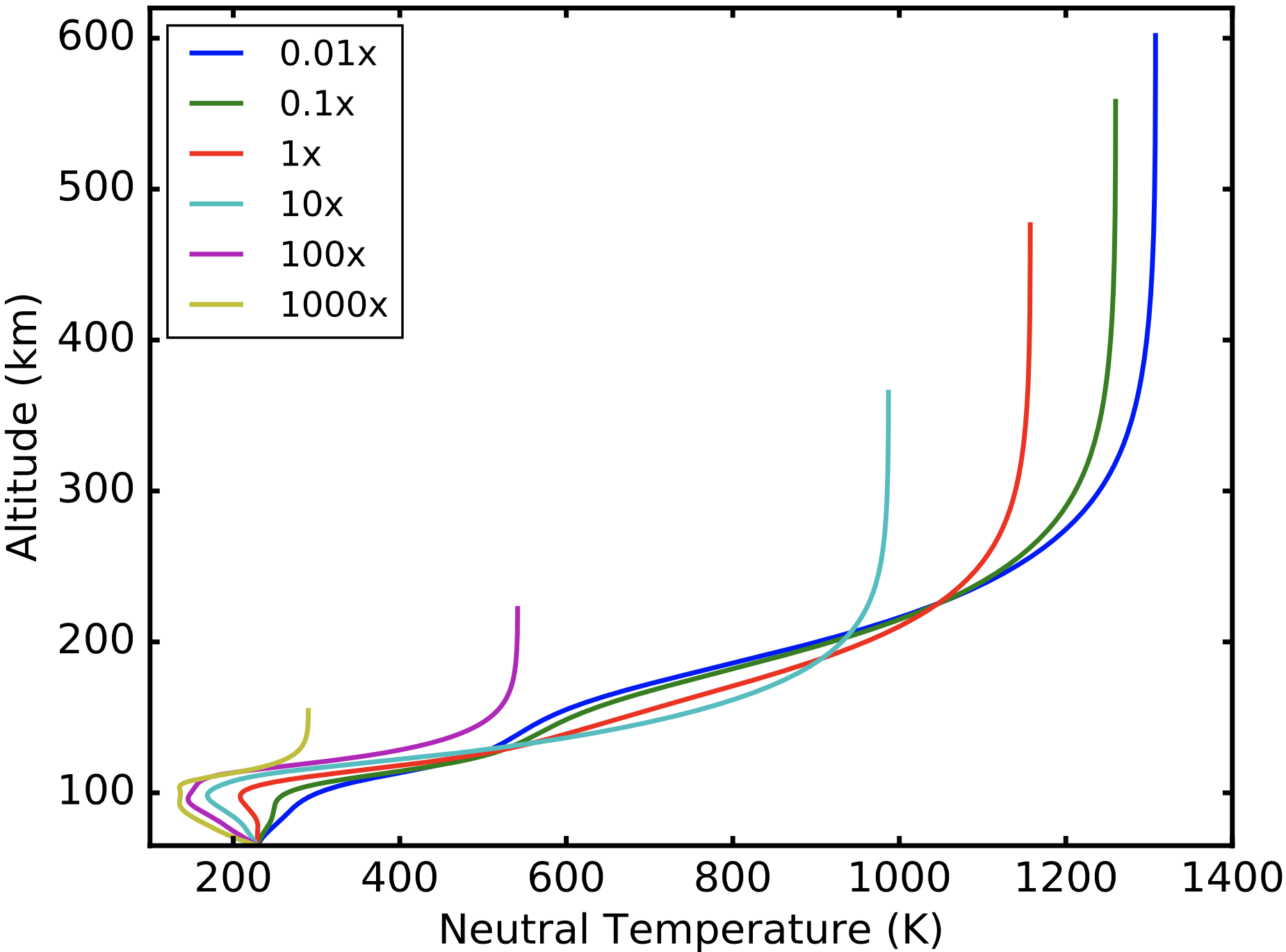} 
\includegraphics[width=0.49\textwidth]{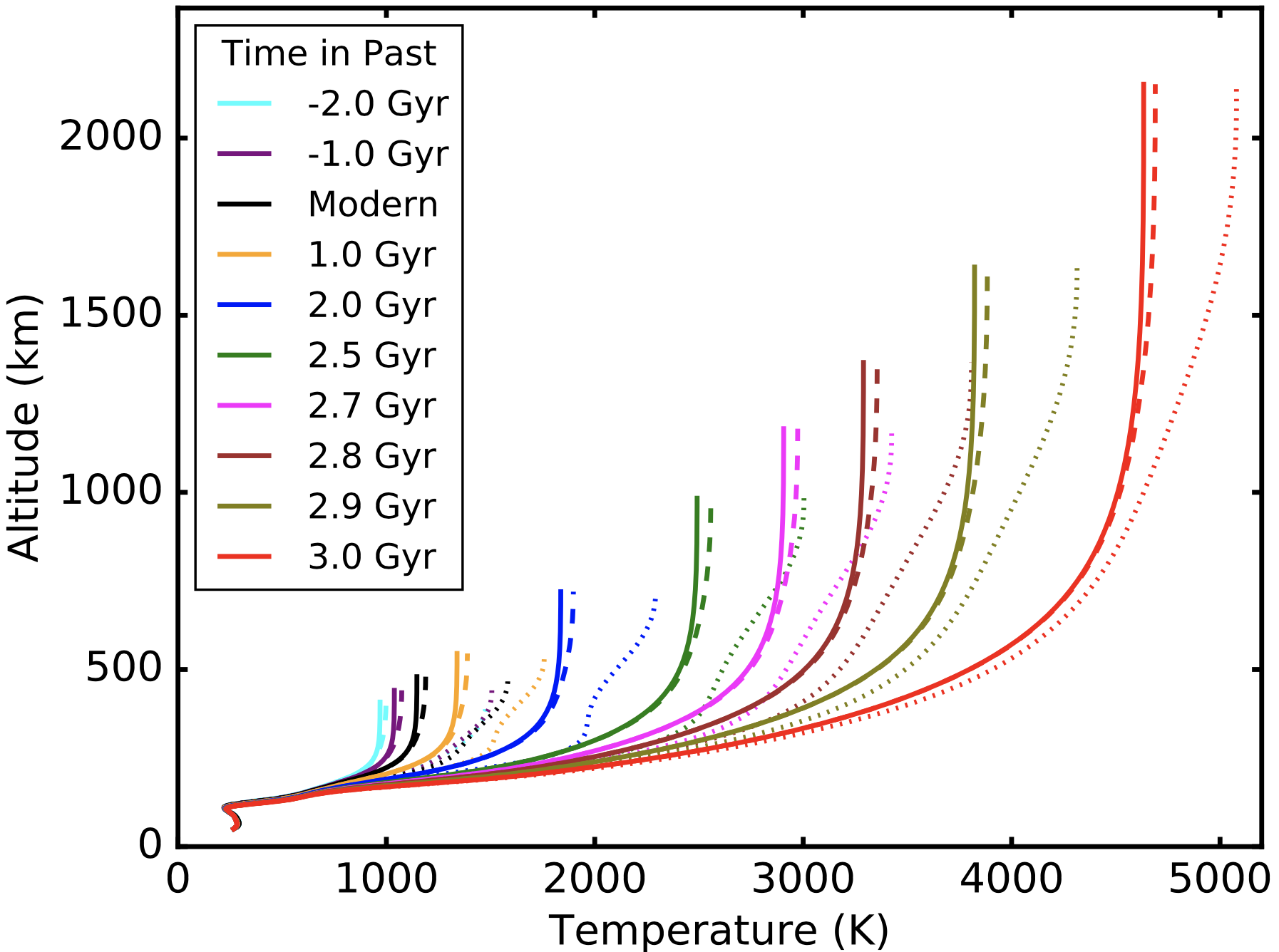} 
}
%\vspace*{-0.2 cm}
\caption{\textit{Left:} Upper-atmospheric temperature profiles for Earth's atmosphere with modified CO$_2$ mixing ratios as given in the inset, relative to the present-day level. The lines refer to the temperature of the neutrals. The upper end of the lines mark the positions of the exobase. Note the temperature decrease with increasing CO$_2$ mixing ratios. -- \textit{Right:} Upper-atmospheric temperature profiles for various XUV irradiation levels, translated here to evolutionary age of the Earth. Sold, dashed, and dotted lines refer to  neutrals, ions, and electrons, respectively. The upper end of the lines mark the positions of the exobase.(From \citealt{johnstone2018}.)}\label{fig:thermosphere}
\end{center}
\end{figure}
Atmospheric loss from secondary atmospheres primarily depends on the XUV flux above the atmosphere and the atmospheric composition. How significantly these two parameters influence the thermal profile of atmospheres is illustrated in \citet{johnstone2018}. Changing the CO$_2$ content of an atmosphere that has otherwise the same composition as Earth's atmosphere dramatically changes the thermospheric temperature (Fig.~\ref{fig:thermosphere}a). While the upper-atmospheric temperature of present Earth is about 1000--1200~K (depending on solar activity), an increase of the CO$_2$ mixing ratio by a factor of 1000 decreases the temperature to $\sim 250$~K, similar to present-day Venus. Decreasing CO$_2$ by a factor of 100 increases $T$ to $\sim 1300$~K. But we also emphasize that the exobase altitude changes with CO$_2$ content, lower CO$_2$/higher temperatures implying larger altitudes (variation from 150~km to 600~km for the above extremes). 

The effect of XUV heating is illustrated on Fig.~\ref{fig:thermosphere}-right. Here, the XUV flux is translated to evolutionary time for Earth using a reasonable solar activity evolution model. In the Archean period about 3~Gyr ago, the exobase temperature reached 4500--5000~K (for electrons, ions, and neutrals), the temperature subsequently dropping to today's $\sim 1200$~K as solar XUV activity was declining. The mass-loss rates derived from these models vary by many orders of magnitude, depending on the atom considered, especially for the 
C and O (about 40 orders of magnitude for Jeans escape alone).

\textbf{Model applications:} \citet{tian2008a} studied Earth's secondary atmosphere under elevated EUV irradiation (e.g., early-Earth conditions) including specifically adiabatic cooling by expansion of a hydrodynamic upper atmosphere. In their model, heating was balanced by downward conduction up to a critical irradiation flux of $\sim 5$ times the present-day level (the latter is $\approx 5$~erg~cm$^{-2}$~s$^{-1}$). Up to that irradiation level, the thermospheric temperature monotonically increases up to its peak at the exobase. At higher irradiation levels, peak temperatures in the thermosphere reach $\sim$8{,}000--14{,}000~K while the exobase is considerably cooler and at much higher altitudes; the exobase level gets progressively cooler for higher irradiation fluxes. But in their simulations, adiabatic cooling counteracts increased heating such that a blow-off state is never reached ($X \leq 1.5$ in Eq.~\ref{Xc}).
The hydrodynamic flow sets in, for an Earth-like atmosphere, when the peak thermospheric temperatures (initially at the exobase) reach 7000--8000~K. In follow-up work applying a more detailed electron transport/energy deposition model, the onset  of hydrodynamic flow and the transition to substantial adiabatic cooling was found already at an irradiation level of  4$\times$ the present solar EUV level \citep{tian2008b}.

For irradiation levels in the habitable zone corresponding to early times of Earth's evolution ($L_{\rm X} = 10^{29}$~erg~s$^{-1}$)\footnote{This is approximately at the [lower] 5th percentile value at 100~Myrs for a solar-mass star, corresponding to an initially slowly rotating Sun; see \citet{johnstone2021}.} for an atmosphere with a present-day Earth's composition (i.e. containing predominantly nitrogen, oxygen, and argon), photodissociation of molecules and partial ionization become substantial, and a transonic wind develops \citep{johnstone2019}. Although H is lost the most efficiently, other atoms and ions are also lost. The entire Earth's atmosphere would be lost within 0.1~Myrs under such irradiation levels, which is essentially instantaneous. The early Earth's atmospheric composition must have been different, e.g., containing significant fractions of efficiently cooling CO$_2$ or CH$_4$.

\begin{figure}
\begin{center}
%\hbox{
\includegraphics[width=0.7\textwidth]{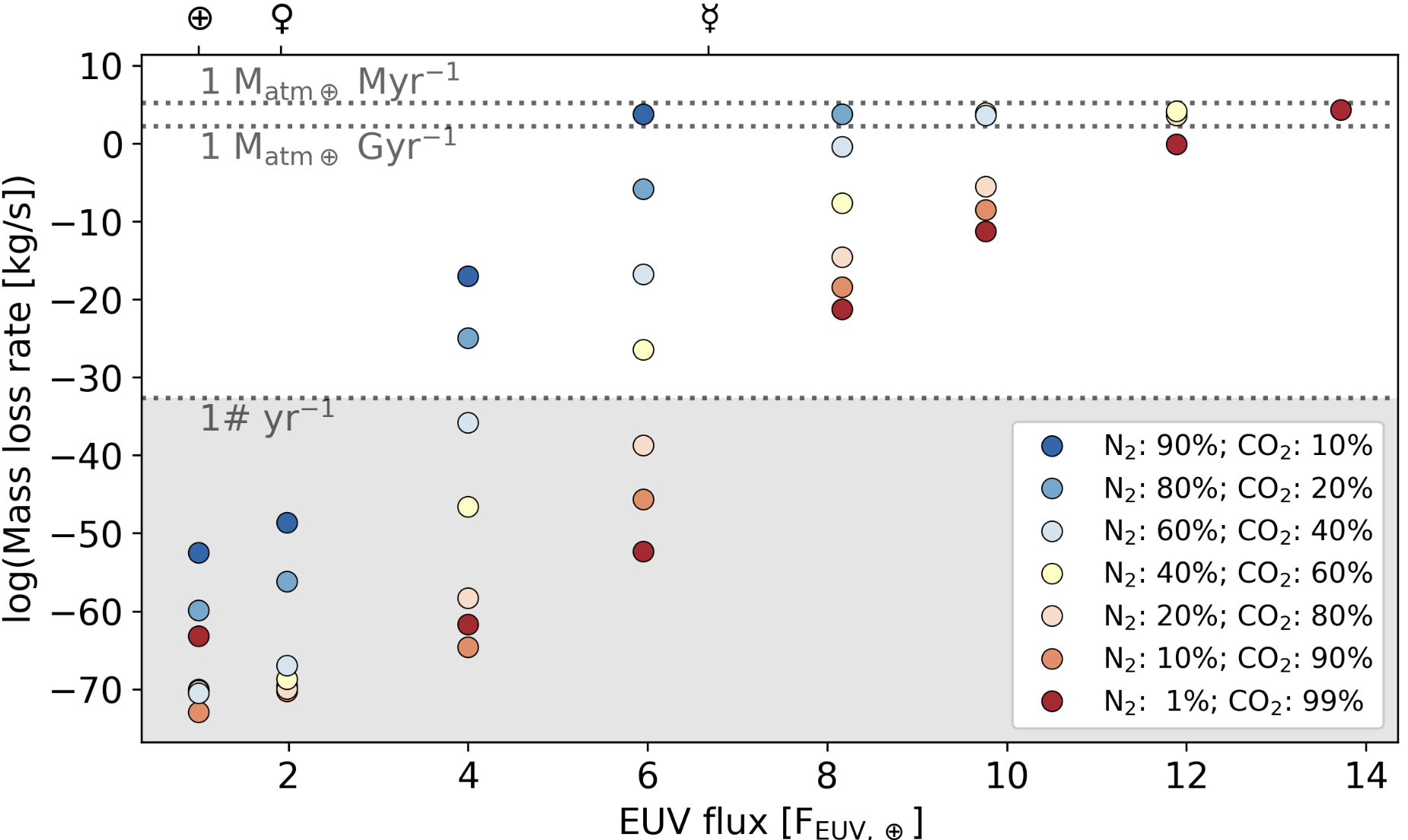} 
%}
\caption{Abundance-weighted average Jeans loss rates for Earth-mass planets with atmospheres composed of CO$_2$ + N$_2$ with various mixing ratios indicated in the inset, as a function of EUV flux at the top of the atmosphere (in units of the present-day flux at Earth). The horizontal dotted lines mark special loss rates of interest; the gray area indicates negligible loss. The planet symbols at the top  show the locations of the three planets Earth, Venus, and Mercury. (From \citealt{vanlooveren2024}.)} %\kristina{I think this plot should be shown in Chapter 4 -- }\manuel{Chapter 4 does deal with such simulations but in an evolutionary context; here, I wanted to show the physical effect of variable irradiation and variable atmospheric composition. I can put the evolutionary aspects into Chapter 4; similar things hold for applications for the early Earth, super-Earths, and Mars described in this section.}
\label{fig:TrappistLoss}
\end{center}
\end{figure}
The combined effects of XUV flux and atmospheric composition was studied for the Trappist-1 system, assuming a wide range of mixing ratios for pure CO$_2$+N$_2$ atmospheres by \citet{vanlooveren2024}. Fig.~\ref{fig:TrappistLoss} shows mass-loss rates in kg~s$^{-1}$ for a range of EUV fluxes in units of the flux level at present-day Earth (the present-day Earth receives of order 5~erg~cm$^{-2}$~s$^{-1}$ in the EUV range).
The figure includes extremely low mass-loss rates only 
for numerical illustration; most models are  compatible with zero mass loss. We clearly see that the mass-loss rate is a very steep function of EUV flux, reaching catastrophic mass-loss rates for fluxes only a few times the solar flux at Earth. However, the figure also shows the profound influence of the cooling CO$_2$ content. High levels of CO$_2$ make the atmosphere significantly more resistant to loss than low levels (but note that all cases shown here have a CO$_2$ content of at least 10\%, implying considerable cooling of the upper atmosphere in all cases, see Fig~\ref{fig:thermosphere}a).

How important planetary mass and consequently the gravitational potential is to retain atmospheres was studied by \citet{tian2009a} for the case of early Mars. A 3-bar CO$_2$-rich atmosphere was assumed, but the EUV irradiation level was $3\times$, $10\times$, or $20\times$ higher than today, 
corresponding to ages $\sim$3--4.3~Gyr ago, depending on the initial solar rotation period \citep{tu2015}. As already discussed, CO$_2$ acts as an efficient coolant, and it does so for  the $3\times$ and $10\times$ EUV level simulations, with Jeans parameters remaining high at 54 and 23, respectively. However, for the $20\times$ EUV case, CO$_2$ gets strongly photodissociated in the upper atmosphere, removing efficient cooling there and raising the exobase to 10,000~km and its temperature to $>2500$~K. The Jeans escape parameter reaches 1.8 for C and 2.4 for O, indicating loss via hydrodynamic flow. One bar of CO$_2$ would have been lost in a matter of a few Myrs; the atmosphere would therefore not have been stable during the early Noachian at $\sim 4$~Gyr ago. However, whether at that time the Sun did reach 20$\times$ the present solar EUV flux is uncertain; for an initially slowly rotating Sun, the EUV level would have been only of order 10$\times$ the present level \citep{tu2015, johnstone2021}. Also, ongoing outgassing or late formation (at about 4~Gyr ago and later) of the atmosphere at times when the solar EUV flux was rapidly declining helps keeping a substantial atmosphere on Mars for some limited time.

At the other extreme, in the case of super-Earths with masses of 6--10 Earth masses and efficiently cooling CO$_2$ atmospheres, the hydrodynamic regime is entered at irradiation levels of 600 times the present-day EUV irradiation, with an exobase at 10,000~km and an exobase temperature of 27,000~K (for a 6 Earth-mass planet; \citealt{tian2009b}). Again, the peak thermospheric temperature increases with the EUV flux but the exobase temperature drops -- the exobase is far above the hottest layers due to the effect of adiabatic cooling in the flow. For higher-mass planets, higher exospheric temperatures are achieved, because it takes more energy to lose an atmosphere in a stronger gravitational field. However, CO$_2$ atmospheres are retained on Gyr time scales for irradiation levels up to order of 1000$\times$ the present-day EUV irradiation of Earth \citep{tian2009b}. Even for a HZ around an active M dwarf, irradiation significantly higher (several 1000 to 10$^4$) is unlikely at planetary/stellar ages beyond $\sim$500~Myr \citep{johnstone2021}. 

The loss behavior completely changes again if the atmosphere is H$_2$O rich.
\citet{johnstone2020} simulated the escape behavior of a pure H$_2$O atmosphere above a water ocean planet.
\citet{garciamunozetal2020} have simulated the escape of a hypothetical H$_2$O atmosphere
around the sub-Neptune $\pi$ Men c, which orbits a solar-type star.
Provided the FUV flux is strong enough, 
Water vapor in the upper atmosphere easily photodissociates as a consequence of stellar irradiation, creating O$_2$, O$_3$, and OH, and eventually H, O, as well as H$^+$ and O$^+$. Hydrogen atoms easily escape into space given their low mass, while oxygen may be left behind, forming an oxygen-rich atmosphere although oxygen may also be absorbed into the surface \citep{wordsworth2018}. On the other hand, at sufficiently high XUV irradiation levels, oxygen will escape into space as well, with little oxygen enrichment left behind. For an Earth-like planet with a pure H$_2$O atmosphere orbiting at 1~au around an active Sun with incident EUV fluxes in the range of $\sim$100--5600~erg~cm$^{-2}$~s$^{-1}$ (X-ray fluxes being $\sim 5-8$ times lower), transonic hydrodynamic oxygen+hydrogen flows develop that reach temperatures of up to 19,000~K at altitudes of several $10^4$~km, and ionization fractions of $\sim10-90$\%. During the first Gyr of a Sun-like star's evolution, the Earth-mass planet at 1~au loses between 1 Earth ocean's worth of water (for the Sun's slow rotator track\footnote{This means that the Sun rotated slowly with a rotation period of $P_{\rm rot} \approx 14$~d after its protoplanetary disk phase and $P_{\rm rot} \approx 6$~d on the zero-age main sequence, see \citet{tu2015}, \citet{johnstone2021}, and \citet{Segura2026}.}) and 40 Earth oceans if such an amount is available (for the fast rotator track\footnote{This corresponds to $P_{\rm rot} \approx 0.75$~d after the disk phase and  $P_{\rm rot} \approx 0.55$~d on the zero-age main sequence.}). An Earth-like planet is therefore at risk to lose any H$_2$O atmosphere in its early evolution, and the situation would be worse for planets around M dwarfs (see \citealt{Segura2026}).

These results make it amply clear that secondary atmospheres cannot be treated with simplistic fluid models including estimates for mass-loss rates as given in Eq.~(\ref{eq:jeans}) or (\ref{eq:energylimited}) that are appropriate at best for (approximately) single fluids like primordial hydrogen envelopes. Each atmosphere requires thermo-chemical simulations including radiative transport, heating and radiative cooling processes, and large chemical reaction networks; the results depend on the planetary mass, the atmospheric composition, and the entire evolution of the stellar irradiation.

\begin{figure}
\begin{center}
%\hbox{
\includegraphics[width=1.0\textwidth]{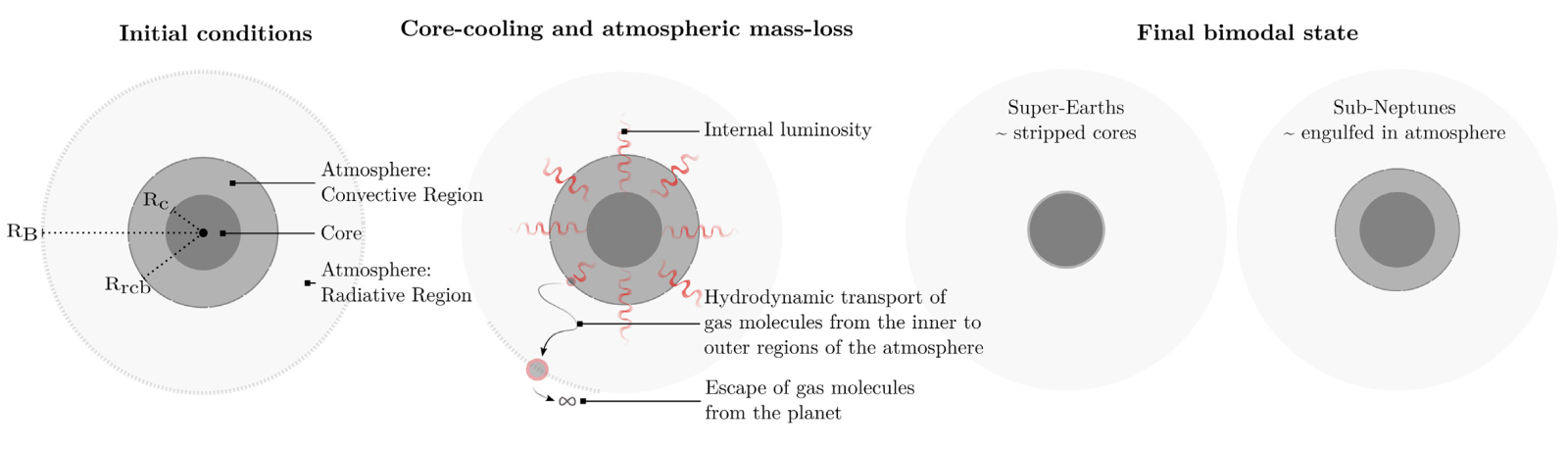} 
%}
\caption{Schematic representation of the core-powered mass-loss from \cite{Gupta2019}. Left panel: initial components of the planet structure: core (dark grey), atmospheric convective (grey), and radiative (light grey) regions. Middle panel: the thermal evolution and atmospheric mass loss at the Bondi radius. Right panel: the two end-member states at the end of 3 Gyr of evolution: super-Earths which are the rocky cores found below the valley, and sub-Neptunes located above the valley and still retaining a significant hydrogen-helium.} 
\label{fig:corepowered}
\end{center}
\end{figure}

\textbf{Core-powered escape:} in addition to stellar radiation, internal luminosities of exoplanets can also power mass-loss rate. Although this mechanism is not directly driven by the star and as such does not technically belong to the family of SPIs, it still plays a very important part in sculpting the planetary population and as such should be mentioned. Naturally, since this mass-loss is powered by internal luminosity of a planet, this mechanism is the most efficient for young planets with high intrinsic luminosities due to their recent formation, and atmospheres composed mostly of light gases accumulated from the gaseous disc \citep{Ginzburg2018,Gupta2019}. Removal of the external pressure from the gas in the protoplanetary disc once it dissipates also leads to loss of accumulated gaseous envelopes \citep{Stoekl2015}, but internal luminosity can accelerate this process. An interesting feature of the core-powered mechanism is that the luminosity of the cooling rocky core can completely erode light envelopes while preserving heavy ones which can produce a deficit of intermediate sized planets influencing the planetary population in a similar manner to radiation driven escape \citep{OwenWu2017}.  Fig.~\ref{fig:corepowered} provides an illustration of the mechanism. The convective region of the atmosphere extends from the core to the radiative–convective boundary, $R_{\rm rcb}$, which is comparable to a few core radii, $R_c$, at the end of the disc dispersal phase (our initial
condition), and the radiative region extends from the $R_{\rm rcb}$ to the Bondi radius, $R_B$. The luminosity of the planet is given by its cooling rate:
\begin{equation}
    L(t) = - \frac{dE_{\rm cool}}{dt} = \frac{64 \pi}{3} \frac{\sigma T_{\rm rcb}^4 R_B}{\kappa \rho_{rcb}},
\end{equation}
where $\sigma$ is the Stefan-Boltzmann constant and $\kappa$ is the opacity at the radiative–convective boundary $R_{\rm rcb}$, and $T_{\rm rcb}$ and $\rho_{\rm rcb}$ are the temperature and the density at the radiative-convective boundary. The radiative-convective boundary is illustrated in Fig.~\ref{fig:corepowered}. Planets accrete their envelopes with $R_{\rm rcb} \approx R_{\rm B}$, where $R_{\rm B}$ is the Bondi radius, which marks the location where the escape velocity equals to the sound speed. The modified Bondi radius is given by \citep{Gupta2019}:
\begin{equation}
    R_{\rm B} = \frac{\gamma - 1}{\gamma}\frac{G M_c \mu}{k_{\rm B} T_{\rm rcb}},
\end{equation}
where $\gamma$ is the adiabatic index of the atmospheres, $\mu$ is its molecular mass, $k_{\rm B}$ is the Boltzmann constant, $G$ is the gravitational constant, and the temperature are the convective-radiative boundary can be approximated by the equilibrium temperature $T_{\rm rcb} \approx T_{\rm eq}$. 

The core-powered mass-loss rate at the Bondi radius can then be estimated as
\begin{equation}
    \dot{M}_{\rm B} = 4 \pi R_{\rm s}^2 c_{\rm s} \rho_{\rm rcb} \exp \left( -\frac{G M_{\rm p}}{c_{\rm s} R_{\rm rcb}}\right).
    \label{eq:corepowered}
\end{equation}
In Eq.~\ref{eq:corepowered}, $R_{\rm s} = G M_{\rm p}/2c_{\rm s}^2$, where $c_{\rm s} = (k_{\rm B} T_{\rm eq}/\mu)^{1/2}$ is the isothermal speed of sound, and $M_{\rm p}$ is the planet's mass. The total mass-loss rate in the core-powered mechanism is the minimum mass loss rate between the mass-loss rate given by Eq.~\ref{eq:corepowered} and the energy limited mass-loss rate defined as $\dot{M}_{\rm E} \approx L(t)/gR_{\rm c}$.

Once the core-powered escape becomes less efficient, primordial atmospheres can still be lost very efficiently through thermal escape in the form of Jeans escape and hydrodynamic escape \citep{erkaev2007,kubyshkina2018a}. Thermal escape mechanisms are the most important for planets with atmospheres dominated by light elements such as hydrogen and helium. These atmospheres can be the atmospheres of terrestrial planets with still present thin hydrogen and helium envelopes at the very early stages of their formation (e.g. \citealp{Lammer2020}), and more massive planets on the volatile-rich side of the radius valley (e.g. \citealp{Affolter2023}). For the latter case, thermal evolution of planetary interiors also plays an important role for atmospheric escape even after the core-powered mass-loss phase \citep{Kubyshkina2022}. Various non-thermal escape mechanisms dominate atmospheric losses for planets with secondary atmospheres (that is, dominated by gases other than hydrogen and helium which are heavier; \citealp{Gronoff2020}).

\subsubsection{Non-thermal atmospheric losses from exoplanets}\label{sec:nonthermalescape}

In addition to thermal mass loss mechanisms described in the previous section, non-thermal mass loss can also be important for planetary evolution. Usually one understands under non-thermal losses the losses that are not driven directly by the stellar irradiation, but are the result of interaction of the planet with the stellar wind, the surrounding particle environment, and photochemistry \citep{Gronoff2020}. In this subsection, we focus on rocky exoplanets and Solar System planets with secondary atmospheres, because the main escape mechanism from planets with hydrogen-dominated atmospheres is thermal escape described above \citep{erkaev2007,Owen2020}. These mechanisms are very diverse and require different tools to study. In today's Solar System, the dominant escape mechanisms from all rocky planets are non-thermal, but it is interesting to note that the three rocky planets with substantial atmospheres have different dominant escape processes: the polar wind for the Earth \citep[e.g.,][]{Abe1993,Yau2007}, the ion pick up for Venus \citep{Gillmann2022}, and the photochemical escape from Mars \citep{Amerstorfer2017,Jakosky2018}. 

Although it is unclear if an intrinsic magnetic field amplifies or reduces the overall atmospheric escape \citep{Gunell2018,Blackman2018}, it is clear that it changes the patterns and influences the dominant loss mechanisms. Throughout its history, Earth has been a magnetized planet, although its magnetic moment has varied over geological times \citep{Biggin2015,Tarduno2025}. Isotopic studies indicate that the Earth has accumulated a small primordial enveloped consisting primarily of hydrogen and helium, which was then lost thermally \citep{Lammer2021}. It is likely that after that, the dominant loss mechanism for the Earth has been the polar outflow for a large period of time \citep{Kislyakova2020,Grasser2023}. Currently, neutral hydrogen is still lost from the Earth due to its light mass, but all other heavier elements are lost through the polar wind. In this mechanism, the ions are accelerated along the magnetic field lines by various physical mechanisms such as the ambipolar (or charge separation) electric field and wave-particle interactions \citep{Yau2007,Lysak2023}. Naturally, this acceleration is the most effective in the polar regions of the planet, where the magnetic field lines are nearly vertical and the wave interactions the most intense. Interestingly, not all ions are lost; observations and theoretical studies indicate that only a small percentage of these particles escape \citep{Seki2001,Peterson2008}. Other escape mechanisms include plasma plumes and the production of energetic neutral atoms (ENAs) by charge exchange, in which energetic ions capture an electron from neutral atoms or molecules; the resulting fast neutrals are no longer bound by the magnetic field and can escape if their velocity exceeds the escape velocity \citep[e.g.,][]{Keika2006}. For exoplanets with lower masses than the Earth, photochemical escape can also be an important non-thermal loss mechanism. 

Photochemical escape is the dominant loss mechanism from the atmosphere of Mars \citep{Jakosky2018}. The main mechanism is photodissociative recombination, which creates hot O atoms with excess energy energetic enough to escape the relatively weak gravity of Mars: $\rm {O_2^+ + e \rightarrow O + O}$ \citep[e.g.,][]{Kim1998,Lee2015,Amerstorfer2017}. At present, Mars lacks an intrinsic magnetic field, but the presence of remnant magnetization in older terrains suggests that up until approximately 4.0 or 4.1 gigayears ago, Mars was a magnetized planet \citep{Acuna1998,Lillis2013}. Measurements of the remnant magnetization of ALH 84001, a Martian meteorite that acquired its magnetization at approximately 4 Ga, indicate that the intensity of the paleomagnetic ﬁeld of Mars was possibly similar to that of the present geomagnetic field \citep{Weiss2008,Weiss2025}. A magnetic field with a dipole strength of approximately 60 nT at equator could form a magnetosphere similar to the current terrestrial one \citep{Kallio2008}, which would mean that early Mars also lost its atmosphere via polar outflow. \cite{Sakata2020,Sakata2022} have investigated loss processes from the early magnetized Mars, and have come to the conclusions that weak magnetic fields can in fact amplify the losses.

Venus is similar to Mars in some ways: it has an atmosphere dominated by CO$_2$, and it currently lacks an intrinsic magnetic field. Similar to Mars, it is possible that it did have an active dynamo in the past \citep{ORourke18}, but due to a lack of geological record, there is no definitive knowledge about Venus' history. For hydrogen, the dominant escape mechanism is in fact photochemical escape \citep[e.g.,][]{Lammer2006,Gu2025}. Since Venus has stronger gravity compared to Mars, for heavier elements such as oxygen, the dominant escape process is the ion pick up \citep{Wei2017,Gillmann2022}. In the ion pick up escape mechanisms, the upper layers of a planet's atmosphere are ionized by stellar radiation or particles; for planets which lack an intrinsic magnetic field, the solar or stellar wind can come into a much closer contact with the atmosphere, because an induced magnetosphere is typically very compressed \citep{Ramstad2021,Gillmann2022}. Therefore, the solar or stellar wind can interact with these ions almost directly. The ions can thus interact with the magnetic field of the wind, which leads to them being picked up by the wind and carried away from the planet until eventually they acquire the speed of the stellar wind. In a very simplified way, this type of mass loss from a non-magnetized planet can be estimated as \citep{Michel1971}

\begin{equation}
    \frac{d M}{dt} = 2 \pi H R \rho V <\sigma \sin \theta>.
\end{equation}
This equation combines the mass flux transported away with the effective cross-section through which this flux flows. Here, $\rho$ is the plasma density of the incoming solar or stellar wind, and $V$ is its speed, $H$ is the atmospheric scale height, $R$ is the radius of the planet, $\sigma$ is the loading fraction, and $\theta$ is the interaction angle between the incoming stellar wind and the atmosphere. Since the mass loading $\sigma$ is large only near the stagnation point where $\theta = 0$, one can assume that the weighted average is of the order of unity or less \citep{Michel1971}. For more precise estimates, MHD and particle codes are currently the state-of-the-art \citep{Terada2009,Kallio2012,Ma2020}  in combination with empirical relations between the power in the upstream solar wind and the power leaving the atmosphere \citep[e.g.,][]{Persson2021}.

Terrestrial exoplanets are notoriously difficult to observe. It is even more difficult to determine if they have intrinsic magnetic fields or not. So far, not only there are no direct detections of magnetic fields in terrestrial exoplanets, but also no unambiguous detections of atmospheres in such planets \citep[e.g.,][]{Zieba2023,Fischer2025}, with possible exception of very close-in super-Earths that are not fully Earth-like \citep{Teske2025}. For this reason, we can only speculate about the dominant escape mechanisms from exoplanets. From out knowledge of the Solar System planets, it seems reasonable to assume that some exoplanets should have magnetic dynamos similar to the one of the Earth \citep[e.g.,][]{Christensen2006,Zhang2022}. Many terrestrial planets orbit stars smaller than the Sun, in particular low mass M dwarfs \citep[e.g.,][]{Ment2023}. These systems are often more compact that the inner Solar system, which makes the star-planet interaction and the consequent non-thermal escape much more efficient \citep[e.g.,][]{Dong2020,2022ApJ...934..189C}. The magnetic field of the star itself is very important; stars with stronger magnetic fields likely significantly amplify non-thermal escape from their planets by compressing their magnetospheres, which raises questions of habitability of planets orbiting M dwarfs \citep{Vidotto2013,Gupta2023}. Planets orbiting lower mass stars also likely experience significant Joule heating of their atmospheres (see section \ref{sec:OhmicHeating} for more details), which also amplifies atmospheric loss. According to our present knowledge, it is possible that there are many more Venus analogues among exoplanets compared to Earth's analogues \citep{Kane2014,Kane2019}, but even they might have difficulties keeping their atmospheres \citep{Scherf2024,vanlooveren2025}.

\begin{figure}
\begin{center}
\hbox{
\includegraphics[width=1.0\textwidth]{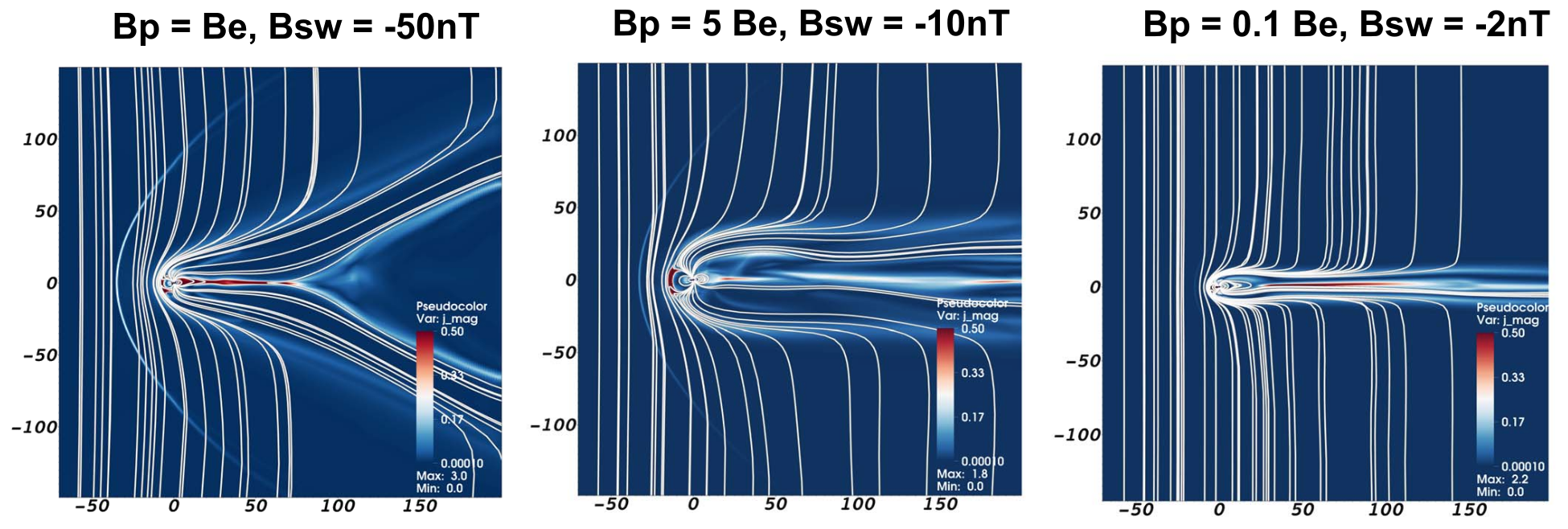} 
}
%\vspace*{-0.2 cm}
\caption{Steady-state planetary magnetospheric conﬁgurations for varying stellar ($B_{\rm sw}$ in nanotesla) and planetary ($B_p$, scaled to the Earth's current dipole strength, $B_e$) magnetic ﬁelds \citep{Gupta2023}. The axes are in planetary radii. One can see how the stellar and the planetary fields interact forming differently shaped magnetospheres.}\label{fig:magnetopsheres}
\end{center}
\end{figure}

In general, numerical simulations show that for exoplanets, one should expect a complex interplay between the stellar magnetic field, stellar wind dynamic pressure, and the intrinsic magnetic field of planets (\citealp{Ramstad2021}; see also Fig.~\ref{fig:magnetopsheres}). The geometry of the magnetosphere certainly plays a role: when the plasma flow around the magnetosphere is sub-Alfv\'enic, an Alfv\'en wing magnetosphere forms, which has a different shape compared to super-Alfv\'enic (and supermagnetosonic) magnetospheres such as the one of the Earth \citep{Baumjohann2010}. In the Solar System, the moon Ganymede is the only body with this type of magnetosphere \citep[e.g.,][]{Volwerk1999,Baumjohann2010}. While the Earth's magnetosphere can rarely form Alfv\'en wings, it only happens under very specific space weather conditions such as a strong CME \citep{Chen2024}. For exoplanets, especially if they orbit close to their host stars or are embedded into very dense stellar winds, such magnetospheres might be much more common \citep[e.g.,][]{Gupta2023,Presa2024}, which further complicates the studies of atmospheric loss. This type of interactions was described in detail in section \ref{sec:magnetic_interactions} of the present review.

Simulations seem to agree that exoplanets similar to Venus and Mars undergo efficient non-thermal loss, especially if they orbit lower mass stars, which expose them to higher particle fluxes and higher XUV fluxes \citep{Airapetian2017,Basak2021,Sakai2026}. \citet{Dong2017} showed that the ion escape rates from Proxima Cen b would be about two orders of magnitude higher than those of the Solar System terrestrial planets if the planet is unmagnetized, and that an intrinsic dipole field lowers the escape rates for some stellar wind conditions, although they remain higher than for the Solar System planets (see also \citealt{Egan2019} for heavy-ion escape from unmagnetized exoplanets). Photochemical escape can also be increased by a present intrinsic magnetic field, because it allows a larger extent of neutral atmosphere \citep{Lee2021}. Hopefully future observations by observatories such as the Habitable Worlds Observatory (HWO; \citealp{Feinberg2024}) and the Large Interferometer for Exoplanets (LIFE; \citealp{Quanz2022}) can provide us with a statistical sampling of atmospheres of terrestrial planet, which will give us a better understanding of the role of the planetary magnetic field for the star-planet interactions. The exact influence of the planet's intrinsic magnetic field on non-thermal losses is still not fully understood and requires further research combined with observational evidence from exoplanets.

\subsubsection{Photochemistry and Disequilibrium Chemistry}

In simple terms, stellar photons (and energetic particles) break molecules apart (photodissociation) and strip electrons from atoms and molecules (photoionization). The resulting radicals, ions, and electrons react with the ambient gas and produce species that would be absent or rare in a chemically equilibrated atmosphere. Photochemistry is therefore a direct pathway by which the host star modifies the composition of a planetary atmosphere and, through this composition, also the atmospheric heating and cooling, the escape processes discussed above, and the spectral features observed in transmission and emission. Below, we summarise the key ingredients of photochemical models and a few recent observational applications; a more detailed treatment in the context of atmospheric evolution and habitability is given by \citet{Segura2026}.

In the atmospheric sciences, the term photochemistry serves as an umbrella that encompasses all gas-phase chemistry driven, at least initially, by the local radiation field within the atmosphere or by particles precipitating from outside the planet. High-energy radiation environment of exoplanets, specifically XUV radiation (X-ray, EUV, and FUV; see Sect.~\ref{sec:intro} for the wavelength ranges), acts as the primary driver of non-equilibrium chemistry and atmospheric escape. Absorbed primarily in the thermosphere and exosphere, this flux governs the stability of the planetary envelope by dissociating stable volatiles (H$_2$, H$_2$O, CO$_2$, CH$_4$) and ionizing the neutral species.

Photochemistry changes the atmospheric composition and, by changing the relative abundances of chemicals, it affects the way the gas interacts with the radiation. Because in the layers of an atmosphere where photochemistry is important the chemical composition naturally departs from the condition of chemical equilibrium (as defined by a minimum of Gibbs' free energy), the terms photochemistry and disequilibrium chemistry are often used interchangeably. 

\paragraph{Stellar XUV Evolution and Saturation}
While photochemistry serves as the mechanism for chemical evolution, the intensity and duration of this evolution are fundamentally dictated by the nature of the input energy source. The chronology of photochemical forcing is therefore inextricably linked to the evolutionary stage of the host star. Multiwavelength observations of young solar analogs demonstrate that Sun-like stars undergo a ``saturation phase'' lasting approximately 100~Myr, during which the X-ray luminosity remains roughly constant at $L_{\rm X}/L_{\rm bol} \approx 10^{-3}$ and the XUV flux is about 100 times the present solar value \citep{Gudel_1997, Ribas2005, Gudel2007}. Following this phase, the flux declines according to a power law, reaching $\sim 10\times$ modern levels about 4~Gyr ago \citep{Ribas2005, Claire_2012}. Crucially for exoplanet demographics, M dwarfs remain in this saturated regime significantly longer than G-type stars, with saturation timescales extending up to a gigayear \citep{Scalo_2007}. Consequently, protoatmospheres orbiting such stars integrate an XUV fluence orders of magnitude higher than that experienced by the early Earth, driving extreme cumulative chemical alteration.

\paragraph{Quantitative Description and Rate Coefficients}
Translating these broad stellar trends into predictive atmospheric models requires a rigorous mathematical framework to calculate how individual photons interact with specific gas-phase species. The quantitative description of photochemistry therefore requires a network of processes that connect reactants to products. Ideally, each forward process, e.g., $A+B \rightarrow C+D$, should be accompanied by its reverse, $C+D \rightarrow A+B$, although the pairing may only become truly important at the high pressures for which the system should lead to chemical equilibrium. For photodissociation and photoionization processes, the rate coefficient $J_{\lambda}$ [s$^{-1}$] is determined by:
\begin{equation}
J_{\lambda}=\frac{4\pi}{hc} \int \sigma(\lambda) \mathcal{J}_{\lambda}(\lambda) \lambda d\lambda,
\end{equation}
where $\mathcal{J}_{\lambda}$ [erg cm$^{-2}$s$^{-1}$sr$^{-1}$] is the average radiation in energy units and $\sigma$ the process cross section. Alternatively, the rate coefficient $J_i(z)$ for a specific process $i$ can be expressed as:
\begin{equation}
J_i(z) = \int_{\lambda_{min}}^{\lambda_{max}} \sigma_i(\lambda) F_\lambda e^{-\tau(\lambda, z)} \, d\lambda
\end{equation}
where $F_\lambda$ is the stellar flux and $\tau$ is the optical depth of the overlying atmosphere. 

At the state-resolved level, the cross sections are independent of temperature. However, $\sigma$ depends on temperature when the cross sections refer to the thermalized population of states in a molecule, reflecting the averaging over the Boltzmann distribution \citep{ranjanetal2020}. $\mathcal{J}_{\lambda}$ is a local property of the radiation field accounting for photons from all directions. Typically, the main contribution is stellar radiation. Stellar photons of XUV wavelengths eject electrons (photoionization), while lower energies (FUV/NUV) photodissociate molecules. Because upper atmospheres are usually dominated by atoms that absorb short wavelengths quickly, photodissociation rates remain large down to much lower altitudes than photoionization rates.

\paragraph{Chemical Networks and State-to-State Kinetics}
Once these individual rate coefficients are defined, they must be integrated into comprehensive reaction networks that account for the collective behavior and interdependence of all atmospheric constituents. Historically, chemical networks have been published for object-specific conditions, e.g., Titan \citep{vuittonetal2019} or Mars \citep{foxetal2015}. The predicted diversity of exoplanet atmospheres calls for more general approaches capable of simulating atmospheres of varying compositions (metal-rich secondary vs. hydrogen-dominated primary atmospheres) \citep{mosesetal2013}. Reliability depends on the implementation of cross sections and rate coefficients \citep{nahar2020, ranjanetal2020}, with ongoing efforts to assess network completeness \citep{garciamunozetal2025, agundez2026} and process variety \citep{loccietal2022, garciamunoz2023}.

While most species refer to ground states, certain instances require state-to-state networks. Metastable states like O($^1D$, $^1S$) act as thermostats for young Earth-like exoplanets \citep{nakayamaetal2022}, while He($2^3S$) provides diagnostic insight into sub-Neptunes \citep{oklopcic2019, ahreretal2025}. Vibrationally excited H$_2$ at moderate-to-high temperatures boosts the reactivity of dissociative charge exchange (He$^+$ + H$_2$ $\rightarrow$ He + H$^+$ + H), controlling He$^+$ and He($2^3S$) abundances in warm exo-Neptunes like GJ~3470~b \citep{garciamunozetal2025}. \citet{garciamunoz2025} proposed a state-to-state network for metal-free atmospheres, showing that rate coefficient inaccuracies have biased inferred He/H ratios from measurements of the He~{\sc i} triplet line at 1.08~$\mu$m.

\paragraph{High-Energy Environments and Photoelectrons}
In the most extreme radiative environments, the photochemical influence extends beyond simple molecular dissociation and includes the potential effects of secondary particles and thermal excitation. In extreme environments like the ultrahot Jupiter KELT-9b ($T_{\rm eq} \approx 4{,}050$~K), self-radiation provides enough XUV photons to photoionize ground-state H atoms to low altitudes \citep{garciamunozschneider2019}. Excited H(2) atoms also represent a major source of electrons. In less extreme atmospheres, photoionization of atoms with lower potentials (C, Mg, Fe, Ca) provides electrons where stellar XUV cannot reach \citep{garciamunoz2007, huangetal2023}. Electrons ejected by energetic photons carry enough energy to trigger further chemistry. The rate coefficient $J_{\rm{e}}$ for photoelectron-driven chemistry is formulated similarly to $J_{\lambda}$, replacing photon flux with particle flux. The details of the energy-degradation cascade are specific to atmospheric composition \citep{garciamunozbataille2024} and affect mass loss rates \citep{Gillet_2023}.

\paragraph{Photochemical Markers: SO$_2$ and Molecular Ions}
The theoretical frameworks of disequilibrium chemistry are no longer purely computational; they have been increasingly validated by the empirical detection of specific chemical 'fingerprints' in exoplanetary spectra. The detection of SO$_2$ in JWST transmission spectra \citep{aldersonetal2023, beattyetal2024, gressieretal2025} has solidified photochemistry as a valid framework in exoplanet science. SO$_2$ mixing ratios of $10^{-5}$--$10^{-6}$ on WASP-39~b result from the multi-step oxidation of sulfur radicals produced from H$_2$S photodissociation \citep{Tsai_2023}. This involves S and SH radicals, with H and OH as free radicals facilitating conversion. Multiple pathways exist, with photodissociation by photons up to 350~nm playing a key role \citep{degruijteretal2025}. 

While SO$_2$ evidences neutral disequilibrium, no molecular ions have been detected yet. Models predict H$_3^+$ as the dominant ion in low-metallicity hydrogen-dominated atmospheres, as CO reactions remove it rapidly \citep{garciamunoz2007}. In water-rich atmospheres, H$_3$O$^+$ is predicted to dominate \citep{garciamunoz2023}, as also suggested by laboratory experiments
\citep{bourgalais2020}. Both are efficient IR radiators that can affect mass loss rates, such as in $\pi$~Men~c \citep{garciamunozetal2020}. Although searches for H$_3^+$ auroral features via high-resolution spectroscopy have been unsuccessful thus far \citep{lenzetal2016}, future instruments may provide the needed sensitivity. Current modeling is limited by scarce information on excitation cross sections for H$_3^+$ and H$_3$O$^+$ in collisions with species like H$_2$.

\section{The Solar System Laboratory}
\label{sec:solar_system_laboratory}

While the above sections provides the necessary theoretical formalism to describe the governing physics of star-planet systems, the predictive utility of these models relies on empirical calibration and the validation of scaling laws. To transition from generalized physical foundations in the diverse parameter space of exoplanets,  the Solar System functions as a primary astrophysical benchmark, a ``ground truth'' laboratory where the fundamental plasma dynamics, radiative forcing, and magnetic topologies can be directly characterized. The following section examines the heliosphere as the definitive prototype for some regimes of these interactions, establishing the empirical baseline required to extrapolate star-planet coupling to more extreme and varied extrasolar environments.

\subsection{The heliosphere}

To quantify how stellar environments affect planetary atmospheres, it is common to adopt the Sun as a baseline and scale its wind and magnetic properties to other stars of different ages and activity levels. This “solar standard” approach is particularly useful because the solar wind and its interaction with planets are observationally relatively well constrained.

The solar wind (SW) is a radially expanding plasma originating from the solar corona. It consists primarily of electrons and protons, with a smaller contribution from helium nuclei. Due to the very large mean free path of particles at large distances from the Sun, the solar wind behaves as a nearly collisionless plasma while interacting with the heliospheric planets. Within this plasma, the interplanetary magnetic field (IMF) is effectively frozen into the outflowing plasma. The solar wind undergoes rapid acceleration within the orbit of Mercury, transitioning from subsonic to supersonic flow as it expands outward from the hot solar corona. This acceleration is primarily driven by thermal pressure gradients in the corona, as originally described by \citet{Parker1958}, and is further influenced by wave-particle interactions and additional magnetic processes in the expanding plasma. The wind typically reaches terminal velocities of about 400–500 km/s in the slow solar wind regime, although faster streams can originate from coronal holes at higher latitudes. At these velocities, the flow is essentially supermagnetosonic, exceeding both the local sound speed and the Alfvén speed, which implies that information cannot propagate upstream and the wind effectively decouples dynamically from the solar surface.

\begin{figure}[htpb]
    \centering
    \includegraphics[width=0.75\linewidth]{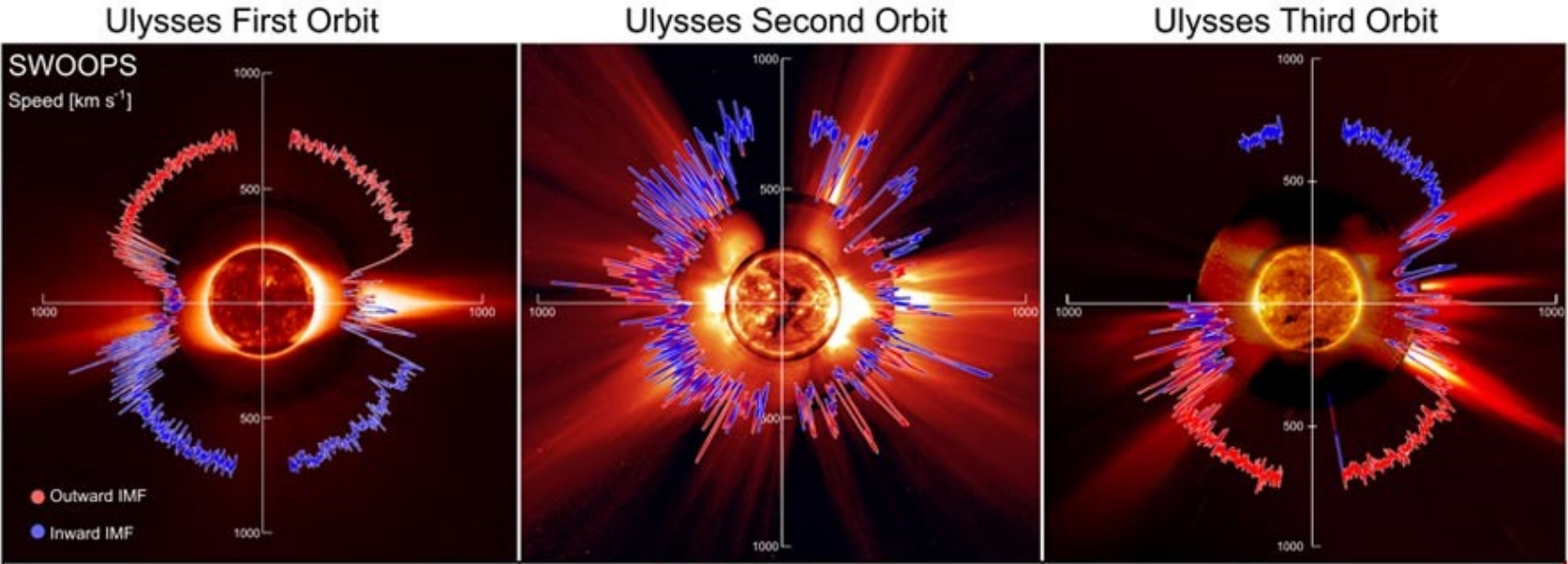}
    \caption{Polar representations of the solar wind velocity measured during Ulysses mission’s three high-latitude orbits, with color coding indicating the polarity of the interplanetary magnetic field (IMF). The solar wind speed distributions are superimposed on representative solar images corresponding to distinct phases of the solar cycle: solar minimum of cycle 22 (17 August 1996), solar maximum of cycle 23 (7 December 2000), and solar minimum of cycle 23 (28 March 2006). The background imagery is constructed by radially combining observations from the Solar and Heliospheric Observatory Extreme Ultraviolet Imaging Telescope (Fe~{\sc xii}, 19.5 nm), the Mauna Loa K-coronameter (700–950 nm), and the SOHO C2 white-light coronagraph \citep{mccomas2008}.}
    \label{fig:ulysses_speed}
\end{figure}

The three-dimensional structure of the solar wind is intrinsically anisotropic and is governed by the large-scale magnetic field configuration of its source, the Sun. In particular, the global magnetic topology determines both the velocity and density distributions of the outflowing plasma. In the vicinity of the solar magnetic equator, where closed magnetic field regions and streamer belts prevail, the solar wind is generally slower and denser. Conversely, at higher latitudes, where open magnetic field lines dominate, the plasma can escape more freely, producing faster and more rarefied wind streams \citep{mccomas2008}. This latitudinal dependence is clearly illustrated in velocity maps obtained during the ULYSSES mission (see the top three panels of Figure \ref{fig:ulysses_speed} corresponding to different levels of solar activity where the middle panel represents the solar activity maxima whereas the left and right represent the neighbouring minima), overlaid on extreme ultraviolet (EUV) images from the Solar and Heliospheric Observatory (SOHO) Extreme Ultraviolet Imaging Telescope. The distinction between fast and slow solar wind streams is most pronounced during periods of low solar magnetic activity, corresponding to the left and the right panels, while it becomes progressively less structured as the system approaches the maximum phase of the solar magnetic cycle denoted by the middle panel.

\begin{figure}
    \centering
    \includegraphics[width=0.9\linewidth]{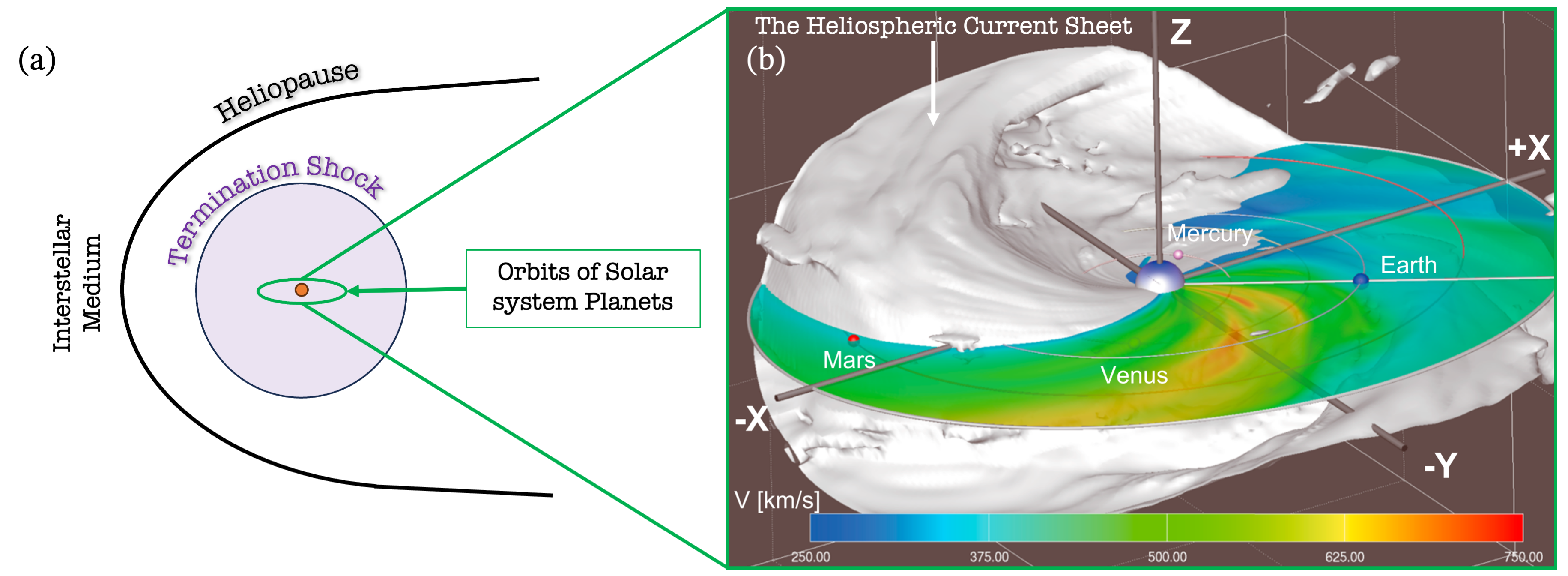}
    \caption{Panel (a) illustrates a large-scale overview of the heliosphere-interstellar interface. This panel marks the termination shock, where the solar wind begins its deceleration, and the heliopause, which defines the outermost limit of the Sun's plasma influence. Panel (b), adapted from \citet{2014SpWea..12..187S}, highlights the wavy white surface of the heliospheric current sheet, the distribution of solar wind speeds across the ecliptic plane, and the relative positions of planetary orbits.}
    \label{fig:heliopause_and_hcs}
\end{figure}

At sufficiently large heliocentric distances, the global solar magnetic field can be approximated as predominantly dipolar for quiet solar conditions. Under these conditions, open magnetic field lines are progressively stretched into a quasi-radial configuration by the outward expansion of the solar wind. Field lines rooted in the two polar regions of the Sun carry opposite magnetic polarities. At the interface where these oppositely directed magnetic flux systems converge, the finite curl of the magnetic field implies the presence of an electric current layer, forming what is known as the heliospheric current sheet (HCS), visualized as the wavy white surface in Figure \ref{fig:heliopause_and_hcs}b. The morphology of this current sheet is strongly modulated by the phase of the solar magnetic cycle: it is relatively well-ordered during solar minimum, whereas it becomes increasingly warped and dynamically variable during solar maximum when the large-scale topology of the solar magnetic field can diverge significantly from the dipolar approximation. Furthermore, the interplay between the radially expanding solar wind and the rotation of the Sun generates the characteristic spiral geometry of the interplanetary magnetic field, known as the Parker spiral; the resulting distribution of solar wind speeds across the ecliptic plane and the relative positions of planetary orbits are illustrated in Figure \ref{fig:heliopause_and_hcs}b. At large heliocentric distances, beyond the orbit of Pluto, the solar wind interacts with the interstellar medium (ISM). As shown in the global schematic in Figure \ref{fig:heliopause_and_hcs}a, the solar wind is first decelerated at the termination shock and is ultimately halted at the heliopause, marking the boundary between the solar and interstellar plasma environments.

The properties of the solar wind described above are not intrinsic constants, but rather emerge from the Sun’s magnetic field and the processes that heat the corona. Extending these considerations to other stars therefore requires understanding the dynamics of their respective stellar magnetic fields and coronal properties, since they ultimately govern the stellar wind properties. In cool stars, magnetic fields are generated by dynamo processes operating in the convective envelope. These dynamos arise from the interplay between rotation and convection, where differential rotation and turbulent plasma motions convert kinetic energy into magnetic energy \citep{Parker1955, 1969AN....291...49S}. As a consequence, the efficiency of the dynamo is intimately linked to stellar rotation. This relationship is parameterized by the Rossby number ($Ro = P_{rot}/\tau_c$), which represents the ratio of the stellar rotation period to the convective turnover time $\tau_c$. Note that $\tau_c$ is not uniquely defined, nor is it a single constant for a star: in stellar models, the convective velocity, and hence the local turnover time, varies strongly with depth in the convection zone, so that $\tau_c$ is either evaluated at a specific location (commonly about half a mixing length above the base of the convection zone) or replaced by a global, volume-averaged turnover time, while observational studies usually adopt empirical calibrations of $\tau_c$ as a function of stellar mass or colour derived from activity--rotation relations \citep[e.g.,][]{Noyes1984, 2011ApJ...741...54C, 2011ApJ...743...48W}. Different prescriptions can differ by factors of a few, so that values of $Ro$, and of the critical Rossby number marking saturation, depend on the adopted definition; here, we follow the definitions used in the cited works. In the ``unsaturated'' regime ($Ro \gtrsim 0.1$), activity scales inversely with the Rossby number; thus, faster rotators sustain stronger fields and enhanced magnetic activity. However, as rotation increases and the Rossby number drops below a critical threshold ($\approx 0.1$), the magnetic flux and high-energy emission enter a saturated regime. In this state, the magnetic activity reaches a plateau, potentially due to the physical saturation of the dynamo or the complete coverage of the stellar surface by magnetic active regions \citep{1984A&A...133..117V,2011ApJ...743...48W}. This distinction can be clearly seen in observations of surface magnetic flux across a range of stellar masses.

\begin{figure}
    \centering
    \includegraphics[width=0.5\linewidth]{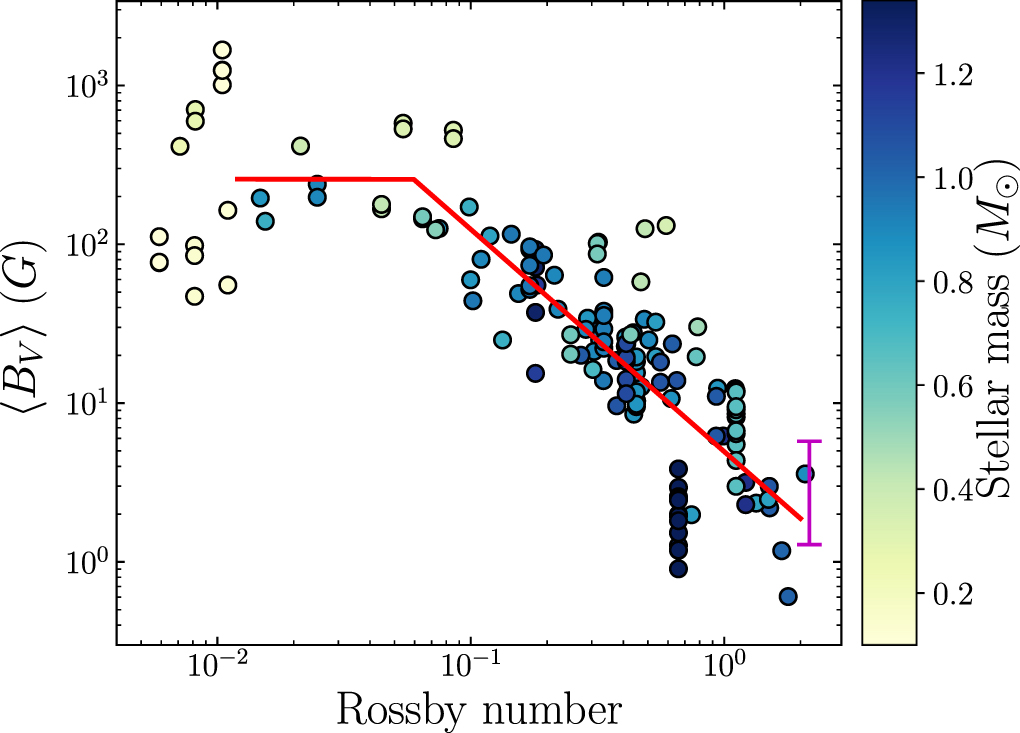}
    \caption{Average unsigned photospheric magnetic flux obtained from ZDI against Rossby number color coded by stellar mass. The three-parameter fit (solid red line) has a saturated field strength, a critical Rossby number of $Ro_{crit} = 0.06 \pm 0.01$, and an unsaturated regime slope value of $\beta = -1.40 \pm 0.10$. The magenta strut represents the range of solar values over cycle 24 (the magnetograms used to calculate this range were truncated to $\ell_{max} = 5$). Adapted from See et al. 2019}
    \label{fig:mag_vs_rossby_see_19}
\end{figure}

The link between rotation and magnetism further extends to the regulation of a star's angular momentum. The magnetized wind carries away angular momentum, which gradually slows the stellar rotation over time \citep{1967ApJ...148..217W,2000BASI...28..707K}. This feedback loop, where rotation powers the dynamo while the magnetized wind brakes the rotation, constrains an evolutionary track for cool stars. To quantify this stellar wind evolution, the Sun serves as a primary benchmark since the solar wind is the only stellar wind measured \textit{in situ}. Consequently, stellar mass-loss rates ($\dot{M}_\star$) are often parameterized as a function of magnetic activity proxies, such as X-ray luminosity ($L_X$) or surface flux ($F_X$) \citep{2004A&ARv..12...71G, 2011ApJ...741...54C} which can be observationally constrained for other stars.

Empirical relations derived from astrospheric absorption measurements of solar-type stars in the unsaturated regime suggest that the mass-loss rate per unit area, $\dot{m}$, scales with the X-ray surface flux as a power law:
\begin{equation}
\dot{m} \propto F_X^{a}
\end{equation}
The value of the exponent $a$ has loose constraints and appear to be sensitive to the stellar sample and the specific activity range considered. Early estimates for solar-like stars \citep{2004LRSP....1....2W} suggested a relatively steep scaling of $a \approx 1.34 \pm 0.18$. This is similar to recent estimates of mass-loss rates for several stars based on their astrospheric X-ray emission \citep{Kislyakova2024}, although so far not enough data has been accumulated with this method to allow for proper statistical estimates. However, more recent multi-wavelength analyses and numerical MHD simulations spanning G to M dwarfs favor a shallower relationship, with $a \approx 0.77 \pm 0.04$ \citep{2021ApJ...915...37W, Chebly2023}. This lower exponent suggests that the efficiency of mass-loss production may decrease as stars become increasingly active.
In the saturated regime ($Ro \le Ro_{\text{crit}}$), magnetic activity reaches the plateau shown in Figure \ref{fig:mag_vs_rossby_see_19}, where $L_X/L_{\text{bol}} \approx 10^{-3}$ \citep{1984A&A...133..117V,2011ApJ...743...48W}. In this state, young, rapidly rotating stars can drive winds significantly stronger than the contemporary solar wind. It has also been suggested that for extremely active stars, the wind-activity relation may fundamentally change or even saturate, as high-density magnetic topologies could potentially suppress plasma outflow despite high X-ray luminosities \citep{2004LRSP....1....2W, johnstone2015}.

\subsection{Planetary Magnetospheres and Solar Wind Interaction}

As described in the preceding section, solar and stellar winds, constitute the dynamic upstream boundary conditions for any planetary body. Planetary magnetospheres arise from a fundamental balance between this incident wind and the localized obstacle presented by the planet. This interaction governs the transfer of mass, momentum, and energy across a wide range of plasma environments, operating across a hierarchy of spatial and temporal scales. 

\subsubsection{Classification of Planetary Obstacles}

The nature of the obstacle varies significantly, even across the Solar System, and comparative studies of planetary magnetospheres reveal both universal processes and system-specific differences \citep[see the review by][]{bagenal2013}. Broadly, the interaction regions can be segregated into three distinct categories based on the intrinsic properties of the planetary body:

\begin{itemize}
    \item \textbf{Intrinsic Magnetospheres:} Magnetized planets such as Earth, Jupiter, and Saturn possess internally generated dynamo magnetic fields. These fields form extended magnetospheres capable of deflecting the solar wind at a significant distance from the planetary surface (typically several planetary radii). In these environments, the direct interaction of the solar wind with the gravitationally bound atmosphere is largely prevented by the magnetic shield.
    \item \textbf{Induced Magnetospheres:} Unmagnetized planets with substantial atmospheres, such as Venus and Mars, lack strong global intrinsic dynamos. Consequently, they rely on their ionospheres to stand off the solar wind. Thermal pressure gradients and currents induced by the motional electric field of the solar wind create an effective barrier, resulting in a well-defined region of perturbed IMF populated by both solar wind and local planetary plasma.
    \item \textbf{Airless Bodies:} At smaller scales, airless bodies such as the Moon or asteroids lack both intrinsic magnetic fields and dense atmospheres. This allows the solar wind to directly interact with, and be partially absorbed by, their bare surfaces or localized crustal magnetic anomalies \citep{bagenal2013}.
\end{itemize}

Despite these structural differences, common features at the obstacles such as bow shocks, magnetopauses (or induced magnetosphere boundaries), and magnetotails consistently emerge from the interaction with the solar wind. The specific characteristics of these boundary layers and the overall magnetospheric geometry are highly sensitive to the upstream solar wind parameters. Table \ref{tab:upstream_parameters} summarizes the primary solar wind and solar radiation parameters and their distinct physical influences on the planetary plasma environment \citep{arridge2011}.

\begin{table}[htpb]
\centering
\caption{Impact of upstream solar wind and radiation parameters on planetary magnetospheric interactions (Adapted from \citealt{arridge2011}).}
\label{tab:upstream_parameters}
\begin{tabular}{p{0.35\linewidth} p{0.6\linewidth}}
\hline
\textbf{Upstream Parameter} & \textbf{Magnetospheric Influence} \\ \hline
Ion gyroradius / Inertial length & Dictates whether the interaction occurs in a collisional or a collisionless regime. \\ 
Dynamic pressure ($\rho v^2$) & Determines the standoff distance and overall size of the magnetosphere (compressibility). \\ 
IMF direction & Controls the orientation of the motional (or `convective') electric field of the solar wind, $\mathbf{E} = -\mathbf{v}_{sw} \times \mathbf{B}_{sw}$, and the efficiency of dayside magnetic reconnection. \\ 
Plasma beta ($\beta$) & Governs the dominance of magnetic field stresses versus thermal particle dynamics. \\ 
Mach number & Defines the properties, strength, and structure of the planetary bow shock. \\ 
EUV flux & Regulates ionospheric density, upper atmospheric heating, and the rate of photoionization for ion pickup. \\ 
Flow velocity ($v_{sw}$) & Influences magnetospheric standoff distance, tail geometry, motional electric field strength, and ion pickup acceleration. \\ 
Electron density and distribution & Drives electron impact ionization processes and influences diffusion regions during reconnection. \\ \hline
\end{tabular}
\end{table}

\subsubsection{Pressure Balance at Planetary Boundaries}\label{sec:pressurebalance}

At the global scale, the size and shape of any magnetosphere are determined by the pressure balance across the magnetopause (or induced boundary). The total pressure includes contributions from dynamic (ram), thermal, and magnetic components. A general form of the pressure balance across the boundary layer can be expressed as:
\begin{equation}
\rho_{sw} v_{sw}^2 + p_{sw} + \frac{B_{sw}^2}{2\mu_0}
\;\approx\;
\rho_{mp} v_{mp}^2 + p_{mp} + \frac{B_{mp}^2}{2\mu_0},
\end{equation}
where the subscripts denote solar wind ($sw$) and magnetospheric or planetary ($mp$) quantities.

For strongly magnetized planets, the intrinsic magnetic field dominates the pressure on the planetary side. Because the bulk magnetospheric flow velocity near the dayside magnetopause is typically small ($v_{mp} \ll v_{sw}$), and the thermal pressure $p_{mp}$ is negligible compared to the magnetic pressure ($B_{mp}^2/2\mu_0 \gg p_{mp}$), the balance reduces primarily to a competition between solar wind ram pressure ($\rho_{sw} v_{sw}^2$) and planetary magnetic pressure. 

In contrast, for non-magnetized planets, the magnetic pressure on the planetary side is negligible ($B_{mp}^2/2\mu_0 \ll p_{mp}$). Here, the solar wind ram pressure is balanced by the thermal pressure of the ionosphere ($p_{mp}$) and the dynamic pressure of escaping planetary plasma ($\rho_{mp} v_{mp}^2$). The inclusion of the magnetospheric mass density term $\rho_{mp}$ is critical for unmagnetized planets with active atmospheric escape, as mass loading dynamically alters the boundary structure \citep{arridge2011, bagenal2013}.

While pressure balance dictates the static geometry, the internal global magnetic and convective dynamics significantly differ among magnetized planets. These dynamics are primarily described by two fundamental paradigms:
\begin{figure}[htpb]
\centering
\includegraphics[width=0.75\linewidth]{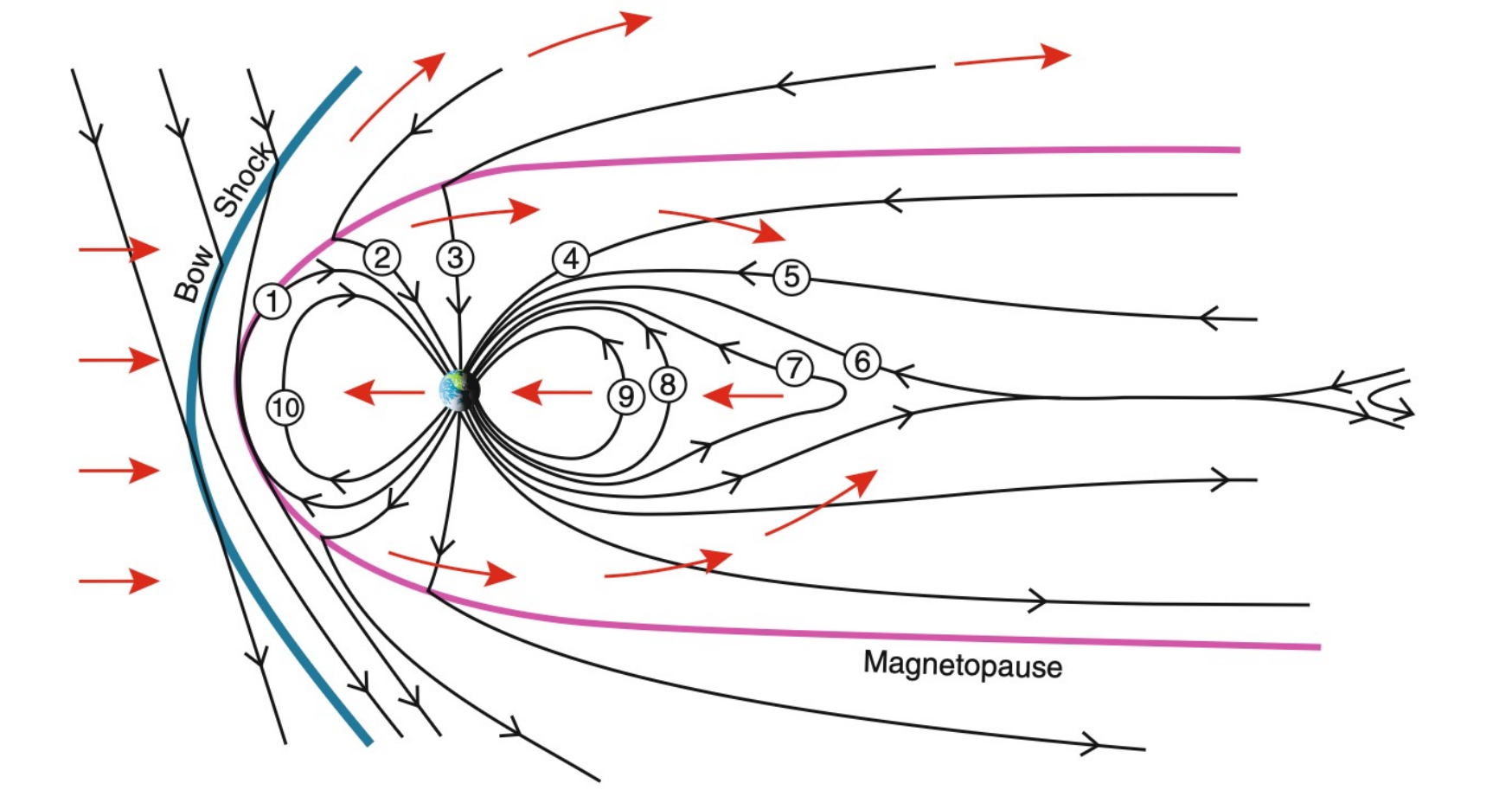}
\caption{The Dungey cycle at Earth, illustrating the solar wind-driven convective flow of magnetic flux \citep{bagenal2013}.}
\label{fig:bagenal3_dungey}
\end{figure}

The Dungey cycle \citep{Dungey1961} governs solar wind-driven magnetospheric convection, which is the dominant mechanism at Earth. In this cycle, the interplanetary magnetic field (IMF) brought by the solar wind reconnects with the planet's intrinsic magnetic field at the dayside magnetopause, a process initiated in step 1 of Fig. \ref{fig:bagenal3_dungey}. Once reconnected, these newly opened magnetic flux tubes are dragged by the solar wind over the polar caps and into the magnetotail, progressing through the spatial stages labeled 2, 3, 4, and 5. Within the magnetotail current sheet, a secondary magnetic reconnection event occurs at an X-line (step 6), which snaps the field lines closed and facilitates the conservation of the planet's total magnetic flux. This explosive reconnection process pinches off a blob of plasma (a plasmoid) which is ejected down the magnetotail (visible on the far right of the figure), while simultaneously driving a sunward convective return flow of mass-depleted flux tubes (steps 7, 8, 9, and 10) back toward the dayside inner magnetosphere to restart the cycle.

\begin{figure}[htpb]
\centering
\includegraphics[width=0.75\linewidth]{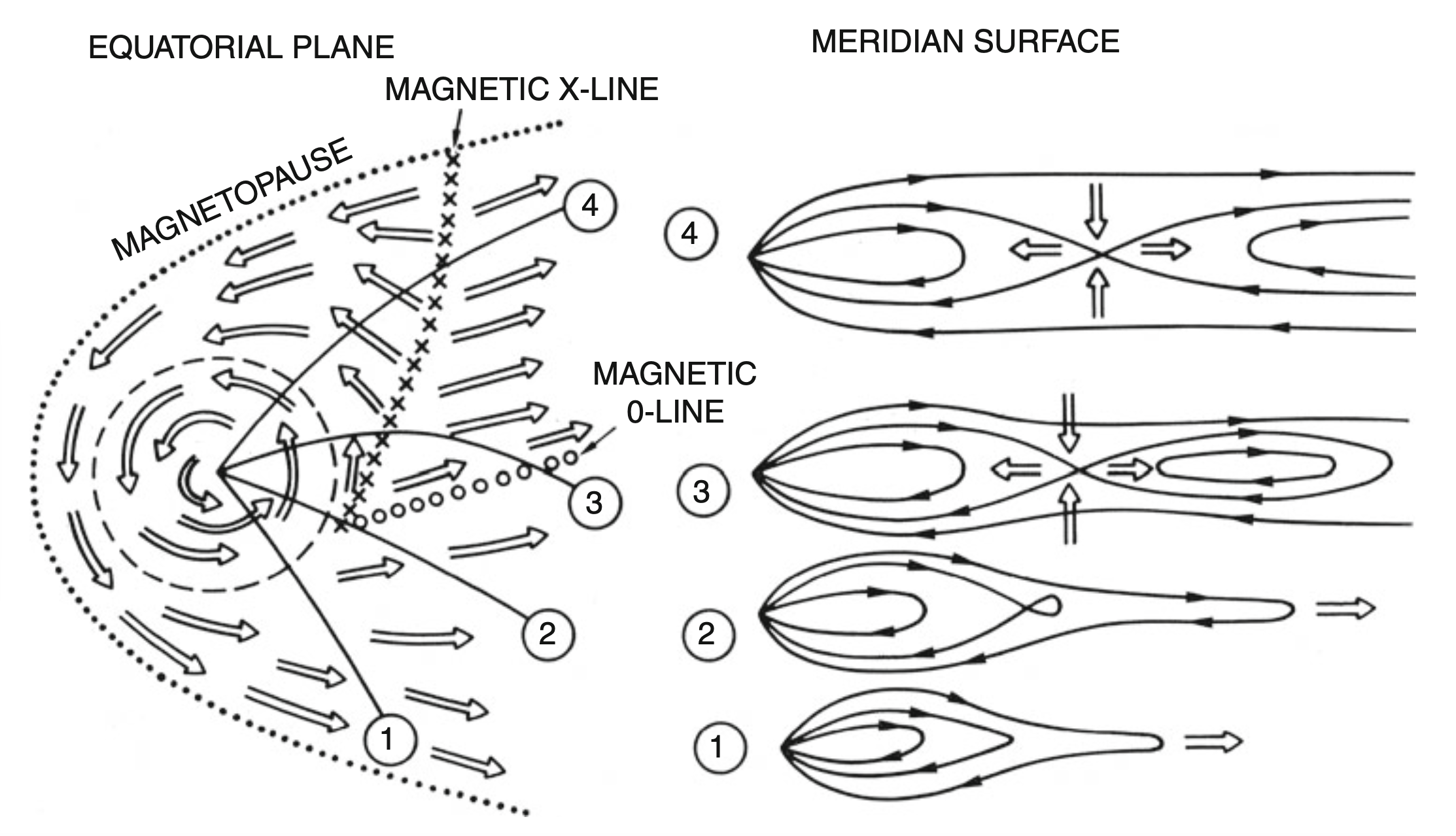}
\caption{The Vasyliunas cycle, depicting rotationally driven plasma transport and nightside plasmoid release. \citep{vasyliunas1983, bagenal2013}.}
\label{fig:bagenal4_vasyliunas}
\end{figure}

Conversely, the Vasyliunas cycle characterizes rotationally driven magnetospheres possessing significant internal plasma sources, such as Jupiter (mass-loaded by Io) and Saturn (mass-loaded by Enceladus). In these systems, locally ionized plasma is trapped on closed magnetic field lines and accelerated to the planet's corotation speed (position 1 in Fig. \ref{fig:bagenal4_vasyliunas}). The resulting centrifugal forces drive the plasma radially outward, causing the closed magnetic field lines to stretch severely into a magnetodisc configuration as they migrate through positions 2 and 3. This extreme radial stretching eventually triggers magnetic reconnection at a nightside x-point (step 4), where the field lines reconnect and detach a plasmoid that escapes down the magnetotail. After the release of this plasma mass, the resulting ``empty'' or mass-depleted flux tubes snap back toward the planet and rotate around to the dayside, where they are eventually re-filled with new plasma from internal sources to maintain the internally driven circulation.

\subsubsection{Multiscale Plasma Dynamics and Reconnection}

\begin{figure}
    \centering
    \includegraphics[width=0.75\linewidth]{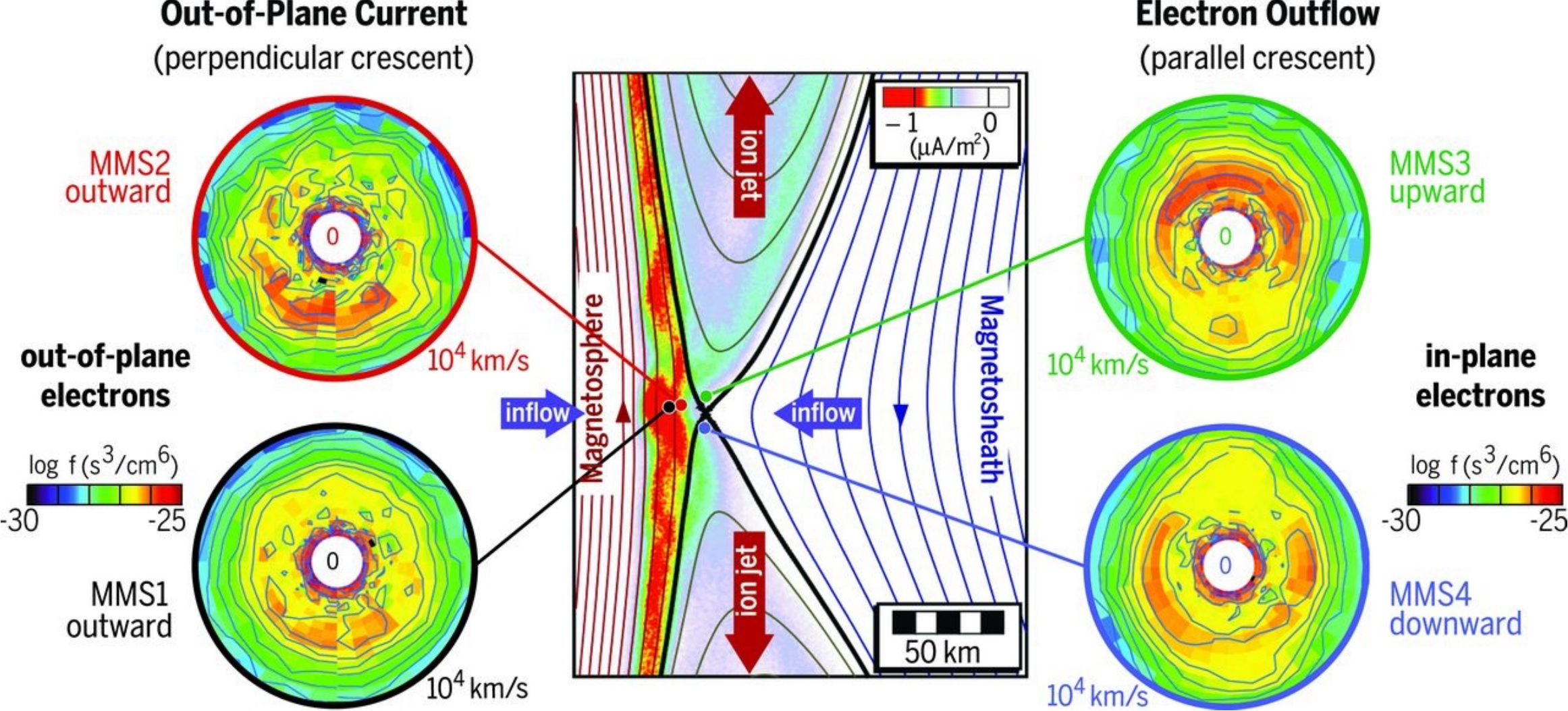}
    \caption{Magnetospheric Multiscale (MMS) mission measurements showing crescent shaped electron distribution functions close to the diffusion region of an active magnetic reconnection site at the dayside magnetopause \citep{burch2016}}
    \label{fig:burch_mms_crescent_distros}
\end{figure}

The interaction between the solar wind and planetary environments within the heliosphere occurs in a collisionless regime where a fluid description is valid only to the lowest order. Therefore, as noted in Table \ref{tab:upstream_parameters}, this approximation breaks down in boundary layers where kinetic scales (ion/electron gyroradii and inertial lengths) become comparable to the characteristic length scales on which the system varies \citep{Treumann2013,Nakamura2025}.  Energy and momentum transfer are therefore mediated by a combination of large-scale electromagnetic stresses and kinetic processes, including collisionless shocks, wave-particle interactions, and magnetic reconnection \citep{arridge2011,Paul_2022}. The process of magnetic reconnection at the dayside of Earth is typically believed to be patchy and intermittent leading to the formation of Flux Transfer Events (FTEs) that have been modelled as well as observed on numerous occasions \citep{Akhavan_Tafti_2018,Paul_2022,Paul_2023}. A direct observational demonstration of such non-MHD behavior was provided by in situ measurements of the electron diffusion region at the Earth's magnetopause by the Magnetospheric Multiscale (MMS) mission \citep{burch2016}. Figure \ref{fig:burch_mms_crescent_distros} shows the characteristic crescent-shaped electron velocity distribution functions observed during active magnetic reconnection. These highly non-Maxwellian distributions arise from the demagnetization of electrons within the diffusion region, driven by the reconnection electric field and meandering orbits across the sharp current layer. These observations confirm that global topological reconfiguration is intrinsically modulated by localized, electron-scale physics.

\subsubsection{Ionization, Mass Loading, and Escape}\label{sec:ionisation_massload_esc}
\begin{figure}[htpb] 
\centering \includegraphics[width=0.85\linewidth]{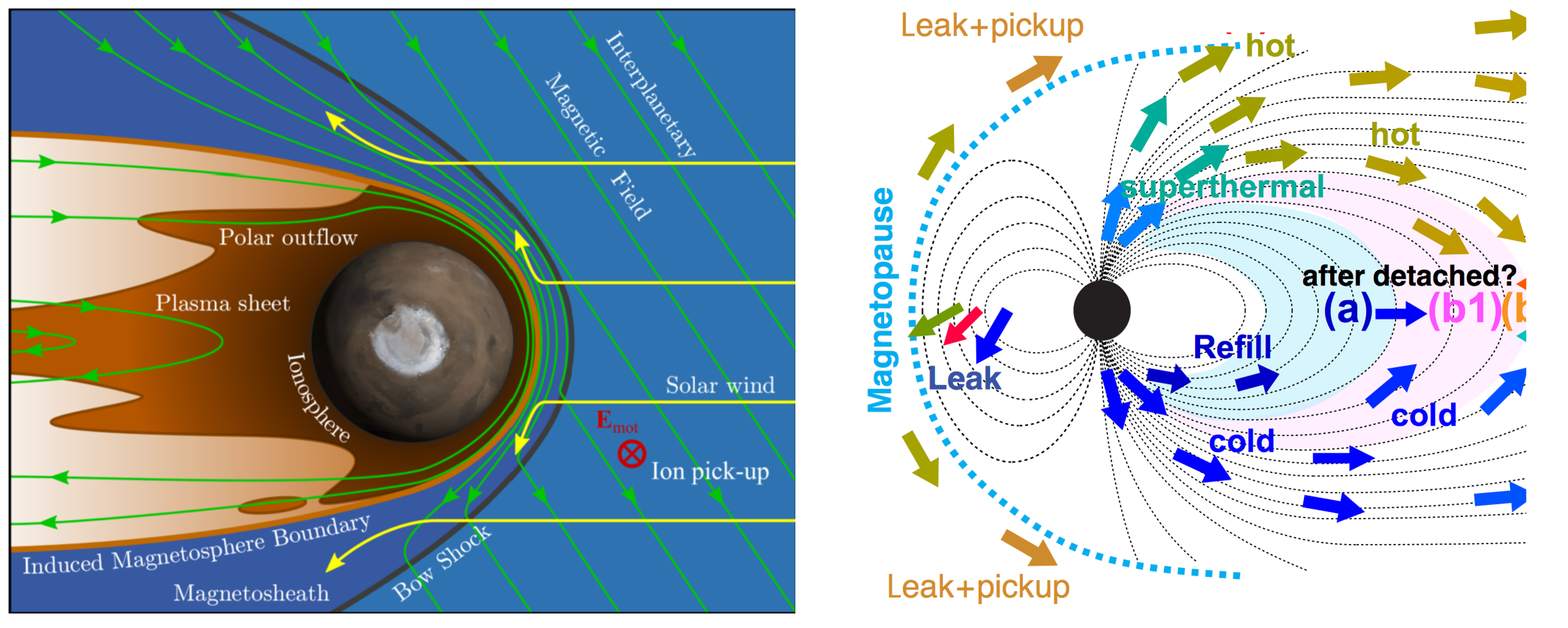}
\caption{The left panel, adapted from \citet{Ramstad2021}, illustrates the atmospheric escape processes of a non-magnetized planet interacting with ambient magnetized plasma, where ion pickup is expected to be the dominant mechanism. In contrast, the right panel, adapted from \citet{Yamauchi_2019}, highlights the processes relevant to a magnetized planet, emphasizing the combined roles of ion pickup, ion leakage, and polar outflows in the overall escape process.} 
\label{fig:escape_mechanisms} 
\end{figure}

The left panel of Figure \ref{fig:escape_mechanisms} illustrates the atmospheric escape processes typical of unmagnetized bodies, such as Venus and Mars, where the solar wind interacts directly with the planetary ionosphere and extended neutral exosphere. In these systems, the lack of a global intrinsic magnetic field allows the solar wind and its embedded interplanetary magnetic field (IMF) to impinge upon the upper atmosphere, where induction currents are generated to deflect the bulk of the flow around the planet. Neutral atoms and molecules that escape the lower atmosphere become ionized through EUV photoionization or charge exchange with solar wind protons. Once ionized, these newly created ions are immediately influenced by the motional electric field of the solar wind plasma, a process known as ``ion pickup'', which entrains them into the flow and causes them to extract momentum from the solar wind. This interaction results in the draping of the IMF around the obstacle, forming a comet-like magnetic tail through which picked-up planetary ions are continuously lost to the interplanetary medium.
In contrast, the right panel of Figure \ref{fig:escape_mechanisms} depicts the more specialized escape landscape of magnetized planets like Earth or Jupiter, where a global magnetic field carves out a magnetospheric cavity that shields the atmosphere from direct solar wind impingement. While the magnetopause acts as a primary barrier, it remains a site for mass and momentum exchange, allowing for ion leakage at the boundaries. In these magnetized systems, atmospheric escape is primarily facilitated through polar outflows, such as the polar wind, where wave-particle energization drives plasma upward along open field lines and into the magnetotail region. Additionally, internal plasma sources (such as volcanic moons) can significantly mass-load the magnetosphere, requiring efficient transport to the magnetotail where reconnection events eventually release the accumulated plasma as plasmoids. Consequently, while a global magnetic field provides a shield, it also creates structured channels like polar outflows and tail reconnection that define the unique escape signatures of magnetized planets.

\subsection{Moons of Solar System giant planets as SPI analogues}

The Galilean moons of Jupiter and to a lesser extent the larger moons of Saturn are perfect examples of analogues of star-planet interactions present in the Solar System. The Jovian system has inspired pioneering research which in many paved the way for consequent application of the same physical mechanisms to exoplanetary systems. 

The Galilean moons are the four largest moons of Jupiter, discovered independently by Galileo Galilei and Simon Marius in 1610, shortly after the invention of the telescope. These moons are important for SPI studies for several reasons: first of all, the inner three (Io, Europa, and Ganymede) are locked in the Laplace resonance, with the gravitational perturbations from neighboring moons supporting their non-zero eccentricities \citep{Peale1979}. This tidal interaction is a perfect analogue for possible influence of tidal forces on exoplanets in similar orbital configurations: Io, the innermost Galilean satellite, is the most volcanically active body in the Solar System due to constant distortion by the tidal forces. 

The Galilean satellites are also a perfect example of a magnetic analogue of SPI in the Solar System. Io is the object for which the formalism for the unipolar inductor model was initially developed \citep{Goldreich1969}. Io and Jupiter are connected by a flux tube similar to the flux tubes possibly connecting exoplanets and their host stars described in section \ref{sec:stellarhotspots}. Io is a constant source of plasma, which is the ionized material outgassed by its volcanoes. This plasma drives radio emission from Jupiter due to precipitation of energetic charged particles along the magnetic flux tubes connecting Io and Jupiter (e.g. \citealp{Saur2004}, which makes it a perfect analogue for planetary-induced activity on stars. In addition to that, energetic particles impacting the Jupiter's atmosphere leave footprints in its ionosphere visible in the UV light which trace the orbital motion of Io, Europa, Ganymede, and Callisto (e.g. \citealp{Hue2023, Rabia2025}). 

Io is also heated by the variable magnetic field along its orbit. The variability of the magnetic field in this case arises due to the tilt of Jupiter's magnetic dipole axis with respect to its rotation axis (and hence to Io's nearly equatorial orbital plane) of approximately 10$^\circ$ \citep{Acuna1976}. Although electromagnetic heating in Io is trumped by tidal heating \citep{Peale1979,Colburn1980}, it is a great analogy for electromagnetic induction heating of planets and their atmospheres considered in section \ref{sec:OhmicHeating}. In Europa, magnetic interaction with Jupiter is believed to power an large current in its salty ocean \citep{Gissinger2019}. In fact, this ocean was discovered due to magnetic interactions between Europa and Jupiter: the magnetic field induced in Europa could be best explained by the presence of a conducting layer of salty liquid water \citep{Kivelson2000}. In Ganymede, the varying magnetic field also induces a dipole moment in addition to the intrinsic magnetic field of the satellite \citep{Kivelson2002}.

\begin{figure}
    \centering
    \includegraphics[width=0.9\linewidth]{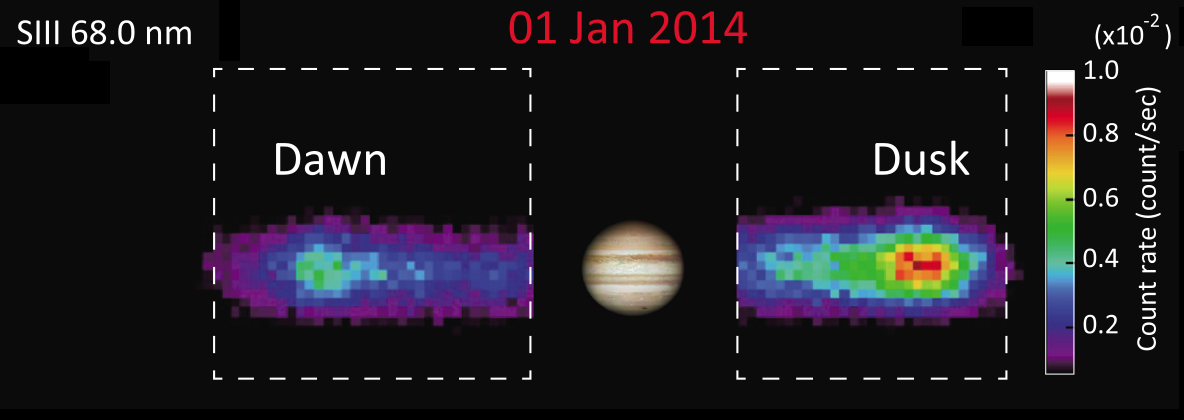}
    \caption{An example spectral image of the Io plasma torus acquired by Hisaki/EXCEED of SIII emission line at 68.0 nm. The Jupiter picture is for scale \citep{Murakami2016}}
    \label{fig:io_torus}
\end{figure}

Constant volcanic eruptions on Io also feed the Io neutral and plasma tori, which are analogues for tori observations around host stars of volcanically active exoplanets \citep{Bagenal2020}). 

On Enceladus, intensity of the water plumes which originate on the satellite's Southern pole correlates with the tidal stress (e.g. \citealp{Hedman2013}). This material then forms a torus somewhat similar to the Io plasma torus (e.g. \citealp{Felici2016, Hadid2026}). Titan also interacts with Saturn both via tidal \citep{Iess2012} and magnetic mechanisms \citep{Snowden2011}, but there exist no stable Io-like flux tube connecting Titan and Saturn.

\section{Applicability of Solar System Analogues to Exoplanetary SPI}
% %%%%%%%%%%%%%%%%%%%%%%%%%%%%%%%%%  Description  %%%%%%%%%%%%%%%%%%%%%%%%%%%%%%%%%%%%%%
% \begin{itemize}
%     \item \texttt{Contents: Which mechanisms scale? Which don’t?}
%     \item \texttt{\textbf{Contributors:} Arghyadeep Paul}
% \end{itemize}
% %%%%%%%%%%%%%%%%%%%%%%%%%%%%%%%%%%%%%%%%%%%%%%%%%%%%%%%%%%%%%%%%%%%%%%%%%%%%%%%%%%%%%%
\begin{figure}
    \centering
    \includegraphics[width=1\linewidth]{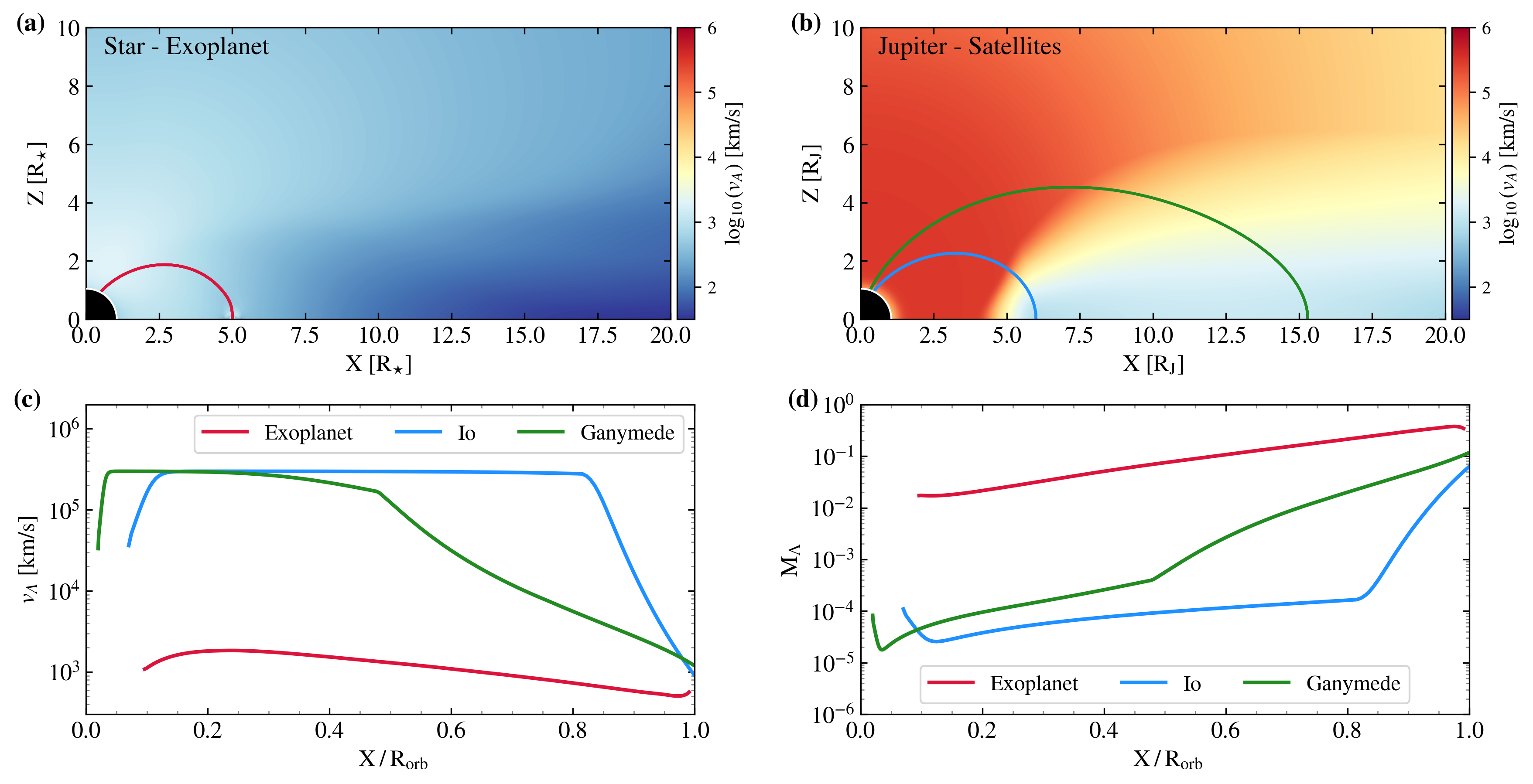}
    \caption{Comparison of Alfvén speed and Alfvénic Mach number profiles in SPMI and the Jovian system. Panel (a) shows the Alfvén speed distribution in the simulated SPMI system of \citet{Paul_2026}. The red curve traces a magnetic field line from an exoplanet at 5 stellar radii to the star. Panel (b) presents a meridional slice of Alfvén speed within Jupiter’s magnetosphere, highlighting the localized influence of the Io plasma torus. The blue and green lines mark magnetic field lines traced from the Io and Ganymede orbits, respectively. Panel (c) plots the Alfvén speed along these field lines and  panel (d) shows the corresponding Alfvénic Mach number profiles along the field lines. Adapted from \citet{Paul_2026}}
    \label{fig:Jupiter_vs_spi}
\end{figure}
\subsection{Are Jovian satellites true analogues of sub-Alfvénic SPMI?}
The Jovian magnetosphere has traditionally been regarded as the foundational archetype for sub-Alfvénic magnetic interactions, with the Jovian moons orbiting their host planet, Jupiter, in a sub-Alfvénic regime. However, it has started to become clearer that while analytical frameworks like the Saur model \citep{Saur2013, Kotsiaros2024} accurately describe satellite-magnetosphere interactions within the heliosphere, they significantly underestimate the power scales expected for tentative Star-Planet Magnetic Interactions (SPMIs). This discrepancy necessitates a critical evaluation of when Jovian analogs remain valid in the exoplanetary context and where the unique environment of the stellar wind dictates a departure from classical planet-satellite interactions.
To evaluate the validity of the analogy, \citet{Paul_2026} compared the Alfvén speed profiles in the Jovian magnetosphere with those characteristic of SPMI environments. In SPMIs, the stellar wind is generally expected to be relatively homogeneous in density on concentric spherical shells. Combined with the smooth radial variation of the stellar magnetic field, this leads to Alfvén speed profiles that typically vary by only a factor of about 4–5 between the planetary orbit and the stellar surface. Although localized inhomogeneities may arise around particularly active stars, these structures are usually spatially confined, and the overall environment remains comparatively uniform. Panel (a) of Figure \ref{fig:Jupiter_vs_spi} illustrates the Alfvén speed distribution in the simulated SPMI system. The red curve traces a magnetic field line connecting a close-in exoplanet at an orbital radius of $r_{\rm orb} = 5R_{\star}$ to the stellar surface. The resulting two-dimensional profile appears largely homogeneous between the planet and the star. The colorbar limits in panel (a) are intentionally set to relatively high values to maintain consistency with the Jovian case shown in the following panel.
By contrast, Jupiter’s magnetosphere exhibits a far more intricate structure. Although inner planetary magnetospheres are generally characterized by low plasma densities, Io’s intense surface activity, at an orbital distance of approximately $6R_J$, injects substantial material into Jupiter’s magnetosphere. This enhanced plasma density subsequently diffuses radially outward, forming the doughnut-shaped Io plasma torus. The torus also supports electrical currents that locally perturb Jupiter’s large-scale magnetic field, introducing pronounced radial and meridional anisotropies in both plasma density and magnetic field strength. Consequently, the Alfvén speed within Jupiter’s magnetosphere varies strongly with position. Panel (b) of Figure \ref{fig:Jupiter_vs_spi} shows a meridional slice of the Alfvén speed distribution in Jupiter’s magnetosphere. The plasma density in the torus is modeled following \citet{Su_2006, Lysak_2020} and \citet{Bagenal_2011}, with a minimum value of 0.01~${\rm cm}^{-3}$, while the magnetic field is represented as a superposition of Jupiter’s dipole field and the residual fields associated with the plasma torus, following \citet{Connerney_1981} with updated corrections from \citet{Connerney_2020}. As shown in panel (b) of Figure \ref{fig:Jupiter_vs_spi}, the Alfvén speed distribution within Jupiter’s magnetosphere is highly structured. The blue and green curves mark magnetic field lines traced from the orbital positions of Io and Ganymede, at $6R_J$ and $15.3R_J$, respectively. For consistency, the color scales in panels (a) and (b) are identical. Panel (c) presents the Alfvén speed profiles along the magnetic field lines shown in panels (a) and (b), plotted in matching colors. The horizontal axis is normalized to the orbital radius of the respective orbiting body. Near the orbital location ($X/R_{\rm orb} \sim 1$), the Alfvén speeds are comparable in the Jovian and SPMI cases. However, moving inward toward the central object, the Alfvén speed along the Jovian field lines diverges dramatically from the SPMI profile, with differences reaching nearly three orders of magnitude.
To estimate the Alfvénic Mach number along these field lines, \citet{Paul_2026} approximated the Alfvén wings as rigidly rotating structures embedded in the ambient plasma, with an angular velocity set by the relative motion of the orbiting body with respect to the surrounding flow. Using this framework, they computed the local Alfvénic Mach number, $M_A$, along each infinitesimal segment of the field lines connecting the orbiting body to its host. They adopted a relative velocity of 195~km~s$^{-1}$ for the exoplanet, corresponding to its orbital speed, while for Io and Ganymede they used relative plasma velocities of 57~km~s$^{-1}$ \citep{Futaana_2015} and 140~km~s$^{-1}$ \citep{Kivelson_2004}, respectively. The resulting Alfvénic Mach number profiles are shown in panel (d) of Figure \ref{fig:Jupiter_vs_spi}, with the horizontal axis defined analogously to panel (c). Near the planetary orbit, $M_A$ is approximately 0.4 in the SPMI case, whereas for the Jovian moons it is significantly lower, ranging from $10^{-2}$ to slightly above $10^{-1}$.
A central assumption of the analytical models is that the majority of the interaction power propagating along the Alfvén wings is generated by dynamics in the immediate vicinity of the planetary obstacle. In the Jovian system, where $M_A$ reaches appreciable values only near the moons and rapidly declines toward Jupiter, supports the expectation that most of the power is produced locally near the moons. In contrast, panel (d) of Figure \ref{fig:Jupiter_vs_spi} shows that in SPMI systems, although $M_A$ decreases along the field lines toward the star, the decline is relatively modest. The minimum $M_A$ values in SPMIs remain comparable to those near the orbits of the Jovian moons. As a result, the Alfvén wings in SPMIs can act as extended obstacles, with power generation distributed along a significant fraction of the star–planet magnetic connection. This interpretation is consistent with simulation results of \cite{Paul_2026}, which reveal a gradual increase in inward-directed Poynting flux with distance from the planet.

\subsection{Influence of Solar Variability on heliospheric planets}

The Solar System offers the only setting in which the response of planetary environments to changes in stellar forcing can be measured \textit{in situ}. In this subsection, we summarise how the magnetospheres, ionospheres, and atmospheric escape of the Solar System planets respond to solar variability, from transient solar wind disturbances to the solar cycle. We first discuss the response of induced magnetospheres, then that of Earth, of the giant planets, and of Venus, Mars, and Titan, and finally turn to the implications for exoplanets orbiting more active stars.

Because the nature of the obstacle to the solar wind is fundamentally different, intrinsic and induced magnetospheres respond differently to solar cycle changes in EUV flux and solar wind dynamic pressure. As discussed by \citet{Bertucci2011}, the transfer of momentum and energy from the solar wind is highly critical to the atmospheric evolution of unmagnetized objects. 

\begin{figure}[htpb]
    \centering
    \includegraphics[width=0.5\linewidth]{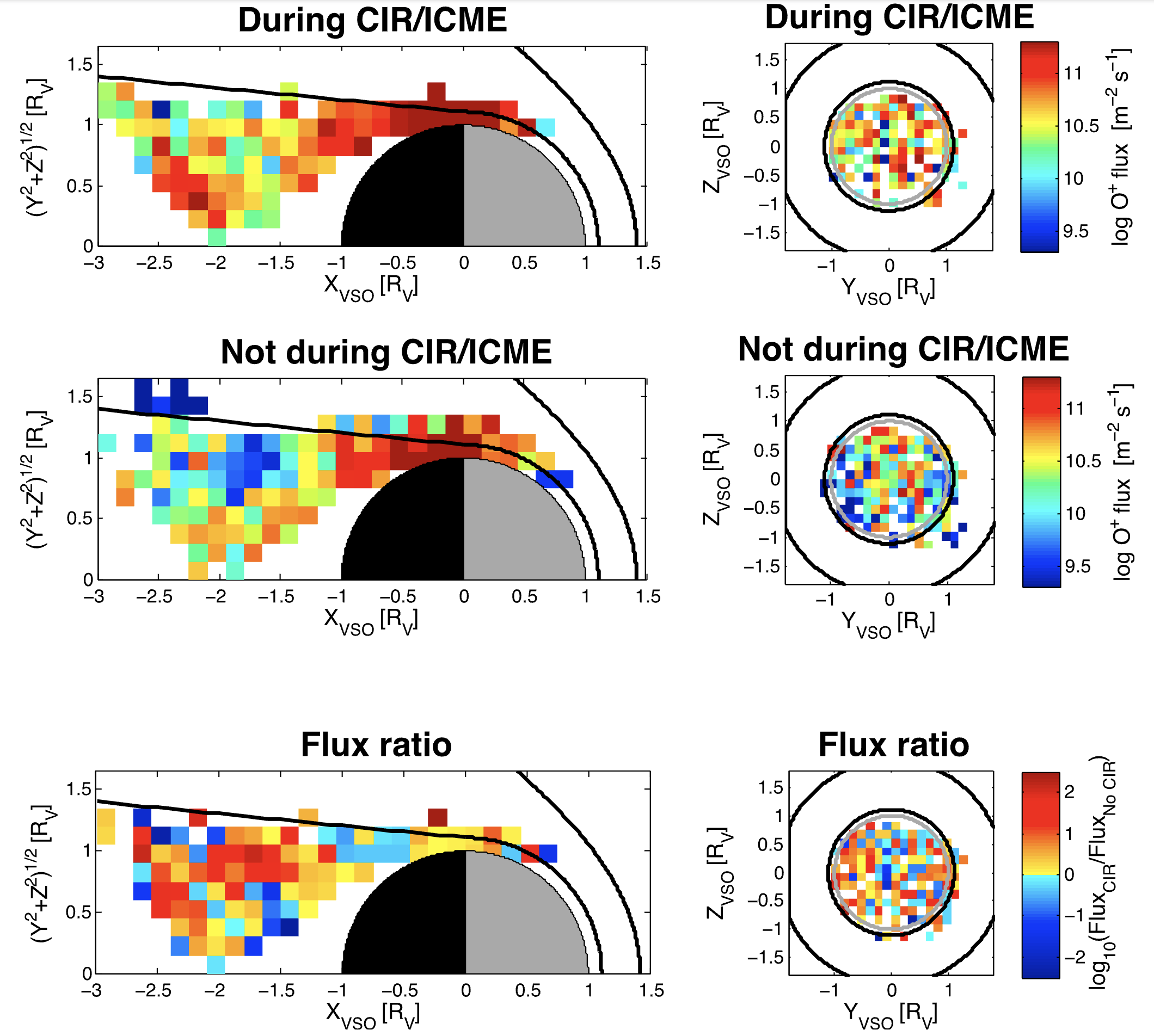}
    \caption{Antisunward planetary O$^+$ fluxes at Venus, measured by the Venus Express spacecraft, during CIR/ICME impacts (top), quiet solar wind (middle), and their ratio (bottom) (adapted from \citet{edberg2011}). Distributions are shown in cylindrical VSO coordinates (left) and the VSO $y-z$ plane for $-3 R_V < x < -1 R_V$ (right); in the Venus Solar Orbital (VSO) frame, the $x$-axis points from Venus towards the Sun, the $z$-axis is perpendicular to the orbital plane of Venus (northward), the $y$-axis completes the right-handed system (opposite to the orbital motion of the planet), and $R_V$ is the radius of Venus. Black lines denote average Bow Shock and Induced Magnetospheric Boundary (IMB) locations \citep{Martinecz_2008}; grey circles mark the planetary limb.}
    \label{fig:O_flux}
\end{figure}

For induced magnetospheres, solar variability strictly governs boundary layer altitudes and directly modulates atmospheric escape rates. As illustrated in Figure \ref{fig:O_flux}, spacecraft observations demonstrate that atmospheric escape rates at Venus and Mars increase significantly during the passage of transient solar wind disturbances, such as Corotating Interaction Regions (CIRs; compressed regions of plasma and magnetic field that form where fast solar wind streams from coronal holes overtake the slower wind ahead of them, and which corotate with the Sun) and Interplanetary Coronal Mass Ejections (ICMEs; the interplanetary counterparts of coronal mass ejections, i.e., large clouds of magnetised plasma expelled from the solar corona) \citep{Bertucci2011, edberg2011}.
The top panels of the figure show a clear enhancement in antisunward planetary O$^+$ fluxes during these solar transients compared to quiet conditions (middle panels). This intensification is particularly evident in the bottom panel, where the flux ratio highlights a significant increase in ion stripping downstream of the obstacle within the wake ($y-z$ plane). During periods of high solar activity, the enhanced EUV flux also leads to the further ionization and expansion of the upper atmosphere; this expands the cross-section of the obstacle, exposing more ionized particles to direct stripping by the dense, fast solar wind associated with solar maximum conditions. Consequently, the planetary O$^+$ ions are accelerated out as they are swept into the magnetotail.

In many cases, the non-thermal escape of planetary ions stoichiometrically limits the retention of water, establishing a direct link between plasma escape and global climate evolution. For exoplanets residing in the habitable zones of highly active host stars, such as the rapidly rotating M dwarfs discussed in the previous section, the intense stellar XUV flux and frequent coronal mass ejections present a severe erosive environment. Under these saturated dynamo conditions (low Rossby numbers), the presence or absence of an intrinsic magnetic shield therefore may become a crucial factor in atmospheric erosion. There are additional pathways for atmospheric loss even for magnetised planets and this has been elaborated further in section \ref{sec:ionisation_massload_esc} and also in the following paragraphs.

\paragraph{The Earth}

At Earth, the magnetospheric dynamics are the archetype of the solar wind-driven Dungey cycle (Figure \ref{fig:bagenal3_dungey}). The energy input from the solar wind strongly controls the efficiency and location of magnetic reconnection at both the dayside magnetopause and the magnetotail \citep{nagai21}. As demonstrated by measurements of the Magnetospheric Multiscale (MMS) mission, electron-scale dynamics regulate this reconnection during highly active periods such as dayside reconnection, geomagnetic substorms and storms, resulting in rapid topological reconfigurations \citep{burch2016}.

% \begin{figure}[htpb]
%     \centering
%     \includegraphics[width=0.75\linewidth]{bagenal3.pdf}
%     \caption{The Dungey cycle at Earth, illustrating the solar wind-driven convective flow of magnetic flux \citep{bagenal2013}.}
%     \label{fig:bagenal3_dungey}
% \end{figure}

\begin{figure}[htpb]
    \centering
    \includegraphics[width=0.6\linewidth]{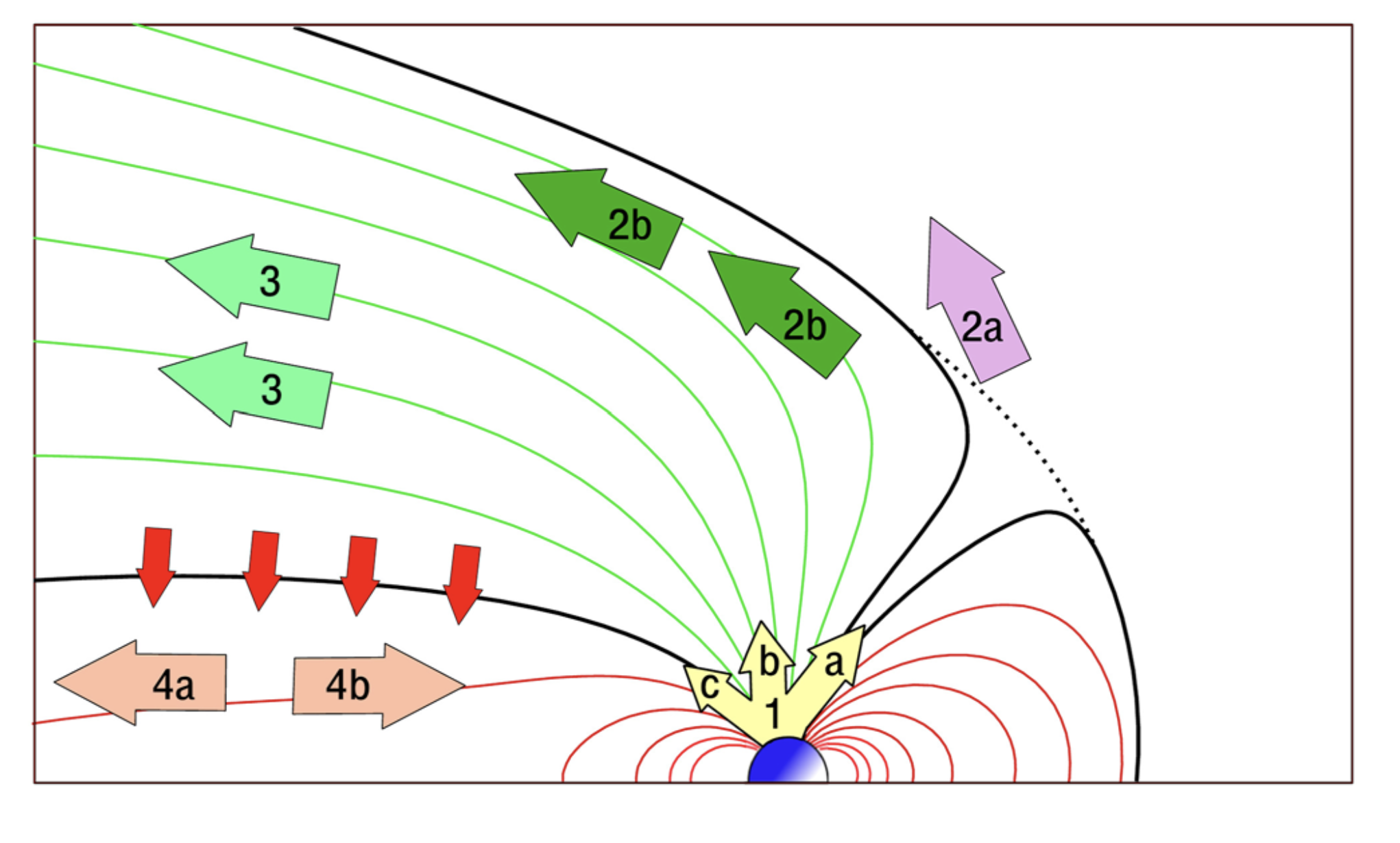}
    \caption{Sketch of the terrestrial magnetosphere and upwelling ionospheric ion transport routes. Green and red lines correspond to open and closed magnetic field lines, respectively. Ionospheric outflow (yellow arrows) occurs via the cleft ion fountain (1a), polar wind (1b), and auroral night zone (1c). Atmospheric escape to the magnetosheath occurs via direct cusp escape (2a) and the plasma mantle (2b). Lower energy ions fill the lobes (3) and feed the plasma sheet (red arrows), undergoing tailward and earthward transport (4a, 4b) \citep{dandouras2021}.}
    \label{fig:dandouras1_escape}
\end{figure}

Crucially, this solar wind-driven convection governs terrestrial atmospheric escape. Ionospheric outflow is distributed across several complex transport routes, as sketched in Figure \ref{fig:dandouras1_escape} as well as the right panel of Figure \ref{fig:escape_mechanisms}. Upwelling ions escape via the cleft ion fountain, polar wind, and auroral night zone outflows. Once in the magnetosphere, atmospheric plasma can escape directly into the magnetosheath (the region of shocked, slowed, and compressed solar wind plasma between the bow shock and the magnetopause) from the cusp (funnel-shaped regions at high latitudes on the dayside, where magnetic field lines converge and solar wind plasma can directly access low altitudes) or via the plasma mantle (a layer of tailward-flowing plasma on open field lines just inside the magnetopause, poleward and tailward of the cusps). Lower energy ions fill the magnetospheric lobes (the low-density regions of open magnetic field lines in the magnetotail) and subsequently feed the plasma sheet (the layer of hotter, denser plasma surrounding the equatorial current sheet of the magnetotail), where both tailward ejection and earthward injection can take place \citep{dandouras2021}. 

\begin{figure}[htpb]
    \centering
    \includegraphics[width=0.5\linewidth]{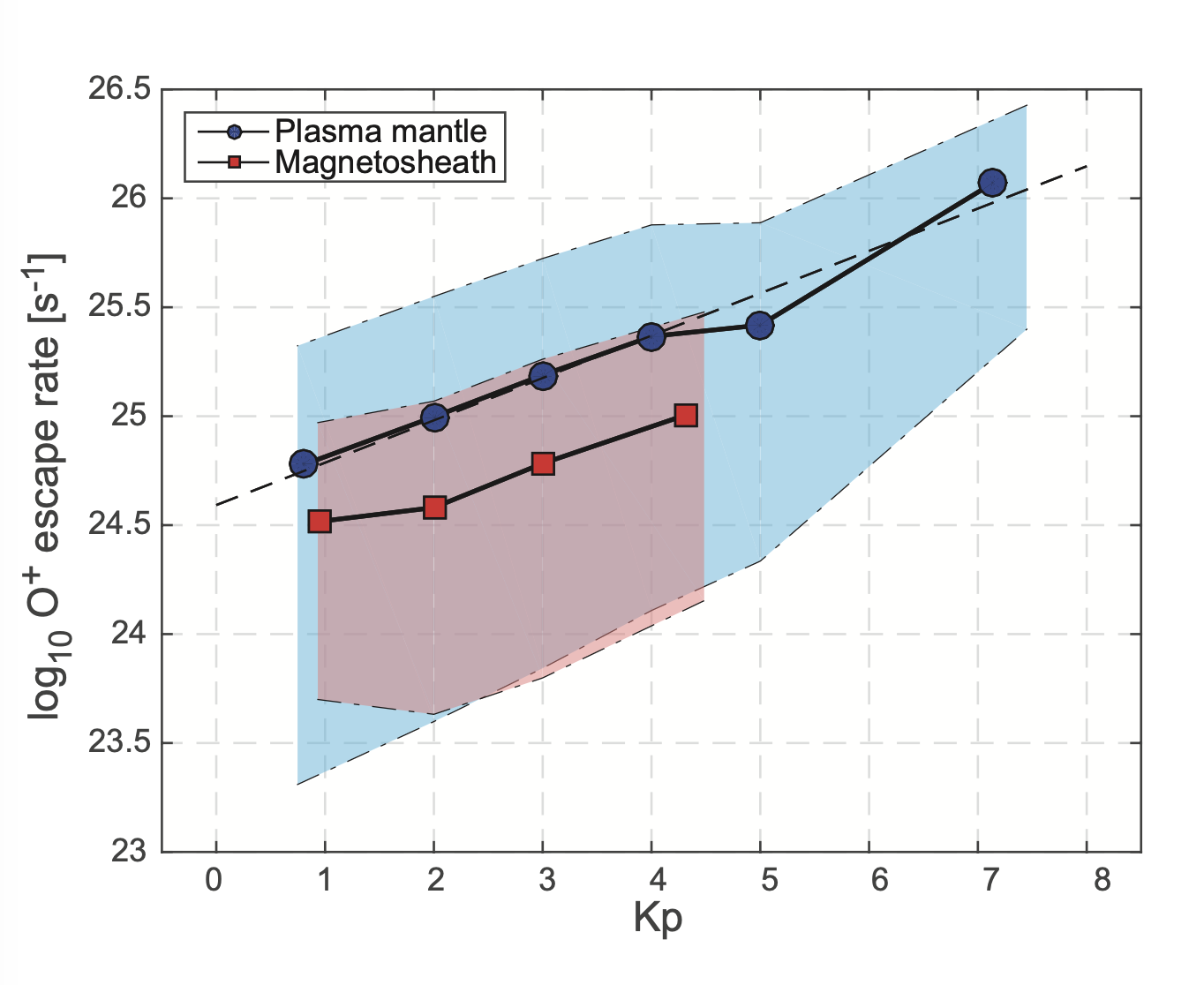}
    \caption{The average O$^+$ escape rates for the plasma mantle (solid blue line/circles) and the dayside magnetosheath (solid red line/squares) as a function of the geomagnetic Kp index \citep{slapak2017}.}
    \label{fig:slapak1_Oplus}
\end{figure}

The rate of this outflow is highly sensitive to solar activity. For instance, the flux of escaping O$^+$ ions in the plasma mantle is nearly three times larger than in the dayside magnetosheath, and both exhibit a strong positive correlation with geomagnetic activity (Kp index) \citep{slapak2017} (Figure \ref{fig:slapak1_Oplus}). Extrapolating these mantle and magnetosheath escape rates to extreme solar wind events yields O$^+$ loss rates on the order of $10^{26}$ s$^{-1}$ (compared to quiet time values of $10^{24}$ s$^{-1}$) \citep{slapak2017}. Assuming this escape is heavily modulated by solar EUV flux, atmospheric stripping would have been significantly stronger in the early Solar System when the young Sun was far more active \citep{dandouras2021}.

\paragraph{Jupiter and Saturn}

Unlike Earth, the giant planets Jupiter and Saturn are dominated by the Vasyliunas cycle (Figure \ref{fig:bagenal4_vasyliunas}). Material originating from their respective moons and rings is ionized and rapidly incorporated into their rotating magnetospheres. The balance between outward centrifugal forces and inward magnetic stresses leads to a highly stretched magnetodisc configuration supported by an equatorial, azimuthal current sheet \citep{achilleos2021}. At Jupiter, periodicities arising from non-axisymmetric momentum transport, and at Saturn, the rotating ``camshaft'' signal (a quasi-uniform magnetic perturbation field in the equatorial inner magnetosphere that rotates with a period close to, but slightly different from, the planetary rotation period, like a rotating camshaft driving periodic motions of the surrounding magnetodisc), strongly govern magnetospheric periodicities \citep{southwood2007}. Interestingly, despite being internally driven, the auroral and kilometric radio emissions (radio emission at kilometre wavelengths, i.e., at frequencies of $\sim$10~kHz to $\sim$1~MHz, generated by the electron cyclotron maser instability above the auroral regions; see Sect.~\ref{sec:emissionnearplanet}) at these planets remain highly sensitive to external solar wind dynamic pressure variations \citep{cecconi2022, burne23}.

% \begin{figure}[htpb]
%     \centering
%     \includegraphics[width=0.75\linewidth]{bagenal4.png}
%     \caption{The Vasyliunas cycle, depicting rotationally driven plasma transport and nightside plasmoid release \citep{bagenal2013,vasyliunas1983}.}
%     \label{fig:bagenal4_vasyliunas}
% \end{figure}

\begin{figure}[htpb]
    \centering
    \includegraphics[width=0.4\linewidth]{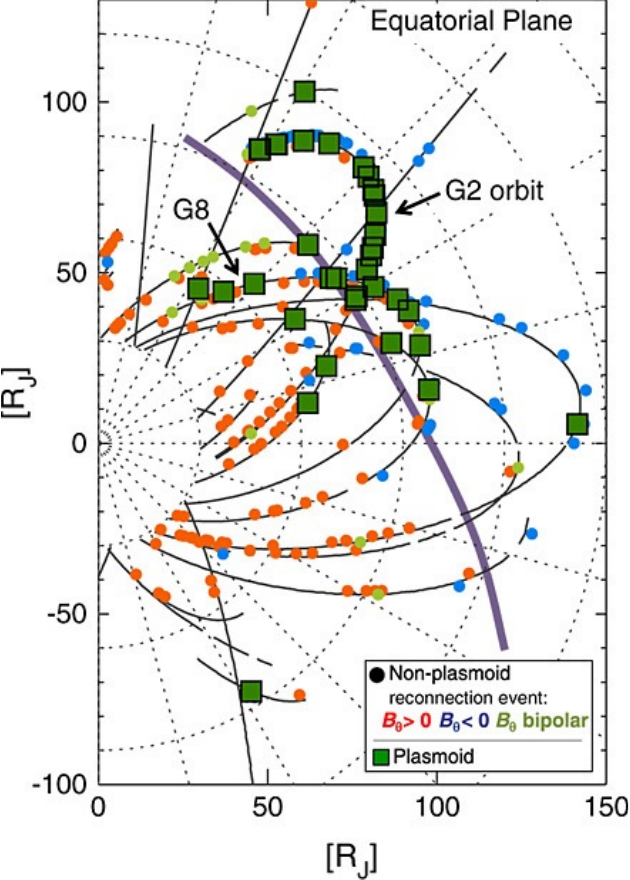}
    \caption{Distribution of tailward-moving plasmoids (green squares) observed in Jupiter’s nightside magnetosphere. Also shown are the locations of other observed reconnection signatures (colored dots, with color indicating the dominant sign of $B_\theta$ during each event) \citep{vogt2014}.}
    \label{fig:vogt1_plasmoid}
\end{figure}

Plasma escape at these giant planets follows this rotationally driven cycle, which depends heavily on the mass-loading rate from internal sources \citep{vogt2014}. The culmination of the Vasyliunas cycle is the pinching off of the distended magnetotail and the release of massive plasmoids down the tail, a process successfully observed during the Galileo mission (Figure \ref{fig:vogt1_plasmoid}). However, the susceptibility to solar variability differs between the two giants. At Jupiter, internal processes overwhelmingly dominate, and only extreme solar wind conditions can significantly perturb the global system \citep{nichols2007}. At Saturn, the solar wind exerts a more measurable impact, with solar wind-driven (Dungey-like) variations sometimes temporarily overriding the Vasyliunas cycle dynamics \citep{thomsen2015}.

\paragraph{Venus, Mars, and Titan: Induced and Transition Magnetospheres}

For unmagnetized planetary bodies, the structure of the induced magnetosphere is also intimately tied to upstream solar conditions. At Venus, the global induced magnetosphere and ionosphere are strictly dependent on the solar cycle \citep{marsvenusagumonograph,lican2015,futaana2017, changvenus2018}. As depicted in the left panel of Figure \ref{fig:escape_mechanisms}, the interaction creates a sharp Induced Magnetosphere Boundary (IMB) that effectively shields the lower atmosphere. However, during the passage of transient solar phenomena such as Interplanetary Coronal Mass Ejections (ICMEs) and Corotating Interaction Regions (CIRs), plasma escape rates are drastically enhanced \citep{edberg2011}. 

% \begin{figure}[htpb]
%     \centering
%     \includegraphics[width=0.75\linewidth]{venusinteraction_albers.jpg}
%     \caption{Conceptual overview of the Venusian induced magnetosphere. Key plasma regions include: 1. the unperturbed solar wind, 2. the interplanetary magnetic field (IMF), 3. the Bow shock, and 4. the induced magnetospheric boundary \citep{albers2024}. \textbf{I will remove this when I include the two regimes in the plot from my slide above}}
%     \label{fig:albers_venus}
% \end{figure}

Similarly, the Martian induced magnetosphere expands in tandem with its exosphere near perihelion, which effectively increases the cross-sectional area of the escaping plasma region \citep{brain2005, clarke24}. Spacecraft observations confirm that, much like Venus, Martian atmospheric escape rates spike during extreme solar events due to the enhanced penetration of the motional (or `convective') electric field of the solar wind, $\mathbf{E} = -\mathbf{v}_{sw} \times \mathbf{B}_{sw}$, which accelerates newly ionised planetary particles away from the planet (ion pickup; Sect.~\ref{sec:ionisation_massload_esc}) \citep{edberg2010,jakosky2015}.

Finally, Saturn's largest moon, Titan, serves as a unique ``transition'' object in the study of planetary plasma interactions. Titan is the only moon in the Solar System to host an optically thick, nitrogen-dominated atmosphere \citep{horst2017}, yet it possesses virtually no intrinsic magnetic field \citep{wei2010}. Because of its orbital distance, Titan is typically embedded within Saturn's oscillating magnetodisc. Consequently, rather than interacting with the supersonic solar wind, Titan's atmosphere interacts with Saturn's corotating, sub-Alfvénic magnetospheric plasma \citep{arridge2011}. 

However, Titan is exposed to highly variable upstream conditions driven by both solar and planetary forcing. Cassini observations revealed that the classical magnetic field ``draping'' pattern, expected for an induced interaction, is present only about 50\% of the time \citep{kaba2017}. This is largely attributed to ``fossil fields'': because the transit time of mass-loaded magnetic flux tubes through Titan's dense induced ionosphere is quite slow (20 minutes to 2 hours), the ionosphere essentially ``remembers'' previous upstream magnetic topologies \citep{bertucci2008}. Furthermore, the timescales for upstream changes range from minutes to hours. Under high solar wind dynamic pressure, Saturn's magnetopause can be compressed inside Titan's orbit, abruptly exposing the moon directly to the supersonic solar wind and forcing it to rapidly transition from a sub-Alfvénic moon interaction to a Venus-like induced magnetosphere \citep{bertucci2015}.

%In the context of exoplanets
\paragraph{Implications for exoplanets beyond the Solar System regime}
While the Solar System represents a ``moderate'' regime of SPI, exoplanets orbiting active M dwarfs (e.g.,  Proxima Centauri b) can operate in a regime where stellar wind dynamic pressures are typically $10^3$ to $10^4$ times higher than those experienced by Earth \citep{Garaffo_2016, Airapetian2017}. Crucially, these environments are characterized by extreme spatial non-uniformity. As a planet traverses along its orbit, it may encounter total pressure variations of a factor of 10 to 1000 within a single orbital period. This induces a violent `breathing' of the planetary magnetosphere, where the standoff distance compresses and expands by a factor of $\sim 3$ on timescales as short as one day. The consequences of this extreme compression are not merely morphological. At Earth, the dissipation of solar wind energy via Joule heating in the polar caps is typically $\sim 10$ erg cm$^{-2}$ s$^{-1}$ during active periods. For Proxima b, however, ionospheric electric currents induced by the compressed magnetic field can dissipate energy at a rate of $\sim 10^4$ erg cm$^{-2}$ s$^{-1}$ \citep{Airapetian2017}. Furthermore, at an orbital distance of $\sim 0.05$ au, massive energy injection from Extreme Ultraviolet (XUV) fluxes, which can be 60 times higher than the Solar flux at Earth, creates steep electron temperature gradients. These gradients generate powerful ambipolar electric fields above the exobase, driving a persistent and heavy outflow of O$^+$ and N$^+$ ions. Under these conditions, it is estimated that a 1-bar oxygen-rich atmosphere could be completely stripped away in as little as 10 Myr, significantly shortening the window for the potential development of life \citep{Airapetian2017}.

While Venus and Mars provide examples of ``ion stripping,'' many close-in exoplanets experience a more extreme form of mass loss known as hydrodynamic escape. In these systems, the intense XUV irradiation (in particular its X-ray and EUV part) from the young or active host star heats the upper atmosphere up to $10^4$ K, causing it to expand as a ``planetary wind'' that is able to overcome the planetary gravity. The interaction between this outgoing planetary wind and the incoming stellar wind creates massive, comet-like tails of neutral and ionized hydrogen. This process has been directly observed via Lyman-$\alpha$ transit spectroscopy for planets like GJ 436b and GJ 3470b, where the absorption depths reach $\sim$35\% (GJ~3470b) to $\sim$56\% (GJ~436b), indicating an exosphere that extends many planetary radii beyond the Roche lobe \citep{Ehrenreich_2015,Bourrier_2018}.

%%%%%%%%%%%%%%%%%%%%%%%%%%%%%%%%%%%%%%%%%%%%%%%%%%%%%%%%%%%%%%%%%%%%%%%%%%%%%%%%%%%%%%

\subsection{Expected SPI Dependence on Heliosphere-analogous Stellar Activity}
%%%%%%%%%%%%%%%%%%%%%%%%%%%%%%%%%  Description  %%%%%%%%%%%%%%%%%%%%%%%%%%%%%%%%%%%%%%
% \begin{itemize}
%     \item \texttt{Contents: Analogues from solar activity}
%     \item \texttt{\textbf{Contributor:} Shyama Narendranath}
% \end{itemize}

Stellar activity manifests through diverse phenomena spanning multiple timescales: from impulsive flares lasting minutes to hours, to coronal mass ejections (CMEs) propagating through interplanetary space over days, to the gradual evolution of high-energy radiation output over billions of years. Stellar flares represent sudden releases of magnetic energy that dramatically increase a star's output across the electromagnetic spectrum, from radio wavelengths to X-rays and gamma rays.

Solar energetic particles (SEPs); high-energy protons, and heavier ions accelerated during flares and CME-driven shocks, can penetrate deep into planetary atmospheres, driving ionization, dissociation, and complex photochemical cascades. The atmospheric effects of SEPs depend on particle energy spectra, atmospheric composition and density, and the presence of magnetic shielding. For planets with thin atmospheres or weak/no magnetic fields, SEPs can reach surface levels, posing direct biological hazards while simultaneously driving atmospheric chemistry relevant to prebiotic synthesis.

Beyond direct radiation effects, flares may also induce interior heating in rocky planets through associated CMEs. Recent modeling suggests that magnetic flux carried by flare-associated CMEs can result in planetary interior heating via Ohmic dissipation, with the majority of heat produced within days or weeks following impact \cite{Cohen_2024,Strugarek_2025}. This heating mechanism may be sufficient to drive geological processes, facilitate volcanism, and promote outgassing, potentially replenishing atmospheres even as direct flare and CME impacts erode them.

Coronal mass ejections represent large-scale expulsions of plasma and magnetic field from stellar coronae. When CMEs impact planetary magnetospheres, they can compress magnetic field lines, enhance particle precipitation into atmospheres, and drive atmospheric erosion through ion pickup processes. The frequency and intensity of CME impacts depend critically on both stellar activity levels and the geometry of stellar magnetic fields. The Solar System provides the only laboratory where we can directly study the outcomes of stellar activity effects on planetary atmospheres over 4.5 billion years of evolution. Observations of the present-day terrestrial planets reflect the integrated effects of stellar activity, planetary properties, and geological processes operating over Solar System history.

\paragraph{Solar Flare Effects}
X-ray fluorescence (XRF) and scattering have been observed from all planets and several small bodies in the Solar System. X-ray fluorescence spectroscopy is a well-established tool to determine the elemental abundances of airless bodies and could in principle also be used for planetary atmospheres \citep{AIKIN1970}. Minor constituents like Argon are observed from planetary atmospheres at the XRF line energies which are distinct from the solar spectrum and could be distinguished even in the presence of a scattered solar X-ray continuum. Airless bodies like the Moon and Mercury exposed to long-term X-ray irradiation suffer from degraded surface organics; such intense irradiation severely limits the possibility of finding complex organic molecules on the surface of airless exoplanets orbiting active M dwarfs. Solar flares modulate the energy input into planetary atmospheres, leading to changes in the neutral density profiles and affecting atmospheric photochemistry. On Earth, flare EUV and soft X‑ray increases cause rapid total electron content (TEC) and E‑region ionization enhancements, localized heating of the lower thermosphere, and increases in nitric oxide affecting thermospheric cooling. These events trigger traveling atmospheric disturbances (TADs) with dayside density perturbations; model–data studies identify EUV in the 15–35 nm range as a primary driver of TEC and show that EUV ``late phases'' can prolong electron‑density recovery by about nine hours for some events \citep{Liu2024}. Comparative modeling and observations show that electron density increases occur across Earth, Venus, and Mars but with different magnitudes, thermal responses, and upward plasma fluxes. While Earth and Venus can share similar ion temperature responses, Mars exhibits distinct electron temperature behavior and potentially enhanced short‑term escape via flare‑driven upward drifts. Venus displays flare‑induced emissions such as changes in the 5577 Angstrom oxygen green line, and Mars often forms pronounced E‑layer peaks during flares, as observed by the MAVEN mission \citep{https://doi.org/10.1002/2015GL065271}.

\paragraph{Solar Wind Interactions and Charge Exchange} 
The solar wind acts as a continuous stream of magnetized plasma that carves out the heliosphere and interacts directly with planetary obstacles. A primary consequence of this interaction is Solar Wind Charge Exchange (SWCX), which occurs when highly charged solar wind ions (e.g., O$^{7+}$, C$^{6+}$) collide with neutral atoms in a planetary exosphere or cometary coma. This results in the capture of an electron into an excited state, followed by the emission of a soft X-ray photon. SWCX was first clearly identified in X-rays from comets, notably Comet Hyakutake in 1996 by ROSAT \citep{Lisse1996}. Cometary comae provide dense neutral targets producing strong, extended X-ray emission. In the context of planets, SWCX provides a unique diagnostic of the ``collision zone'' between the solar wind and the neutral exosphere. For Earth, this emission is concentrated in the magnetosheath and cusps, acting as a tracer for the magnetopause position. For unmagnetized bodies like Mars and Venus, SWCX is a significant contributor to the X-ray halo and serves as a direct proxy for atmospheric escape rates, as the emission intensity correlates with the neutral gas density extending beyond the ionopause \citep{Cravens1997}.

\paragraph{Surface Interactions and In-Situ Chemistry} 
Beyond atmospheric stripping, the solar wind drives chemical evolution on the surfaces of airless bodies. A significant discovery in the last decade is the in-situ production of hydroxyl (OH) and water (H$_2$O) on the Moon and Mercury via solar wind interaction. Protons (H$^+$) from the solar wind implant themselves into the oxygen-rich silicate minerals of the regolith. These protons can break metal-oxide bonds and bond with the freed oxygen to form hydroxyl, which may further evolve into water molecules \citep{Pieters2009}.This mechanism suggests that ``water'' on airless exoplanets may not solely be delivered by comets but can be continuously ``baked'' into the surface by the stellar wind itself. This flux is highly variable, depending on the stellar wind density and the shielding provided by any crustal magnetic fields.

\paragraph{Solar Energetic Particles (SEPs) and Radiation Environments}
The near-planetary radiation environment is highly sensitive to SEPs, which dictate the habitability of the upper atmosphere and surface. SEPs drive ``Ozone depletion'' events on Earth-like planets by producing odd nitrogen (NO$_x$) and hydrogen (HO$_x$) species that catalytically destroy O$_3$ \citep{Thomas2007}. In exoplanetary systems, particularly those around M dwarfs where flare frequency is orders of magnitude higher than the Sun, the cumulative effect of SEPs could lead to the permanent removal of protective ozone layers, exposing the surface to lethal UV radiation even during ``quiet'' periods of the star.Furthermore, solar X-rays trigger X-ray fluorescence (XRF) emission from the surfaces of airless bodies. XRF emission is strongly dependent on both the flux and the spectral hardness of the solar X-ray spectrum. This allows for the remote sensing of surface elemental abundances (such as Mg, Al, and Si). In exoplanetary SPI studies, the detection of specific X-ray fluorescence lines from a planet could theoretically reveal its surface composition, provided the stellar flares provide a sufficient excitation flux; to our knowledge, however, such XRF diagnostics have so far been explored only for Solar System bodies (see, e.g., the review by \citealt{2007P&SS...55.1135B}), and dedicated feasibility studies for exoplanets remain to be carried out.

\paragraph{Magnetic cycle of the host star}
Stellar magnetic cycles drive periodic, large-scale topological reorganizations of the global stellar magnetic field, transitioning from largely ordered, dipoles at activity minimum to complex multipoles at maximum \citep{Fares2013}. This dynamo-driven structural evolution modulates the spatial distribution of stellar wind velocity, mass-loss rates, and magnetic configurations throughout the surrounding astrosphere \citep{Vidotto2015, Chebly2023}. Consequently, the boundary conditions governing star-planet interactions on the planetary side fluctuate cyclically in synchronisation with the host star's cycle. During periods of peak multipolar complexity, planets may encounter highly variable magnetospheric compression, enhanced magnetopause reconnection rates, as well as accelerated atmospheric erosion \citep{Maggio2022, Finociety2023}. For short-period exoplanets, these periodic reconfigurations subject the upper atmosphere to chronic thermodynamic and magnetohydrodynamic stresses, fundamentally regulating long-term volatile retention and habitability.

\section{Multi Wavelength Diagnostics and Observations of Star-Planet Interactions}\label{sec:obs_diags}
Robust detections of star-planet interactions remain a significant observational challenge to this date. Because these interactions depend on the dynamic magnetic alignment and physical properties of all the bodies involved, the resulting signals are often intermittent. This has led to the common characterization of SPI observations as an ``on-off'' phenomenon; a signal may be clearly detectable during one observing epoch and completely absent in the next, depending on the stellar magnetic cycle or the orbital geometry. To account for this inherent variability, a comprehensive multi-wavelength strategy is required, as no single diagnostic provides a complete picture of the system's energetics. While X-ray and optical emission lines reveal the response from the stellar atmosphere due to a planetary companion, radio and infrared observations allow us to potentially probe the planet's own magnetosphere and the morphology of its escaping atmosphere. Table~\ref{tab:spi_summary} provides a roadmap of the diversity of these diagnostics, summarizing their physical origins and the primary technical hurdles associated with their detection.

\begin{table}[ht]
\centering
\caption{Summary of Multi-Wavelength SPI Diagnostics}
\label{tab:spi_summary}
\small
\setlength{\tabcolsep}{4pt}
\renewcommand{\arraystretch}{1.2}
\begin{tabular}{l l p{0.22\textwidth} p{0.38\textwidth}}
\toprule
\textbf{Diagnostic} & \textbf{Wavelength} & \textbf{Origin} & \textbf{Status \& Challenges} \\
\midrule
\textbf{Ca~{\sc ii} H\&K} & Optical / UV & Chromospheric heating & Transient, on--off signals; difficult to separate from stellar rotation. \\
\addlinespace
\textbf{X-ray} & Soft X-ray & Coronal heating; tidal spin-up & Robust in wide binaries; biased toward active hosts; needs high cadence. \\
\addlinespace
\textbf{Radio (ECMI)} & Meter / Decameter & Particle dynamics in strong magnetic fields & Promising candidates, but no confirmed detections; narrow beaming limits visibility. \\
\addlinespace
\textbf{He Triplet} & Near-IR & Hanle effect in escaping exosphere & Emerging; requires high-resolution spectropolarimetry and high SNR. \\
\addlinespace
\textbf{Lyman-$\alpha$} & Far-UV & Resonant scattering in H-rich tails & Robust escape tracer; depends strongly on stellar wind models. \\
\addlinespace
\textbf{White Light} & Optical & SPI-induced flare clustering & Seen in young systems; needs long baselines to beat stochastic variability. \\
\bottomrule
\end{tabular}
\end{table}

The diverse observational markers summarized in Table~\ref{tab:spi_summary} can be broadly organized into two primary categories based on the location of the emission source. The first part of our discussion focuses on signatures originating from or near the host star, where the planetary companion acts as an external driver of chromospheric and coronal variability. The second part addresses emissions originating from the exoplanet itself or its immediate environment, including both coherent magnetospheric radio bursts and spectroscopic signatures within the escaping upper atmosphere.

\subsection{Emissions from/near The Stellar Surface}

The conceptual framework of observations of SPI emerged relatively early in the era of exoplanet discovery. Five years after the discovery of the first exoplanet around a solar-type star \citep{Mayor1995}, \cite{Schaefer2000} reported nine superflares on solar analogs of spectral types F8-G8. These stars were single (or in wide-binaries) and were characterized as neither excessively rapid rotators nor young stars, exhibiting low rotational velocities, low Ca~{\sc ii} H\&K emission, and low X-ray luminosities. While the studied stars did not host known exoplanets at the time, the authors discussed the potential effects of such superflares on orbiting planets. To explain the energies and duration of these superflares, \cite{Rubenstein_2000} suggested that, similar to observations of large flares in RS-CVn systems (detached late-type binaries that are very active as a result of their tidally enforced rapid rotation), magnetic reconnection between the fields of the star and that of a close-in Jovian exoplanet could be at play. They argued that this model could be tested by efforts to confirm the presence of extrasolar giant planets around these stars and by studying their stellar magnetic fields, which should be high enough to explain the observed energy.

\begin{figure}[h]
    \centering
    \includegraphics[width=0.60\linewidth]{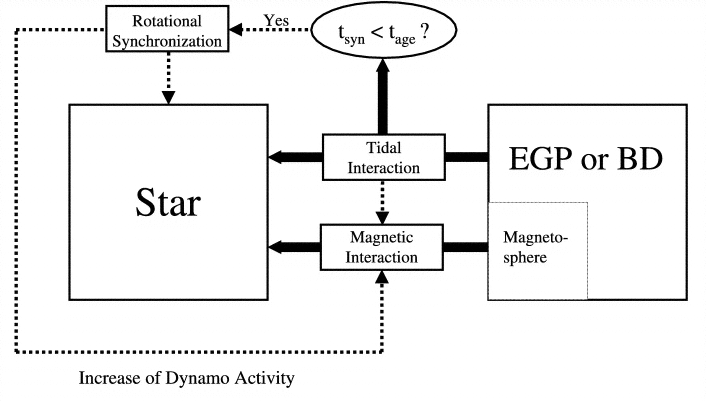}
    \caption{The different scenarios of interactions between an extrasolar giant planet (or Brown Dwarf) and its hosting star, as presented in \cite{Cuntz_2000}.}
    \label{Fig:Cuntz2000}
\end{figure}

In the same year, \cite{Cuntz_2000} suggested that interactions between extrasolar giant planets (or brown dwarfs) and their host stars could fundamentally affect stellar activity. During the first years of exoplanet discovery, observations were biased toward massive planets on close-in orbits, known as hot Jupiters \citep[as of 2011, approximately 25\% of all discovered exoplanets were Hot Jupiters; see][]{Fares2011PhDT}. \cite{Cuntz_2000} discussed two possible interaction scenarios for such effects (see Fig.~\ref{Fig:Cuntz2000}): tidal interactions and magnetic interactions. Tidal interactions between a massive giant planet and its host star could affect the outer convective envelope and the flow in its atmospheric layers. This could either affect the dynamo generation of the stellar field or cause stellar activity enhancement. In contrast, magnetic interaction between the stellar field and the planetary magnetosphere could lead to a direct enhancement of stellar activity. This activity enhancement is predicted to be modulated by periods related to the planetary orbital period ($P_{\rm orb}$). In the case of tidal interactions, the modulation would occur at $P_{\rm orb}/2$ because two tidal bulges would form and cross in front of the observer twice per orbit. For magnetic interaction, the reconnection would be seen once per orbit, leading to an activity enhancement modulated by the full planetary orbital period \citep{Saar2001}.

While a wider variety of interaction scenarios is now recognized \citep[see, e.g., Fig. 1 of][and other sections of this review]{Vidotto2025}, these early papers drove the field forward, resulting in increased observational efforts to detect SPI in activity indices (particularly Ca~{\sc ii} H\&K and $H_{\alpha}$), as well as in X-rays \citep[e.g.,][]{Kashyap2008, Poppenhaeger2010, Pillitteri2010, Pillitteri2011, Pillitteri2014, Poppenhaeger2014, Foster2020} as well as in radio \citep[e.g.,][]{Bastian2000, Lazio2004, Lecavelier2009, Smith2009, Vedantham2020, Callingham2021, Turner2021, Route2023, Tasse2026}. This was accompanied by efforts to study stellar magnetic fields to model the stellar atmosphere, the magnetic environment at the planetary orbit, and the expected emissions \citep[e.g.,][]{Ip2004, Preusse2006, McIvor2006, Lanza2008, Fares2009, Fares2010, Cohen2011, Fares2012, Fares2013, Vidotto2015, Alvarado-Gomez2015, Alvarado-Golmez2016a, Alvarado-Golmez2016b, Garraffo2017, Strugarek2019, Alvarado-Gomez2020, Strugarek2022, Bellotti2023, Vidotto2023, Chebly2023, Paul_2025, Chebly2026, Paul_2026}.

\subsubsection{SPI Signatures in Ca~{\sc ii} H\&K Emission Lines}
In this particular section, we focus specifically on observational efforts to detect SPI in activity indices. Soon after their initial paper, \cite{Saar2001} analyzed the periodicity in activity indices for seven hot Jupiter-hosting stars. They used the Ca~{\sc ii} infrared triplet (IRT), a reliable indicator for cool F-G-K stars, and defined an activity index by comparing emission inside the lines to the continuum in nearby, line-free bands. Using period search methods (see \citealt{Figueira2026}), they found no modulation by either the orbital period or half of it for any star in their sample. This negative result could have resulted from low signal-to-noise (SNR) data or the fact that the non-radiative heating in the Ca~{\sc ii} line due to SPI was relatively small.

\begin{figure}[h]
    \centering
    \includegraphics[width=0.45\linewidth]{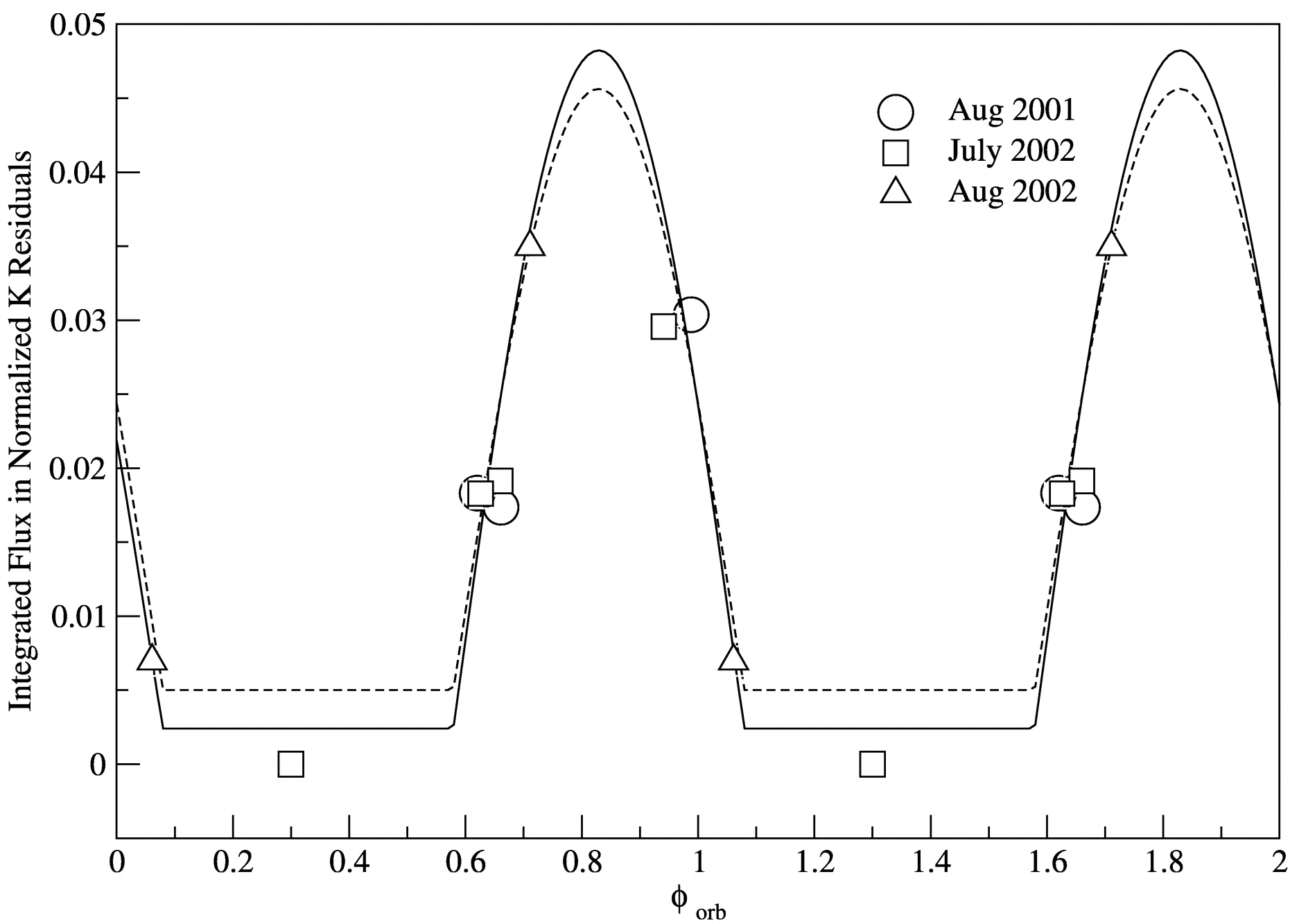}
    \includegraphics[width=0.45\linewidth]{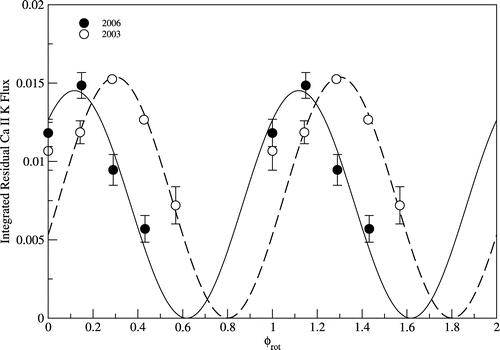}
    \caption{Left panel: Residual activity variation in Ca~{\sc ii}~K for HD~179949 as a function of the orbital period \citep{Shkolnik2003}. Right panel: Variation for the same star in other epochs as a function of rotation period \citep{Shkolnik2008}, described as the ``on-off nature'' of SPI.}
    \label{Fig:Shkolnik}
\end{figure}

\cite{Shkolnik2003} followed up on this work by observing five exoplanet-hosting stars in Ca~{\sc ii} H\&K using the high-resolution Gecko spectrograph at CFHT to collect high-SNR spectra. In their analysis, they normalized the line cores for Ca~{\sc ii}~H, Ca~{\sc ii}~K, and Al~{\sc i}, and calculated the variability in the line cores (see \citealt{Figueira2026}). They verified if the residual activity modulation corresponded to stellar rotation or the planetary orbital period. This provided the first hint of SPI: HD~179949's residual activity in the Ca~{\sc ii}~K line showed modulation with the planet's orbital period (see Fig.~\ref{Fig:Shkolnik}, left panel).

\cite{Shkolnik_2005} extended these studies to 13 stars, 10 of which hosted close-in planets, exploring both long- and short-term variability across multiple observing epochs and telescopes. They confirmed their previous results and found orbital-related variability in $\upsilon$~And. Long-term monitoring revealed that activity could be modulated by the planetary orbit in some epochs and by stellar rotation in others for the same system \citep{Shkolnik2008}. For example, in HD~179949, orbital modulation was found in four out of six observing epochs, with maximum variability leading the subplanetary point by about $70^{\circ}$. This led to the classification of SPI as having an ``on-off nature.'' Conversely, in other epochs, activity was modulated purely by stellar rotation, indicating a low or absent SPI effect (see Fig.~\ref{Fig:Shkolnik}, right panel). Other systems reported to show SPI signatures included HD~189733 and $\tau$~Boo, a system in which the planet's orbit is synchronized with stellar rotation. They also found that Ca~{\sc ii}~K correlated with Ca~{\sc ii}~H and the Ca~{\sc ii} IRT, while showing a weaker correlation with $H_{\alpha}$.

\begin{figure}[h]
    \centering
    \includegraphics[scale=0.35]{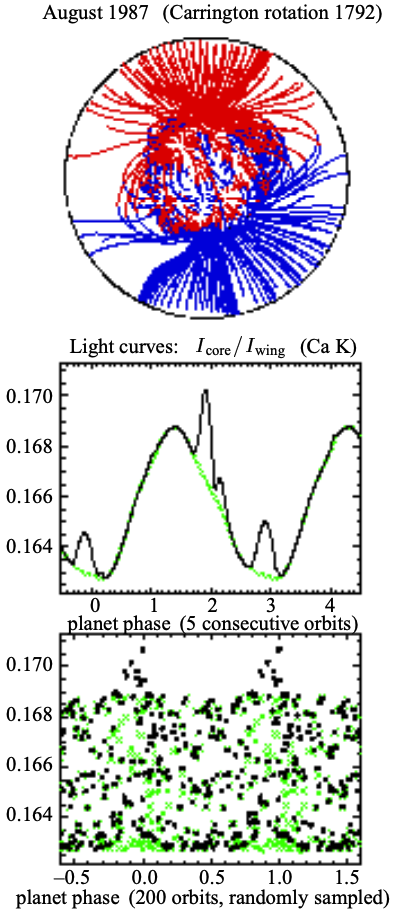}
    \includegraphics[scale=0.35]{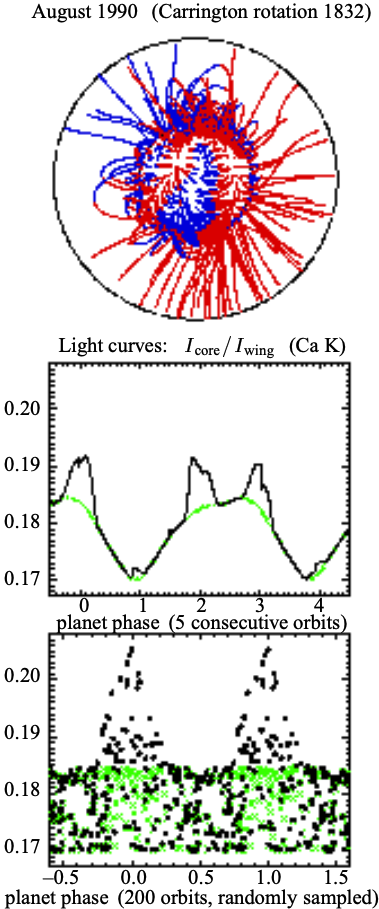}
    \caption{Modeling of SPI based on solar magnetograms by \cite{Cranmer2007} for two different solar magnetograms, plotting solar magnetic field, activity enhancement, and overall enhancement over 200 planetary orbits.}
    \label{Fig:Cranmer}
\end{figure}

Theoretical studies have modeled SPI to explain the phase shifts and the on-off nature of these detections \citep[e.g.,][]{Ip2004, Gu2005, McIvor2006, Lanza2008}. \cite{Cranmer2007} used solar magnetograms to model interactions through the solar cycle, showing that activity enhancement varies along the planetary orbit because the planet crosses different magnetic field configurations. They demonstrated that for a cyclic magnetic field, SPI observations could be missed if the effect on activity is small in certain field configurations. Figure \ref{Fig:Cranmer} shows the modeling of an SPI for a close-in planet around the Sun, where the configuration leads to different interactions from one orbit to the next and varying enhancement between epochs.

It became increasingly clear that the stellar magnetic field plays the major role in shaping these interactions. Stellar magnetic field reconstructions using Zeeman-Doppler Imaging (ZDI) followed in a series of papers exploring planet-hosting stars \citep[e.g.,][]{Catala2007, Moutou2007, Donati2008, Fares2009, Fares2010, Fares2012, Fares2013, Alvarado-Gomez2015, Moutou2016, Hussain2016}. These studies reconstructed surface magnetic fields, investigated activity modulation, and introduced the first modeling of stellar radio emission based on real magnetic maps. 

We focus here on three systems reported to have on-off SPIs. $\tau$~Boo, a massive hot Jupiter host, has shown polarity reversals of its magnetic field, the first magnetic cycle by polarity reversal ever observed in a star other than the Sun. \cite{Donati2008} reported a polarity flip compared to \cite{Catala2007} in just one year. Dense campaigns confirmed short cycles of 8 months to 2 years, with the star flipping polarity every 4 months \citep{Fares2009, Mengel2016, Jeffers2018}. While the star's rotation is synchronized with the planet, disentangling a planetary signal from a stellar signal is not trivial. Although \cite{Shkolnik2008} reported activity enhancement at a specific planetary phase, \cite{Fares2009} found that such effects were not significant once nightly variability and long-term trends were removed \citep{Evensberget2026}.

In HD~189733, an active K dwarf, \cite{Moutou2007} and \cite{Fares2010} found activity primarily modulated by stellar rotation, with no detected modulation at the planetary orbital period in Ca~{\sc ii} H\&K or $H_{\alpha}$. Residual analysis after cleaning the stellar signal yielded no planetary signal, except for a possible detection in one epoch of $H_{\alpha}$ observations. In the HD~179949 system, \cite{Fares2012} observed two epochs; one was modulated by stellar rotation, and the other showed low-amplitude variability at the synodic (beat) period, which occurs when a planet crosses the same magnetic configuration only once per beat period. Notably, \cite{Gurdemir2012} analyzed line cores separately for HD~179949 (April 2006) and found that Ca~{\sc ii}~K variability could be attributed to SPI while Ca~{\sc ii}~H did not show the same behavior. Conversely, \citet{Scandariato2013} collected simultaneous optical and X-ray data and found both lines to be modulated purely by stellar rotation.

More recently, \cite{Cauley_2019} found SPI signatures in the Ca~{\sc ii}~K line for four systems ($\tau$~Boo, HD~189733, HD~179949, and $\upsilon$~And) and predicted tentative planetary magnetic field strengths based on these detections. In \cite{Cauley_2018}, analysis of six epochs for HD~189733 showed SPI signatures in only one epoch. \cite{Strugarek2022} used the magnetic map from \cite{Fares2017} to model this specific epoch (August 2013), demonstrating that observational cadence plays a major role, with detections likely possible only 12\% to 23\% of the time. This is consistent with the 1 out of 6 detection rate found by \cite{Cauley_2018}. A more recent campaign of HD 189733 carried out in July 2023 showed again a clear SPI modulation (Yang et al., in prep). Additionally, \cite{Klein2022} found modulation by the planetary period in the He~{\sc i}~D3 line for the M dwarf host AU~Mic using optical HARPS spectra. Recently, \cite{Revilla2026} found a planet-induced periodic modulation in GJ~436 system, by analyzing 17 years worth of data. GJ 436 has an activity cycle of about 7-8 years observed both in spectroscopy \citep{kumar2023} and photometry \citep{Lothringer2018}. The authors find an increased activity at the same phase of the cycle, that is modulated by a combination of the stellar rotation and planetary orbital period. The use of long data span of observation and finding hints of SPI is promising for the use of archival data in such type of studies.

\subsubsection{X-ray and White-Light Observations of SPI}\label{sec:xray_whitelight_obs}
% \begin{itemize}
% \item \texttt{\textbf{Contributors:} Mayank Narang, Katja Poppenhaeger}
% \end{itemize}
% (\katja{I wrote this X-ray part and also added something on white-light observations, because they are expected to root in similar physical processes as the X-ray observables, but there's not that much work done on that yet that it would require its own sub-section.})
Observable manifestations of SPI, whether of tidal or magnetic origin, are hypothesized to enhance or simulate intrinsic stellar activity features across multiple spectral regimes. This phenomenological overlap is particularly evident in the high-energy regime, where soft X-ray emission serves as a primary diagnostic of the stellar corona. Given that coronal existence is fundamentally coupled to magnetic activity, SPI-driven energetic enhancements are expected to manifest as increased coronal luminosity and elevated plasma temperatures.
Tidal SPI, specifically, may induce secular angular momentum transfer from the planetary orbit to the host star, resulting in a ``tidal spin-up'' that counteracts the natural rotational deceleration driven by magnetic braking \citep[see][]{Penev2012, Jackson2016, Isik2026}. This mechanism effectively mitigates the age-dependent decay of stellar rotation, thereby maintaining $L_{\rm X}$ and coronal thermal profiles exceeding the baseline expectations for isolated stars of equivalent chronological age. Empirical investigations into this elevated activity have persisted for two decades; early population analyses by \cite{Kashyap2008} indicated that stars hosting hot Jupiters exhibit statistically higher X-ray activity than those with lower-mass or longer-period companions—a correlation reinforced by subsequent studies \citep{Poppenhaeger2010, Miller2015}.
However, these population-level results are susceptible to systematic selection biases inherent in exoplanet detection methodologies. High stellar jitter and spot modulation characteristic of fast-rotating, active stars facilitate the discovery of hot Jupiters while masking lower-mass or longer-period companions, potentially leading to an overrepresentation of active host stars in the known exoplanet census \citep{Kashyap2008, Poppenhaeger2011}. To decouple genuine SPI effects from these selection effects, researchers have utilized wide stellar binaries where only one component hosts an exoplanet. As the stellar components are co-eval, any significant divergence in coronal activity can be attributed to the planetary companion rather than age-related evolution \citep{Poppenhaeger2014}. Comparative studies of such systems confirm that hot Jupiter hosts are intrinsically over-active in X-rays relative to their co-eval companions, supporting (although not conclusively validating) tidal SPI as a driver of stellar activity evolution and ruling out detection biases as the sole causative factor \citep{Ilic2022}.
Beyond secular trends, X-ray observations facilitate the search for short-term variations indicative of magnetic SPI. In the HD~189733 system, high-resolution X-ray timing revealed marginal evidence for orbital periodicity in coronal brightenings, potentially signifying the accretion of evaporated planetary material onto the host star \citep{Pillitteri2011}. Conversely, MHD modeling suggests that genuine magnetic SPI features may remain undetected in current datasets due to insufficient observational cadence \citep{Strugarek2022}. Similar periodicities have been investigated in the HD~179949 system, though existing data remain statistically inconclusive \citep{Acharya2023}. Eccentric systems provide additional diagnostic leverage, as SPI effects are predicted to peak near periastron passage. While \cite{Maggio2015} detected an activity increase in HD~17156 at periastron, subsequent chromospheric observations of the similarly eccentric HD~80606 found no significant enhancement \citep{Figueira2016}.
Complementary to these high-energy diagnostics, long-baseline optical photometry from space-based observatories such as \textit{Kepler} and TESS has facilitated the systematic search for SPI-induced flaring. Such events are theoretically predicted for planets occupying sub-Alfv\'enic orbital regimes. While numerical estimates of dissipated energy vary across modeling frameworks \citep{Lanza2012, Lanza2018}, the dissipation required to trigger premature flaring in existing coronal loops may be energetically modest \citep{Loyd2023}.
\begin{figure}[h]
\centering
\includegraphics[width=0.39\linewidth]{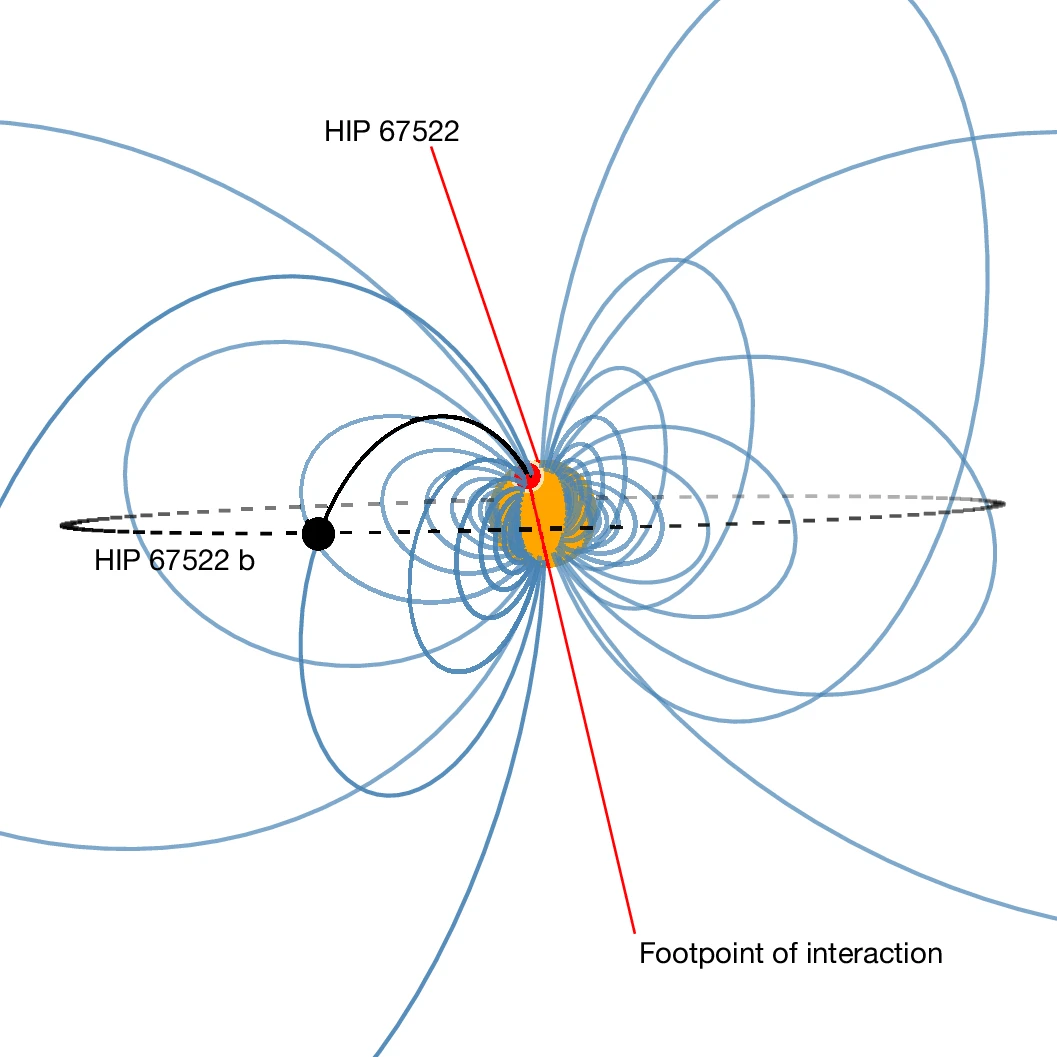}
\includegraphics[width=0.59\linewidth]{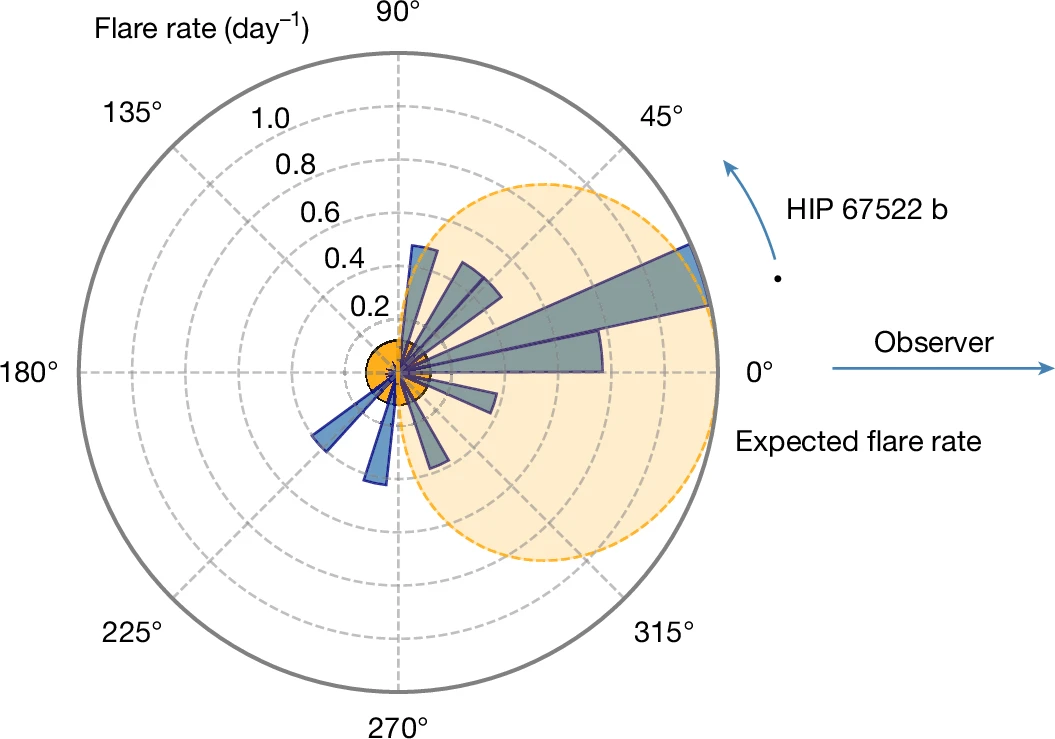}
\caption{The HIP~67522 system, characterized by a gas giant in a close-in orbit around a young G-type host. The observed stellar flares in broad-band optical light curves exhibit significant clustering as a function of the planetary orbital phase. The orange dashed region (right panel) indicates the predicted flare rate relative to an interaction footpoint at the sub-planetary longitude. Figures from \cite{Ilin2025}.}
% \katja{NOTE: I HAVE ASKED EKATERINA ILIN FOR HER PERMISSION TO USE THESE FIGURES AND SHE IS FINE WITH IT; BUT: THESE FIGURES WERE ORIGINALLY PUBLISHED IN NATURE AND SOMEONE FROM ISSI NEEDS TO GET LEGALLY VALID PERMISSION FOR THIS.}
\label{Fig:Ilin2025}
\end{figure}
Initial investigations of the TRAPPIST-1 system yielded no evidence of SPI-correlated flare clustering \citep{Fischer2019}. However, subsequent studies have characterized deviations in flare frequency from purely stochastic baselines for various sub-Alfv\'enic candidates \citep{Ilin2022, Klein2022, Ilin2024}. This research culminated in the empirical detection of strong flare clustering in the young, active HIP~67522 system. Analysis of TESS and CHEOPS data by \cite{Ilin2025} revealed a significant correlation between flare occurrence and planetary orbital phase (see Fig.~\ref{Fig:Ilin2025}). The observed clustering at the orbital period implies that the planetary companion induces excess flaring at a consistent magnetic footpoint, which is subsequently modulated in observability by stellar rotation. These findings underscore that for stars with lower intrinsic flaring rates, resolving SPI-induced signals from the stochastic background requires extended observational baselines. Space-based photometry can also reveal SPI signatures other than flares. In the HD~118203 system, which hosts a close-in Jupiter-sized planet on an eccentric orbit ($e \simeq 0.32$, $P_{\rm orb} \simeq 6.1$~d), \citet{CastroGonzalez2024} detected TESS photometric variability modulated at the planetary orbital period, which they interpret as likely evidence of magnetic SPI, while discussing in detail the caveats associated with stellar rotation; the evolved nature of the host star and the orbital eccentricity make synchronous stellar rotation unlikely. The authors further note that the pseudo-synchronous rotation expected for an eccentric planet could lead to a stronger planetary magnetic moment than in an equivalent circular, synchronized system, making short-period eccentric systems promising targets for magnetic SPI searches.

\subsection{Emissions from/near the Planetary Magnetosphere}\label{sec:emissionnearplanet}
While the preceding sections detailed the star's observable response to a close-in companion, potentially some of the most direct empirical constraints on exoplanetary properties, specifically exoplanetary magnetism, are derived from emissions originating within the planetary vicinity itself. In particular, non-thermal radio signatures provide a window into the particle acceleration processes and field topologies that define the planetary magnetic environment or lack thereof. The interpretation of non-thermal, coherent radio signatures from exoplanetary systems necessitates a robust microphysical framework to account for the conversion of particle kinetic energy into electromagnetic radiation. While the macroscopic observability of these signals is governed by the global magnetic topology and the stellar wind environment, the fundamental amplification of such waves is expected to occur on kinetic scales through the process of the electron cyclotron maser instability (ECMI), which provides the physical mechanism for the high-brightness temperature, polarized emissions observed in both Solar System planets and their exoplanetary analogues.

\subsubsection{The electron cyclotron maser instability (ECMI)}
The electron cyclotron maser instability is a coherent emission mechanism that directly amplifies electromagnetic radiation near the electron cyclotron frequency $\omega_{c}$ and its harmonic frequencies in magnetized plasmas (\cite{Twiss} and \cite{Schneider}). The ECMI theory successfully explains the generation of auroral radio emissions from the magnetized planets of the Solar System (\cite{Zarka1998}, \cite{WuLee1979}) and it could be responsible for similar radio emissions of exoplanets (\cite{Zarka_2001}, \cite{Zarka2007}). ECMI radiation can attain extremely high brightness temperatures of up to $10^{24}$\,K \citep{Treumann2006}, is polarized and beamed along the surface of a cone with a large opening angle and an axis parallel to the ambient magnetic field. The ECMI instability generally requires a non-thermal anisotropic particle distribution \citep{Treumann2006}. In the context of SPI, such supra-thermal electrons can be created by wave-particle interactions in the Alfv\'en wing, as is thought to be the case for Io-induced ECMI emission in the Jovian magnetosphere \citep{Zarka1998}. An essential requirement for ECMI emission is $\omega_{\rm c}>\omega_{\rm p}$ (i.e., cyclotron frequency exceeds plasma frequency). This corresponds to a magnetic field strength that exceeds $B>3.2\times 10^{-3}\,(n_e/{\rm cm}^{-3})^{1/2}\,{\rm G}$. The condition is usually met in the tenuous magnetospheres of planets and could also be met in a substantial fraction of the coronae/magnetospheres of stars with strong large-scale magnetic fields. These theoretical and phenomenological expectations have led to several searches for SPI induced ECMI radiation in the radio band (see Sect.~\ref{sec:radio_obs}). Because the ECMI emission directly originates from non-thermal charges, it provides a direct probe of particle acceleration in Alfv\'en wings on kinetic scales. 

\begin{figure}[t]
    \centering
    \includegraphics[width = 0.48\linewidth]{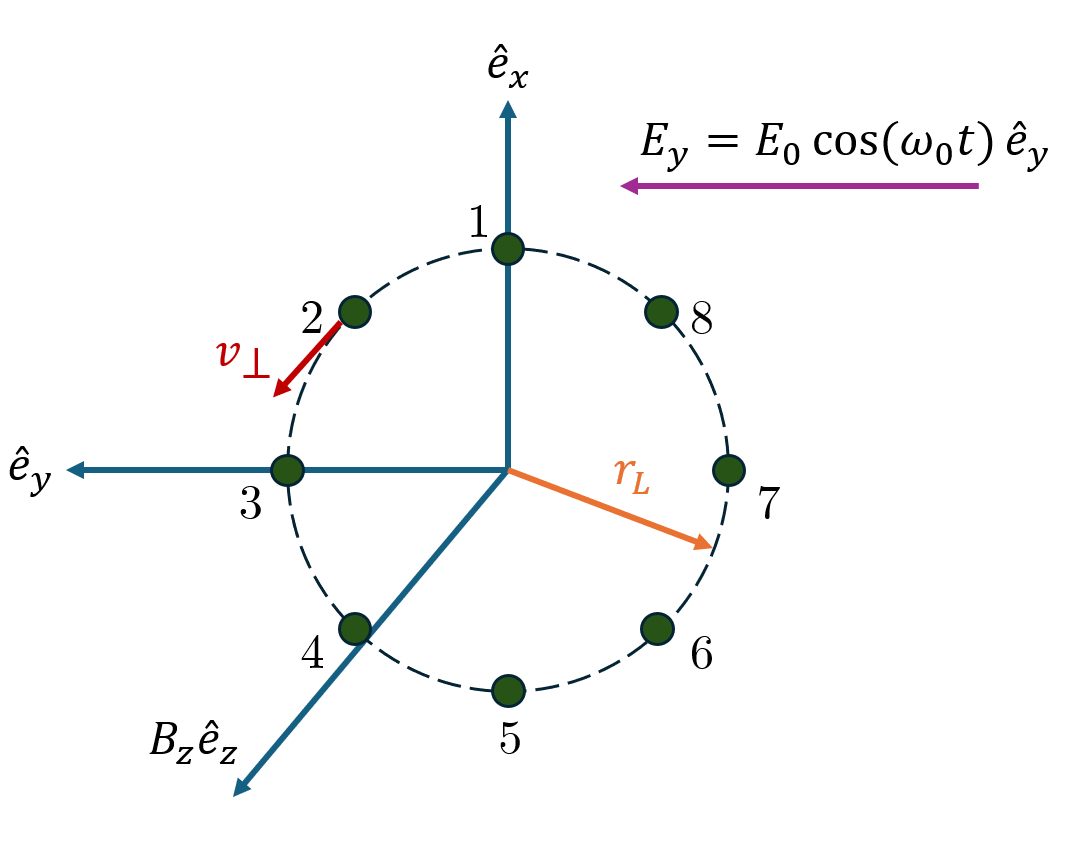}
    \includegraphics[width = 0.48\linewidth]{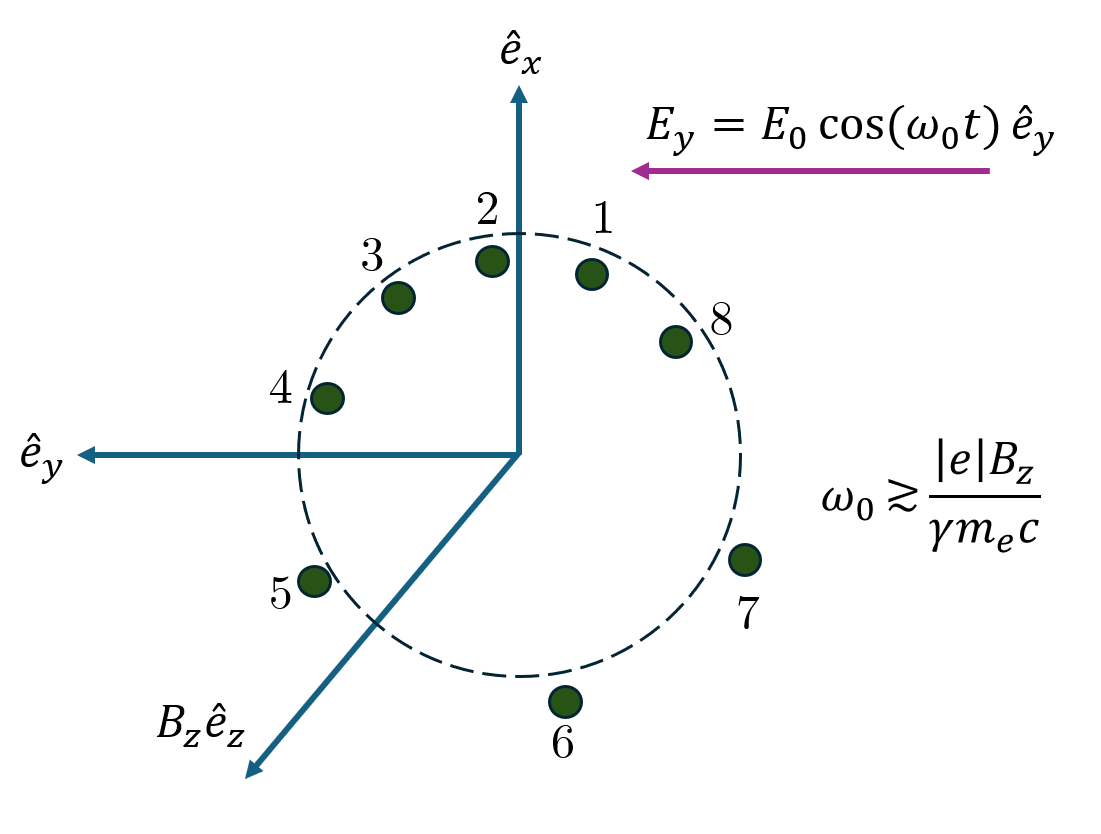}
    \caption{Left panel: Initial configuration for the electron cyclotron maser instability (ECMI). Right panel: bunched particles after several cycles \cite{CMI2}.}
    \label{fig:CMI}
\end{figure}

To understand the physical mechanism responsible for ECMI, consider the orbits of eight test electrons, labeled 1-8, in velocity space (see the left panel of Figure \ref{fig:CMI}), under the influence of a small external field. Starting with the electrons uniformly distributed along a gyro-orbit, with zero $z$-velocity component, a constant and uniform magnetic field is directed along the $z$-axis, so that $\mathbf{B}=(0,0,B_z)$. The electrons rotate in the counter-clockwise direction due to the magnetic force and the initial radius of the electron ring is the gyration radius $r_{L}$:
\begin{equation}
    r_{L} = \frac{\gamma_{\perp} m_{e}cv_{\perp}}{|e|B_{z}},
\end{equation}
where $v_{\perp}$ is the initial perpendicular velocity, $m_{e}$ is the mass of the electron, and $\gamma = (1-v_{\perp}^{2}/c^{2})^{-1/2}$ is the Lorentz factor. By applying a small electric field of constant amplitude, with frequency $\omega_{0}$, in the $y$-direction: $\mathbf{E} = (0,E_{0}\cos(\omega_{0}t),0)$, the electrons feel an additional force $-e\mathbf{E}$ and their orbits are modified so that some electrons cluster together due to the difference in acceleration depending on the relative phase of the electric field. In the case where the cyclotron frequency is equal to the frequency of the electric field, $\omega_{0} = \omega_{c} = \frac{|e|B_{z}}{\gamma m_{e}c}$, the rate of change of the particle energy with respect to time is found by taking the dot product of the total Lorentz force with the electron velocity, $\mathbf{v}(t)$:
\begin{align}
    \frac{d\mathbf{p}(t)}{dt}\cdot \mathbf{v}(t) &= -e(\mathbf{E}+\mathbf{v}(t)\times \mathbf{B})\cdot \mathbf{v}, \\
    \frac{d\varepsilon_{p}}{dt} &= -|e|v_{y}(t)E_{0}\cos(\omega_{0}t),
\end{align}
where $\mathbf{p}$ is the electron's momentum and $v_{y}(t)$ is the component of the electron's velocity along the $y$-direction. Particles 1, 2, and 8 lose energy because their velocity vector has a component in the direction of $\mathbf{E}$, causing them to spiral inward. Their Lorentz factor $\gamma$ decreases, the relativistic cyclotron frequency increases, and the gyration radius decreases, leading to their phase slipping ahead of the wave. Conversely, particles 4, 5, and 6 gain energy because their velocity vector has a component opposite to $\mathbf{E}$; their cyclotron frequency decreases, they spiral outward, and their phase falls behind the wave. Particles 3 and 7 move perpendicularly to the electric field and do not experience any change in their energy; their instantaneous cyclotron frequency and gyration radius remain constant, and they remain in phase with the rotating field. 

The particles cluster around the positive $y$-axis after an integral number of wave periods \cite{CMI2} due to relativistic effects, since the rotational frequency of the electrons depends on their energy. In order to obtain a net exchange of energy between the electrons and wave, the frequency of the electric field must be slightly greater than the relativistic cyclotron frequency: $\omega_{0} \gtrsim \frac{|e|B_{z}}{\gamma m_{e}c}$. In this case, during a wave period of $2\pi/\omega_{0}$, the electrons traverse a coordinate space angle less than $2\pi$. All particles slip behind the wave and cluster in the upper half-plane after an integral number of wave periods. This phase slippage causes the net kinetic energy of the ensemble to decrease. Due to conservation of total energy, the field amplitude must increase, promoting more bunching and eventually producing an instability (see the right panel of Figure \ref{fig:CMI}). Electromagnetic waves can be amplified when the resonance condition between waves and energetic electrons is met:
\begin{equation}
    \gamma -\frac{s\omega_{c}}{\omega_{q}} - N_{q}\mu \cos\theta\frac{u}{c}=0,
\end{equation}
where $u$ denotes the momentum per unit mass with parallel component $u_{||}$, $\mu = u_{||}/u$ is the cosine of the pitch angle, $\gamma$ is the Lorentz factor, $s$ is the harmonic number, $\omega_{c}$ the electron cyclotron frequency and $\omega_{q}$ the frequency of the excited wave, with all frequencies normalized by the plasma frequency $\omega_{p}$. $N_{q}$ is the refractive index of the excited wave propagating with phase angle $\theta$ to the magnetic field, and $q = \pm$ designates the wave modes for the extraordinary mode (X, $q=+$) and the ordinary mode (O, $q=-$). 

Fast electrons lead to maser instabilities in magnetized plasmas. Considering that the non-thermal electrons occupy a small component in the plasma compared to thermal ones, the dispersion relation of the excited wave can be approximated by cold-plasma theory \cite{Wu2002}: $N_{q}^{2}=1-\frac{1}{\omega_{q}(\omega_{q}+\tau_{q}\omega_{c})}$, where $\tau_{q} =-s_{q}+q\sqrt{s_{q}^{2}+\cos^{2}\theta}$ and $s_{q}=\omega_{q}\omega_{c}\sin^{2}\frac{\theta}{2}(\omega_{q}^{2}-1)$. Discussion of the ECMI growth rate can be found in \cite{Wu2002}, \cite{Wu2009}, and \cite{CMI1}. The ECMI instability grows efficiently due to special-relativistic effects, even in the mildly relativistic plasma in the Alfv\'en wings of Io and main-sequence stars. A non-relativistic plasma cannot support ECMI since it requires a particle population with a significant ratio of $v/c$ \citep{Mottez2014}. At high growth rates, harmonic components tend to be generated due to nonlinear effects, causing broadening of the emitted frequency spectrum.

\subsubsection{Auroral radio emission in the Solar System}
Star–planet magnetic interactions can produce intense radio emission, much like the auroral radio bursts seen in our own Solar System. Well-studied planetary auroral emission, including Jupiter’s decametric (DAM, $\sim10$-40\,MHz) and hectometric (HOM, $\sim0.3$-3\,MHz) emissions, Saturn’s kilometric radiation (SKR, $\sim10$--800\,kHz), and Earth’s auroral kilometric radiation (AKR, $\sim50$-800\,kHz), are all powered by the electron cyclotron maser instability (ECMI). Despite spanning different frequency ranges, these emissions share a common origin: energetic electrons are accelerated along converging magnetic field lines, creating anisotropic velocity distributions that trigger ECMI and produce intense, beamed coherent radiation. These radio bursts are ultimately driven by the interaction between a magnetized body and an external energy source. At Earth and Saturn, the solar wind governs the auroral current system and modulates AKR and SKR. At Jupiter, however, a substantial fraction of the auroral power is generated internally by electrodynamic coupling with its moons. Io and Ganymede act as unipolar inductors, exciting Alfv\'en waves that propagate along magnetic field lines and form large-scale Alfv\'en wings \citep{Bigg1964, Neubauer1980, Zarka1998, Paul_2026}. These wings channel energy and momentum into Jupiter’s auroral acceleration regions, where ECMI emission originates.

Electrons within the Alfv\'en wings are accelerated along magnetic field lines by the electric fields induced as a moon or planet moves through the magnetized plasma. When these electrons enter regions of stronger magnetic field, they experience mirroring, which reflects electrons whose pitch angles exceed a critical value. This process produces a loss-cone velocity distribution, in which only electrons with sufficiently small pitch angles escape along the field lines toward the planet or star, while the remaining electrons are reflected back. Magnetic reflection can temporarily trap electrons, modulating both the intensity and direction of the emission, and creating quasi-periodic features in the observed radio flux. This anisotropic electron distribution provides free energy for the electron cyclotron maser instability (ECMI), which efficiently converts electron kinetic energy into coherent, highly beamed, circularly polarized radio emission. The characteristic emission frequency is near the local electron gyrofrequency: $f_\mathrm{ce} = \frac{e B}{2 \pi m_\mathrm{e}}$, where $B$ is the local magnetic field strength, $e$ is the electron charge, and $m_\mathrm{e}$ is the electron mass \citep{Treumann2006, Hess2010}. Typical electron energies for auroral-type ECMI are $\sim 10$--$20\,\mathrm{keV}$, consistent with Jupiter's and Saturn's auroral regions. 

The ECMI emission forms a hollow conical beam along the magnetic field lines. The half-opening angle depends on the pitch-angle distribution and energy of the electrons: $\theta_\mathrm{beam} \approx \arccos \frac{v_\parallel}{v}$, where $v_\parallel$ is the electron velocity parallel to the field and $v$ is the total electron speed \citep{Treumann2006}. Higher-energy electrons produce narrower beams, while lower-energy electrons yield broader emission cones. Magnetic reflection affects $v_\parallel$, further modulating the effective beam angle and the duration of emission. In addition to loss-cone distributions, ECMI can also be driven by shell-type (or ring) distributions, which arise during rapid magnetospheric compression or reconfiguration. Here, electrons move predominantly perpendicular to the magnetic field, generating ECMI emission with broader spectral coverage, larger beaming angles, and distinct polarization properties compared to loss-cone emission \citep{WuLee1979, Bingham2013}. Shell-type emission is particularly relevant during sudden reconnection events or strong solar wind compressions. 

Once generated, radio waves propagate through the surrounding plasma, where refraction, reflection, scattering, and absorption influence their detectability. Propagation is limited to regions where the emission frequency exceeds the local plasma frequency: $f_\mathrm{pe} = \frac{1}{2 \pi} \sqrt{\frac{n_\mathrm{e} e^2}{\varepsilon_0 m_\mathrm{e}}}$, where $n_\mathrm{e}$ is the electron number density and $\varepsilon_0$ is the vacuum permittivity. Plasma inhomogeneities can reflect or scatter the waves, while the intrinsic ECMI beaming controls which emissions are observable. Weaker electrodynamic interactions observed between Jupiter and its moons Ganymede and Europa demonstrate that such moon–planet coupling is a general phenomenon in magnetized plasma environments \citep{Louis2017, Zarka2018, Jacome2022}. Overall, the combination of electron acceleration, magnetic mirroring, pitch-angle anisotropy, and energy-dependent beaming shapes the intensity, spectral properties, and observability of radio emission. These mechanisms observed in the Solar System provide a robust framework for understanding SPI in exoplanetary systems. While these Solar System examples clarify the basic physics of Alfv\'en-wave propagation, electron acceleration, and ECMI-driven radio emission, the associated magnetic fields, plasma densities, and energy budgets remain modest compared to those expected in close-in exoplanetary systems. Hot Jupiters orbiting within a few stellar radii of their host stars encounter denser winds, stronger magnetic fields, and substantially higher Poynting fluxes. As a result, star–planet interactions can operate in far more extreme regimes, potentially producing radio emission orders of magnitude more powerful than any Solar System analogue. We will come back to SPI induced radio emission in a bit. 

\subsubsection{Radio emission from star--planet interactions}
For in-depth discussions of Solar System radio emissions and the underlying plasma physics, we refer the reader to dedicated reviews \citep[e.g.][]{Zarka1998, Zarka2004, Treumann2006, Saur2021}. This Solar System framework provides a natural basis for studying star–planet interactions (SPI) in exoplanetary systems. Close-in star–planet systems often operate in the sub-Alfv\'enic regime, where the Alfv\'enic Mach number $M_\mathrm{A} < 1$. A useful Solar System analogue is the Jupiter–Io or Jupiter–Ganymede interaction: although no solar wind penetrates Jupiter's magnetosphere, the dense, corotating, and slightly outward-drifting magnetospheric plasma constitutes an effective sub-Alfv\'enic flow in the satellite frame, generating stationary Alfv\'en wings that transport Poynting flux toward the auroral source region \citep[e.g.,][]{Saur2013, Zarka2018}. Similarly, close-in exoplanets orbiting within a few stellar radii may reside inside the sub-Alfv\'enic region of their host star's wind \citep{Zarka2007, Strugarek2016}. In this regime, the planet can maintain a magnetic connection with the star via Alfv\'en wings, which channel energy along magnetic field lines and potentially power coherent radio emission. 

\begin{figure}[htb]
    \centering
    \includegraphics[width=\linewidth]{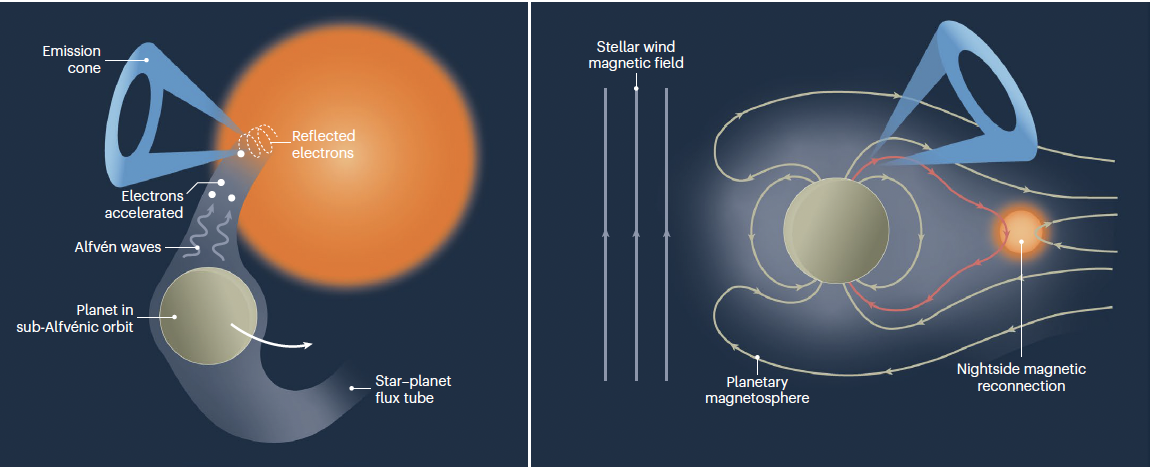}
    \caption{
Schematic of the two sources of ECMI emission in exoplanetary systems. 
\textbf{Left:} Star-planet interaction. A close-in planet inside the stellar Alfv\'en surface perturbs the stellar magnetic field, generating Alfv\'en waves that accelerate electrons along field lines. Magnetic mirroring produces a loss-cone distribution, resulting in ECMI emission in a hollow cone. 
\textbf{Right:} Planetary auroral emission. Stellar wind compresses the planet’s magnetosphere, causing nightside magnetic reconnection. Accelerated electrons follow field lines to the poles, where mirroring produces ECMI emission. Emission cones are shown only in the northern hemispheres for clarity. Image taken from \cite{Callingham2024}.
}
    \label{fig:ecm_sources}
\end{figure}

The interaction produces two primary sources of emission. First, perturbations of the stellar magnetic field by the planet launch Alfv\'en waves that accelerate electrons toward the star. These electrons undergo magnetic mirroring, forming a loss-cone distribution that drives ECMI emission in a hollow conical beam originating from the stellar surface (see Fig.~\ref{fig:ecm_sources}, left panel). Second, the incident stellar wind compresses the planetary magnetosphere, leading to magnetic reconnection on the nightside. Electrons accelerated along planetary field lines toward the poles are reflected, producing auroral ECMI emission from the planet itself (Fig.~\ref{fig:ecm_sources}, right panel). The efficiency, location, and observability of these emissions depend critically on the local Alfv\'en speed and the detailed topology of the stellar magnetic field, emphasizing the importance of characterizing the stellar Alfv\'en surface for predicting SPI-induced radio signals. By understanding electron acceleration, ECMI generation, and radio wave propagation in the Solar System, we can interpret exoplanetary radio signals, anticipate emission timing, and extract key physical parameters such as magnetic field strength, plasma density, electron energy distributions, and beaming angles \citep[e.g.,][]{Zarka1998, Treumann2006, Hess2010, Strugarek2016, Louis2017}. Observing these emissions provides rare indirect measurements of exoplanetary magnetic fields, informs on planetary atmospheric protection and habitability, and constrains stellar wind properties and star–planet magnetospheric interactions. Beyond the framework developed by Zarka and collaborators, the expected radio signatures of magnetic star--planet interactions have been investigated by several other groups, e.g., through radiometric scaling-law predictions of radio fluxes from known exoplanets \citep{Lazio2004, Griessmeier2007}, analytical models of the energetics of the Alfv\'en-wing interaction in M dwarf systems \citep{Saur2013, 2018ApJ...854...72T}, stellar wind models used to assess the propagation and visibility of the emission \citep{Kavanagh2019, Kavanagh2023}, and numerical simulations of the interaction \citep{Strugarek2019, Strugarek2022}; on the observational side, coherent low-frequency radio emission from M dwarfs has been interpreted in this context \citep{Vedantham2020, Callingham2021, Pineda2023}.

\subsubsection{Observational challenges and current facilities}\label{sec:radio_obs}
Observing star-planet interaction (SPI)-induced radio emission is intrinsically challenging. The expected signals are typically faint, highly directional due to electron cyclotron maser instability (ECMI) beaming, and often occur at low frequencies ($\lesssim 10$~MHz), which are blocked by Earth's ionosphere, necessitating space-based instruments for direct detection. Stellar activity further complicates observations: stochastic flares, bursts, and intrinsic variability can mask or mimic SPI signatures, making it difficult to disentangle planetary contributions from the stellar background \citep{Dulk1985, Weber2017, Weber2017b, Kavanagh2019, Ilin2025}. Additional limitations arise from the geometry of the system. The orbital configuration and magnetic orientation of the exoplanetary system relative to the observer strongly affect detectability. For example, if the system is viewed along its rotational axis (a “pole-on” orientation), much of the emission may be directed away from the observer, reducing visibility, whereas equator-on configurations increase the likelihood that the emission cone crosses the line of sight \citep{Lamy2023}. The inclination of the stellar rotation axis and planetary orbital plane relative to Earth, combined with the phase of the planet in its orbit, further modulate the probability of observing emission. 

Radio-wave propagation is also constrained by the plasma environment. High plasma densities near the star can prevent low-frequency waves from escaping, effectively setting a minimum observable frequency. Density gradients and inhomogeneities in the stellar wind can partially reflect, refract, or scatter the emitted waves, further reducing detectability. Systems exhibiting diverse radio emission patterns—such as separate low- and high-frequency components or complex spectral structures—are preferred observational targets, as these features increase the probability that at least one emission component will intersect the observer’s line of sight. Recurrent emission, modulated by the orbital period of the planet, further enhances detectability by raising the chance of capturing emission during any given observing session. Practical constraints of the observatories also affect detection. The maximum elevation of a star above the local horizon during transit determines how long it can be observed and influences sensitivity due to atmospheric attenuation. Telescope sensitivity limits the detection of weaker signals: some systems may be detectable across a wide range of planetary magnetic field strengths, while others are only visible if the planet’s field exceeds a certain threshold, and still others remain undetectable with current instrumentation. 

Recent advancements in radio astronomy are transforming the study of star-planet interactions (SPI). Modern observatories now offer unprecedented sensitivity and frequency coverage for stellar radio observations. These include the Low-Frequency Array (LOFAR; \citealt{vanHaarlem2013}), the Giant Metrewave Radio Telescope (GMRT; \citealt{Swarup1991, Gupta2017}), the Karl G. Jansky Very Large Array (JVLA; \citealt{Perley2011}), the Five-hundred-meter Aperture Spherical Radio Telescope (FAST; \citealt{Nan2006, Nan2011}), the Australian Square Kilometre Array Pathfinder (ASKAP; \citealt{Johnston2008}), and NenuFAR (\citealt{Zarka2020}). These facilities enable detailed studies of low-frequency radio emission, significantly enhancing the search for SPI signatures in both the Solar System and exoplanetary systems. Several systems have emerged as promising SPI candidates. These include low-frequency emission from the M~dwarf GJ~1151 \citep{Vedantham2020}, bursts from $\tau$~Boo \citep{Turner2021}, orbital-phase-coherent bursts from YZ~Ceti \citep{Pineda2023}, and fine-structured bursts from AD~Leo \citep{Zhang2023}. NenuFAR detected a circularly polarized burst from HD~189733 consistent with SPI \citep{Zhang2025}. Next-generation instruments such as the Square Kilometre Array (SKA; 50~MHz-15~GHz) promise even greater sensitivity and frequency coverage, enabling more detailed studies. 

\subsubsection{Modelling SPI-induced radio emission}
Progress in observational capabilities is paralleled by significant advances in modeling and numerical simulations. One of the pioneering tools in this domain is \texttt{MASER}~\citep{Kavanagh2023}, which computes synthetic radio emission produced at the planet (see Figure~\ref{fig:ecm_sources}, right panel) for specific star–planet systems. \texttt{MASER} has enabled detailed exploration of key SPI properties—including expected emission frequencies, beaming geometries, and phase-dependent variability—providing valuable guidance on which systems are most likely to be detectable under given observational conditions. However, \texttt{MASER} relies on several simplifying assumptions, such as uniform stellar wind conditions, idealized dipolar or axisymmetric magnetic fields, and constant plasma densities. These assumptions do not capture the local variations in wind speed, density, or the complex magnetic topologies inferred from Zeeman–Doppler Imaging (ZDI), including higher-order multipolar components. Consequently, while \texttt{MASER} is well suited for first-order estimates and parameter studies, it cannot fully reproduce realistic phase-dependent variability, fine-scale beaming structures, or the effects of stellar wind inhomogeneities on Alfv\'en-wing formation and ECMI-driven radio emission. 

To capture the full complexity of star-planet interactions and provide quantitatively reliable predictions, accurate modeling of SPI-driven radio emission requires 3D~MHD simulations that self-consistently account for stellar wind acceleration, magnetic topology, and plasma environment. The \texttt{ExPRES} code \citep{Louis2019a} incorporates realistic physics by using detailed 3D inputs for the stellar magnetic field and plasma environment. Unlike \texttt{MASER} \citep{Kavanagh2023}, which relies on simplified magnetic and wind geometries, \texttt{ExPRES} can ingest magnetic field maps reconstructed from ZDI and plasma densities from 3D MHD simulations. This enables self-consistent computation of ECMI-driven radio emission visibility, frequency, and beaming patterns as a function of orbital phase, capturing complex magnetic topologies, local plasma variations, and Alfv\'en-wave propagation. For Alfv\'en-wing-mediated star-planet interactions to occur, the planet must orbit within the stellar Alfv\'en surface (introduced in Sect.~\ref{sec:magnetic_interactions}), i.e., the surface where the stellar wind speed equals the local Alfv\'en speed: $v_\mathrm{sw}(r_\mathrm{A}) = v_\mathrm{A}(r_\mathrm{A}) = \frac{B(r_\mathrm{A})}{\sqrt{\mu_0 \rho(r_\mathrm{A})}}$ with $v_\mathrm{sw}$ the stellar wind velocity, $v_\mathrm{A}$ the Alfv\'en speed, $B$ the magnetic field strength, $\rho$ the plasma mass density, and $\mu_0$ the vacuum permeability. Inside this surface, perturbations from the planetary motion can propagate upstream along Alfv\'en characteristics, forming Alfv\'en wings that channel energy and momentum between the star and planet. Planets beyond the Alfv\'en surface are in the super-Alfv\'enic regime, where no direct magnetic coupling occurs. 

\begin{figure}[hbt]
    \centering
    \includegraphics[width=0.5\linewidth]{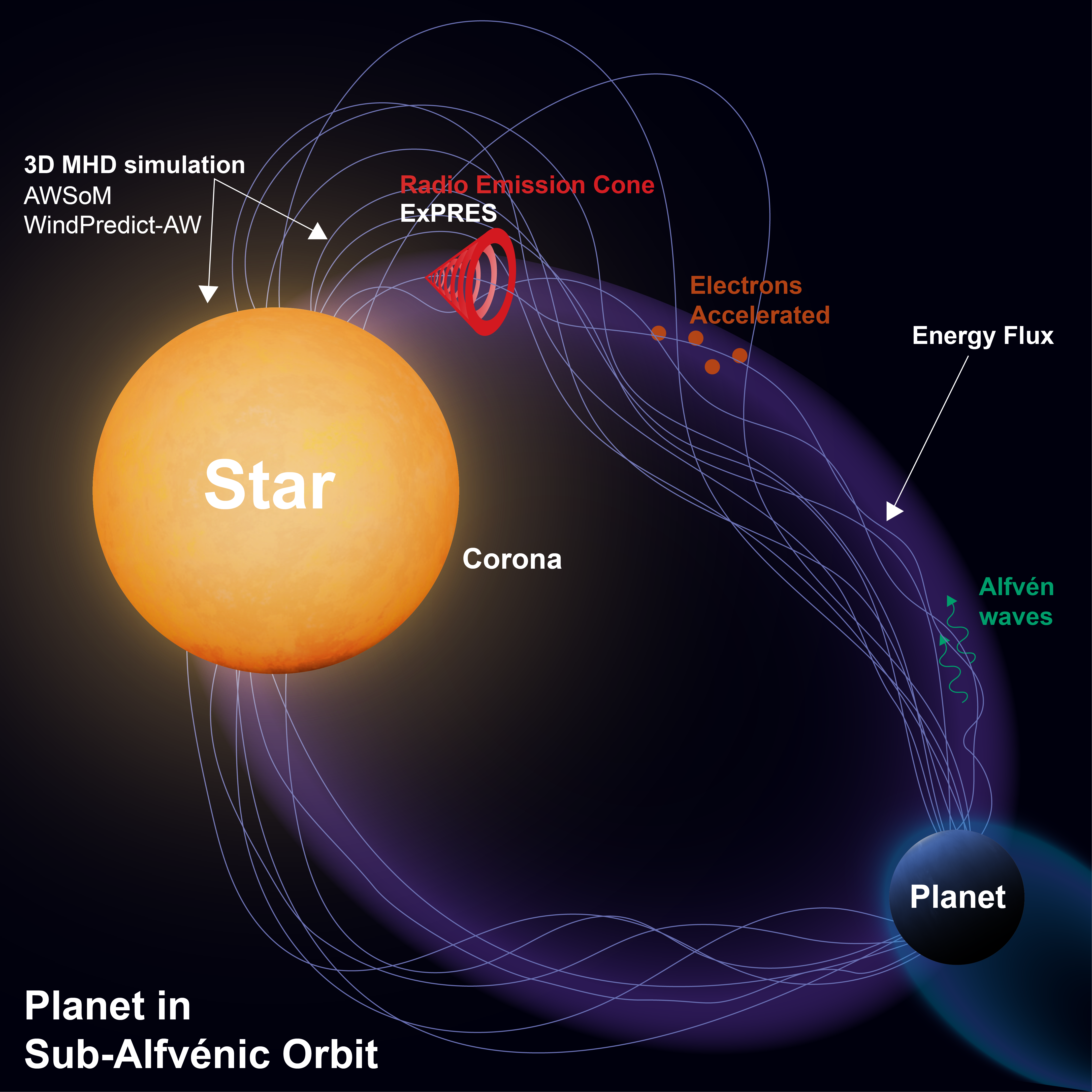}
    \caption{Artist’s rendering of star-planet interaction-induced radio emission. The illustration summarizes the physical processes leading to radio emission, from stellar activity to the transfer of energy mediated by Alfv\'en waves (green), which accelerate electrons to relativistic velocities (red circles). These electrons generate cyclotron radio emission, depicted as a red emission cone. The figure also highlights the use of three-dimensional MHD stellar wind simulations, including the Alfv\'en Wave Solar Model (AWSoM) and WindPredict\_AW. These wind models are coupled with the ExPRES code to simulate and predict radio emission arising from star-planet interactions~(adapted from \citealt{Chebly2026}).}
    \label{fig:SPI_wind_coupling}
\end{figure}

\texttt{ExPRES} has been validated against Solar System cases (Jupiter, Saturn, Io, Ganymede, Europa) and extended to exoplanetary SPI \citep{Hess2011} and stellar radio bursts \citep{Zarka2025}. Using the local magnetic field, \texttt{ExPRES} also predicts the characteristic radio emission frequency via the electron cyclotron frequency: $f_\mathrm{ce} = \frac{e B}{2 \pi m_\mathrm{e}}$ where $e$ is the electron charge and $m_\mathrm{e}$ the electron mass. This allows direct connection between simulated emission and observable radio signals from the star. Recent studies have strengthened this framework by coupling \texttt{ExPRES} with data-driven 3D~MHD Alfv\'en wave driven stellar wind models such as Alfv\'en wave solar model applied to other cool stars (see Figure \ref{fig:SPI_wind_coupling}). 

These simulations self-consistently compute wind acceleration, plasma density, and the large-scale stellar magnetic topology along the planetary orbit, thereby providing realistic boundary conditions for the propagation of Alfv\'en waves and the generation of ECMI emission. A detailed description of the wind models employed by \citet{Chebly2026}, including their numerical setup, magnetic field reconstruction, and sensitivity to coronal base parameters, is provided by \citet{AlvaradoGomez2026}.

\begin{figure}
    \centering
    \includegraphics[width=\linewidth]{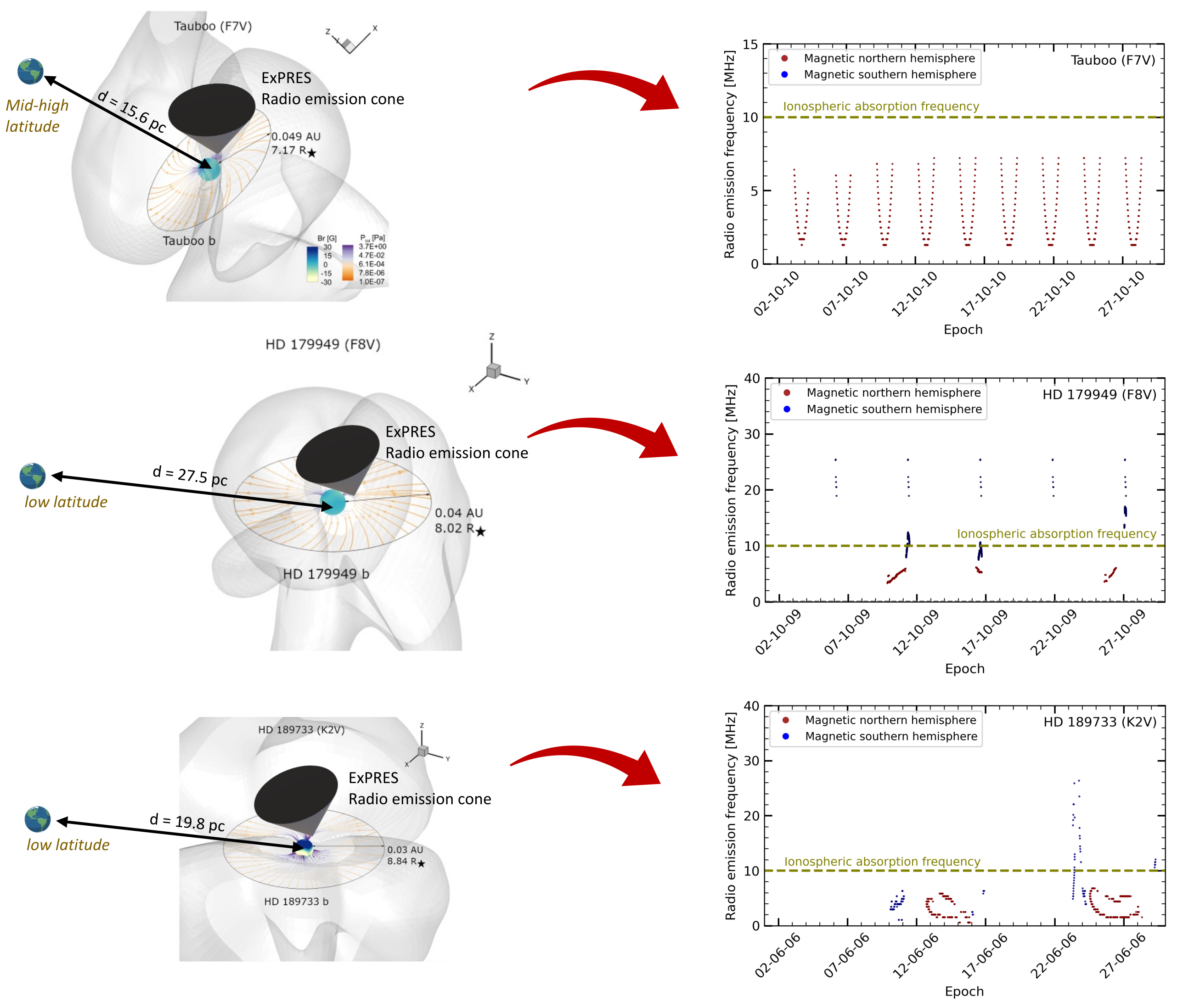}
      \caption{Predicted star-planet interaction (SPI) radio emission for three systems ($\tau$~Boo, labelled `Tauboo' in the figure, HD~179949, and HD~189733) based on 3D MHD stellar wind models driven by ZDI maps of the host stars, which are used as the inner boundary condition (here the model used is the Alfv\'en wave solar model), coupled with ExPRES simulations. \textbf{Left panels:} 3D view of the stellar wind environment and magnetic field topology. The gray translucent iso-surface represents the Alfv\'en surface, while the star is located at the center of the domain. Selected magnetic field lines connecting the star to the planetary orbit are color-coded by the total wind pressure, with higher values shown in purple. The stellar surface is color-coded by the unsigned radial magnetic field. The planetary orbit is indicated by a black circle, and the 3D orientation axes are shown for reference in all panels. The resulting radio emission cones (black) are generated near the star. The cone shown here is illustrative and not intended to represent the modeled emission beams. The observer geometry (Earth line of sight) is indicated for each system, highlighting the role of viewing angle (from mid- to low-latitudes) in determining detectability. \textbf{Right panels:} Time-dependent radio emission frequency as a function of epoch. The first epoch corresponds to that of the ZDI map used as the inner boundary condition in the stellar wind simulation. The emission is computed with ExPRES for both northern (red) and southern (blue) magnetic hemispheres, corresponding to radiation along magnetic field lines directed from the planet to the star (northern hemisphere) and from the star to the planet (southern hemisphere). The dashed horizontal line marks the ionospheric cutoff frequency ($\sim 10$~MHz), below which ground-based observations are not possible due to absorption in the Earth's ionosphere. The results exhibit strong variability with magnetic field strength and topology, with detectable emission occurring primarily when the emission cone intersects the observer’s line of sight. Variations in stellar wind conditions, orbital distance, and magnetic field strength, which together control the efficiency of electron acceleration and ECMI emission. \textbf{Note:} Adapted from \cite{Chebly2026}, with modifications to provide a clearer summary of the modeling framework.}

    \label{fig:SPI-induced_radioemission}
\end{figure}

This provides a realistic description of the local environment, including magnetic field strength, plasma density, and wind velocity—parameters that are critical for Alfv\'en-wing formation and electron acceleration efficiency. In particular, \cite{Chebly2026} used this framework to predict ECMI emission properties such as characteristic frequencies and beaming geometry from 3 well known exoplanetary systems; $\tau$~Boo (F7V), HD~179949 (F8V) and HD~189733 (K2V). However, it is important to note that ExPRES does not directly compute the absolute radio power or intensity of the emitted signal. Instead, it models the emission geometry and visibility based on the distribution of accelerated electrons~(see Figure \ref{fig:SPI-induced_radioemission}). To estimate the radio power, one must rely on empirical or semi-empirical scaling laws, such as those proposed by \cite{Zarka2018}, which relate the emitted radio power to the incident Poynting flux from the stellar wind. These estimates depend sensitively on the underlying stellar wind properties (e.g., magnetic field strength, density, and velocity) provided by the stellar wind MHD simulations.

By incorporating the pitch-angle distribution of accelerated electrons, the simulations account for both loss-cone and shell-type emission, determining the observer-dependent visibility of the signal. This predictive modeling framework serves two complementary purposes. First, it identifies the most favorable conditions for SPI-induced radio emission by constraining orbital phases, emission frequencies, and beaming directions that maximize detectability, thereby enabling targeted observational campaigns with instruments such as LOFAR, GMRT, JVLA, and the upcoming Square Kilometre Array (SKA). Second, it supports the interpretation of observed radio signals to extract key physical parameters. In particular, the emission frequency constrains the local magnetic field via $f_\mathrm{ce} = \frac{e B}{2 \pi m_\mathrm{e}}$, while the inferred radio power—estimated through scaling laws—combined with modulation and beaming properties, provides constraints on the Poynting flux, stellar wind conditions, and planetary magnetic field strength. Together, these diagnostics enable detailed inference of planetary magnetospheric properties, orbital geometry, and the surrounding stellar wind environment.

\subsubsection{Caveats and outlook}
However, there are some caveats. Current models typically assume steady-state stellar wind conditions and neglect temporal variability in magnetic field topology, coronal density, and temperature. These factors influence the location of the Alfv\'en surface, the efficiency of Alfv\'en-wing formation, and the resulting radio emission. In addition, uncertainties in magnetic field reconstructions from ZDI and in the initial conditions of stellar wind simulations propagate into predictions of emission frequency, intensity, and beaming. Future improvements should incorporate time-dependent stellar wind and magnetic field models to capture variability over rotational and magnetic activity cycles. Dynamical coronal conditions would improve the accuracy of sub-Alfv\'enic region predictions and help identify additional promising targets. Moreover, tailoring wind simulations and magnetic reconstructions to individual stars will enhance the reliability of predicted emission characteristics. With these enhancements, the framework offers a scalable, physically consistent approach for detecting, interpreting, and characterizing magnetic star-planet interactions across a broad range of exoplanetary systems. Detecting SPI-induced radio emission is scientifically crucial because it provides rare indirect measurements of exoplanetary magnetic fields, which are otherwise extremely difficult to probe. Magnetic fields influence planetary atmospheric retention and surface radiation environments, directly affecting habitability. Moreover, observed SPI signals constrain the local stellar wind conditions and reveal how close-in planets interact magnetically with their host stars, informing models of magnetospheric dynamics, angular momentum transfer, and energy deposition in exoplanetary systems \citep{Yantis1977, Lazio2004, Zarka2007, Zarka2015, Lazio2024}.

\subsection{Inference of Exoplanetary Magnetic Fields}\label{sec:obs_magfields}
% \begin{itemize}
%     \item \texttt{\textbf{Contributor:} \textbf{Kristina Kislyakova} (nobody else has contributed)}
%     \item \textbf{AP: I will Add in to this section next, specifically on abservations of the He Triplet}
% \end{itemize}

Magnetic fields of exoplanets, as highlighted before, are very difficult to measure remotely. Although theoretical models estimating the strength of magnetic fields of exoplanets do exist, they sometimes predict diverging strengths of the planetary magnetic fields (e.g. \citealp{Christensen2006, Griessmeier2007}). A direct method of measuring magnetic fields of exoplanets is detecting radio emission from them as elaborated in the preceding section. Such observations are difficult and so far, no reliable detections have been reported, although some systems such as the $\tau$~Boo system seem like promising targets \citep{Vidotto2012, Sirothia2014, Turner2021, Tasse2026}. 

Another, more indirect, method was originally suggested by \cite{Vidotto2010,Vidotto2011}. The method allows to estimate magnetic moments of exoplanets from the UV observations of close-in exoplanets. In this method, the magnetopause stand-off distance (i.e., the distance from the centre of the planet to the magnetopause along the star--planet line, set by the pressure balance between the stellar wind and the planetary magnetic field; Sect.~\ref{sec:pressurebalance}) can be found based on Lyman-$\alpha$ observations of planetary transits performed with the Hubble Space Telescope. Close-in gaseous exoplanets often exhibit much deeper and longer transits in the UV light compared to visible observations, due to inflated hydrogen-dominated atmospheres efficiently absorbing the Lyman-$\alpha$ photons. In this method, the estimate of the magnetic moment of an exoplanet is tangled with assumptions on stellar wind parameters, which one has to make to interpret the observations. Later, \cite{Kislyakova2014} has used this method to estimate both the magnetic moment and the stellar wind in the vicinity of an exoplanet HD~209458b. \cite{Vidotto2017} have also used the method to estimate the mass-loss-rate from from a red dwarf star GJ~436. 

%In a similar manner, one can try to estimate exoplanetary magnetic fields based on polarization signatures in the He I 1083~nm triplet (e.g. \citealp{Oklopcic2020,Strugarek2025Heline}. However, such detections are difficult with the current facilities and might require the capabilities of new generation missions such as the Habitable Worlds Observatory. 

In a similar manner, one can try to estimate exoplanetary magnetic fields based on polarization signatures in the He~{\sc i} 1083~nm triplet. This method, as described by \citet{Oklopcic2020}, relies on the presence of metastable helium atoms ($2^3S_1$) in the extended or escaping atmospheres of hot, close-in exoplanets. These atoms are optically pumped by the anisotropic radiation from the host star, which creates a population imbalance (alignment) among different atomic sublevels. In the presence of a planetary magnetic field, this alignment is modified through the Hanle effect, which results in linearly polarized absorption at 1083 nm that traces the field's direction. More precisely, the atomic alignment is modified by Larmor precession over the lifetime of the polarized level, which alters both the amplitude and the orientation of the linear polarization signal (Stokes $Q$ and $U$). Unlike the Zeeman effect, which is primarily sensitive to the line-of-sight component of the field (Stokes $V$) and is often difficult to detect in broadened exoplanetary lines, the Hanle effect is sensitive to field strengths comparable to the critical Hanle field (set by the inverse lifetime of the relevant level) and depends on the full vector geometry of the field. In principle, this allows the 3D orientation of the field within the escaping atmospheric wind to be reconstructed, providing a direct map of the magnetospheric cavity. This technique is particularly versatile because it is sensitive to a broad range of magnetic field strengths, from as low as a few $\times 10^{-4}$ G up to 800 G, covering the field strengths typically found in Solar System planets. Furthermore, the line-of-sight component of the magnetic field can induce a slight circular polarization via the Zeeman effect. While the predicted signals are relatively weak, with linear polarization levels ranging from approximately $10^{-3}$ in optimistic cases to $10^{-5}$ in less favorable geometries, the method provides a direct way to probe the magnetic environment of the upper atmosphere. Such detections are challenging for current facilities but may be achievable with high-resolution infrared spectropolarimeters such as SPIRou or future large-scale observatories like the Extremely Large Telescope (ELT). Atmospheric escape from exoplanets and their implications on habitability has been covered in considerably more detail by \citet{Segura2026}. 

\subsection{Interior Signatures and Volcanism}
Magnetic fields also provide a key observational probe of the interior structure and thermal evolution of planets and satellites. Across the Solar System, observed surface magnetic field strengths $B_p$ span several orders of magnitude, from $\sim 300$ nT at Mercury \citep{1975JGR....80.2708N,2010SSRv..152..307A} to $\sim 4 \times 10^{5}$ nT at Jupiter \citep{2018GeoRL..45.2590C}. These fields originate either from self-sustaining magnetohydrodynamic (MHD) dynamos operating within electrically conducting layers of planetary interiors or, in some satellites, from electromagnetic induction driven by time-varying external fields. In both cases, magnetic fields are direct consequences of internal structure and energy transport processes.

A self-sustaining large-scale dynamo requires four fundamental ingredients: an electrically conducting fluid, sustained fluid motion, rotation, and a persistent internal energy source. These conditions are intimately linked to interior structure and thermal evolution. Conducting regions vary widely across planetary bodies, ranging from liquid iron alloys in terrestrial planet cores, to metallic hydrogen in gas giants, and ionic or partially degenerate fluids in ice giants \citep{Kivelson2002, 2003E&PSL.208....1S, 2015MPLB...2930018N, 2017PNAS..11411873Z}. The existence and extent of such layers are set by internal pressure and temperature profiles, which in turn reflect planetary composition and evolutionary history.

Convective motion within these regions is driven by thermal and compositional buoyancy, arising from secular cooling, latent heat release, and differentiation processes such as core formation or inner-core solidification \citep{2000Sci...288.2007B,2007cody.book...31N}. These processes are fundamentally tied to a planet’s ability to transport internal heat, which also influences geological and volcanic activity in rocky bodies. Planetary rotation further shapes convection through Coriolis forces, organizing turbulent flows into columnar structures that are favorable for large-scale magnetic field generation \citep[e.g.,][]{Christensen2006}. Sustained dynamo action ultimately requires a long-term energy source, typically provided by continued cooling or radiogenic and gravitational energy release.

In addition to intrinsic dynamos, several icy satellites exhibit induced magnetic fields generated by the interaction between external planetary fields and subsurface conducting layers. These induced responses provide some of the strongest geophysical evidence for present-day subsurface oceans, linking magnetic observations directly to internal structure and thermal state \citep{1998Natur.395..777K,2025AGUA....601237C}.

Despite the complexity of dynamo processes, planetary and satellite magnetic fields exhibit empirical correlations with bulk properties. This behavior is often referred to as magnetic Bode’s law, which relates magnetic dipole moments (or surface field strengths) to planetary mass and rotation rate (or angular momentum) \citep{1978Natur.272..147R, 1999JGR...10414025F, Christensen2006, 2006E&PSL.250..561O}. These correlations reflect the coupling between interior structure, convective power, and rotation.

The physical basis of this scaling lies in convective dynamo theory, where more massive planets tend to develop larger volumes of electrically conducting material due to higher internal pressures and temperatures. This increases the size of the dynamo region, which is ultimately controlled by interior structure. Rotation governs the organization of convection: rapidly rotating planets favor low-Rossby-number flows that produce stable, dipole-dominated fields.

A heuristic expression for the planetary magnetic dipole moment is given by
\begin{equation}
\mu_p \propto L_p \sim \omega_p M_p R_p^{2},
\end{equation}
where $M_p$, $R_p$, and $\omega_p$ denote planetary mass, radius, and rotation rate. Using mass–radius relations \citep[e.g.,][]{2017ApJ...834...17C}, gas giants exhibit approximately constant radii with increasing mass, while rocky planets follow $R_p \propto M_p^{1/3}$.

The surface magnetic field then scales as
\begin{equation}
B_p \propto \frac{\mu_p}{R_p^{3}} \propto \frac{M_p}{R_p},
\end{equation}
leading to $B_p \propto M_p$ for gas giants and $B_p \propto M_p^{2/3}$ for rocky planets.

More physically grounded descriptions are based on energy flux scaling within the convecting interior. Geodynamo simulations show that magnetic field strength is controlled by buoyancy-driven convective power \citep{Christensen2006}, which in turn is closely related to a planet’s thermal evolution. This connection links magnetic field generation directly to interior cooling and energy transport processes, which also govern volcanic and tectonic activity in rocky planets.

It has been shown that buoyancy power correlates with surface luminosity across a wide range of objects, including giant planets and low-mass stars \citep{2009Natur.457..167C}. This leads to a general scaling for the magnetic field at the dynamo region \citep{2010A&A...522A..13R},
\begin{equation}
    B_{\text{dyn}} \approx 4.8 \,\text{kG}
    \left( \frac{M}{M_{\odot}} \right)^{1/6}
    \left( \frac{L}{L_{\odot}} \right)^{1/3}
    \left( \frac{R}{R_{\odot}} \right)^{-7/6}.
\end{equation}

For giant planets, the observable dipole field is related to the dynamo field through
\begin{equation}
  B_{\mathrm{dip}}^{\mathrm{eq}} = \frac{B_{\mathrm{dyn}}}{2\sqrt{2}} \left(1 - \frac{0.17}{M_p/M_{\mathrm{J}}}\right)^3,
  \label{Bdip}
\end{equation}
with the polar field given by $B_{\rm dip}^{\rm pol} = 2 B_{\rm dip}^{\rm eq}$.

\begin{figure}[htb]
    \centering
    \includegraphics[width=0.5\linewidth]{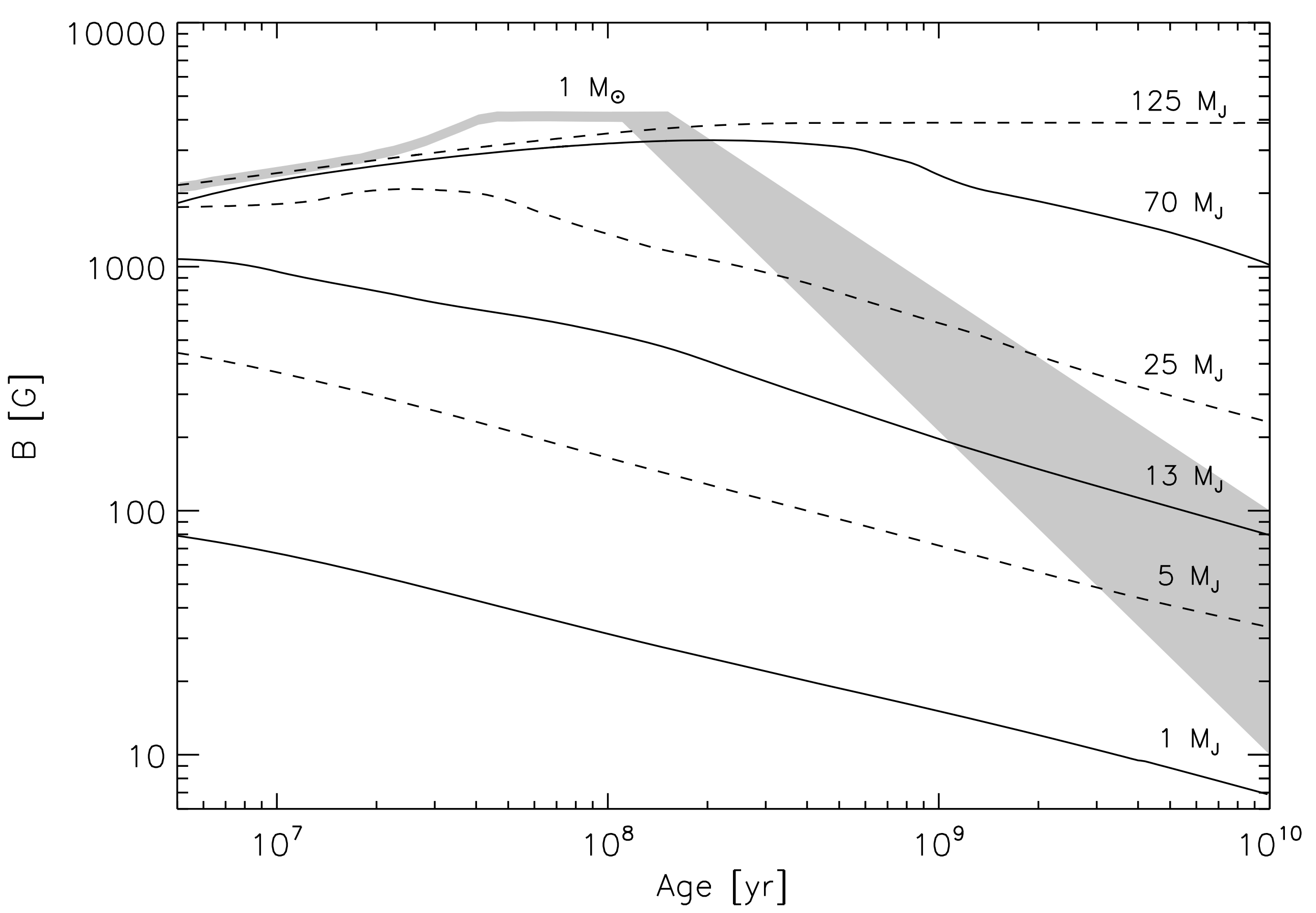}
    \caption{The average surface magnetic field $B_{\mathrm{dyn}}$ for objects with $M > 13 M_{\mathrm{J}}$, and the dipole field $B_{\mathrm{dip}}^{\mathrm{pol}}$ for objects with $M \le 13 M_{\mathrm{J}}$, are shown as a function of age for giant planets, brown dwarfs, and a very low-mass star with $M = 125\,M_{\mathrm{J}}$. All low-mass objects are assumed to be rapidly rotating. For comparison, an estimate of the Sun’s average magnetic field is also shown as a shaded gray region (Adapted from \citet{2010A&A...522A..13R}).}
    \label{fig:mag_model}
\end{figure}

Using evolutionary models for substellar objects \citep{2003A&A...402..701B}, the mass, luminosity, and radius can be tracked as a function of age, allowing the temporal evolution of magnetic fields to be inferred. Figure~\ref{fig:mag_model} shows the predicted magnetic field strengths of exoplanets and brown dwarfs as a function of age. Importantly, these scaling relations refer to the dynamo region rather than the visible surface. In most planets, the dynamo operates deep in the interior, near the transition to a liquid metallic phase at pressures of order $\sim$1~Mbar \citep{2017ApJ...849L..12Y,2019ApJ...872...51C}, which is itself determined by interior structure and composition.

The observational detection of active volcanism relies on the synthesis of chemical, thermal, and magnetospheric signatures that distinguish geologically active bodies from inert ones. The presence of short-lived chemical species in the upper atmosphere provides a primary diagnostic for active outgassing, as volcanic discharge injects volatiles such as sulfur dioxide (SO$_2$), atomic sodium (Na), and potassium (K) into the environment \citep{Kaltenegger2010, Misra2015}. Because these species are rapidly sequestered by surface mineralogy or destroyed via stellar UV photolysis, their persistent detection in a terrestrial-sized atmosphere necessitates a continuous replenishment rate consistent with high-flux volcanic recycling. Complementing these chemical tracers, thermal phase curves and secondary eclipse mapping provide direct constraints on the planetary internal heat budget. In systems dominated by intense tidal dissipation, such as the super-Earth 55 Cancri e, observed brightness temperatures often exceed those predicted by radiative equilibrium with the host star \citep{Demory2016, Kreidberg2016}. Such thermal anomalies, which may manifest as significant phase shifts in thermal emission or localized surface ``hotspots,'' are indicative of large-scale magma oceans or transient volcanic plumes that dominate the planetary heat flux. Furthermore, volcanic activity exerts a profound influence on the circumplanetary environment through the injection of neutral material into the magnetosphere. Upon ionization, this material forms a dense plasma torus that facilitates the ECMI, leading to the generation of coherent radio emissions \citep{Zarka1998, Saur2013}. Consequently, the detection of periodic, high-frequency radio bursts provides a dual probe of the planetary magnetic field strength $B_{\rm p}$ and the mass-loading rate from volcanic sources \citep{2018ApJ...854...72T, Oza2019}.

\section{Extreme Frontiers and Edge Cases of SPI}
\subsection{SPI Mediated by Energetic Transients}

The studies by \citet{2022ApJ...928..147A} and \citet{2022ApJ...934..189C} jointly frame Coronal Mass Ejections (CMEs) as central agents in mediating star-planet interactions in magnetically active systems. Rather than focusing on stellar irradiation or steady winds as the dominant pathways for SPI and its forcing on planetary atmospheric evolution, these works explore how energetic magnetic transients can fundamentally regulate the coupling between a young stars and close-in planets. Using the AU Microscopii (AU Mic) system as a case study (see \citealt{2020Natur.582..497P, 2021A&A...649A.177M}), the combined analysis establishes how CME initiation, magnetic confinement, and fragmentation propagate through the stellar environment and imprint themselves directly on planetary magnetospheres, atmospheric escape processes, and observational diagnostics.

\citet{2022ApJ...928..147A} investigate the stellar side of this interaction by modeling the quiescent and eruptive magnetized environment of AU Mic with fully 3D MHD simulations. The stellar corona and wind are computed using the Alfvén Wave Solar Model (AWSoM; \citealt{2014ApJ...782...81V}), part of the Space Weather Modelling Framework (SWMF; \citealt{2021JSWSC..11...42G}), constrained by magnetic field information obtained with Zeeman Broadening and Zeeman-Doppler Imaging (ZDI) observations of the star \citep{ 2020ApJ...902...43K,2021MNRAS.502..188K}. Energetic magnetic transients are introduced through the eruption of a Titov-Démoulin flux rope \citep{1999A&A...351..707T} whose parameters are chosen to represent one of the most powerful CME candidates inferred for AU Mic (see \citealt{1994ApJ...435..449C, 1999ApJ...510..986K}). This framework assumes that the global magnetic topology dominates CME evolution and that large-scale dynamics can be captured without explicitly resolving flare reconnection or energetic particle populations. By holding the eruption properties fixed and varying the background magnetic field, the study isolates the role of stellar magnetism in shaping CME-driven star-planet coupling.

\begin{figure*}[!t]
\centering%  left, bottom, right and top
\vspace{-0.36cm}
\includegraphics[trim=0.0cm 0.0cm 0.0cm 0.0cm, clip=false, width=0.99\textwidth]{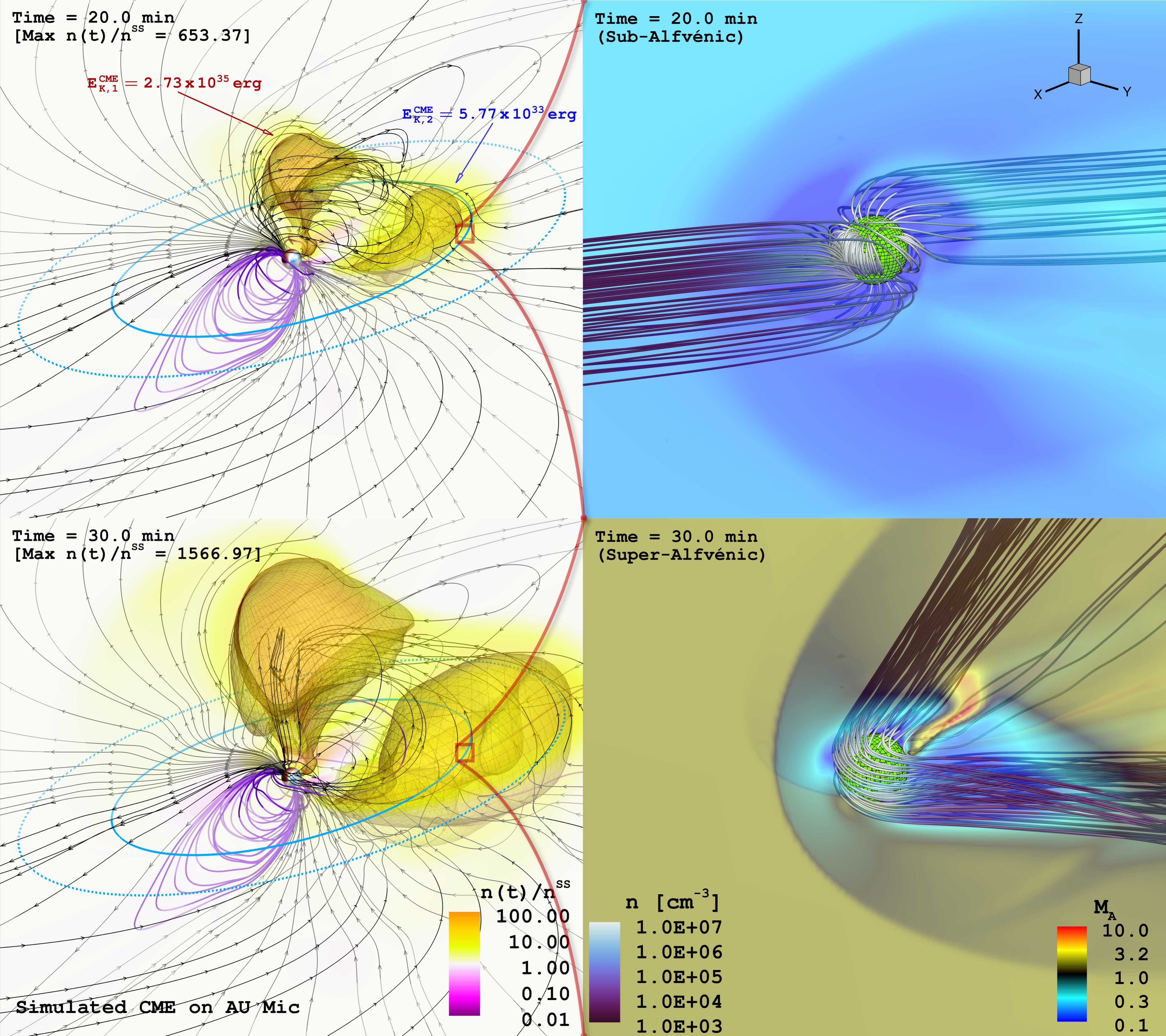}\vspace{-0.4cm}
\caption{Snapshots from a CME simulation of the AU Mic system illustrating how energetic magnetic transients modulate star-planet interactions and their potential radio signatures. Prior to the arrival of the CME (top panels), the planet orbits within a sub-Alfvénic stellar wind, giving rise to a magnetospheric configuration characterized by Alfvén wings around a planet, assumed to host an Earth-like magnetic field (top-right panel). In this regime, Alfvénic coupling between the planet and the star may enable SPI-driven radio emission via the Electron Cyclotron Maser Instability (ECMI) in the stellar corona. During CME passage (bottom panels), large perturbations in plasma density, flow speed, and magnetic field strength drive the local environment into a super-Alfvénic state. The Alfvén wing geometry is consequently disrupted and replaced by a compressed, quasi-closed magnetosphere bounded by a bow shock (bottom-right panel). This CME-driven transition suppresses the Alfvénic coupling required for SPI-ECMI emission, leading to a temporary shutdown of the associated radio signal until pre-eruption conditions are re-established. Detecting such transient modulation of SPI-induced radio emission offers a potential indirect pathway for identifying highly energetic CMEs in close-in exoplanetary systems. \textbf{Adapted from:}~\citet{2022ApJ...928..147A} and \citet{2022ApJ...934..189C}.}\label{Fig:SPI_CMEs}
%\vspace{-0.4cm}
\end{figure*}

The simulations reveal that CMEs in the AU Mic system do not propagate as simple, coherent disturbances but instead undergo strong magnetic confinement and fragmentation (see also \citealt{2018ApJ...862...93A, 2019ApJ...884L..13A}). As the erupting flux rope interacts with the complex stellar magnetic field, it breaks into multiple magnetized plasma structures with distinct trajectories, masses, and velocities. This fragmentation is a critical mediator of star-planet interactions, as it determines which portions of the eruption intersect the planetary orbital plane and on what timescales. Consequently, the impact of a CME on AU Mic b is governed not by the total eruptive energy but by the properties of the CME fragments that will eventually become magnetically connected to the planet (Fig.~\ref{Fig:SPI_CMEs}, left panels). When these fragments reach the planetary orbit, they drive transient but extreme enhancements in stellar wind pressure that dominate the local plasma environment.

A defining feature of this CME-mediated interaction is the transition of the stellar wind at the planetary orbit from a sub-Alfvénic to a super-Alfvénic regime during eruption passage (Fig.~\ref{Fig:SPI_CMEs}, right panels). Under quiescent conditions, AU Mic b is expected to resides largely in a sub-Alfvénic flow, enabling strong magnetic coupling between the planet and the star (see also \citealt{2021MNRAS.504.1511K}). In such regimes, star-planet interactions are expected to generate Alfvén waves that propagate upstream toward the star and power coherent stellar radio emission via the electron cyclotron maser instability, as described in Section \ref{sec:emissionnearplanet} (see also \citealt{Saur2013, 2018ApJ...854...72T}). The simulations show that CME-driven magnetic transients disrupt this coupling by lifting the sub-Alfvénic condition, thereby suppressing Alfvén wave transmission and quenching any planet-induced ECMI radio emission.

This CME-induced disruption leads to a counterintuitive manifestation of star-planet interactions at radio wavelengths. Rather than producing enhanced emission, the passage of a CME may result in a possibly detectable transient radio dimming, provided that a baseline star-planet interaction signal can be identified. This phenomenon is analogous to coronal dimmings observed during solar and stellar CMEs at EUV and X-ray wavelengths \citep{2016ApJ...830...20M, 2021NatAs...5..697V, 2025ApJ...988..167M}, but it arises from the temporary shutdown of magnetic coupling rather than plasma evacuation. If the CME is associated with a stellar flare, the time delay between the flare onset and the disappearance of the ECMI signal could offer a means of estimating the CME propagation speed, given knowledge of the planetary orbit. Although the detectability of such signals is not quantified, the study highlights CME-driven radio dimming as a novel diagnostic of energetic magnetic transients, particularly relevant in light of the limited number of stellar CME candidates and detections \citep{2019NatAs...3..742A, 2019ApJ...877..105M, 2022SerAJ.205....1L, 2025ApJ...993...80N}) and the growing capabilities of low-frequency radio observations of cool stars \citep{Vedantham2020, Callingham2021, 2025Natur.647..603C}.

\citet{2022ApJ...934..189C} extend this CME-centric view of star-planet interactions to the planetary environment by coupling the time-dependent stellar wind and CME solutions to a global magnetospheric model of AU Mic b that includes atmospheric escape. The planetary atmosphere is represented by a hydrodynamic outflow of neutral hydrogen imposed at the inner boundary, allowing the simulations to focus on how external magnetic and dynamic pressures associated with CMEs regulate atmospheric structure and evolution. This approach assumes that the dominant control on the escaping atmosphere arises from stellar magnetic transients rather than from internal thermodynamic feedback or detailed ion chemistry.

In the absence of CMEs, the steady stellar wind already imposes strong magnetic and ram pressure on the planet, shaping the outflow and producing highly asymmetric and orbit-dependent structures. Synthetic Ly$\alpha$ transit signatures derived from these conditions vary significantly with orbital phase, indicating that even quiescent star-planet interactions are intrinsically time dependent. The introduction of CME-driven transients amplifies this variability dramatically. As CME fragments engulf the planetary magnetosphere, the external pressure overwhelms the atmospheric escape flow, suppressing it entirely and, in some cases, reversing it into an inflow. The CME plasma displaces neutral atmospheric material and erodes the magnetospheric cavity, directly altering the observable Ly$\alpha$ absorption.

These results demonstrate that energetic magnetic transients can dominate both the dynamics and observability of star-planet interactions in young, active systems. The simulations show that high-velocity Ly$\alpha$ absorption observed during exoplanet transits is unlikely to reflect the intrinsic speed of atmospheric escape. Instead, such signatures are more plausibly produced by the interaction between stellar wind transients and planetary material, including sweeping and charge-exchange processes (see also \citealt{2008Natur.451..970H, murrayclay2009, 2019ApJ...873...89M, 2022MNRAS.517.1724D}). In this framework, CMEs act as episodic regulators that intermittently suppress atmospheric escape while reshaping the magnetospheric interface between the star and the planet.

Both studies \citep{2022ApJ...928..147A, 2022ApJ...934..189C} acknowledge limitations inherent to this modeling approach, including the omission of explicit flare reconnection, energetic particles, and self-consistent ionization and heating in the planetary atmosphere. Nevertheless, the results robustly establish CMEs as key mediators of star-planet interactions, capable of altering magnetic connectivity, shutting down interaction-driven emissions, and temporarily erasing canonical observational signatures of atmospheric escape. Both investigations underscore the need for time-dependent, fully coupled stellar-planetary MHD models to interpret observations of young, magnetically active exoplanetary systems.

One should note that the general role played by the CMEs for the atmospheric escape is not understood yet. While the studies cited above show that CMEs can suppress atmospheric escape, other works show that CMEs can temporarily increase non-thermal escape both from exoplanets \citep{Lammer2007,Khodachenko2007,Hazra2025}  and Solar System planets \citep{jakosky2015,Dimmock2018,Ma2020,Sakata2022}.

\subsection{Exoplanets around Pulsars}
% \begin{itemize}
%     \item \texttt{\textbf{Contributors:} Miljenko Cemeljic, Sergio Joya}
% \end{itemize}
%\section{Introduction} \label{sec_intro}
%Intro by Sergio and Miki
An edge case for SPI is the pulsar environment. The magnetosphere of a pulsar is one of the most extreme places in the universe due to its properties like strong magnetic fields and highly relativistic plasma. With the planets observed around pulsars, new open questions arise, as the possibly observable effects are in the form of radio emission, facilitated by such magnetosphere and pulsar winds.

Pulsar magnetosphere is a plasma-rich region outside a rapidly rotating neutron star (NS), dominated by its magnetic field. NSs are the end product of the main-sequence stellar evolution of massive stars (more than $\approx 10 M_{\odot}$; \citealp{Iliadis}). The magnetosphere acts as a mediator through which the rotational energy of the NS is conveyed to accelerating charged particles and producing multi-band EM emission, which is observed as pulsars \citep{Lorimer}.

The first mention of the possibility of a planet around pulsar was related to the timing-method observations of the bright pulsar PSR B0329+54 \citep{DemPro79}. The suggested planet was later not confirmed \citep{1985ApJS...59..343C}, however, this system still remains a tentative case \citep[and references therein] {StaRod17}. A serendipitous discovery of planets around the millisecond pulsar PSR 1257+12 \citep{WolFra92} came as a surprise, amplified by the fact that the three planets, one with 0.02 and other two with 4 Earth masses, respectively \citep{Wol94, Konacki_2003}, were the first confirmed exoplanets. Nobody expected them around a pulsar \citep{Wolszczan2012}.

The fraction of pulsar systems that have planetary or planet-like companions is less than 0.5 \% \citep{Nitu}. Such scarcity suggests that survival of planets around stars forming the NS is either rare, or our methods of observation are not sufficiently sensitive for their detection. The large number of NS in the Galaxy ($10^{9}$) still guarantees existence of a relatively large number of such planetary systems. It is important to assess the effects of the high energy pulsar emission on such planets, that might be detected using the (pulsed) emission, scaled to much higher energies, due to the interaction of the pulsar wind and the planet. It could lead to the observable aurora events on exoplanets orbiting around pulsars \citep{MishraR}.

%At distances of the order of an astronomical unit, pulsar winds are expected to still be Poynting-flux-dominated. The EM energy density of such waves is $B_{0}^{2}/\mu_{0}$, where $\mu_{0}$ is the magnetic permeability of vacuum, is much higher than the plasma energy density $\gamma_{0}\rho_{0}c^{2}$, with $\gamma_{0}$ being the Lorentz factor and $\rho_{0}$ the rest-mass density of a supposedly cold wind. These quantities are measured in the observer's frame. Alfv\'{e}nic perturbations in such conditions propagate at a phase velocity close to the speed of light, so that the flow might remain sub-Alfv\'{e}nic, rendering a planet unscreened from the wind by a bow shock. When the wind crosses the Alfv\'{e}n wing, it encounters a rotation of the ambient magnetic field, causing electromagnetic wave instabilities. In the observer’s reference frame, such waves of extremely high intensity are collimated in a very narrow range of directions. 

Radio emission from planets and small objects in the pulsar wind was first studied in \cite{Mottez1, Mottez2, Zarka2020}, extending the theory of Alfv\'{e}n wings to the relativistic regime. In \cite{Mottez2014} and \cite{Zarka2020}, a relativistic aberration narrowing the beam of the emission from such wings near the smaller body orbiting a pulsar was evoked as a possible explanation for fast radio bursts. 

%Because of the extremely harsh conditions in the environment near the pulsar, planetary objects found there could provide insights about novel forms of planetary environment. Pulsar planets are the perfect probes to investigate such conditions. Numerical simulations involving pulsar planets could provide information needed for a direct observation of a larger number of such objects whose existence is, until now, known only indirectly, from the timing measurements of pulsars. Indicating EM observational signatures of planets from simulations, would render them subject to further investigation.

%When a magnetized flow in the stellar wind is obstructed by the presence of an obstacle such as a planet or its moon, it leads to a partial dissipation of this energy flow. A fraction of the energy is dissipated in different ranges of the electromagnetic (EM) spectrum. The study of star-planet magnetospheric interaction can be used to analyze the radiation emanating from the process of dissipation. This interaction will depend on the stellar wind, interplanetary magnetic field (IMF, the component of the stellar wind carried by the stellar wind into the interplanetary space), and the intrinsic planetary magnetic field. In the case of planets orbiting non-compact stars, such interaction is observed as auroral radio emission \citep{Desch}. It is due to the energetic electrons traveling along the magnetic field lines which are formed in the reconnection region between the planetary magnetic field and IMF \citep{WuLee}.

In the Solar System, aurora-like emissions are observed from most of the planets and also some of their moons \citep{Badman}. In the case of planets orbiting non-compact stars, such interaction is observed as auroral radio emission \citep{Desch}. It is due to the energetic electrons traveling along the magnetic field lines which are formed in the reconnection region between the planetary magnetic field and IMF \citep{WuLee1979}. The radiation from magnetized planets in low-frequency radio range is only one to two orders of magnitude less intense than the radiation produced by the Sun \citep{Zarka2007}. Magnetized planets of the Solar System are sources of non-thermal radio emission. The source region is close to the auroral field lines at distances of two to four times the planetary radii. For the total emitted power, simple scaling laws have been obtained, leading to expected radio signals of the orders of hundreds of mJy \citep{Winterhalter}. As for the possibility of observing an aurora on a planet around pulsar, this is still a novelty, first introduced by \cite{MishraR}, and there is yet no observational confirmation. Magnetospheric interaction in the vicinity of a pulsar, producing analogous emission to an aurora, is a potential channel for obtaining observational information about the pulsar's nearest physical environment and in particular the pulsar wind.

\begin{figure*}
    \centering
    \includegraphics[width=0.60\linewidth]{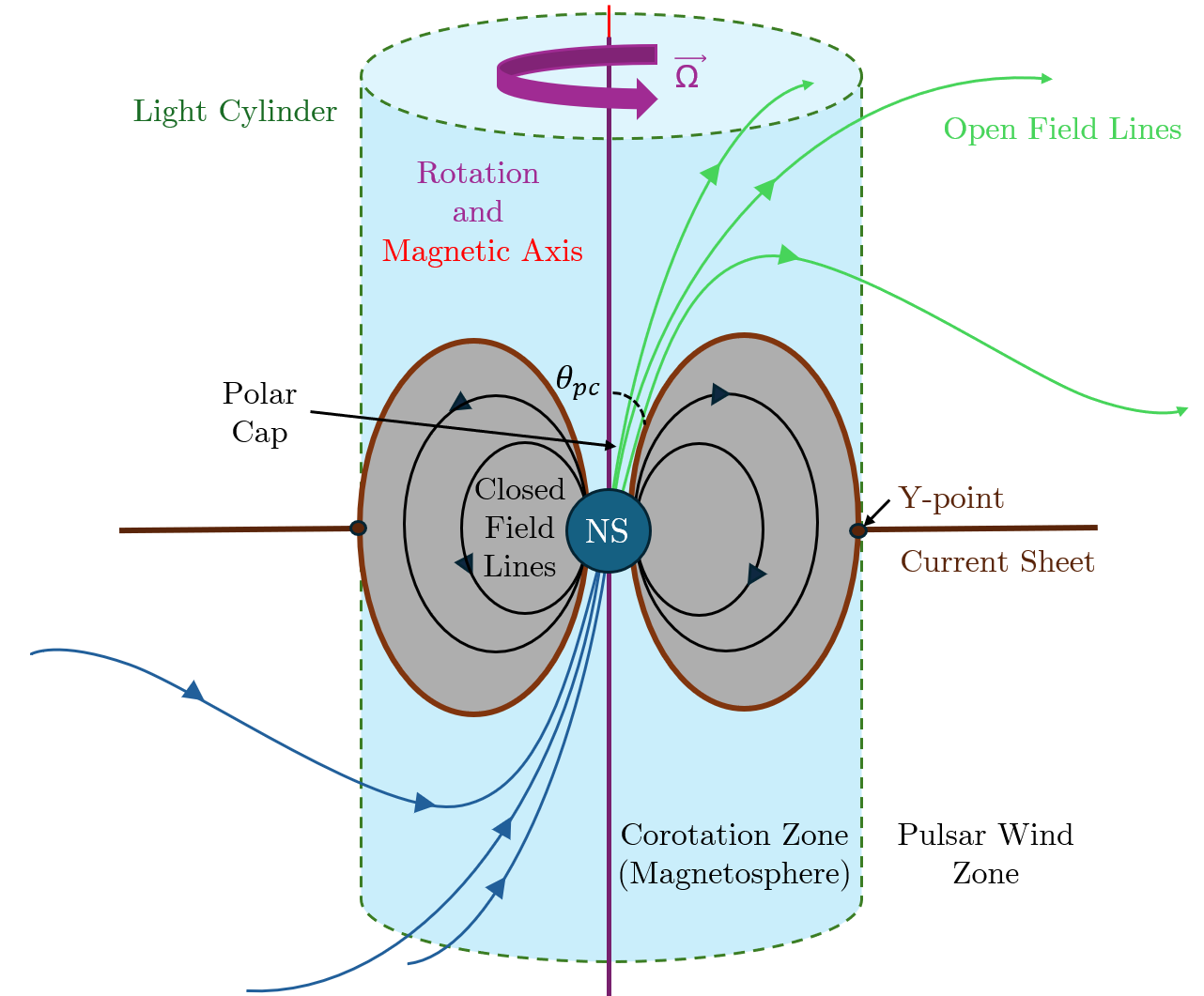}
    \caption{A schematic of the canonical pulsar magnetosphere in the simplified aligned-rotator configuration (magnetic and rotation axes parallel); real pulsars are generally oblique rotators. Adapted from \cite{Cerutti_2016}.}
    \label{fig:AlignedRotator}
\end{figure*}

Pulsars behave as unipolar inductors as they are highly conductive, rotate at high velocities, and are provided with an intense magnetic field (see section \ref{sec:unipolar_inductor}). In general, they are oblique rotators, i.e., their magnetic axis is inclined with respect to the rotation axis, that may be seen as magnetic dipoles rotating at high angular velocities. Beyond the light-cylinder zone and as a result of the rotation of the pulsar, the magnetic field lines become spiral-shaped and wound into a Parker spiral \citep{Bogovalov}. Due to the large angular velocity, the spirals are tightly wound up and the magnetic field beyond the light cylinder $r_{LC}$ can be approximated as purely azimuthal \citep{Kirk}. The approximate geometry of a canonical pulsar and its magnetic field is depicted in Fig.~\ref{fig:AlignedRotator} for the simplified case of an aligned rotator (magnetic and rotation axes parallel), which already captures the main structural elements relevant here: the corotating region of closed field lines, the light cylinder, the open field lines, the equatorial current sheet, and the outflowing pulsar wind.

\textbf{Location of Pulsar Companion.}
In the course of a pulsar formation, any planets orbiting the initial star will be exposed to several hazards and catastrophic threats. Hence a key question to ask is whether any planets can survive the subsequent supernova. Consider a binary system, consisting of a planet and  a massive star, that will become supernova. As the star evolves, the planet could be engulfed and spiral into the supergiant envelope, or it could be destroyed by the supernova ejecta. When the explosion is symmetric in mass ejection, the considerable and fast loss of mass might cause the remaining system to become unbound. There is no kick velocity imparted to the remnant star and the planet is sent away.  If the planet still remains in its orbit after these events, it is not totally safe, as it could be destroyed by the high energy pulsar wind. \cite{Planets1} describes several scenarios of planetary evolution during various post-MS evolutionary stages of their host star. 

\textbf{Pulsar wind characteristics.}
Pulsar winds are not completely understood and there exists several models, such as those of \cite{PWC1}, \cite{PWC2}, \cite{PWC3}, and \cite{PWC4}. However, they all agree on the fact that the wind is dominated by the Poynting flux at distances of the order of an astronomical unit. Hence, the electromagnetic energy density dominates the plasma energy density. To use the theory of \cite{Neubauer_1980} for the formation of Alfvénic wings and generalize it, it is necessary to prove that the flow is sub-Alfvénic, which is shown by models such as \cite{Mottez1} and \cite{Kirk} for aligned pulsars. This is valid for planets which are found at distances larger than $r_{LC}$, meaning that they orbit in a sub-Alfv\'{e}nic relativistic wind around the pulsar.

\textbf{Relativistic Alfv\'{e}n Wings.}
Following the study of \cite{Mottez1, Mottez2, Zarka2020}, about radio emission from planets and small objects in the relativistic pulsar wind, one extends the theory of Alfvén wings to the relativistic regime.  The treatment of the non-relativistic regime closely parallels the derivation provided by \cite{Neubauer_1980}. The idea is to use the equations of special-relativistic ideal MHD and the Lorentz transformations in order to find an analogous relation  for the relativistic case. 

The maximum power involved in an Alfvén wing is:
\begin{equation}
    \dot E_{J}= \frac{\pi c R_{ob}^2(B_{0}^{\phi})^{2}}{\mu_{0}},
\end{equation}
where $R_{ob}$ is the obstacle radius, $B_{0}^{\phi}$ is the azimuthal magnetic field of the pulsar, $\mu_0$ is the vacuum magnetic permeability, and $c$ is the speed of light. A specific instance of exoplanet interaction with highly magnetized winds occurs in pulsar systems. Outside the light cylinder, pulsar winds are typically relativistic. Therefore, the flow sweeps both Alfvén wings away from the planetary body, regardless of whether the planet itself is magnetized \citep{Mottez1}.

\textbf{Numerical simulations of pulsar-planet interaction.}
The results in numerical simulations of pulsar–planet magnetospheric interaction for quantifying the emitted electromagnetic radiation from pulsar planets were presented in \cite{MishraR}. They implemented the PLUTO code \citep{Mignone2007} setup in 3D ideal MHD by \cite{Varela2016a, Varela2018, Varela2022}, which was previously used in a pilot study for pulsar planets in Newtonian version in \cite{cemvar23}. Now it was adopted to use with the special relativistic module.  The pulsar itself was assumed far away, out of the computational box.
\begin{table}
\caption{The predicted intensity of the radio emission flux $\Phi$ for a ground observer in simulations with conducting and ferromagnetic planetary surfaces. The planets are positioned at 750~pc, the distance to the planets around PSR 1257+12. For comparison, the corresponding values with the distances of many known pulsars, at 250~pc and 100~pc, are also shown. With expected larger value of $B_{\rm sw}$ for pulsars, equal to 7.4 G and 13 G in each of the cases, the $\Delta\nu$ would be 20.1 and 36.4, respectively, rendering ground observations possible (shown in brackets).}.
\centering                          % used for centering table
\begin{tabular}{ c c c c c c c }        % centered columns
\hline    % inserts single horizontal line
 Set-up & $\Phi_{\mathrm a}(750)$ & $\Phi_{\mathrm b}(250)$ & $\Phi_{\mathrm c}(100)$ & $P_{\rm radio}$ & $B_{\rm sw}$ & $\Delta\nu$ \\
 & (mJy) & (mJy) & (mJy) & (W) & (G) & MHz  \\
\hline\hline
 Pulsar-planet (cond.) & 0.60 & 5.4 & 33.75 & $3.65\times 10^{12}$ & 0.0025 (7.4) & 0.007 (20.1) \\
\hline 
Pulsar-planet (ferrom.) & 0.47 & 4.23 & 26.43 & $1.14\times 10^{13}$ & 0.01 (13) & 0.028 (36.4) \\
\hline
\hline
\end{tabular}
\label{observs}
\end{table}
%-------------------
%The magnetic Reynolds number in the simulation with characteristic length of the order of planet radius $L\sim 10^8$~cm and characteristic velocity V$\sim 10^{10}$~cm~s$^{-1}$, of the order of pulsar wind velocity, is $R_{\mathrm m} = VL/ \eta \sim 10^3$, with the magnetic diffusivity in the code $\eta\sim 10^{12}$~cm$^2$~s$^{-1}$. It was evaluated from numerical experiments with the same grid resolution in a simpler setup, as described in \cite{Varela2018}.

Compared to the magnetic field of Sun-like stars, millisecond pulsar magnetic field of $B\sim10^{7-9}$~G, \citep{Wijnands98} is large, but it is still much weaker, for the orders of magnitude, from the field of newly born pulsars, which is $B\sim10^{12}$~G \citep{BackerMilis82}, or from even stronger field on magnetars $B\sim10^{14}$~G \citep{KaspiMagn17}.

In the setup without a planetary atmosphere, if the pulsar wind is reaching the planet surface, the flow can be super-Alfv\'{e}nic and still produce the Alfv\'{e}n wings \citep{Mottez20cor}. In addition, as shown in \cite{Varela2016a} in the case of Mercury, a rocky planet under heavy bombardment of particles from the stellar wind can emanate enough material from the surface to produce a visible aurora.

Auroras are observed in most of the planets in the Solar system, even in the non-magnetic planets, like Venus and Mars \citep{Gray2022}. Pulsar magnetic field is much stronger than the solar one, so we can expect that any aurora there would be the enhanced version of the Solar System aurora events. Mechanism for pulsar planet aurora could be different, but we still could learn from the ones we observe closer to us. A tentative example is the green glow observed from Venus magnetotail - produced by the solar magnetic field facing the non-magnetized obstacle \citep{ZhangVenus2012}. The similar event in a much stronger magnetic field of a pulsar could provide observable signature. 

Taking into account that pulsar magnetic field is much stronger than the planetary field, and the fact that non-magnetized planets also show aurora, in the setups in \cite{cemvar23, MishraR} the planetary field was neglected. Two cases of planetary surface as the inner boundary condition in the simulations were considered: a perfectly conducting and a ferromagnetic. The same method was used as in the previous simulations for magnetized planets in \cite{Varela2018}, with modified boundary conditions at the planet surface: the radial component of the planetary magnetic field was set to zero. Polar and azimuthal components were smoothly absorbed, copied from the values from the last active zone to the boundary ghost zone with zero gradient. The azimuthal component of the magnetic field in the ferromagnetic case was set with the changed sign at the inner boundary.

\textbf{Non-magnetized planets around pulsars.}\label{plplan}
In the case with conducting planet surface, the pulsar wind magnetic field lines connect with the planetary surface and the electric field lines remain close to it. In the case of the ferromagnetic planet surface, the electric field near the planet forms a dipole-like structure. Simulations of such a case were performed by \cite{MishraR}.

\begin{figure}[tbp]
\centering    \includegraphics[width=0.8\columnwidth]{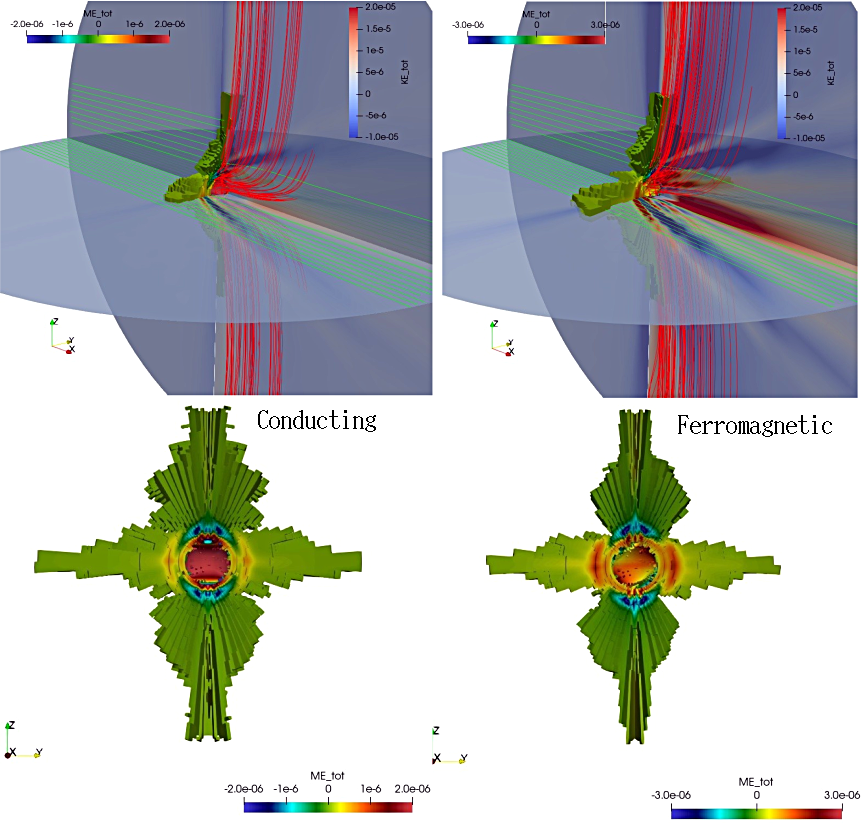}
      \caption{{\it Top panels}: with the color graded isocontours are shown the divergence of the Poynting flux {\bf ME}\_{\bf tot}$=\mathbf{E}\times\mathbf{B}/\mu_0$ and the kinetic energy flux {\bf KE}\_{\bf tot}$=0.5~\rho\mathbf{v}|\mathbf{v}|^2$ in the cases with conducting and ferromagnetic planet surfaces, in the {\it left} and {\it right} panels, respectively. Note that the plotted quantities are not fluxes through a surface but the divergences of the corresponding energy-flux vectors, i.e., local volumetric rates of energy conversion (in code units); positive (red) values of $\nabla\cdot${\bf ME}\_{\bf tot} mark regions where electromagnetic energy is released and can be converted into radiation. The semi-transparent slices are colour-coded by the divergence of the kinetic energy flux (right colour bar, labelled KE\_tot), and the isosurfaces by the divergence of the Poynting flux (top colour bar, labelled ME\_tot); red lines show the magnetic field lines and green lines the velocity streamlines of the pulsar wind. {\it Bottom panels}: zoom into the radiative patterns in the same cases, viewed along the $x$-axis (i.e., facing the planetary dayside); the extended green structures are isosurfaces tracing the Alfv\'{e}n wings, and the red regions on and around the planetary disc mark the largest positive values of $\nabla\cdot${\bf ME}\_{\bf tot}. Locations with the maximum radiated power are in the middle of the planet dayside and in the base of the Alfv\'{e}n wings. Figure adopted from \cite{MishraR}.}
         \label{nonmagrad}
   \end{figure}

The emission from the two simulated cases is shown in Fig.~\ref{nonmagrad}. Geometry of the radio emission is similar in both cases, with the intensity about three times lower with a conducting planetary surface, probably caused by the slight difference in the setup. Local maxima are distributed differently: in the conducting case, the maximum is located at the centre of the planetary dayside, with negative values above and below the poles, while in the ferromagnetic case the maxima form two lobes on either side of the planetary disc, close to the equatorial limbs. The local maxima are higher in the ferromagnetic than in the conducting case (note the different colour-bar ranges, $\pm 3\times 10^{-6}$ versus $\pm 2\times 10^{-6}$ in code units). In the conducting case, the resulting integrated radio emission is $3.65 \times 10^{12}$~W and in the ferromagnetic case it is $1.14 \times 10^{13}$~W. Those values enable prediction of the observed signal intensity.

 The empirical generalized radiometric Bode's law (RBL), which is derived from the observations of magnetized planets in the Solar system, can be used for an estimate of the radio emission intensity \citep{Desch1984U,Zarka2007}. The radiated power $P_{\mathrm{rad}}$ is computed as the energy transferred  from the flow around the solid obstacle in the pulsar wind. The divergence of the magnetic Poynting flux is
\begin{equation}
P_{\mathrm{rad}}=2\times 10^{-3}\int_V\mathbf{\nabla}\cdot{\mathbf{ME}}_{\mathbf{tot}}
dV=2\times 10^{-3}\int_V\mathbf{\nabla}\cdot\frac{(\mathbf{v}\times\mathbf{B})\times\mathbf{B}}{\mu_0}dV,
\label{pflux}
\end{equation}
where the factor $2\times 10^{-3}$ comes from the efficiency of dissipated power to radio emission conversion \citep{Zarka2018} and V is the volume enclosed between the center of the dayside of the planet and the magnetotail.

The density of the radio flux from a planet at a distance d is
\begin{equation}
\Phi=\frac{P_{\mathrm{rad}}}{\Omega d^2 \Delta\nu},
\label{phi1}
\end{equation}
where $\Omega=0.16$~sr is the solid angle of the beam of emitted radiation. It was set as ten times smaller than in the case of Jovian emission from \cite{Zarka04}, following a rough estimate that in the relativistic flow the emission region will be more narrow \cite{Mottez2014,Mottez20cor}. The emission bandwidth $\Delta\nu$ is assumed equal to the maximum emission frequency \citep{Lynch18}.

The two frequencies, $\nu_{\mathrm{max}}$ and $\nu_{\mathrm{min}}$, which are the maximum emission and the characteristic plasma frequencies, respectively \citep{Zarka2007,Varela2016a}, are:
\begin{equation}
\nu_{\mathrm{max}}=\frac{e B_{\mathrm{sw}}}{2\pi m_{\mathrm e}}\sim 2.8~{\mathrm {MHz}}B_{\mathrm{sw}},\ \nu_{\mathrm{min}}=\sqrt{\frac{ne^2}{\pi m_{\mathrm e}}}\sim 8.98~{\mathrm {kHz}}\sqrt{n}.
\label{nus1}
\end{equation}
Here $e$ and $m_{\rm e}$ are the electron charge and mass. $B_{\rm sw}$ is the pulsar wind magnetic field in the vicinity of the planet and ${\mathrm n}$ (in cm$^{-3}$) is the plasma number density. The obtained radio emission fluxes are, as shown in  Table~\ref{observs}, more than enough for observations with LOFAR, MeerKAT and SKA, with minimal sensitivities of the order of 0.1, 0.01 and 0.001 mJy, respectively. However, any emission is absorbed by the plasma below a lower limit of the frequency $\nu_{\mathrm{min}}$. Near the planetary orbit around a pulsar, the results are well above this limit, but for the terrestrial observations, the ionosphere, with a number density of $10^6$~cm$^{-3}$, absorbs all the frequencies below 10~MHz, and in practice even higher, about 30~MHz, and the obtained results are all well below this limit.

The simulations in \cite{MishraR}, because of the numerical reasons, could be performed only with an order or two smaller magnetic field than expected in the pulsar wind. Since the cut-off frequency $\nu_{\rm max}$ is proportional with the magnetic field strength, they argued that for the realistic values of $B_{\rm sw}$ the obtained frequencies of the emission would increase to the more realistic values. Their conclusion is that auroras on planets around pulsars could be observable even with current instruments.

Another caveat in \cite{MishraR} study is that they reached only the 0.87~c velocity of the pulsar wind, or $\gamma=2$, when the realistic values are much higher, reaching $\gamma$ of $10^4-10^5$ \citep{Aharon12}. In such ultra-relativistic flow, geometry of the flow would probably change significantly, modifying the predicted emission. It remains to be seen in the future, more realistic simulations, if this moves the predicted intensities and frequencies towards the more realistic observables, or renders them even more forbidding

\subsection{SPI Edge Cases and Non-Detections}
%\begin{itemize}
%    \item \texttt{\textbf{Contributors:} Silva Järvinen, Julian Alvarado-Gomez [New numerical simulations]}
%\end{itemize}

The tightest star-planet systems are expected to be the most profound to show SPI. The tidal and magnetic force roughly depend on the distance from the planet to the star as $1/d^3$ and $1/d^2$ (or sometimes as $1/d^3$), respectively \citep{Shkolnik_2005}. Further, planets with masses comparable to Jupiter’s are expected to have also comparable magnetic fields (4.2 G). Thus, the searches for SPI have been concentrated on systems that have hot Jupiters within semimajor axes of $\sim$0.1~au.

\cite{Shkolnik2003} reported the first evidence for planet-induced chromospheric activity on HD\,179949, one out of five stars in that study. In a subsequent study \cite{Shkolnik2008} had already a sample of 13 stars with hot Jupiters and reported that five of them, HD\,179949 included, appeared to be actively engaging in SPI. They suggested that an observable magnetic interaction between a star and its hot Jupiter appears as a cyclic variation of stellar activity synchronized to the planet’s orbit. However, since the synchronicity was not always detected, it was suggested that the on/off nature of SPI is likely a function of the changing stellar magnetic field structure throughout its activity cycle.

When \cite{Hellier2009} detected 10-Jupiter mass planet in a 0.02 au orbit around WASP-18, it became one of the most promising systems to detect SPI. The host star itself has a spectral type F6V star with rotation period of 5.6\,d \citep{Salz2015} whereas the planet has orbital period of 0.94\,d \citep{Hellier2009,Southworth2009}. At the time of the planet detection nothing was known about the magnetic field of the host star. Later studies have shown a magnetic field detection rates of 27--32 per cent and longitudinal field strengths from 0.3 to 8.3\,G (in case of fossil field it can be much stronger, several hundreds of gauss) in F-type spectral range \citep[e.g.,][]{Marsden2014,Seach2020}. However, according to these studies non-detections of magnetic field are more common in earlier than later spectral types and there is no clear correlation between longitudinal magnetic field strength and age.

The first attempts to detect magnetic field and SPI in WASP-18 during 2010 and 2011 failed. Several factors, but mostly short exposure times and thus low signal-to-noise ratio of the data, could have caused the non-detection. However, subsequent studies where SPI signatures were searched via multiple magnetic proxies like Ca~{\sc ii} H\&K \citep{Miller2012}, coronal X-ray emission \citep{Pillitteri2014}, and transition region far UV lines \citep{Fossati2018} revealed an extremely anomalous behavior for WASP-18. Furthermore, deep \emph{Chandra} exposure yielded a non-detection of its X-ray corona \citep{Pillitteri2014}. All things considered, WASP-18 seems to have abnormally low activity and variability levels for its spectral type and age \citep{Pillitteri2023}.

The abnormal behavior of WASP-18 has inspired also theoretical interpretations. Both \cite{Pillitteri2014} and \cite{Lanza2024} suggested that tidal SPI in the system could be strong enough to disrupt the required plasma flows to sustain a dynamo, particularly given the shallow convection zone of the host star. Similarly, \cite{DelSordo2021} proposed that magnetic SPI, combined with evaporating outflows from the planet, could modify the photosphere/corona boundary conditions for the field, resulting in an effective quenching on the dynamo action inside the star. However, while these hypotheses involve SPI and the magnetic field generation, the evidence motivating them relies only on proxies of magnetism and not on the stellar magnetic field itself.

Recently, Alvarado-G\'omez et al.\ (2026) obtained new spectropolarimetric observations of WASP-18 which were optimized to get higher signal-to-noise than previous observations. The obtained data confirmed that the star shows constantly very low levels of activity. They also reported the S-index value as 0.13, which indicates a significantly low activity regime. For comparison, the solar value, which is above 0.16 \citep{Egeland2017}, and previously studied F-type stars also have higher values \citep{Seach2020}. Furthermore, this study reported a non-detection of magnetic field in WASP-18. The authors brought up the possibility of close-in planet suffocating stellar activity instead of enhancing it.

\section{Open Questions and Future Directions}
%%%%%%%%%%%%%%%%%%%%%%%%%%%%%%%%%  Description  %%%%%%%%%%%%%%%%%%%%%%%%%%%%%%%%%%%%%%

The most contentious debate in modern SPI centers on whether a global intrinsic magnetic field acts as a ``shield'' for habitability or a ``conduit'' for energy input. Traditionally, magnetospheres were viewed as simple deflectors of stellar wind \citep{Tarduno2010}. However, recent cross-comparative studies show that mass loss rates across Earth, Venus, and Mars are clustered around $10^3\text{--}10^4\text{ g s}^{-1}$, suggesting the presence of a dynamo is not a primary inhibitor of atmospheric loss \citep{Gunell2018} and a much in depth exploration of the role of intrinsic magnetic fields in atmospheric escape is imperative taking into account the chemical evolution and the reaction chains that are dependent on the energy input.
The magnetospheric interaction with the stellar wind is currently parameterized by the enhancement factor $Q$, defined as the ratio of magnetized to unmagnetized non-thermal atmospheric mass loss. For terrestrial planets, this interaction is governed by a competition between the local deflection of plasma and the increase in the effective cross-section for energy and momentum capture. Theoretical frameworks suggest that when the magnetopause standoff distance ($R_{mp}$) exceeds a critical threshold of approximately $3.75$ planetary radii ($R_p$), the magnetosphere intercepts a disproportionate fraction of the incident stellar wind’s Poynting flux, resulting in $Q > 1$, or a net acceleration of atmospheric loss \citep{Blackman2018}. This geometric threshold is particularly significant for Earth-sized and super-Earth planets orbiting within the close-in habitable zones of M dwarf stars ($\sim 0.05$ AU). In these high-pressure environments, a planet with an Earth-like magnetic moment may experience magnetospheric compression that keeps it near the $Q \approx 1$ boundary. However, the presence of an exceptionally strong dynamo can push the system into the energy-capture regime. Recent high-pressure mineral physics indicates that in super-Earths, specifically those exceeding $3\text{--}6 M_\oplus$, internal pressures and insulating mantle conditions facilitate the persistence of a Basal Magma Ocean (BMO) that remains stable for several billion years \citep{Nakajima2026}. At pressures exceeding $100$ GPa and temperatures above $4000$ K, liquid silicates, particularly iron-enriched (Mg,Fe)O, exhibit semi-metallic electrical conductivities ($\sigma \gtrsim 10^4$ S/m), which may be sufficient to sustain a self-exciting dynamo \citep{Dragulet2025}. Whether such BMO dynamos actually operate in super-Earths, however, remains a hypothesis: these conductivities suggest that the thresholds for dynamo action could be exceeded, but BMO dynamos have not yet been demonstrated with self-consistent dynamo models. If they do operate, they could generate magnetic moments up to an order of magnitude stronger than typical core-driven dynamos. By expanding the magnetospheric cross-section well beyond the $3.75 R_p$ limit, such dynamos might then inadvertently act as an oversized energy harvester. The subsequent deposition of Poynting flux into the upper atmosphere would drive intense Joule heating and facilitate potentially catastrophic polar ion outflow, potentially creating an evolutionary crisis where a robust magnetic ``shield'' paradoxically accelerates the depletion of the planetary volatile inventory.

A critical refinement in future SPMI studies must also address the self-consistent energy transport, deposition, and repartition across a range of wavelengths for the stellar hotspots tentatively expected to be produced by SPMI. Future studies must address the consistency of this scenario from a multi wavelength observational perspective. Furthermore, the current state-of-the-art remains limited in its understanding of how much energy deposited into stellar chromospheres can be converted into observable emission through the dissipation of SPMI energy typically expected to be in the form of Alfv\'en waves. Future studies should systematically quantify this process to determine whether the enhanced emissions proposed as the basis of SPMI-induced stellar hotspots are indeed physically plausible in magnitude as well as mechanism through the dissipation of Alfvén waves in the stellar chromosphere. 

A significant scientific open end in tidal interactions is the understanding of which close-in exoplanets are slowly spiralling into their host stars and which ones remain in stable orbits. At present, there are very few confirmed cases of orbital decay beyond WASP-12b, making it difficult to establish clear trends. Transit-timing variations are frequently obscured by degeneracies, including apsidal precession, magnetic activity cycles, and barycentric acceleration, which impede the precise quantification of the orbital period variation \citep[see, e.g.,][]{WatsonMarsh2010, Patra2017, Bouma2020, 2022AJ....163..175W}. A rigorous observational census is therefore required that contrasts actively spiralling systems with stable counterparts across a diverse stellar parameter space. Such population-level constraints are vital for breaking the theoretical degeneracies between competing tidal dissipation pathways.

Another key challenge in SPI research is the characterization of the thermospheric energy budget, specifically the balance between XUV-driven heating and infrared radiative cooling. While the detection of sulfur dioxide (SO$_2$) in WASP-39b \citep{Tsai_2023} confirms the presence of active photochemistry in highly irradiated atmospheres, diagnostic molecular ions such as H$_3^+$ and H$_3$O$^+$, which trace the ionization state and thermal gradient, remain elusive in current transmission and emission spectroscopy. Most Global Circulation Models (GCMs) and 1-D hydrodynamic models currently parameterize stellar energy deposition using a simplified ``heating efficiency'' factor ($\eta$), which assumes a fixed fraction of incident XUV energy is converted into local kinetic heat. This approach, however, fails to account for the non-Maxwellian energy degradation of primary photoelectrons. As demonstrated by \citet{garciamunoz2023}, high-energy photoelectrons initiate complex secondary ionization and electronic excitation cascades that partition a significant portion of absorbed energy into chemical potential and internal molecular states rather than kinetic heating. Neglecting these cascades leads to a systematic misallocation of the thermospheric energy flux, resulting in hydrodynamic escape rate estimates that may be biased by tens of percent \citep{Gillet_2023}. This accurate partitioning is critical for predicting the column density of H$_3^+$, the primary infrared coolant in hydrogen-dominated upper atmospheres. The current lack of definitive H$_3^+$ $v_2$ rovibrational signatures in high-sensitivity observations from the James Webb Space Telescope (e.g., NIRSpec and MIRI) highlights a significant discrepancy between predicted cooling rates and observational upper limits. This tension is further compounded by the fact that in low-density regimes, molecular ion populations are governed by non-Local Thermodynamic Equilibrium (non-LTE) conditions, where the cooling rate is highly sensitive to uncertain neutral-ion collisional cross-sections and local electron densities. Additional discrepancies may arise from electrodynamic effects: if models overestimate the Pedersen conductivity ($\sigma_P$), they will also overestimate Joule heating. This can produce temperature profiles that are too hot and atmospheres that are too extended, while simultaneously reducing the density in the regions that contribute most to observable emission. The net effect is an atmosphere that favors expansion but produces weaker molecular ion signals than expected. To help reconcile the gap between GCM predictions and the non-detections from the James Webb Space Telescope, several research priorities emerge. First, kinetic solvers should be incorporated into 3D GCMs to self-consistently capture non-thermal electron energy distributions and the resulting species-dependent heating efficiencies. Second, improved mapping of the spatial and vertical structure of Pedersen and Hall conductivities is needed to better constrain Joule heating, particularly across the dayside and terminator regions. Current GCM prescriptions of magnetic drag, often implemented as spatially uniform or weakly depth-dependent terms, are likely unphysical and fail to capture the inherently anisotropic and conductivity-dependent nature of Lorentz forces throughout the atmosphere. Developing more realistic magnetohydrodynamic drag formulations that vary with ionization fraction, field geometry, and flow regime is therefore essential. Adding to this complexity, it is well plausible that the significance of the aforementioned heating and cooling processes may evolve over time as the bulk composition of the planet's atmosphere evolves. This highlights the importance of the evolutionary context of exoplanets, especially those of lower mass. Finally, targeted deep-integration observations of close-in gas giants should aim to place tighter upper limits on H$_3^+$ $Q$-branch emission near 3.96 micron, providing a direct probe of thermospheric cooling rates.

A fundamental limitation in current SPI power-scaling laws is their reliance on analytical frameworks derived from the Jupiter-Io interaction \citep{Saur2013}, which assume the planet behaves as a localized perturbation in a spatially homogeneous flow. However, for exoplanets orbiting in the sub-Alfvénic regime of their host star, the star-planet system is coupled via a stable magnetic tether. Recent 3D magnetohydrodynamic simulations \citep{Paul_2026} demonstrate that as the Alfvénic Mach number ($M_A$) approaches finite values, the interaction undergoes a regime transition from a localized point-source to an Extended Obstacle regime. In this regime, Poynting flux is tapped from the ambient stellar wind continuously along the entire 3D structure of the Alfvén wings. This may also, in turn, alter the predicted torque on the host star and suggest that constraining any resulting emissions, whether chromospheric or radio, should consider the complete global geometry of the magnetic tether rather than a discrete point source. Current observational cadences, however, typically optimized for point-source transits, are insufficient to resolve these spatially extended, time-varying structures. 

Spectropolarimetry of the He~{\sc i} $2^3S$--$2^3P$ triplet (Sect.~\ref{sec:obs_magfields}) is a promising probe in this context, because the metastable helium population preferentially traces the extended, escaping thermosphere. Through the Hanle effect, it could in principle provide the 3D orientation of the magnetic field within the escaping atmospheric wind, allowing a direct mapping of the magnetospheric cavity and helping to constrain the reconnection efficiency at the magnetopause \citep{Oklopcic2020}. 

It is also important to note that when stellar wind stripping and radiation-driven photoevaporation act in concert for exoplanetary atmosphere, the resulting mass loss likely exceeds the sum of its parts. Existing research underscores that far-out SPI is a physically rich regime where the interplay of magnetized winds, radiation, and stellar storms in concert determines the overall evolutionary pathway for the exoplanet's atmosphere. Future challenges include modeling these interactions over billion-year timescales, understanding the impact of episodic CME frequency versus secular losses, and developing observational diagnostics, such as radio emissions and UV transit signatures.

% \textbf{Antoine's suggestion: Energy transport, deposition and repartition in various wavelenghth for SPMI, other few points that would be worth mentionning I think: effect of magnetic field in different depths of the 'atmosphere', GCMs still prescribe globally simplistic (unphysical?) magnetic drags and it's not satisfactory
% how would the triggering scenario (à la Illin) work on physics-grounds? How is that happening, and how ofen will it happen?}

% \subsection{Key Observational and Modelling Requirements}
% \begin{itemize}
%     \item \texttt{Contents: Measurement and mission needs}
% \end{itemize}

\subsection{Strategies to Address Open Questions}
A critical step forward lies in establishing a hierarchy of coupled models that systematically bridge the gap between microphysical plasma processes and global atmospheric dynamics. Rather than attempting to directly embed fully kinetic treatments into large-scale simulations, future efforts should prioritize intermediate descriptions based on reduced multi-fluid and resistive formulations of magnetohydrodynamics. These models can resolve ion-neutral drift, anisotropic conductivity, and current closure in partially ionized flows while remaining computationally compatible with Global Circulation Models. Within this framework, machine learning techniques can be deployed in a constrained manner to emulate unresolved terms, provided they are trained on physically consistent datasets and explicitly regularized to preserve conservation laws. As usual, it remains important to trace the physical consistency of each of the models and their built-in modules.

A first key frontier is the development of fully coupled interior–atmosphere evolution frameworks, in which the planetary dynamo, mantle convection, and atmospheric escape are treated as a single co-evolving system rather than independent components. In such an approach, core cooling rates and mantle rheology should be dynamically linked to surface volatile loss, allowing atmospheric erosion to feed back onto interior thermal evolution through changes in surface temperature and boundary heat flux. This is particularly important for super-Earths, where long-lived high-pressure silicate phases and deep magma reservoirs can modify the efficiency of convective heat transport and, in turn, regulate dynamo longevity. By coupling magnetic moment evolution directly to volatile depletion history, these models would move beyond static prescriptions of planetary magnetism and instead treat magnetic field strength as an emergent property of coupled interior–exterior evolution. This would enable a physically consistent treatment of how atmospheric stripping, radiogenic heating, and core solidification jointly determine long-term planetary habitability. This of course as a first step requires resolving well the effects of planetary magnetism itself on atmospheric loss and habitability which remains an active area of research.

In that light, an equally critical advancement would be the extension of Earth-like magnetosphere-ionosphere-thermosphere-atmosphere (M-I-T-A) coupling frameworks to exoplanetary environments. Current terrestrial models resolve tightly coupled current systems, ion-neutral drag, and electrodynamic feedbacks across these regions, but analogous treatments for exoplanets remain largely parameterized. Future efforts should, at the very least, elucidate under which conditions these feedbacks might be important within the large diversity of known exoplanets and their host stars. For strongly irradiated exoplanets, especially those orbiting active M dwarfs, the interaction between stellar wind forcing, ionospheric conductivity, and thermospheric expansion must be solved self-consistently, as each layer actively modifies the boundary conditions of the others in ways that may be specific to the chemical make-up of the gas. In this regime, magnetospheric currents can drive significant Joule heating in the thermosphere, which in turn alters scale heights and ionization profiles, feeding back into conductivity and current closure. Extending coupled M-I-T-A models beyond Earth therefore requires generalized formulations that can accommodate extreme stellar wind pressures, non-Earth-like compositions, and highly variable ionization states, enabling a unified description across diverse exoplanet classes.

\section{Acknowledgements}

The authors gratefully acknowledge all the contributions from the participants of the International Space Science Institute (ISSI) Workshop on Stellar Magnetism and its Impact on (exo)Planets, held in Bern, Switzerland, June 2 - 6, 2025. We thank the ISSI and its staff for hosting and supporting the workshop. AP and KK would also like to extend their special thanks to Dr. Harish Vedantham for the discussions during the workshop, which improved the overall structure of the paper.

\section{Funding}

% I am adding the disclaimer text for the ERCs as a footnote with a special symbol (2 for dagger), so that we don't have to write it out for each ERC grant.
\renewcommand{\thefootnote}{\fnsymbol{footnote}}

AP gratefully acknowledges the MERAC Foundation for research funding and the academic support provided by CEA Paris-Saclay. KGK acknowledges the support from the European Research Council (ERC) under grant agreement 101123041 (ERC-CoG EASE).\footnotemark[2] AS and JC acknowledge support from the European Research Council (ERC) under grant agreement 101125367 (ERC-CoG ExoMagnets).\footnotemark[2]
  RF acknowledges support from the United Arab Emirates University (UAEU) UPAR grant number G00005451. M\v{C}  acknowledges the Czech Science Foundation (GAČR) grant No. 21-06825X. KP acknowledges support from the European Research Council (ERC) under grant agreement 101170037 (ERC-CoG Evaporator).\footnotemark[2]

\footnotetext[2]{Views and opinions expressed are however those of the author(s) only and do not necessarily reflect those of the European Union or the European Research Council Executive Agency. Neither the European Union nor the granting authority can be held responsible for them.}
%%%%%%%%%%%%%%%%%%%%%%%%%%%%%%%%%%%%%%%%%%%%%%%%%%%%%%%%%%%%%%%%%%%%%%%%%%%%%%%%%%%%%%

\section{Conflict of interest}

The authors declare no competing interests as defined by the journal.

%\newpage
%\noindent \textbf{For information (from Manuel): planned structure of Chapter 4 (e-mail from Antigona)}

%\bigskip

%Topics to be covered:

%\begin{itemize}
%\item 1.    Atmospheric evolution:
%    \begin{itemize}
%    \item a.    Stellar evolution: long term stellar changes from pre-main sequence to MS    
%    \item b.    The impact of the star on primordial atmospheres
    %\item c.    Secondary atmospheres: atmospheric escape vs. outgassing
    %\item d.    Planetary magnetic fields and atmospheric escape: lessons from the solar system
    %\end{itemize}
%\item 2.    Atmospheric photochemistry
 %   \begin{itemize}
  %  \item a.    Prebiotic chemistry
   % \item b.    Biosignatures
    %\item c.    Hazes 
    %\end{itemize}
%\item 3.    Planetary habitability
 %   \begin{itemize}
  %  \item a.    The habitable zone
   % \item b.    Atmospheric retention
    %\item c.    Stellar UV and particles: damage vs prebiotic chemistry
    %\item d.    The stellar input for habitability models
    %\item e.    Exoplanet characterization: escape, habitability and life detection
    %\end{itemize}
%\end{itemize}

%\bibliographystyle{alpha}  % original version
\bibliographystyle{aasjournal-3}   % temporarily added by MG for clarity 
\bibliography{main}

\end{document}